\documentclass[11pt,letterpaper]{mythesis}
\usepackage[nottoc,notlot,notlof]{tocbibind}
\usepackage{sastyle}

\usepackage{acro} 
\usepackage{tabularx} 
\acsetup{first-style=short}
\DeclareAcronym{LP}{
	short = LP,
	long  = Linear Programming,
	tag = abbrev
}

\DeclareAcronym{SDP}{
	short = SDP,
	long  = Semi-definite Programming,
	tag = abbrev
}

\DeclareAcronym{OPE}{
	short = OPE,
	long  = Operator Product Expansion,
	tag = abbrev
}

\DeclareAcronym{GN}{
	short = GN model,
	long  = Gross-Neveu model -- the theory of $N$ fermions with quartic interaction,
	tag = abbrev
}

\DeclareAcronym{GNY}{
	short = GNY model,
	long  = Gross-Neveu-Yukawa model -- a scalar field coupled to the Gross-Neveu model via Yukawa interactions,
	tag = abbrev
}

\DeclareAcronym{IR}{
	short = IR ,
	long  = Infrared -- used to indicate low energies,
	tag = abbrev
}

\DeclareAcronym{UV}{
	short = UV ,
	long  = Ultraviolet -- used to indicate high energies,
	tag = abbrev
}

\DeclareAcronym{TQFT}{
	short = TQFT ,
	long  = Topological Quantum Field Theory,
	tag = abbrev
}

\DeclareAcronym{MFT}{
	short = MFT ,
	long  = Mean Field Theory,
	tag = abbrev
}

\DeclareAcronym{QFT}{
	short = QFT ,
	long  = Quantum Field Theory,
	tag = abbrev
}

\DeclareAcronym{CFT}{
	short = CFT ,
	long  = Conformal Field Theory,
	tag = abbrev
}

\DeclareAcronym{SCFT}{
	short = SCFT ,
	long  = Supersymmetric Conformal Field Theory,
	tag = abbrev
}

\DeclareAcronym{CFTdp}{
	short = CFT$_{d^+}$ ,
	long  = Conformal Field Theory in $d$ dimensions  with Euclidean signature,
	tag = abbrev
}

\DeclareAcronym{CFTdm}{
	short = CFT$_{d^-}$ ,
	long  = Conformal Field Theory in $d$ dimensions  with Lorentzian signature,
	tag = abbrev
}

\DeclareAcronym{d}{
	short = \ensuremath{d} ,
	long  = Spacetime dimension: we consider either \ensuremath{\R^{d}} or \ensuremath{\R^{d-1,1}},
	sort  = a,
	tag = nomencl
}

\DeclareAcronym{T}{
	short = \ensuremath{T^{\mu\nu}} ,
	long  = Stress energy tensor of a given field theory,
	sort  = a,
	tag = nomencl
}

\DeclareAcronym{phi}{
	short = \ensuremath{\f} ,
	long  = A local operator in the trivial representation of the rotation subgroup of conformal group,
	sort  = b,
	tag = nomencl
}

\DeclareAcronym{cO}{
	short = \ensuremath{\cO} ,
	long  = A local operator in the generic representation of the rotation subgroup of conformal group,
	sort  = c,
	tag = nomencl
}

\DeclareAcronym{psi}{
	short = \ensuremath{\psi} ,
	long  = A local spin\ensuremath{-\half} operator of conformal group in \ensuremath{\R^{2,1}},
	sort  = c,
	tag = nomencl
}

\DeclareAcronym{Delta}{
	short = \ensuremath{\Delta} ,
	long  = Scaling dimension of a local operator: \ensuremath{\cO_i(\lambda x)=\lambda^{-\De_i}\cO_i(x)},
	sort  = d,
	tag = nomencl
}

\DeclareAcronym{Delta3}{
	short = \ensuremath{\Delta_{abc}} ,
	long  = Shorthand notation for the frequently used combination \ensuremath{\De_a+\De_b-\De_c},
	sort  = d2,
	tag = nomencl
}

\DeclareAcronym{spin}{
	short = \ensuremath{l} ,
	long  = Spin of a local operator in three spacetime dimensions,
	sort  = e,
	tag = nomencl
}

\DeclareAcronym{z}{
	short = \ensuremath{z} ,
	long  = Holomorphic cross ratio of four points in the standard conformal frame,
	sort  = f,
	tag = nomencl
}

\DeclareAcronym{zb}{
	short = \ensuremath{\zb} ,
	long  = Anti-holomorphic cross ratio of four points in the standard conformal frame,
	sort  = g,
	tag = nomencl
}

\DeclareAcronym{mu}{
	short = \ensuremath{\mu(\cO)} ,
	long  = Plancherel measure of the conformal group,
	sort  = h,
	tag = nomencl
}

\author{Soner Albayrak}
\title{Analytic Studies of Fermions in the Conformal Bootstrap}
\date{June  2021}
\advisorline{Dissertation Director: Professor David Poland}

\newcommand{\Laplacian}{\mathop{}\!\mathbin\Box}

\newcommand{\dDisct}[1]{\text{dDisc}_{t}\left[#1\right]}
\newcommand{\DD}{\mathrm{D}}
\newcommand{\pre}{\mathsf{p}}
\newcommand{\shift}[1]{\Pi_{#1}}
\newcommand*{\uniq}{\raisebox{-0.7ex}{\scalebox{1.8}{$\cdot$}}}
\newcommand\sixj[6]{\ensuremath{\left\{\begin{array}{ccc} #1 & #2 & #6 \\ #3 & #4 & #5\end{array}\right\}}}
\newcommand\sixjBlock[6]{\ensuremath{\left(\begin{array}{ccc} #1 & #2 & #6 \\ #3 & #4 & #5\end{array}\right)}}

\newcommand{\sixjDecomp}[9]{
	\newcommand{\sixjDecompTemp}[4]{
		\ensuremath{
			\mathcal{J}^{#1}
			\footnotesize
			\left(\begin{array}{cccccc} #2 & #3 & #4 & #5 & #6 & #7 \\ #8 & #9 & ##1 & ##2 & ##3 & ##4\end{array}\right)
			\normalsize
		}
	}
	\sixjDecompTemp
}

\newcommand{\sixjDecompp}[9]{
	\newcommand{\sixjDecomppTemp}[5]{
		\ensuremath{
			\mathcal{J}^{#1}_{#2}
			\footnotesize
			\left(\begin{array}{cccccc} #3 & #4 & #5 & #6 & #7 & #8 \\ #9 & ##1 & ##2 & ##3 & ##4 & ##5\end{array}\right)
			\normalsize
		}
	}
	\sixjDecomppTemp
}

\newcommand{\opeFuncDecomp}[9]{
	\newcommand{\opeFuncDecompTemp}[5]{
		\ensuremath{
			\mathcal{K}^{#1}_{#2}
			\footnotesize
			\left(\begin{array}{cccccc} #3 & #4 & #5 & #6 & #7 & #8 \\ #9 & ##1 & ##2 & ##3 & ##4 & ##5\end{array}\right)
			\normalsize
		}
	}
	\opeFuncDecompTemp
}

\newcommand{\parity}[1]{\mathord{\vtop{\ialign{##\crcr
				$\hfil\displaystyle{#1}\hfil$\crcr\noalign{\kern1.5pt\nointerlineskip}
				$\hfil\dot{}\hfil$\crcr\noalign{\kern-5.5pt}}}}}

\begin{document}
	
	\frontmatter
	
	\begin{abstract}
	In this thesis, we analyze unitary conformal field theories in three dimensional spaces by applying analytic conformal bootstrap techniques to correlation functions of non-scalar operators, in particular Majorana fermions. Via the analysis of these correlation functions, we access several sectors in the spectrum of  conformal field theories that have been previously unexplored with analytic methods, and we provide new  data for several operator families. In the first part of the thesis, we achieve this by the large spin expansions that have been traditionally used in the conformal bootstrap program for scalar correlators, whereas in the second part we carry out the computations by combining several analytic tools that have been recently developed such as weight shifting operators, harmonic analysis for the Euclidean conformal group, and the Lorentzian inversion formula. We compare these methods and demonstrate the superiority of the latter by computing nonperturbative correction terms that are inaccessible in the former. A better analytic grasp of the spectrum of fermionic conformal field theories can help in many directions including making new precise analytic predictions for supersymmetric models, computing the binding energies of fermions in curved space, and describing quantum phase transitions in condensed matter systems with emergent Lorentz symmetry.
\end{abstract}

\maketitle
\makecopyright{2021}
\hbox{\hfil}\vspace{4in}
\begin{center}
	To my family
\end{center}
\tableofcontents


\listoffigures
\addcontentsline{toc}{section}{\listfigurename}%
\listoftables
\addcontentsline{toc}{section}{\listtablename}%
\clearpage

\printacronyms[include=noSuchClass,name=Acknowledgments, heading =chapter*]

\addcontentsline{toc}{section}{Acknowledgments}%
I would like to thank my supervisor, David Poland, for his invaluable support throughout my Ph.D. years. I appreciate his mentoring, his academic guidance, and his patience with me: he is a great collaborator and a tremendous advisor to have, and I am extremely grateful to have been his PhD student. I  sincerely hope that one day I can be an advisor  like him, with such an encouraging approach towards people and an insightful approach towards physics.

I also would like to thank my dear friend Savan Kharel for practically being a second mentor for me. I have learned substantially from him about a lot of things, not least the academic world, and I am utterly thrilled that our casual discussions have led to several collaborations and diverse publications. 

There are numerous friends and colleagues that I have crossed paths with in Yale particle theory group and in Walter Burke Institute at Caltech; and,  I would like to acknowledge their support through my Ph.D. years (they are simply too many to name). In particular, I thank David Meltzer who has been a great friend and a treasured collaborator: I'm indebted to him for his contributions to the work presented in this thesis.  Additionally, I'd like to thank Chandramouli Chowdhury: I certainly enjoyed our collaborations. Lastly, I would like to give my special thanks to Taylan Nurlu, who help me forget a theoretical physicist's inevitable doom of being never understood hence never sincerely appreciated by their friends and family, as he has always been there to listen and try to understand what I do throughout this thesis. This is in fact the second thesis I'm thanking him with that exact sentence!

Acknowledgment is also due to my thesis committee: I thank Witold Skiba, Thomas Appelquist, Keith Baker, and Simon Caron-Huot for reading this thesis and for giving their much appreciated feedback.

Finally, I owe my sincerest thanks to my family for their everlasting love. My father Halim, my mother Münevver, and my brother $\dot{\text{I}}$lker have always been there for me whenever I needed them. In particular, I'd like to express my gratitude to my wife Bengisu for her endless mental support. I could not have completed my Ph.D. without my family’s help and this thesis is dedicated to them.

\clearpage

\printacronyms[include=abbrev,name=Abbreviations, heading =chapter*]

\addcontentsline{toc}{section}{Abbreviations}%
\clearpage

\printacronyms[include=nomencl,name=Nomenclature, heading =chapter*]

\addcontentsline{toc}{section}{Nomenclature}%

	\mainmatter
	
\chapter{Introduction}

It would not be an overstatement to say that quantum field theories (\ac{QFT}s) with conformal symmetry are tremendously important in modern theoretical physics. Understanding such theories may shed light on numerous areas, including but not limited to
\begin{enumerate}
	\item the study of quantum gravity through the gauge gravity correspondence,
	\item beyond the standard model physics via phenomenological applications,
	\item condensed matter systems through critical phenomena,
	\item effective field theories perturbed around CFTs via renormalization flow.
\end{enumerate} 
Such relevance of conformal field theories (\ac{CFT}s) has stimulated an intensive amount of research into analytic and numerical tools to understand the underlying principles and restrictions of conformal symmetries in field theories. In two dimensional spacetimes, the restrictions of conformal symmetry are enhanced to those of the \emph{Virasaro} symmetry, hence two dimensional CFTs enjoy a greater number of constraints which in turn enable their classification and in some cases their complete solution. In contrast, CFTs in general dimensions are harder to solve analytically and the analysis of these theories has been relatively idle in the $20^{\text{th}}$ century.

In the past decade, there has been a revival in the analytic study of conformal field theories, particularly through nonperturbative tools and general constraints such as unitarity and causality. This program, coined the \emph{conformal bootstrap} in analogy to the s-matrix bootstrap program of 1960s, has yielded novel insight into the analytic structure of CFTs, both in Euclidean signature --- convergence of the operator product expansion (\ac{OPE}), analytic expansions for the conformal blocks, the existence of the operators that shift the conformal weights of local operators, etc. --- and in Lorentzian signature --- the existence of infinite double twist families, analyticity in spin, light-ray operators, etc.

Despite the stupendous progress of the last decade, the analytic studies in CFTs have been mostly focused on the direct study of the \emph{scalar} operators, i.e.  local operators in the trivial representation of the rotation subgroup of the conformal group.\footnote{Among some of the exceptions for the analytic bootstrap with external spinning operators, we can name the lightcone bootstrap \cite{Albayrak:2019gnz,Albayrak:2020rxh,Li:2015itl,Hofman:2016awc,Li:2017lmh,Elkhidir:2017iov,Chowdhury:2018uyv}, Mellin space techniques \cite{Sleight:2018epi,Sleight:2018ryu}, and mean field theory computations \cite{Karateev:2018oml}.} Conformal symmetry tells us that symmetric traceless representations of the $\mathrm{SO}(d)$ group appear in the OPE of scalar operators, so the analysis of scalar operators is deemed sufficient if one is only interested in such operators. This is especially a natural choice as the direct analysis of spinning operators through correlation functions of such operators is far more complicated for a variety of technical reasons. However analysis of correlation functions of scalar operators alone (or correlation functions of any bosonic operators for that matter) does not give access to the fermionic operators in the spectrum of CFTs. 

There are various reasons as to why CFTs with fermionic operators are of great relevance. In \mbox{$d=4$}, we expect non-trivial non-supersymmetric fermionic CFTs because of the \ac{UV} Lagrangian descriptions based on matter coupled to gauge theories that flow in the \ac{IR} to weakly coupled Caswell-Banks-Zaks fixed points \cite{Caswell:1974gg,Banks:1981nn}. The situation is similar in $d=3$ where we can consider IR fixed points of QED coupled to matter or the Gross-Neveu-Yukawa (GNY) models. Furthermore, there is a fermionic sector in any supersymmetric conformal field theory (\ac{SCFT}) in any $d$,\footnote{We note that superalgebras actually only exist for $d\le 6$ \cite{Nahm:1977tg}.} and a better analytic grasp of such theories is not only important for their theoretical relevance but also for their utility in the description of many condensed matter systems (such as a critical point on the boundary of topological superconductors \cite{Grover:2013rc}).

In this thesis we address this missing link in the literature by extending the analytic progress achieved for bosonic CFTs to $3d$ CFTs with fermionic sectors. We will do this by considering correlation functions of Majorana fermions with scalars and with each other; this allows us to reach various new sectors such as the fermionic double twist operators.\footnote{The irreducible representations of the $\mathrm{so}(2,1)$ algebra are real, so we do not lose any generality by considering Majorana spinors of the $\mathrm{Pin}(2,1)$ group. This is in contrast to $d=4$ case where the irreducible representations of the $\mathrm{so}(3,1)$ algebra are complex hence Majorana fermions are special cases (Technically, in both $d=3,4$, we have pseudo-Majorana spinors instead of Majorana spinors but we will gloss over this distinction \cite{RauschdeTraubenberg:2005aa}).} Our work has two main consequences: firstly we provide explicit new data that is applicable in any unitary $3d$ CFT; secondly we detail and discuss two different technologies to achieve this --- an old style method that has been used for scalars for some time now, and a new method that combines several analytic tools that have been recently developed in the bootstrap literature. We believe that our results can be utilized in various concrete CFTs such as the GNY models and that our method for the analysis of the fermionic correlators be extended to higher-spin operators, which would ultimately enrich our understanding of conformal field theories even further.

\subsection*{Outline of the thesis}
We start with a general introduction of conformal symmetry and the conformal bootstrap program in Chapter~\ref{chapter:preliminaries}. In particular, we review the historical development of both, alongside with a brief mathematical description of conformal symmetry. We then specialize to fermions in three dimensional spacetime in Chapter~\ref{chapter: perturbative}: we set our conventions, define the spectrum of fermionic theories, review the traditional so-called lightcone bootstrap approach, and carry out the computations in this scheme to obtain the CFT data.

The results obtained in Chapter~\ref{chapter: perturbative} are \emph{not} analytic in the spin $l$ of the relevant CFT operators; in particular, they should be regarded as asymptotic results around $l\sim\infty$. At finite spin, we may get spurious poles which ruin the analyticity: such spurious poles are actually canceled by what we call the \emph{non-perturbative} corrections, i.e. terms that are exponentially suppressed in $l$ --- hence are invisible to the framework of Chapter~\ref{chapter: perturbative}. In Chapter~\ref{sec:nonperturbative}, we detail this difference and demonstrate its importance in concrete scalar CFTs such as the $3d$ Ising and $O(N)$ models. In these theories, we obtain the non-perturbative corrections through a recently developed analytic tool (the Lorentzian inversion formula). In the rest of the chapter, we discuss how to extend this approach to spinning operators, review necessary ingredients, and apply the derived relations to obtain the non-perturbative corrections for various fermionic operators. The backbone of this approach relies on the $6j$ symbols of the conformal group; in Appendix~\ref{qppandix: 6j symbols}, we provide further details for these objects.

\subsection*{Citations to Previous Work}
Chapter 3 is essentially identical to
\begin{quotation}
	S. Albayrak, D. Meltzer, and D. Poland, \emph{“More Analytic Bootstrap: Nonperturbative Effects and Fermions,”} \href{https://inspirehep.net/literature/1727557}{JHEP 08 (2019) 040, arXiv:1904.00032 [hep-th]}
\end{quotation}

\noindent Chapter 4 is excerpted from
\begin{quotation}
	S. Albayrak, D. Meltzer, and D. Poland, \emph{“More Analytic Bootstrap: Nonperturbative Effects and Fermions,”} \href{https://inspirehep.net/literature/1727557}{JHEP 08 (2019) 040, arXiv:1904.00032 [hep-th]}
	\par
	S. Albayrak, D. Meltzer, and D. Poland, \emph{“The Inversion Formula and 6j Symbol for 3d Fermions,”} \href{https://inspirehep.net/literature/1801105}{JHEP 09 (2020) 148, arXiv:2006.07374 [hep-th]}
\end{quotation}

\noindent Appendix A is excerpted from
\begin{quotation}
	S. Albayrak, D. Meltzer, and D. Poland, \emph{“The Inversion Formula and 6j Symbol for 3d Fermions,”} \href{https://inspirehep.net/literature/1801105}{JHEP 09 (2020) 148, arXiv:2006.07374 [hep-th]}
\end{quotation}

\subsection*{Other PhD Work of the Author}

\begin{quotation}
	S.~Albayrak, { S.~Kharel} and {D.~Meltzer}, \emph{``On duality of color and kinematics in (A)dS momentum space,''} \href{https://arxiv.org/pdf/2012.10460.pdf}{arXiv:2012.10460 [hep-th]}
\end{quotation}
\begin{quotation}
	S.~Albayrak, C.~Chowdhury and  { S.~Kharel}, \emph{``On loop celestial amplitudes for gauge theory and gravity,''}
	\href{https://inspirehep.net/literature/1807927}{ Phys.Rev.D 102 (2020)}
\end{quotation}
\begin{quotation}
	S.~Albayrak and  { S.~Kharel}, \emph{``Spinning loop amplitudes in Anti-de Sitter space,''} \href{https://inspirehep.net/literature/1802534}{Phys.Rev.D 103 (2021) 2} 
\end{quotation}
\begin{quotation}
	S.~Albayrak, C.~Chowdhury and  { S.~Kharel}, \emph{``Study of momentum space scalar amplitudes in AdS spacetime,''} \href{https://inspirehep.net/files/950d888f8dbaf39b96ea1bee25d6bda0}{Phys.Rev.D 101 (2020) 12}
\end{quotation}
\begin{quotation}
	S.~Albayrak and  { S.~Kharel}, \emph{``Towards the higher point holographic momentum space amplitudes. Part II. Gravitons,''} \href{https://inspirehep.net/files/eb148d7ad84e19945017f5ab2aa85393}{JHEP 12 (2019) 135}
\end{quotation}
\begin{quotation}
	S.~Albayrak, C.~Chowdhury and  { S.~Kharel}, \emph{``New relation for Witten diagrams,''} \href{https://inspirehep.net/files/44b75a5187c14730a849c04e3b909bd3}{JHEP 10 (2019) 274}
\end{quotation}
\begin{quotation}
	S.~Albayrak and  { S.~Kharel}, \emph{``Towards the higher point holographic momentum space amplitudes,''} \href{https://inspirehep.net/files/fdfe40b4cfec540b04df94d4cb9bc1c4}{JHEP 02 (2019) 040}
\end{quotation}

\chapter{Preliminaries}
\label{chapter:preliminaries}
\section{Historical background}
\subsection{Brief account of conformal symmetry in the past centuries}
\label{sec: brief history of conformal symmetry}
In this section, we will review the evolution of the conformal bootstrap. A more appropriate history should actually start around $19^{\text{th}}$ century, encompassing the story of conformal transformations themselves. Such a review would take us from the times of Joseph Liouville and Sophus Lie, and would include all the relevant topics along the timeline of conformal symmetries. As it happens, there is a such a review and interested reader is directed there \cite{Kastrup:2008jn}. Below, we will recap the developments in the last century and set the scene for the emergence of the conformal bootstrap as we understand it today.\footnote{See \cite{Poland:2018epd} for a thorough analysis along with an extensive historical account.}

The end of the $19^{\text{th}}$ century and the first half of $20^{\text{th}}$ century have witnessed exiting advancements in the area of scale invariance among which the first observation of critical opalescence and the exact solution of $2d$ Ising model come to mind \cite{JS8702300074,Onsager:1943jn}. Conformal invariance also entered into physics around this time through the Maxwell equations. Later with the advancement of the Wilsonian renormalization group, it was realized that CFTs have a much broader usage than their mere application in statistical and classical physics because any QFT connects to a CFT in the UV and another one in the IR via renormalization group flow.\footnote{Technically endpoints of renormalization group flow are either \ac{TQFT}s or \emph{scale invariant} theories, which usually tend to be conformally invariant as well.}

The greater appreciation of CFTs lead to their involvement in more and more physical theories. However, the conformal bootstrap differs from other employment of conformal symmetries in both conceptual and practical senses. To understand that, we need to appreciate the term \emph{bootstrap} in physics.

The 1960's have seen rapid proliferation of strongly interacting particles detected in high energy collisions. The intuitive expectation that we can explain all particles in terms of elementary ones and that local fields should be used to calculate observables started to be doubted. In this environment, Geoffrey Chew proposed what he called \emph{the nuclear democracy} \cite{Chew:1963zza}. It is an approach fundamentally different in the sense that it rejects standard reductionism in high energy physics and proposes that one should focus on fundamental laws and how they constrain observable particles while treating them equally instead of finding a minimal set of elementary particles to which all others can be reduced \cite{Chew:1961ev,Chew:1962mpd}.

Chew's work later turned into what we call today \emph{the s-matrix bootstrap} because bootstrappers were trying to obtain the scattering matrix starting from the fundamental laws such as Lorentz invariance and unitarity. The program was not particularly successful at that time, besides some work on Regge trajectories such as that of Dolen, Horn, and Schmid \cite{Dolen:1967zz,Dolen:1967jr}; and, the interest in s-matrix bootstrap plummeted once strong interactions were understood in the framework of Quantum Chromodynamics. Veneziano utilized the advancements in the Regge theories \cite{Veneziano:1968yb} however this lead to a different direction, i.e. the rise of string theory \cite{Vecchia2008}.

The following year of Veneziano's work, Wilson \cite{Wilson:1969zs} --- followed by Kadanoff \cite{Kadanoff:1969zz} --- initiated the analysis of current algebras in the lines of the s-matrix bootstrap\footnote{Like the s-matrix bootstrap, Wilson's approach was also bypassing conventional local fields. Nonetheless, his work with operator product expansions were proved in the realm of QFT, initially perturbatively by Zimmerman \cite{deser1970lectures}. One can also show it nonperturbatively via radial quantization with the path integral formalism \cite{Weinberg:1996kr}.} where the authors promoted the usage of \emph{operator product expansion} (OPE) associativity as a consistency condition, which was demonstrated to be sufficient to produce an alternative solution for $2d$ Ising model in addition to that of Onsager's original solution \cite{Onsager:1943jn}.

We can see the appeal of these advancements: the main problem with the s-matrix bootstrap was the lack of sufficient constraints to extract a large set of concrete results. However, this can be circumvented if one also includes the constraints of CFTs and the consistency of OPEs. This was first realized by Ferrara, Grillo, and Gatto \cite{Ferrara:1973yt}, and independently by Polyakov \cite{Polyakov:1974gs}. We can safely take these papers to be the birth of the conformal bootstrap, though it was not until 1984 that the term bootstrap was used for this approach \cite{Belavin:1984vu}.

Despite the appreciation of its importance, progress on the conformal bootstrap has been rather slow until the past two decades, except with $2d$ CFTs which have seen rapid development due to their mathematical simplicity and relevance in string theories. The stagnancy in higher dimensional CFTs changed in 2008 when Rattazzi, Rychkov, Tonni, and Vichi introduced a numerical method into the conformal bootstrap \cite{Rattazzi:2008pe}. They argued that one can use linear programming to extract useful bounds on physical quantities such as scaling dimensions of the operators in the CFT, \emph{without} analytically solving them. Since then, the conformal bootstrap attracted much attention and created a revival in the area. This revival also led to a nice amount of progress on the analytical side as well, and we will discuss this in the next section.

\subsection{Recent developments in the field}
\subsubsection{Numerics}
\label{sec: state of art-numerics}
The conformal bootstrap flourished since the introduction of \emph{linear programming} (\ac{LP}) into the area with the seminal paper of Rattazzi et al. \cite{Rattazzi:2008pe}. The method they introduce is a simple yet elegant application of ``proof by contradiction'': one assumes some properties about the spectrum of a CFT, such as the scaling dimension of the lowest lying operators. One then checks whether this spectrum is consistent with crossing symmetry and unitarity; if it is self-consistent, we do not obtain any new information. If not, we conclude that the assumed spectrum cannot be realized in any unitary CFT. This way, we can put bounds on the spectrum of any unitary CFT without the need for a full analytical solution.

Several works were published in the literature within the next three years after \cite{Rattazzi:2008pe} appeared, applying this numerical technique in numerous ways; e.g. to bound the scaling dimension of the $\phi^2$ operator via the analysis of $\<\f\f\f\f\>$ \cite{Rattazzi:2008pe,Rychkov:2009ij,Poland:2011ey}, to bound the OPE coefficient of two scalars with a traceless symmetric tensor $\cO$ \cite{Caracciolo:2009bx,Poland:2011ey}, to obtain similar bounds for $\cN=1$ supersymmetric theories \cite{Poland:2010wg,Poland:2011ey}, and to bound the central charge of generic CFTs \cite{Poland:2010wg, Rattazzi:2010gj, Vichi:2011ux,Poland:2011ey}. In those papers, conserved currents of global symmetries and stress tensors are also studied in (non-)supersymmetric theories.\footnote{See \cite{Rattazzi:2010yc} as well.}

This revival in the conformal bootstrap gained an acceleration with the realization of an alternative to LP in \cite{Poland:2011ey}. In that paper, Poland, Simmons-Duffin, and Vichi introduced \emph{semi-definite programming} (\ac{SDP}) into the conformal bootstrap, which they used to update the previously computed bounds significantly by carving out various $4d$ CFTs with(out) global symmetries and supersymmetry.

The advantages of SDP over LP are discussed in \cite{Poland:2011ey}; for a more detailed review, one can refer to \cite{doi:10.1137/1038003}. Here we will only mention one of these advantages: semi-definite optimization is a very standard problem in engineering hence there are various sophisticated implementations to address such problems.\footnote{For a review of SDP applications, see \cite{parrilo2000structured}.}

Even after the introduction of SDP into the conformal bootstrap, there were some analyses which kept relying on LP. Most of these papers employed the simplex algorithm of LP implemented via the \textsf{ILOG CPLEX} optimizer  through a \textsf{Mathematica} interface \cite{ElShowk:2012hu,Liendo:2012hy,ElShowk:2012ht,Shimada:2015gda,El-Showk:2014dwa,Bobev:2015vsa,El-Showk:2013nia}. In contrast, the bootstrap community gradually switched to SDP: during the next 4 years after the introductions of SDP, the conformal bootstrap analyses were mostly implemented via the algorithm \textsf{SDPA-GMP} \cite{Chester:2014gqa,Beem:2014zpa,Bae:2014hia,Chester:2014mea,Nakayama:2014sba, Caracciolo:2014cxa,Kos:2014bka,Chester:2014fya,Nakayama:2014yia,Nakayama:2014lva,Kos:2013tga,Friedan:2013cba,Friedan:2013bha}. Some papers simply used both LP and SDP in the same work \cite{Beem:2015aoa}.

In 2015 David Simmons-Duffin introduced \textsf{SDPB}, a SDP implementation specialized for the conformal bootstrap applications \cite{Simmons-Duffin:2015qma}. This algorithm boosted the numerical analyses even further and has become the de-facto optimizer used in the conformal bootstrap community and has been actively improved thanks to \emph{Simons Collaboration on the Nonperturbative Bootstrap}.\footnote{See
	\hyperref{https://github.com/davidsd/sdpb}{}{}{https://github.com/davidsd/sdpb} for further details.} Some of the papers among this intense research program can be grouped as follows:
\begin{enumerate}	
	\item \emph{Various $3d$ scalar models}: Precision islands in the Ising model \cite{Kos:2016ysd}, $\cN=1$ Ising and WZ models \cite{Li:2017kck}, $O(N)$ models \cite{Kos:2015mba,Kos:2016ysd}, CFTs with a continuous global symmetry \cite{Dymarsky:2017xzb}, stress-tensor bootstrap \cite{Dymarsky:2017yzx}, and the random-bond Ising model \cite{Komargodski:2016auf}.
	
	\item $4d$ \emph{supersymmetric theories}: Minimial $\cN=1$ SCFTs \cite{Poland:2015mta}, mixed correlators in  $\cN=1$ SCFTs \cite{Li:2017ddj}, chiral correlators in $\cN=2$ SCFTs \cite{Lemos:2015awa,Chester:2015qca}, general $\cN=3$ SCFTs \cite{Lemos:2016xke}, stress tensor supermultiplets  in $\cN=4$ SCFTs \cite{Beem:2016wfs}, and $\cN=4$ $O(N)$ Vector Models  \cite{Chester:2015lej}.
	
	\item \emph{Models in $d>4$ dimensions}: Mixed correlators in the $5d$ critical $O(N)$ Models \cite{Li:2016wdp}, and the $6d$ superconformal bootstrap \cite{Chang:2017xmr}.
	
	\item \emph{Fermionic theories}: Identical Majorana fermions in $3d$ \cite{Iliesiu:2015qra}, Dirac fermions in QED$_3$ \cite{Li:2018lyb}, and Weyl fermions in $4d$ \cite{Karateev:2019pvw}.
\end{enumerate}
Of course, this is only a small fraction of the research using numerical bootstrap techniques. In addition to these, there are a variety of different theories where numerical bootstrap has been extensively used. These include tetragonal CFTs \cite{Stergiou:2019dcv}, the half-BPS line defect \cite{Liendo:2018ukf}, universality of BTZ spectral density \cite{Collier:2017shs}, Virasoro
minimal models \cite{Behan:2017rca,Lin:2016gcl}, M-theory \cite{Agmon:2017xes}, Argyres-Douglas theory \cite{Cornagliotto:2017snu}, flavored $2d$ CFT partition functions \cite{Dyer:2017rul}, conformal multi-flavor QCD on lattice \cite{Nakayama:2016knq}, pure quantum gravity in AdS$_3$ \cite{Bae:2016yna}, many-flavor gauge theories \cite{Iha:2016ppj},
long multiplets of $\cN=2,3$ SCFTs \cite{Cornagliotto:2017dup},
modular constraints on CFTs with currents \cite{Bae:2017kcl},
high-precision bootstrap of a non-unitary CFTs \cite{El-Showk:2016mxr},
and K3 CFT \cite{Lin:2015wcg}.

\subsubsection{Analytics}
\label{sec: state of art-analytics}
The rapid progress in the numerics was accompanied by the development of various analytical tools, allowing us to discover the rich structure of conformal theories.  Below we will briefly review this progress.

The first point that we should mention is the convergence of OPE. Being at the heart of the conformal bootstrap program, this convergence was extensively analyzed in \cite{Pappadopulo:2012jk} where the authors show the exponential suppression of operators with higher and higher scaling dimensions. In that paper, the authors make use of the radial quantization and the mapping between $\R^d$ and $\R\x S^{d-1}$; it is also realized that extensive usage of radial coordinates has various advantages over standard Dolan and Osborn coordinates $\{z,\zb\}$, especially for the convergence of the conformal blocks written as series expansions \cite{Hogervorst:2013sma,Rychkov:2015lca}.

Such series expansions are particularly important since the absence of compact and explicit expressions for the conformal blocks in odd dimensions poses a serious practical problem.\footnote{
	Even though the OPE of two scalars and a spin $l$ operator was known for years, it was first in \cite{Dolan:2000ut} that the summation was carried out and explicit conformal blocks were derived! In that and the following papers, Dolan and Osborn gave the closed form expressions for conformal blocks of external scalars in even dimensions, and provided several expressions such as integral representations for conformal blocks of external scalars in odd dimensions \cite{Dolan:2000ut,Dolan:2003hv,Dolan:2011dv}.
} There are some special cases where the conformal blocks are known even in odd dimensions \cite{Hogervorst:2013kva,Fitzpatrick:2013sya}, however the generic case is yet to be derived.\footnote{
	There is a new research program where the calculations are carried out in the embedding space, including the application of OPE \cite{Fortin:2016dlj,Comeau:2019xco, Fortin:2019fvx, Fortin:2019dnq}. The authors provide the compact and complete results for the necessary ingredients for any conformal block in any dimension.
} In the past decade, this problem was partially circumvented by going to the specific kinematical limits. The most intensively studied one is the \emph{lightcone limit} for which the conformal blocks in any dimension reduce to the collinear conformal blocks whose compact form is known. In \cite{Fitzpatrick:2012yx, Komargodski:2012ek} the authors showed that every CFT admits a large spin expansion and the lightcone limit probes this sector. Using the bootstrap equations, they show the existence of infinitely many operators which organize into families of almost the same twist.\footnote{We define twist $\tau\coloneqq\De-l$ for the scaling dimension $\De$ and spin $l$ for the symmetric traceless representations --- all representations are such in three dimensions. For a general mixed representation in other dimensions, $l$ is taken to be the length of the first row of the Young tableau.} In mean field theory, such operators are composite objects built out of fundamental fields and they have exactly the same twist; i.e., the operators $\phi \partial_{\mu_1}\cdots\partial_{\mu_n}\phi$ all have the twist $2\Delta_\phi$ for the scalar $\phi$ with twist $\Delta_{\phi}$.\footnote{The operator $\phi \partial_{\mu_1}\cdots\partial_{\mu_n}\phi$ is neither in an irreducible representation of the rotation group nor a conformal primary. One can ensure these properties by symmetrizing open indices and subtracting the traces. Furthermore, one needs to replace $\partial$ with $\frac{\overset{\rightarrow}{\partial}-\overset{\leftarrow}{\partial}}{2}$ because $\partial= \frac{\overset{\rightarrow}{\partial}-\overset{\leftarrow}{\partial}}{2}+ \frac{\overset{\rightarrow}{\partial}+\overset{\leftarrow}{\partial}}{2}$ and second term simply creates descendants.} Following \cite{Komargodski:2012ek}, such operators are called \emph{double twist operators} in the bootstrap literature.

In a strongly interacting theory, we can no longer interpret double twist operators as in given the schematic forms; furthermore, those operators start to develop anomalous dimensions. However, \cite{Fitzpatrick:2012yx, Komargodski:2012ek} proved that double twist operators exist in any CFT, and they asymptotically have the same properties as composite MFT operators at large spin.\footnote{The inverse proportionality of anomalous dimension with spin justifies viewing the large spin sector as a perturbation around the generalized free theory. Via AdS/CFT such operators correspond to widely separated weakly interacting particles \cite{Alday:2007mf,Fitzpatrick:2014vua}}

Despite being an asymptotic expansion,\footnote{
	It is shown in \cite{Alday:2015ewa} that the large spin expansion is asymptotic but Borel-summable. The validity of this expansion is partially solidified in \cite{Qiao:2017xif}.} analysis of the large spin sectors via the lightcone bootstrap received a great deal of attention \cite{Kaviraj:2015xsa,Alday:2016mxe,Alday:2016njk,Alday:2016jfr,Kaviraj:2015cxa,Alday:2015eya,Alday:2015ota,Li:2015rfa} where mostly the CFT data of double twist operators are investigated. The generalization to multi-twist operators and eikonalization are discussed in \cite{Fitzpatrick:2015qma}, and David Simmons-Duffin systematized the lightcone bootstrap in \cite{Simmons-Duffin:2016wlq}: he introduced an $\SL(2,\R)$ expansion as a means to go beyond leading order in spin, discussed the sensitivity of large spin expansion to finite spin effects, addressed the issue of mixing, and compared the predictions of analytics with the numerics in the case of the $3d$ Ising model.

Concurrently, Simon Caron-Huot introduced in 2017 an inversion formula which calculates the OPE data using the correlator as an input, much like the Froissart-Gribov formula yields the angular momentum partial wave coefficients using the amplitude as an input in relativistic s-matrix theory
\cite{Caron-Huot:2017vep}. More importantly, Caron-Huot's inversion formula establishes the analyticity in spin in CFTs just as the Froissart-Gribov formula established the analyticity in spin in s-matrix theory \cite{Gribov:1961ex,Collins:1977jy,Donnachie:2002en}. This is an intrinsically Lorentzian phenomena as this inversion formula uses specific causal orderings of the operators. This Lorentzian inversion formula was further analyzed \cite{Simmons-Duffin:2017nub}, generalized to arbitrary Lorentz representations \cite{Kravchuk:2018htv}, and utilized in various works \cite{Liu:2018jhs,Albayrak:2019gnz,Kologlu:2019mfz}. 

We stated above that the conformal blocks have been poorly understood even though the OPE of two scalars has been known for a very long time, and that the explicit conformal blocks were derived only less than two decades ago \cite{Dolan:2000ut}. However, this recently changed quite dramatically. Firstly, the conformal blocks of traceless symmetric operator exchange for external bosonic spinning operators were derived by a judicious use of differential operators on scalar conformal blocks \cite{Costa:2011dw}. Then a procedure to calculate any conformal block was introduced in \cite{SimmonsDuffin:2012uy}, and is employed, among others, to derive superconformal blocks \cite{Khandker:2014mpa,Fitzpatrick:2014oza} and conformal blocks of identical external fermions in $3d$ \cite{Iliesiu:2015qra}.\footnote{The procedure is conceptually quite simple and elegant. One uses the well known shadow formalism \cite{Ferrara:1972xe,Ferrara:1972ay, Ferrara:1972uq, Ferrara:1973vz} to derive the \emph{partial waves} in the embedding space, and then obtains the conformal blocks using a monodromy projection. Even though the process is quite general, it does not immediately provide an explicit compact result.} There has been numerous other progress in the derivation of the conformal blocks in a variety of dimensions as well \cite{Osborn:2012vt,Behan:2014dxa,Echeverri:2015rwa,Bissi:2015qoa,Bobev:2015jxa,Iliesiu:2015akf,Penedones:2015aga,Costa:2016xah,Costa:2016hju}. However, the most comprehensive approach to get spinning conformal blocks was introduced in \cite{Karateev:2017jgd}. The authors use group theoretical arguments and manage to construct a formalism to relate different spinning three point structures to each other by exchanging a finite dimensional representation of the conformal group. One can use these relations inside Euclidean pairings of three point structures and relate partial waves of different operators. Using monodromy arguments explicitly discussed back in \cite{SimmonsDuffin:2012uy}, one can then relate spinning conformal blocks to each other, and as an important subset of that fermion conformal blocks to bosonic ones. Alternatively, one can directly use these \emph{weight shifting operators} with the Lorentzian inversion formula, bypassing partial waves and directly working with conformal blocks \cite{Kravchuk:2018htv}.\footnote{In this thesis, we refer to the single-valued solutions of the conformal Casimir equation as \emph{partial waves}. They decompose into conformal blocks and the \emph{shadow} conformal blocks. The main intuitive difference between Euclidean and Lorentzian pairings is the integration range: the integration is over all spacetime in the Euclidean inversion formula hence it yields both conformal blocks and their shadows; in other words, partial waves are the main objects. In contrast, the integration range is a causal diamond in Lorentzian formula, and Lorentzian pairing of two three point function is proportional to a conformal block. The reason for the difference in the integration range follows from the necessity to work with the so-called \emph{light-ray operators} which are continuous-spin generalization of local operators. Being analytic in spin, Lorentzian inversion formula can be conceptually explained only by these operators, even though it was not originally derived through them \cite{Caron-Huot:2017vep,Simmons-Duffin:2017nub}.}

The advancement in extracting spinning conformal blocks went hand in hand with the analysis of  correlators of external spinning operators. In fact, the authors of \cite{Costa:2011dw} cited above started their analysis by constructing a formalism to analyze the spinning correlators in great generality \cite{Costa:2011mg}.

The problem of classifying three point functions of arbitrary spin $l$ has been discussed several times in the literature \cite{Sotkov:1976xe,Mack:1976pa,Osborn:1993cr,Weinberg:2010fx,Giombi:2011rz}. What Costa, Penedones, Poland, and Rychkov did in \cite{Costa:2011mg} was to reintroduce the embedding space into the conformal bootstrap and to combine it with an index-free formalism, hence developing necessary tools to write down any bosonic spinning three point correlator as well as the conditions on conserved tensors.\footnote{Usage of embedding space started with Dirac and has been on and off since then \cite{Dirac:1936fq,Cornalba:2009ax,Weinberg:2010fx,Ferrara:1973eg,Boulware:1970ty,Ferrara:1973yt} --- it is extensively discussed in \cite{Costa:2011mg}. Likewise, derivation of the conservation conditions on three point structures goes back to the last century \cite{Schreier:1971um,Osborn:1993cr}.
}

The extension to fermionic correlators is a little bit more complicated. Individual cases of fermionic analysis can be found in the literature; e.g. Weinberg used the embedding formalism to analyze $4d$ fermions back in 2010 \cite{Weinberg:2010fx}, see also \cite{Iliesiu:2015qra,Elkhidir:2017iov,Ji:2018yaf} for more recent studies. However, a more general treatment only started relatively recently \cite{Kravchuk:2016qvl, Cuomo:2017wme, Kravchuk:2017dzd, Karateev:2017jgd}. In these papers, Kravchuk and his collaborators analyzed classification of conformal correlators, generalized Casimir recursion relations, discussed generic $4d$ bootstrap equations, and introduced the weight shifting operators.

There are a variety of other areas which benefited from and contributed to the analytical bootstrap program. Among others, we can list some of them:
\begin{enumerate}
	\item \emph{supersymmetric theories}: analyses of $\cN=1,2,4$ superconformal blocks \cite{Dolan:2001tt,Poland:2010wg,Li:2016chh}, discussion of  chiral superconformal primary OPE \cite{Poland:2010wg,Vichi:2011ux,Fortin:2011nq}, study of superconformal theories with global symmetries \cite{Berkooz:2014yda}, and many others  \cite{Goldberger:2011yp,Goldberger:2012xb,Alday:2014qfa,Xie:2016hny,Maruyoshi:2018nod,Manenti:2018xns}.
	
	\item \emph{gauge gravity duality}: analyses of unitarity and analyticity of holographic s-matrix \cite{Fitzpatrick:2011dm,Fitzpatrick:2011hu}, study of AdS$_3$/CFT$_2$ \cite{Benjamin:2016fhe,Qualls:2015bta,Jackson:2014nla,Qualls:2014oea}, establishment of a holographic connection to Witten and geodesic Witten diagrams \cite{Heemskerk:2009pn,Heemskerk:2010ty,Sleight:2017fpc,Castro:2017hpx,Nishida:2018opl}.
	
	\item \emph{various other applications}: finite temperature CFTs \cite{Iliesiu:2018zlz,Iliesiu:2018fao}, Mellin space bootstrap\footnote{The Mellin representation of conformal correlators actually goes all the way back to Mack \cite{Mack:1976pa}, see also \cite{Paulos:2011ie,Fitzpatrick:2011ia,Penedones:2010ue,Mack:2009mi}.} \cite{Dey:2016mcs,Gopakumar:2016cpb}, multi-point conformal blocks \cite{Rosenhaus:2018zqn,Parikh:2019ygo}, defect CFTs \cite{Gaiotto:2013nva,Guha:2018snh}, complex CFTs \cite{Gorbenko:2018ncu}, global symmetries \cite{Sen:2015doa, Dey:2016zbg}, boundary and crosscap CFTs \cite{Hogervorst:2017kbj}, fractal Ising model \cite{Golden:2014oqa}, and so on  \cite{Friedan:2012jk,Perlmutter:2013gua,Green:2013rd,Rychkov:2015naa,Rejon-Barrera:2015bpa,Hofman:2016awc,Bae:2016jpi,Li:2017agi,Hogervorst:2017sfd}.
\end{enumerate} 

\section{Primer on conformal field theories}
In this section, we will review the basics of  conformal field theory. For concreteness, we will consider the Euclidean conformal group though the discussion can straightforwardly be extended to the Lorentzian conformal group.

\subsection{Conformal transformations and conformal algebra}
\label{sec: conformal transformations and conformal algebra}

We define the conformal transformations as the transformations that leave the metric invariant upto a local rescaling, that is
\be 
g'_{ij}(x')=\Lambda(x)g_{ij}(x)\;.\label{E: metric transformation}
\ee
As a second rank covariant tensor, the metric transforms under an infinitesimal transformation $x^i\rightarrow x^i+\e^i(x)$ as
\be 
g'_{ij}=\frac{\partial x^k}{\partial x'^i}\frac{\partial x^l}{\partial x'^j}g_{kl}=g_{ij}-(\partial_i\epsilon_j+\partial_j\epsilon_i)+O(\epsilon^2)\;,
\ee  
which in turn restricts conformal transformations to satisfy the condition
\be 
\label{eq: conformal killing equation}
\partial_i\epsilon_j+\partial_j\epsilon_i=\lambda(x)g_{ij}(x)\;.
\ee 
Contracting both sides with $g^{ij}(x)$, we can solve for $\lambda(x)$ and insert it back to obtain
\be 
\partial_i\epsilon_j+\partial_j\epsilon_i=\frac{2}{\ac{d}}(\partial\.\epsilon)g_{ij}(x)\;.
\ee 
By acting on both sides with $\partial^i$ and $\partial^i\partial^j$, we further obtain the conditions
\bea 
\frac{d}{2}\Laplacian \e_j=&\frac{2-d}{2}\partial_j (\partial\.\e)+\de g_j\;,\\
(d-1)\Laplacian (\partial\.\e)=&\partial\.\de g
\eea 
for $\delta g_j\coloneqq (\partial^i g_{ij})(\partial\.\epsilon)$. In flat spacetimes, which will be the focus of this thesis, we have $\delta g_i=0$, hence we can solve the equations above for the most general form of $\epsilon_i$ in $d>2$:
\be 
\label{eq: most general conformal transformation}
\epsilon_i = a_i+cx_i+\theta_{ij}x^j+2(b\. x)x_i -x^2 b_i
\ee 
for the antisymmetric matrix $\theta_{ij}=-\theta_{ij}$.

We see that conformal transformations in $d$-dimensional flat space are parametrized by the scalar $c$, vectors $a_i$ \& $b_i$, and the antisymmetrix matrix $\theta_{ij}$, leading to the conclusion that the conformal group has  dimension $\frac{(d+2)(d+1)}{2}$.

To get further insight into the generators of the conformal group, let us first consider the translation group: in QFT, this group is generated by the the momentum operator $P^\mu$, which by Noether's theorem should be the conserved charge of a current. The relevant current here is the stress tensor \ac{T}, and we have the relation
\be 
P^\mu(t)=-\int d^{d-1}x T^{0\mu}(x)\;,
\ee 
where the integration is over a constant time slice of the spacetime. In fact, $P^\mu$ is a topologically conserved charge and we can rewrite the integration over any hypersurface $\Sigma$ as $\int_\Sigma dS_\mu T^{\mu\nu}(x)$; in a similar fashion, we can define a new set of conserved charges $Q_\e$ as 
\be 
Q_\e(\Sigma)\coloneqq -\int_\Sigma dS_\mu \e_{\nu}(x)T^{\mu\nu}(x)
\ee 
if $\partial_\mu \left(\e_{\nu}(x)T^{\mu\nu}(x)\right)=0$. Conservation of stress tensor, $\partial_\mu T^{\mu\nu}=0$, then implies
\be 
\left(\partial_\mu\e_\nu+\partial_\nu\e_\mu\right)T^{\mu\nu}=0\;.
\ee 
For a generic stress tensor this implies 
\be 
\partial_\mu\e_\nu+\partial_\nu\e_\mu=0\;.
\ee 
This is the \emph{Killing equation}, and it has the solutions $\e=\e^\mu\partial_\mu$ as
\bea[eq: Poincare generators]
p_\mu\coloneqq&\partial_\mu \label{eq: differential representation of p}\;,\\
m_{\mu\nu}\coloneqq&x_\mu\partial_\nu-x_\nu\partial_\mu\;,
\eea 
which generate the usual Poincar\'e algebra with $p$ and $m$ generating translations and rotations respectively.

If we impose that the stress tensor is \emph{traceless}, i.e. $g_{\mu\nu}T^{\mu\nu}=0$, we instead have the \emph{Conformal Killing equation}
\be 
\partial_\mu\epsilon_\nu+\partial_\nu\epsilon_\mu=\lambda(x)g_{\mu\nu}(x)
\ee 
which is precisely \equref{eq: conformal killing equation}. Therefore, in the field theory context, conformal field theory is a local field theory with a \emph{traceless} stress tensor. For this equation, we obtain two new set of vector fields for $\e=\e^\mu\partial_\mu$:
\bea[eq: Conformal generators]
d\coloneqq&x^\mu\partial_\mu\;,\\
k_\mu\coloneqq&2x_\mu(x\cdot\partial)-x^2\partial_\mu\;,
\eea 
where $d$ generates dilations and $k$ generates so-called special conformal transformations. By comparing the generators in \equref{eq: Poincare generators} and \equref{eq: Conformal generators} with the most general form of the conformal transformation in \equref{eq: most general conformal transformation}, we conclude that the parameters $a,\; b,\; c,\;\theta$ parameterize translations, special conformal transformations, dilations, and rotations respectively.

It is illustrative to investigate the algebras constructed with these generators, which we summarize in \tabref{\ref{table: algebras}}. We can also collect the full set of commutation relations as
\bea[eq: commutation relations of Euclidean conformal group] 
\comm{m_{ij}}{m_{kl}}=&\left(\de_{il}\de_{kr}\de_{jp}+\de_{jk}\de_{ir}\de_{lp}+\de_{ik}\de_{lp}\de_{jr}+\de_{jl}\de_{ip}\de_{kr}\right)m_{pr}\;,\\
\comm{p^\pm_i}{m_{jk}}=&\left(\de_{ij}\de_{kl}-\de_{ik}\de_{jl}\right)p^\pm_l\;,\\
\comm{p^\pm_i}{d}=&\pm p^\pm_i\label{eq: ladder operators of d}\;,\\
\comm{p^+_i}{p^-_j}=&2\de_{ij}d-2m_{ij}\;,
\eea 
for $p_i^+\coloneqq p_i$ and $p_i^-\coloneqq k_i$.

\begin{table}
	\caption[Conformal algebra and its subalgebras]{\label{table: algebras}Conformal algebra and its subalgebras.\footnotemark}
	\begin{tabularx}{\textwidth}{%
			>{\hsize=.5\hsize\linewidth=\hsize}X%
			>{\hsize=.25\hsize\linewidth=\hsize}X%
			>{\hsize=.33\hsize\linewidth=\hsize}X%
			>{\hsize=.54\hsize\linewidth=\hsize}X%
		}
		\hline\\[-.15in]\hline\\[-.15in]
		\textbf{Name of the algebra}&\textbf{Denotation}&\textbf{Decomposition}&\textbf{Commutation relations}
		\\[.02in]\hline\\[-.15in]
		Translation&$\mathrm{t}(N)$&--&$[p^i,p^j]=0$
		\\[.02in]\hline\\[-.15in]
		Rotation&$\mathrm{so}(N)$&--&$\begin{aligned}
			[m^{ij},m^{kl}]=m^{il}g^{jk}-m^{jk}g^{il}\\+m^{ik}g^{jl}-m^{jl}g^{ik}
		\end{aligned}$
		\\[.02in]\hline\\[-.15in]
		Euclidean&$\mathrm{iso}(N)$&$\mathrm{t}(N)\oplus_s\mathrm{so}(N)$&$[p^k,m^{ij}]=g^{ki}p^j-g^{kj}p^i$
		\\[.02in]\hline\\[-.15in]
		Dilation&$\mathrm{d}(1)$&--&--
		\\[.02in]\hline\\[-.15in]
		Translation \& Dilation&--&$\mathrm{t}(N)\oplus_s\mathrm{d}(1)$&$[p^i,d]=p^i$
		\\[.02in]\hline\\[-.15in]
		Euclidean \& Dilation&--&$\mathrm{iso}(N)\oplus_s\mathrm{d}(1)$&$	[m^{ij},d]=0$
		\\[.02in]\hline\\[-.15in]
		Special conformal&$\mathrm{c}(N)$&--&$	[k^i,k^j]=0$
		\\[.02in]\hline\\[-.15in]
		General conformal&$\mathrm{gc}(N)$&--&$\begin{aligned}
			{}[k^k,m^{ij}]=&g^{ki}k^j-g^{kj}k^i
			\\
			[k^i,d]=&-k^i
			\\
			[p^i,k^j]=&2g^{ij}d-2m^{ij}
		\end{aligned}
		$
		\\[.02in]\hline\\[-.15in]\hline\\[-.15in]
	\end{tabularx}
\end{table}

The interpretation of these relations is straightforward. The first equation simply states that $m$ generates a rotation algebra. The second equation and the absence of nonzero $\comm{d}{m_{ab}}$ indicate that $p^\pm$ and $d$ transform as a vector and a scalar under rotations respectively. The third equation shows that $p^\pm$ behave like \index{Ladder operators}ladder operators with respect to the dilation $d$; and finally, the last equation gives their non-commutative nature.

\subsection{Conformal group and its linear realizations}
\label{sec: preliminaries of conformal group}

\footnotetext{In the table, the symbol $\oplus_s$ denotes a semi-direct sum. We remind the reader that an algebra $g$ can be written as a semi-direct sum of its subalgebras $g_1$ and $g_2$ as \mbox{$g=g_1\oplus_s g_2$} if the conditions $\comm{\bar g_1}{\bar g_1}\in g_1$, $\comm{\bar g_2}{\bar g_2}\in g_2$, and $\comm{\bar g_1}{\bar g_2}\in g_1$ are satisfied for all elements $\bar g_i\in g_i$. We also would like to note that one can construct further subalgebras by replacing $t(N)$ with $c(N)$ in the table; for example, we can define a second Euclidean subalgebra of the conformal algebra as $\mathrm{c}(N)\oplus_s\mathrm{so}(N)$.}

In \secref{\ref{sec: conformal transformations and conformal algebra}} we reviewed the conformal symmetry and the differential representation of the conformal algebra. By an exponential map, one can extend the analysis from the conformal algebra to the conformal group --- or rather its connected component around the identity element. Indeed, this way, we can write the action of the conformal group under a \emph{finite} transformation by starting with the infinitesimal transformation in \equref{eq: most general conformal transformation}; here, we simply  present the final results:
\bea[eq: action of conformal group in d dimensions] 
P:&\;x^i\rightarrow x^i+a^i\;,\\
M:&\;x^i\rightarrow e^{\theta_{ij}} x^j\;,\\
D:&\;x^i\rightarrow e^c x^i\;,\\
K:&\;x^i\rightarrow \frac{x^i-b^i x^2}{1-2b\cdot x+b^2x^2}\;.\label{eq: special conformal transformation action}
\eea 
The transformation under the first three generators is fairly intuitive: we translate, rigidly rotate, or scale the coordinates. The last transformation on the other hand can be explained intuitively only through an \emph{inversion} operation:
\be 
I: x^i \rightarrow\frac{x^i}{x^2}\;.
\ee 
Even though this operator is not an element of the conformal group, conjugation by an inversion is actually the outer authomorphism of the conformal group and one can show that
\be 
K=-I\. P\. I\;.
\ee 
In other words, the special conformal transformation in \equref{eq: special conformal transformation action} is simply an inversion followed by a translation followed by another inversion!

The nonlinear nature of the transformation in \equref{eq: special conformal transformation action} complicates the construction of the representations of the conformal group in $\R^{d}$. However, there actually exists another space in which the realization of the conformal group is linear. To see that, we define the generators $M_{AB}$ 
\be 
M_{ij}=&m_{ij}\;,\\
M_{0,-1}=&d\;,\\
M_{i,-1}=&\frac{1}{2}\left(p_i^++p_i^-\right)\;,\\
M_{i,0}=&\frac{1}{2}\left(p_i^--p_i^+\right)\;,
\ee  
where $A=-1,0,1,\dots,d$ and $i=1,\dots,d$. In terms of $M_{AB}$, we can rewrite the comutation relations in \equref{eq: commutation relations of Euclidean conformal group} as 
\be 
\label{eq: embedding algebra}
\comm{M_{AB}}{M_{CD}}=\left(g_{AD}M_{BC}+g_{BC}M_{DA}+g_{AC}M_{DB}+g_{BD}M_{AC}\right)
\ee 
for the metric
\be 
g_{AB}\doteq\mathrm{diag}\left(-1,1,\dots,1\right)\;.
\ee 

We realize that \equref{eq: embedding algebra} is the commutation relations for the rotation algebra which indicates that the Euclidean group $\mathrm{gc}(d)$ is isomorphic to an orthogonal algebra in $d+2$ dimensions, i.e.
\begin{subequations}
	\be 
	\mathrm{gc}(d)\simeq \mathrm{so}(d+1,1)\;.
	\ee 
	Similarly, we can identify the algebra of the Lorentzian conformal algebra with another orthogonal algebra:
	\be 
	\mathrm{gc}(d-1,1)\simeq \mathrm{so}(d,2)\;.
	\ee 
\end{subequations}

This isomorphism of the algebras suggests that we can work in $\R^{d+1,1}$ ($\R^{d,2}$) instead of $\R^{d}$ ($\R^{d-1,1}$) where the action of the conformal group is realized linearly. This \emph{embedding space} approach was first suggested by Dirac, and has been extensively used ever since \cite{Dirac:1936fq,Weinberg:2010fx,Costa:2011mg}. Below, we will quickly review this approach and illustrate how to construct invariant/covariant objects of the conformal group (such as correlation functions) via this approach.

The nonlinear action of the conformal group in \equref{eq: action of conformal group in d dimensions} can be mapped to the linear action of the special orthogonal group in the embedding space, once the extra degrees of freedom are taken care of: this is achieved by mapping \emph{null rays} in $\R^{d+1,1}$ to points in $\R^{d}$. The intuitive explanation for this is that null rays are the objects in $\R^{d+1,1}$ that are mapped to themselves under the action of $\so(d+1,1)$ --- similar to $\R^d$ being mapped to itself under $\gc(d)$ --- and that they have the correct degrees of freedom.

A straightforward procedure to map from the embedding space back to the physical space is by \emph{fixing the gauge} of the freedom $X_A\sim \lambda X_A$; a common choice is the \emph{Poincar\'e} section where we scale $X_A$ to $P_a$ defined as
\be 
\label{eq: Poincare section position}
P_A\coloneqq(1,x^2,x_i)
\ee 
in the lightcone coordinate $X_A=\left(X_+,X_-,X_i\right)$ for $X_\pm=X_0\pm X_{-1}$. With this choice, we see that $\so(d+1,1)$ invariants map as 
\be 
\label{eq: mapping down invariants from embedding space}
X_AY^A\quad\rightarrow\quad -\half(x_i-y_i)^2\;.
\ee 
Similarly, one can show that $\so(d+1,1)$ tensors map as 
\be 
\label{eq: mapping down tensors from embedding space}
T_{A_1\cdots A_l}(X^A_1,\dots)\quad\rightarrow\quad t_{a_1\cdots a_l}(P^A_1,\dots)\equiv\frac{\partial P^{A_1}}{\partial x^{a_1}}\cdots \frac{\partial P^{A_l}}{\partial x^{a_l}}T_{A_1\cdots A_l}(X^A_1,\dots)\;.
\ee 
We can also write down a prescription to map fermionic representations of $\so(d+1,1)$ but it is actually far easier to work in an \emph{index-free} fashion where auxiliary spinors and auxiliary vectors are used in the embedding space to construct $\so(d+1,1)$ invariants. However, before illustrating that, we will briefly discuss the stabilizer subgroup of the conformal group to review the labeling of the operators in the irreducible representations.

The stabilizer subalgebra $h$ of the conformal algebra  consists of the generators $m_{ij},\; p_{i}^-,\; d$ if we choose it to fix the point $x=0$. From \equref{eq: commutation relations of Euclidean conformal group} we then see that
\be 
h=\left(\mathrm{c}(N)\oplus_s\mathrm{d}(1)\right)\oplus_s\mathrm{so}(N)\;.
\ee 
We realize from \equref{eq: ladder operators of d} that $p_i^-$ acts a lowering operator for $d$, hence we can define \emph{primary operator} which satisfies
\be 
\comm{p_i^-}{\cO(0)}=0\;,
\ee 
which is in the highest weight representation of $\mathrm{d}(1)$. As these operators are invariant under the action of $c(N)$,\footnote{Under the action of $c(N)$, $\cO(0)$ transforms as $e^{-b^ip^-_i}\cO(0)e^{b^ip^-_i}=\cO(0)+b^i[\cO(0),p^-_i]+\dots=\cO(0)$ by the Hausdorff formula.} the stabilizer algebra for primary operators is effectively \mbox{$\mathrm{d}(1)\oplus_s\mathrm{so}(N)$} for which the operators are labeled by a continuous parameter $\De\in\R$ for $\mathrm{d}(1)$ and a representation $\rho\in\so(d)$ for the conformal algebra in $\R^d$. When we map from algebra to the group, we need to take the fermionic representations into account as well, hence we denote
\begin{subequations}
	\be 
	\text{primary operators:} \quad\cO_{\De,\rho}\quad\text{with}\quad \De\in\R\;,\quad\rho\in\mathrm{Spin}(d)\;.
	\ee 
	As \equref{eq: ladder operators of d} indicates that $p^+_i$ is a raising operator for $d$, operators $p^+_i\.\cO$, $p^+_i\.p^+_i\.\cO$, and so on are not primary operators. Such operators are called \emph{descendant operators} and are given by \equref{eq: differential representation of p} as
	\be 
	\text{descendant operators:} \quad\frac{\partial^n\cO_{\De,\rho}}{\partial x^{\mu_1}\dots \partial x^{\mu_n}}\quad\text{with}\quad n\in\Z^+\text{ and }\cO=\text{primary}\;.
	\ee 
\end{subequations}

With the little detour of the labeling of operators under the conformal group finished, we can get back to their construction in the embedding space, and in particular, the index-free construction of conformal correlators which bypass technical computations in \equref{eq: mapping down tensors from embedding space}. Without dwelling on the details, we will simply review the procedure
\begin{itemize}
	\item For the bosonic representations in any $d$, introduce an auxiliary vector $Z_A$ for each row of the Young diagram of the representation $\rho$ for the operator $\cO_{\De,\rho}$. In $d=3$, we can instead introduce a single auxiliary spinor $S_I$ for both bosonic and fermionic representations. In $d=4$, we instead introduce two auxiliary spinors $S_I$ and $\bar S_I$ for all representations.
	\item We construct the most general function of the auxiliary vectors/spinors and position vectors which satisfy following constraints:
	\begin{enumerate}
		\item Invariant under the action of $\SO(d+1,1)$.
		\item Homogeneous in both the auxiliary vectors/spinors and position vectors with degrees of homogeneity fixed for a given representation.
		\item Restricted to the null cone $X^2=0$.
		\item Transverse, i.e. position and auxilary vector/spinor are orthogonal. In $3d$, this means $S_I X^I_{\; J}=0$.\footnote{Here $X^I_{\; J}$ is the position vector written in spinor indices.}
	\end{enumerate}
	\item The resultant expression is projected to the Poincar\'e section by $X_{A}\rightarrow P_{A}$ for $P_A$ in \equref{eq: Poincare section position} and by projecting auxiliary vectors/spinors accordingly. For example, in $3d$ we project the $5d$ spinor $S_I$ as 
	\be 
	\label{eq: mapping down invariants from embedding space 2}
	S_I\rightarrow\begin{pmatrix}
		s_\a\\x^\a_{\;\b}s^\b
	\end{pmatrix}
	\ee 
	in terms of the position $x^\a_{\;\b}\equiv x_\mu(\gamma^\mu)^\a_{\;\b}$ and the $3d$ spinor $s_\a$.
\end{itemize}

We can illustrate this procedure for the construction of the two point function $\<\cO(x_1,s_1)\cO(x_2,s_2)\>$ in $3d$, where we use the shorthand notation
\be 
\ac{cO}(x,s)\coloneqq s_{\a_1}\dots s_{\a_{2l}}\cO^{\a_1\dots \a_{2l}}(x)
\ee 
for the spinor indices $\a_i$. In embedding space, we are then trying to construct $\<\cO(X_1,S_1)\cO(X_2,S_2)\>$ which is homogeneous in $X_i$ and $S_i$ with degrees $-\De-l$ and 2\ac{spin}. We then write down the only such $\SO(d+1,1)$ invariant:
\be 
\<\cO(X_1,S_1)\cO(X_2,S_2)\>=c\frac{\left((S_1)_I(S_2)^I\right)^{2l}}{\left((X_1)_A(X_2)^A\right)^{\De+l}}\;,
\ee 
where we note that $(S_1)_I(X_1)^I_{\; J}(S_2)^J=(S_1)_I(X_2)^I_{\; J}(S_2)^J=0$ by the transverness condition. With \equref{eq: mapping down invariants from embedding space} and \equref{eq: mapping down invariants from embedding space 2}, we then obtain
\be 
\<\cO(x_1,s_2)\cO(x_2,s_2)\>\propto \frac{\left[(s_1)_\a(x_{12})^\a_{\;\b}(s_2)^\b\right]^{2l}}{x_{12}^{2\De+2l}}\;.
\ee 

\section{Lightning review of the conformal bootstrap}

In this section, we will briefly review the historical philosophy of the conformal bootstrap and its modern applications in practice.

As we noted in \secref{\ref{sec: brief history of conformal symmetry}}, the bootstrap approach  historically bypasses \emph{constructionism}: one starts with some fundamental conditions and derive the theory with a bottom-up approach, using these conditions as constraints on the landscape of the theory. Conformal symmetry by itself is sufficient to fix the form of two and three point correlation functions, and one aims to use other conditions such as unitarity and operator product expansion associativity to entirely determine the form of all correlation functions, importantly, without resorting to any input from a microscopic theory.

This method is in stark contrast to the traditional top-down approach where one usually starts a theory by writing down a Lagrangian. This is an old habit of physicists and it served them extremely well for the past few centuries. However there is an important philosophical difference between the Lagrangian approach and that of the bootstrap: starting in the UV by writing a Lagrangian, one inherently assumes that physics is \emph{constructive}, i.e. one can not only reduce all physics to some fundamental laws and particles but also obtain useful information regarding phenomena observed in the IR by starting with fundamental laws and particles in the UV and by \emph{constructing their way back}. In contrast, the main objects of the conformal bootstrap are correlation functions, hence there is no need for constructionism as one already works with the theoretically closest objects to the observables.

Despite this philosophical difference between Lagrangian and bootstrap approaches, most physicists including the bootstrappers already accept both a reductionist and a constructionist approach towards nature --- and hence combine the bootstrap approach with the Lagrangian theories --- and we will assume in the rest of the thesis that both top-down and bottom-up approaches are equivalent in this sense.\footnote{Interested reader can refer to the infamous paper of Anderson \cite{Anderson:1972pca} and a more recent relevant talk, $27^{\text{th}}$ Occam lecture of Slava Rychkov --- see \hyperref{https://www.merton.ox.ac.uk/event/27th-ockham-lecture-reductionism-vs-bootstrap-are-things-big-always-made-things-elementary}{}{}{https://www.merton.ox.ac.uk/event/27th-ockham-lecture-reductionism-vs-bootstrap-are-things-big-always-made-things-elementary}.}

Let us switch gears and review the modern employment of the conformal bootstrap; for simplicity, we will consider the analytic bootstrap for scalar 4-point functions $\<\phi_1\phi_2\phi_3\phi_4\>$, which take the general form
\be
\label{eq: scalar four point s channel expansion}
\<\ac{phi}_1(x_1)\phi_2(x_2)\phi_3(x_3)\phi_4(x_4)\> = \left(\frac{x_{23}}{x_{13}}\right)^{\Delta_{12}}\left(\frac{x_{13}}{x_{14}}\right)^{\Delta_{43}}\frac{G(z,\bar{z})}{x_{12}^{\Delta_1+\Delta_2}x_{34}^{\Delta_3+\Delta_4}}\;,
\ee
where $x_{ij} = x_i - x_j$ and $\Delta_{ij} = \Delta_i - \Delta_j$, with \ac{Delta}$_i$ being the scaling dimension of $\phi_i$. The conformal cross-ratios (\ac{z},\ac{zb}) are given by
\begin{equation}
	z\zb=\frac{x_{12}^2x_{34}^2}{x_{13}^2x_{24}^2}\quad,\quad (1-z)(1-\zb)=\frac{x_{14}^2x_{23}^2}{x_{13}^2x_{24}^2}\;.
\end{equation} 
The function $G(z,\bar{z})$ can be expanded in conformal blocks for the $\f_{1}\f_{2}\rightarrow\f_{3}\f_{4}$ OPE as
\be
\label{eq: conformal blocks in four scalars}
G(z,\bar{z}) = \sum_{\cO} f_{12\cO} f_{43\cO} g_{h,\bar{h}}^{r,s}(z,\bar{z})\;,
\ee
where we have introduced the variables
\begin{subequations}
	\begin{alignat}{2}
		r&=\frac{\Delta_{1}-\Delta_{2}}{2}\;,\qquad   &&s=\frac{\Delta_{3}-\Delta_{4}}{2}\;,
		\\
		h&=\frac{\Delta-\ell}{2}\;,  
		&&\bar{h}=\frac{\Delta+\ell}{2}\;,
	\end{alignat}
\end{subequations}
where $\Delta$ and $\ell$ are the scaling dimension and spin of the exchanged operator. This parameterization is convenient for the lightcone expansion, where $\hb$ is the natural expansion parameter.

For a spacelike configuration of the operators, we have \mbox{$\<\phi_1\phi_2\phi_3\phi_4\> =\<\phi_3\phi_2\phi_1\phi_4\> $}
which with \equref{eq: scalar four point s channel expansion} and \equref{eq: conformal blocks in four scalars} indicate
\be
\label{eq: scalar crossing equation}
\sum_{\cO\in\f_1\x\f_2} f_{12\cO} f_{43\cO} \frac{g_{h,\bar{h}}^{r,s}(z,\bar{z})}{\left((1-z)(1-\zb)\right)^{(\De_1+\De_3)/2}}
=
\sum_{\cO\in\f_3\x\f_2} f_{32\cO} f_{41\cO} \frac{g_{h,\bar{h}}^{r,s}(1-z,1-\bar{z})}{(z\zb)^{(\De_1+\De_3)/2}}
\;.
\ee
This is the crossing equation: it constraints the dynamical data --- OPE coefficient $f$'s --- in $\C$ spanned by $z,\zb$ where the conformal blocks, $g(z,\zb)$, are theory-independent kinematic objects.

In general dimensions the conformal blocks are not known in a simple closed form,\footnote{See~\cite{Poland:2018epd} for a review of various methods to calculate the blocks.} but in the limit where two operators become light-like separated they display a universal behavior which makes analytic results possible. In terms of the conformal cross-ratios, if we take the limit $z \rightarrow 0$, then the leading order behavior of conformal blocks in any dimension is
\begin{equation}
	g_{h,\hb}^{r,s}(z,\zb)\underset{z\rightarrow 0}{\simeq} z^h \zb^{\hb}
	\pFq{2}{1}{\hb-r,\hb+s}{2\hb}{\zb}\;,
	\label{2.39 of Dolan:2011dv}
\end{equation}
where ${}_2F_1$ is the standard hypergeometric function. This approximation is sufficient to compute the leading large-$\ell$ corrections to the spectrum of double-twist operators using analytic bootstrap techniques, as was first demonstrated in~\cite{Komargodski:2012ek,Fitzpatrick:2012yx}.

In this thesis we will be interested in higher order corrections in the large-$\ell$ expansion \mbox{\cite{Alday:2015ewa,Alday:2016njk,Simmons-Duffin:2016wlq}}, hence we will need subleading terms in the above expansion. In three dimensions, one can use dimensional reduction to expand the conformal block in terms of the 2d conformal blocks~\cite{Hogervorst:2016hal}, or equivalently use an $\SL(2,\mathbb{R})$ expansion~\cite{Simmons-Duffin:2016wlq}. These two methods are equivalent since the 2d blocks are a simple combination of 1d, or $\SL(2,\mathbb{R})$, blocks.

The conformal blocks in any dimension can be expanded as
\bea[eq:SL2R]
g_{h,\hb}^{r,s}(z,\zb) ={}&{}\sum\limits_{n=0}^\infty\sum\limits_{j=-n}^n A_{n,j}^{r,s}(h,\hb)z^{h+n} k^{r,s}_{2(\bar{h}+j)}(\bar{z})\;,\label{eq:SL2R - expansion}
\\
k^{r,s}_{2\bar{h}}(\bar{z})\coloneqq{}&{}\zb^{\hb}\pFq{2}{1}{\hb-r,\hb+s}{2\hb}{\zb}\;,\label{eq:SL2R - definition of k}
\eea 
where $k^{r,s}_{2\bar{h}}(z)$ is the $\SL(2,\mathbb{R})$ block. Using the decomposition mentioned above, or by solving the Casimir differential equation, the first two levels are straightforward to work out and are given by
\bea 
A^{r,s}_{0,0}(h,\hb)&=1\;,\\
A^{r,s}_{1,-1}(h,\hb)&=\frac{h-\bar{h}}{2 h-2 \bar{h}+1}\;,\\
A^{r,s}_{1,0}(h,\hb)&=\frac{1}{2} \left(\frac{r s \left(h-2 \bar{h}^2+2 \bar{h}-1\right)}{(2 h-1) (\bar{h}-1) \bar{h}}-h-r+s\right)\;,\\
A^{r,s}_{1,1}(h,\hb)&=\frac{(h+\bar{h}-1) (\bar{h}-r) (\bar{h}+r) (\bar{h}-s) (\bar{h}+s)}{4 \bar{h}^2 (2 \bar{h}-1) (2 \bar{h}+1) (2 h+2 \bar{h}-1)}\;,
\eea 
for $d=3$. By using such an expansion one can solve \equref{eq: scalar crossing equation} order by order in $z$.

\chapter{Fermions in Conformal Field Theories: A Perturbative Approach}
\label{chapter: perturbative}
\section{Fermions in $d=3$}
There are important and relevant $3d$ CFTs with fermionic operators such as the $3d$ supersymmetric Ising and \ac{GNY}s, and they are invoked to explain several phenomena such as the phase transitions in graphene and the time-reversal symmetry breaking in d-wave superconductors. We refer to \cite{Poland:2018epd} for details; we'll only briefly review \ac{GNY} below as an example.

The \ac{GNY} can be described by the action \cite{Moshe:2003xn,Hasenfratz:1991it,ZinnJustin:1991yn}
\be 
S=\frac{1}{2}\int d^dx \left(\bar\psi_i(\slashed{\partial}+g\f)\ac{psi}^i+\partial_\mu\f\partial^\mu\f+m^2\f^2+\lambda\f^4\right)
\ee 
for the flavor index $i=1,\dots,N$. In $d=4-\e$, one can show by working perturbatively at large $N$ that this theory has an IR fixed point which coincides with the UV fixed point of the \ac{GN} \cite{Karkkainen:1993ef,Focht:1995ie}. This is similar to the familiar relation between the Wilson- Fisher fixed points of the $\f^4$ theory and the UV fixed points of the non-linear sigma model in $d=2+\e$ \cite{Giombi:2016ejx}. The critical point is believed to survive up to $\e=1$ (hence $d=3$), where large $N$ studies have been carried out \cite{Gracey:1992cp,Derkachov:1993uw,Gracey:1993kc}. Reference \cite{Petkou:1996np} furthermore studies the conformal OPE expansion alongside the large $N$ expansion of this model.

The numerical bootstrap has addressed fermions in $d=3$ --- along with the contact to GNY models --- in \cite{Iliesiu:2015qra} and \cite{Iliesiu:2017nrv}: we refer the reader to those papers for further details.

\subsection{Notations \& Conventions}
\label{sec:EmbeddingReview}

In this thesis, we follow the same conventions with \cite{Albayrak:2019gnz} for $3d$ fermions, so we will be brief here.  We use the Minkowski metric in mostly plus signature $\eta_{\mu\nu}=\text{diag}(-1,1,1)$, and as our focus are fermions we will be considering the double cover of the $3d$ conformal group $\SO(3,2)$, which is isomorphic to $\Sp(2,\mathbb{R})$. We will hence use representations of the symplectic group; in particular, the smallest fundamental representation will act on a real two dimensional vector space, describing Majorana fermions.

The conformal group $\SO(3,2)$ can be realized linearly in two  higher dimensions  as we reviewed in \secref{\ref{sec: preliminaries of conformal group}}, therefore we embed $\Sp(2,\mathbb{R})$ representations as the projective null representations of $\Sp(4,\R)$ in this 5d embedding space, use the fact that conformal symmetry acts as linear isometries in this formalism, and project back to physical structures by fixing the extra degrees of freedom in the embedding space.

In practice, one removes the extra degrees of freedom of the embedding space by going to the so-called Poincar\'e section,
\be 
X^A \rightarrow (x^\mu,1,x^2)\;,
\ee 
where we are working in the lightcone coordinates $X^A=(X^\mu,X^+,X^-)$ and $X^\pm$ are related to the Cartesian coordinates as $X^\pm=X^4\pm X^3$. One can reverse this projection and lift any point to the embedding space in this Poincar\'e section.\footnote{The only exception is the point $\infty$: one cannot use the same Poincar\'e section $(x^\mu,1,x^2)$ used for finite points for the point at infinity as well. A simple way to see this is as follows. We start with $X^a=(x^\mu,X^+,X^-)$ and impose nullness to obtain $X^A=(x^\mu,X^+,x^2/X^+)$. If we now consider an inversion as $x^\mu_R=x^\mu/x^2$, we see that $X^A_R=(x^\mu/x^2,X^+,(x^2X^+)^{-1})\sim(x^\mu,x^2X^+,1/X^+)$ where we use projectiveness of the representation. For $x^2\ne 0$, we can choose $X^+=1/x^2$ for $X_R^A$ and $X^+=1$ for $X^A$, which means we can write both $x^\mu$ and $x_R^\mu$ using the same Poincar\'e section $(x^\mu,1,x^2)$. For $x^2=0$, this is no longer possible, and the reflected point $x_R^\mu=x^\mu/x^2=\infty$ should be instead in the Poincar\'e section $(x^\mu,x^2,1)$.}

As developed in \cite{Costa:2011mg} and generalized to 3d spinors in \cite{Iliesiu:2015qra}, one can encode spinors by polynomials with auxiliary spinor fields:
\be \cO(x,s)={}&s_{\a_1}s_{\a_2}\dots s_{\a_{2l}} \cO^{\a_1\a_2\dots \a _{2l}}(x)\;,\\
\cO^{\a_1\a_2\dots \a _{2l}}(x)={}&\frac{1}{(2l)!}\frac{\partial^{2l}}{\partial_{s_{a_1}}\partial_{s_{a_2}}\dots \partial_{s_{a_{2l}}}}\cO(x,s)\;.
\label{eq: spinor convention}
\ee 
In this formalism, we can relate the spinor field $\Psi(X,S)\coloneqq S_I \Psi^I(X)$ in the embedding space and the spinor field  $\psi(x,s)\coloneqq s_\alpha \psi^\alpha(x)$ in the physical space as
\begin{equation}
	\Psi(X,S)=\frac{1}{(X^+)^{\Delta_\psi}}\psi(x,s)\;,
\end{equation}
where $S_I$ is taken to the Poincar\'e section via setting
\begin{equation}
	S_I=\sqrt{X^+}\begin{pmatrix}
		s_\alpha\\ x^\alpha_{\;\;\beta}s^\beta
	\end{pmatrix}\;.
\end{equation}
Here we use the matrices $\gamma$ and $\Gamma$ to convert the indices,
\begin{equation}
	X^I_{\;\;J}\equiv X^A(\Gamma_A)^I_{\;\;J}\;,\qquad x^\alpha_{\;\;\beta}\equiv x^\mu(\gamma_\mu)^\alpha_{\;\;\beta}\;.
\end{equation}
The $3d$ gamma matrices and $\Sp(2,\mathbb{R})$ invariant $\e$ are defined as
\be 
(\gamma_0)^\a _{\;\;\b}=i(\sigma_2)_{\a \b}\;,\quad
(\gamma_1)^\a _{\;\;\b}=(\sigma_1)_{\a \b}\;,\quad
(\gamma_2)^\a _{\;\;\b}=(\sigma_3)_{\a \b}\;,\quad
\e_{\a\b}=\e ^{\a \b}=i(\sigma_2)_{\a \b}\;,
\ee 
where $\sigma_i$ are the standard Pauli matrices:
\bea 
\sigma_1=\left(
\begin{array}{cc}
	0 & 1 \\
	1 & 0 \\
\end{array}
\right)
\quad,\quad\sigma_2=\left(
\begin{array}{cc}
	0 & -i \\
	i & 0 \\
\end{array}
\right)
\quad,\quad\sigma_3=\left(
\begin{array}{cc}
	1 & 0 \\
	0 & -1 \\
\end{array}
\right).
\eea 
The embedding space gamma matrices and $\Sp(4,\mathbb{R})$ invariant $\Omega$ are then defined as:\footnote{We note that  gamma matrices are defined such that the generators of $\Sp(2,\mathbb{R})$ and $\Sp(4,\mathbb{R})$ are given as $-\frac{i}{4}[\gamma^\mu,\gamma^\nu]$ and $ -\frac{i}{4}[\Gamma^A,\Gamma^B]$, respectively.}
\be 
\Gamma_0=\gamma_2\otimes\gamma_0
\;,\quad
\Gamma_1=\mathbbm{I}\otimes\gamma_1
\;,\quad
\Gamma_2=\mathbbm{I}\otimes\gamma_2
\;,\quad
\Gamma_3=\gamma_0\otimes\gamma_0
\;,\quad
\Gamma_4=\gamma_1\otimes\gamma_0
\;,\quad
\Omega=\epsilon\otimes\mathbbm{I}
\ee 
with the embedding space metric $g_{IJ}=\text{diag}\left(-,+,+,+,-\right)$. In lightcone coordinates, the gamma matrices take the form
\be 
(\G^\mu)^I_{\;\;J}=\begin{pmatrix}
	\left(\gamma^\mu\right)^{\a}_{\;\;\b} & 0 \\
	0 & \left(\gamma^\mu\right)_{\a}^{\;\;\b}
\end{pmatrix}\;,\quad 
(\G^+)^I_{\;\;J}=\begin{pmatrix}
	0 & 2\e^{\a\b}\\ 0 & 0
\end{pmatrix}\;,\quad 
(\G^-)^I_{\;\;J}=\begin{pmatrix}
	0 & 0\\
	2\e_{\a\b} & 0
\end{pmatrix}.
\ee 
Finally, we raise (lower) the spinor indices with $\mathrm{SP}$ invariants by acting from right (left), e.g. $x_\a=\epsilon_{\a\b}x^\b$ and $x^\a=x_\b\epsilon^{\b\a}$. 

To describe the embedding space spinor structures, we define the shorthand notation
\be
\<S_1X_2X_3\dots X_{n-1}S_n\>\equiv&-S_1\cdot X_2\cdot X_3\cdots X_{n-1}\cdot S_n\\=& -(S_{1})_I(X_2)^I_{\;\;J}(X_3)^J_{\;\;K}\cdots(X_{n-1})^L_{\;\;M}(S_n)^M.
\ee
In this convention, we diagonalize the spectrum and normalize the operators $\cO^{\De,l}$ such that they satisfy
\be
\<\cO^{\De,l}(X_1,S_1)\cO^{\De,l}(X_2,S_2)\>= i^{2l}\frac{\<S_1S_2\>^{2l}}{X_{12}^{2\De+2l}}. \label{eq: 2pt convention}
\ee 
For example, two-point functions of  fermions in embedding and physical spaces read as
\begin{equation}
	\<\Psi(X_1,S_1)\Psi(X_2,S_2)\>=i\frac{\<S_1S_2\>}{X_{12}^{\Delta_\psi+\frac{1}{2}}}\;,\qquad \<\psi^\alpha(x_1)\psi_\beta(x_2)\>=i\frac{(x_{12})^\alpha_{\;\;\beta}}{x_{12}^{2\Delta_\psi+1}}\;.
\end{equation}

Conformal symmetry also fixes the form of three point correlators albeit non-uniquely unless two of the operators are scalars. Therefore, in general, one writes down there point correlator $\<\cO_1\cO_2\cO_3\>$ in terms of three point structures $\<\cO_1\cO_2\cO_3\>^{(i)}$ as 
\be 
\label{eq: definition of OPE coefficients}
\<\cO_1\cO_2\cO_3\>=\sum\limits_i \lambda^{i}_{\cO_1\cO_2\cO_3}\<\cO_1\cO_2\cO_3\>^{(i)}\;,
\ee 
where $\lambda$ are the OPE coefficients.  We stress that $\<\cO_1\cO_2\cO_3\>^{(i)}$ are \emph{not} physical three point correlator; in particular, they are formal expressions and do not necessarily satisfy Fermi-statistics if operators are interchanged. Throughout this paper, we will refer to these objects as three-point structures, for which we will follows  the conventions of \cite{Iliesiu:2015qra,Iliesiu:2015akf} and choose three point structures as:
\bea[eq: 3pt structures]
\<\phi_1\phi_2\cO_3\>^{(1)}&=\frac{\<S_3X_1X_2S_3\>^l}{X_{12}^{(\De_{123}+l)/2}X_{23}^{(\De_{231}+l)/2}X_{31}^{(\De_{312}+l)/2}},\label{eq: 3pt structure scalar scalar spin l}
\\
\<\psi_1\phi_2\cO_3\>^{(1)}&=\frac{\<S_1S_3\>\<	S_3X_1X_2S_3\>^{l-\frac{1}{2}}}{X_{12}^{(\De_{123}+l-\half)/2}X_{23}^{(\De_{231}+l-\half)/2}X_{31}^{(\De_{312}+l+\half)/2}},
\\
\<\psi_1\phi_2\cO_3\>^{(2)}&=\frac{\<S_1X_2S_3\>\<	S_3X_1X_2S_3\>^{l-\frac{1}{2}}}{X_{12}^{(\De_{123}+l+\half)/2}X_{23}^{(\De_{231}+l+\half)/2}X_{31}^{(\De_{312}+l-\half)/2}},
\\
\<\psi_1\psi_2\cO_3\>^{(1)}&=
\frac{\<S_1S_2\>\<	S_3X_1X_2S_3\>^{l}}{X_{12}^{(\De_{123}+l+1)/2}X_{23}^{(\De_{231}+l)/2}X_{31}^{(\De_{312}+l)/2}},
\\
\<\psi_1\psi_2\cO_3\>^{(2)}&=
\frac{\<S_1S_3\>\<S_2S_3\>\<	S_3X_1X_2S_3\>^{l-1}}{X_{12}^{(\De_{123}+l-1)/2}X_{23}^{(\De_{231}+l)/2}X_{31}^{(\De_{312}+l)/2}},
\\
\<\psi_1\psi_2\cO_3\>^{(3)}&=
\frac{\left(X_{23}\<S_1S_3\>\<S_2X_1S_3\>+X_{13}\<S_2S_3\>\<S_1X_2S_3\>\right)\<	S_3X_1X_2S_3\>^{l-1}}{X_{12}^{(\De_{123}+l-1)/2}X_{23}^{(\De_{231}+l)/2}X_{31}^{(\De_{312}+l)/2}},
\\
\<\psi_1\psi_2\cO_3\>^{(4)}&=
\frac{\left(X_{23}\<S_1S_3\>\<S_2X_1S_3\>-X_{13}\<S_2S_3\>\<S_1X_2S_3\>\right)\<	S_3X_1X_2S_3\>^{l-1}}{X_{12}^{(\De_{123}+l-1)/2}X_{23}^{(\De_{231}+l)/2}X_{31}^{(\De_{312}+l)/2}},
\eea 
where we define for brevity
\begin{subequations}
	\be 
	X_{ab}\coloneqq-2X_a\. X_b\;,
	\ee 
	and 
	\be 
	\label{eq: definition of delta with three indices}
	\ac{Delta3}\coloneqq\De_a+\De_b-\De_c\;.
	\ee 
\end{subequations}

For integer spin we can also convert to vector notation,
\be 
\label{eq: spinor-vector transition}
\cO^{\a_1\dots \a _{2J}}=&\cO^{\mu_1\dots\mu_J}\gamma_{\mu_1}^{\a _1 \a _2}\cdots \gamma_{\mu_J}^{\a _{2J-1} \a _{2J}},
\\
\cO^{\mu_1\dots\mu_J}=&\left(-\half\right)^J\gamma^{\mu_1}_{\a _1 \a _2}\cdots \gamma^{\mu_J}_{\a _{2J-1} \a _{2J}}\cO^{\a_1\dots \a _{2J}},
\ee 
where $\cO^{\mu_1\dots\mu_J}$ is a symmetric traceless tensor. If we introduce the auxiliary polarization vectors $z_\mu$, and insist that
\be 
z_{\mu_1}\dots z_{\mu_l}\cO^{\mu_1\dots\mu_l}=s_{\a_1}\dots s_{\a_{2l}}\cO^{\a_1\dots\a_{2l}}
\ee 
holds, we find the following relation:
\be 
z_\mu=s_\a s_\b \gamma^{\a\b}.
\ee 
We can now use these relations to convert two and three-point functions to vector notation for integer spin.  In particular, \equref{eq: 2pt convention} and \equref{eq: 3pt structure scalar scalar spin l} become
\bea
\<\cO^{\De,l}(X_1,Z_1)\cO^{\De,l}(X_2,Z_2)\>=&\frac{1}{2^l}\frac{H_{12}^l}{X_{12}^{\De_l}},\\
\<\f(X_1,Z_1)\f(X_2,Z_2)\cO(X_3,Z_3)\>=&\frac{V_3^l}{X_{12}^{(\De_{123}-l)/2}X_{23}^{(\De_{231}+l)/2}X_{31}^{(\De_{312}+l)/2}},
\eea 
where we define
\be 
H_{12}\equiv{} &{}-2\left[(Z_1\.Z_2)(X_1\.X_2)-(X_1\.Z_2)(Z_1\.X_2)\right],\\
V_{3}\equiv {}& {}\frac{(Z_3\.X_1)(X_2\.X_3)-(Z_3\.X_2)(X_1\.X_3)}{X_1\.X_2},
\ee 
in the conventions of \cite{Costa:2011mg}.

Associated to every representation $\cO$ is a shadow representation $\widetilde{\cO}$ which has dimension $\tl\De$ and the same $\SO(d)$ representation $\rho$\footnote{In even dimensions we actually need the reflected representation $\rho^{R}$, but in odd dimensions the two are equivalent so to simplify the discussion we will ignore the distinction.} where we define
\be 
\tl\De=d-\De
\ee 
for brevity. Then there exists a conformally-invariant pairing:
\be 
\label{eq: conformally invariant pairing}
\left(\tl\cO^\dagger,\cO\right)=\int d^dx\tl\cO^\dagger_{\alpha_1\cdots\alpha_{2J}}(x) \cO^{\alpha_1\cdots\alpha_{2J}}(x),
\ee 
where $\tl \cO^{\dagger}$ has dimension $\widetilde{\Delta}$. We will work with real operators, so we can drop the $\dagger$ in the expressions which follow.

In the rest of the thesis we will implicitly contract indices when an operator and its shadow are being integrated over, we will always use spinor indices, and suppressed spinor indices always go from southwest to northeast. In particular, this means that order of correlators in an expression matters in our conventions, for instance
\be 
\int dx  \<\cO_1 \cO(x) \cO_2 \>\<\cO_3 \tl \cO(x) \cO_4 \>  =(-1)^{2l_{\cO}}
\int dx  \<\cO_3 \tl \cO(x) \cO_4 \>\<\cO_1 \cO(x) \cO_2 \>\ .
\label{eq: pairing convention} 
\ee

By using the conformally-invariant pairing in \equref{eq: conformally invariant pairing}, we can define \emph{shadow transformation} which maps a representation $\cO$ to $\widetilde{\cO}$:
\be 
\label{eq: shadow definition}
\mathbf{S}[\cO](x)\equiv \int dy \cO(y)\<\tl\cO(y)\tl\cO(x)\>.
\ee 
We will adopt the convention that the shadow transform always acts by multiplying the two-point function from the left. The distinction is immaterial for bosonic correlators, but when we introduce fermions the ordering does matter. 

\subsection{Double twist sector of $3d$ fermions}
It is shown in \cite{Fitzpatrick:2012yx} that every CFT has a double twist sector which asymptottes to \ac{MFT} operators at large spin. This spectrum is important for various reasons; for example, it is an open question whether several relevant operators such as stress tensor lie in these trajectories. Likewise, these operators correspond to two particle states in the bulk via the AdS/CFT correspondence. Finally, double twist operators are the relevant spectrum in a channel to reproduce low lying operators such as  exchange of the identity operator and stress tensor in the crossed channel in the light-cone limit.

In the free theory limit, double twist operators can be written down as composite operators of the form
\be 
\label{eq: double twist definition}
:A_{\a_1\dots \a_p}\partial_{\b_{11}\b_{12}}\cdots\partial_{\b_{q1}\b_{q2}}(\partial^2)^rB_{\rho_1\dots \rho_s}:-\text{ total derivatives}\;,
\ee 
where total derivative parts are extracted from the normal-ordered product to ensure that the resultant operator is a primary. In addition, the open indices above are symmetrized or antisymmetrized (contracted) dependent on the representation.

We group such operators into towers which has the same twist $\tau=\De-l$. In particular, we observe that inclusion of a partial derivative with open indices does not change the twist hence we define and denote the double twist towers as
\be 
{}[AB]_n\xrightarrow{\text{free theory limit}} \{A(\partial^2)^nB,\;A\partial_{\a_1\b_1}(\partial^2)^nB,\;A\partial_{\a_1\b_1}\partial_{\a_2\b_2}(\partial^2)^nB,\;\cdots \}\;,
\ee 
where $A$ and $B$ may have open or contracted indices. In a free theory, the double twist tower $[AB]_n$ is a collection of composite operators schematically given in the equation above\footnote{We emphasize that the expression $A(\partial^2)^nB$ and so on are schematic as the correct form should be normal-ordered with total derivative pieces removed as defined in \equref{eq: double twist definition}.} and all these operators have the same twist $\tau_{[AB]_n}=\tau_A+\tau_B+2n$. Once we move away from the free theory, we can no longer describe double twist operators with such simple composite structures; nevertheless, they are still in the spectrum and they satisfy
\be 
\lim\limits_{l\rightarrow\infty}\tau_{[AB]_n}=\tau_A+\tau_B+2n\;,
\ee 
hence we can define a double twist tower as a collection of operators with the same twist accumulation point as spin goes to infinity. At finite spin, we have
\be 
\tau_{[AB]_n}=\tau_A+\tau_B+2n+\gamma_{[AB]_n}(n,l)\;,
\ee 
where $\gamma$ is the \emph{anomalous dimension}. One of the successes of the lightcone bootstrap was to show that for any CFT\footnote{We restrict ourselves to local CFTs with twist gap. In $d=3$, any unitary CFT does have the required twist gap.}
\be 
\lim\limits_{l\rightarrow\infty} l^{\tau_0}\gamma_{[AB]_n}(n,l)=\text{ constant}\;,
\ee 
where $\tau_0$ is the twist of lowest twist non-identy operator in the spectrum of the crossed channel, which is usually the stress tensor. For example, we expect a double twist family
\be 
{}[\psi_\a\psi^\a]_0\xrightarrow{\text{free theory limit}} \{\psi_\a\psi^\a ,\;\psi_\a\partial_{\b_{11}\b_{12}}\psi^\a,\;\psi_\a\partial_{\b_{11}\b_{12}}\partial_{\b_{21}\b_{22}}\psi^\a,\;\cdots  \}
\ee 
in the spectrum of fermionic CFTs: one aim of the conformal bootstrap is to find an analytic form for the anomalous dimension of this tower.

Let's now step back and briefly discuss what kind of representations we expect for double twist operators for $3d$ CFTs. We know that the irreducible representations of the $\mathrm{so}(3)$ algebra are labeled by one index $j\in\half \Z_{\ge 0}$, where tensor products of two irreps have the well known reduction
\be 
j_1\otimes j_2 =(j_{1}+j_2)\oplus (j_1+j_2-1)\oplus (j_1+j_2-2)\oplus \cdots \oplus \abs{j_1-j_2}\;.
\ee
For example, $\half \otimes \half = 1\oplus 0$ indicates that we have both spin-1 and spin-0 operators constructed out of two fermions, i.e. $\psi_{\{\a}\psi_{\b\}}$ and $\psi_{[\a}\psi_{\b]}=-\half\e_{\a\b}(\psi_\sigma\psi^\sigma)$. In what follows, we will assume that all open $\Sp(2,\R)$ indices are symmetrized as any antisymmetric pair can be rewritten in terms of contracted indices and $\e_{\a\b}$. So, we say that $\half \otimes \half = 1\oplus 0$ indicates that $\psi_\a\x\psi_\b$ can be decomposed into $\psi_\a\psi_\b$ and $\psi_\a\psi^\a$.

In some cases, we are interested in CFTs with parity symmetry where the relevant algebra is enlarged from $\so(3)$ to $\mathrm{o}(3)$ which are labeled by the previous index $j$ and a sign $p=\pm 1$. We will denote the representation as $j^p$, where $p$ indicates if the corresponding irrep is parity even or parity odd. The decomposition of products of two irreps now becomes
\be 
j_1^+\otimes j_2^+ =(j_{1}+j_2)^+\oplus (j_1+j_2-1)^-\oplus (j_1+j_2-2)^+\oplus \cdots \oplus \abs{j_1-j_2}^{+|-}\;,
\ee
where parity alternates between different irreps. Likewise, if some of the parities are flipped on left hand side, so are those in the right hand side, e.g. $j_1^+\otimes j_2^- =(j_{1}+j_2)^-\oplus \cdots$.

We now turn to the bispinors we can construct out of $\psi^\alpha$ and $\psi^\beta$.  The representation of the Majorana fermion can be taken as either $\half^+$ or $\half^-$: this ambiguity reflects the fact that it is not the $\mathrm{O}(2,1)$ but the $\mathrm{Pin}(2,1)$ group which acts on the spinors. Nevertheless, as we will always be interested in composite operators or correlators with even number of fermions, our results will be ambiguity-free. Let us choose that $\psi_\a$ transforms under the $\mathrm{Pin}(2,1)$ with the representation $\half^+$. We then have the decomposition
\be 
\half^+\otimes\half^+=1^+\oplus 0^-\;,
\ee 
which means we can construct a parity even vector and a parity odd scalar out of two $\psi$.\footnote{One can also check this explicitly. If we choose parity transformation as a reflection \mbox{$x^1\rightarrow -x^1$} and \mbox{$x^{0,2}\rightarrow x^{0,2}$}, the Majorana fermion transforms as \mbox{$\psi^\a\rightarrow\pm(\gamma^1\psi)^\a$} where the ambiguity is the same one discussed above. By using \mbox{$\psi_\a=\e_{\a\b}\psi^\b$}, we can show that \mbox{$\psi_\a\rightarrow\mp(\psi\gamma^1)_\a$}, hence \mbox{$\psi_\a\psi^\a\rightarrow -(\psi\gamma^1)_\a(\gamma^1\psi)^\a=-\psi_\a\psi^\a$} indicating that $\psi_\a\psi^\a$ is a parity-odd scalar. One can similarly check the parities of other operators.} These are exactly $\psi_\a \psi_\b$ and $\psi_\a \psi^\a$, which should then appear in the double twist towers $[\psi_\a\psi_\b]_n$ and $[\psi_\a\psi^\a]_n$ .

We can now  include a partial derivative, which is parity even spin$-1$ object. That means
\be
\frac{1}{2}^+\otimes\frac{1}{2}^+\otimes 1^+=2^+\oplus 1_2^-\oplus 0^+\;,
\ee
where $a$ denotes the multiplicity of the representation $j$ in the notation $j_a$. These objects are
\be
\psi_\a \partial_{\g\l}\psi_\b\;,\quad \psi_\a \partial_{\b\g}\psi^\a \;,\quad \psi_\a \partial_{\b\g}\psi^\g \;,\quad \psi_\a \partial^{\a}_{\;\;\b}\psi^\b\;.
\ee  
First two objects already appear in $[\psi_\a\psi_\b]_n$ and $[\psi_\a\psi^\a]_n$ double twist towers. We are left to conclude that inclusion of a derivative yields only two new towers: a parity odd tower $[\psi_a(\partial\psi)_\b]_n$ and a parity even tower $[\psi_a(\partial\psi)^\a]_n$.

Considering more derivatives will not produce new towers as we expect a total of four different double twist towers. To see that, we first note that out of the irreducibly representations in the decomposition of p-derivatives, i.e. $1^+\otimes 1^+\cdots \otimes 1^+$, only the fully symmetric representation is relevant as contracted derivatives can be rewritten in terms of $(\partial^2)$ and fewer number of derivatives with open indices due to the identity $\partial^\a_{\;\g}\partial^\g_{\;\b}=(\partial^2)\de^\a_\b$. Hence, the independent towers  in the decomposition of two fermions with $p-$derivatives  appear in $\half^+\otimes\half^+\otimes p^+\in \half^+\otimes\half^+\otimes 1^+\otimes 1^+\cdots \otimes 1^+$ which is decomposed as 
\be 
\half^+\otimes\half^+\otimes p^+=(p+1)^+\oplus p_2^-\oplus (p-1)^+\;,
\ee 
which are the towers
\be 
{}[\psi_\a\psi_\b]_n\;,\quad [\psi_\a\psi^\a]_n\;,\quad [\psi_\a(\partial\psi)_\b]_n\;,\quad [\psi_\a(\partial\psi)^\a]_n\;.
\ee 

With the double twist towers constructed as above, we finally turn to the \emph{selection rules} on spin imposed by the fact that double twist operators satisfy certain symmetries as they are defined modulo total derivatives. For example, via integration by parts we see that
\be 
\psi_\a \partial_{\mu_1}\dots \partial_{\mu_l} \partial^{2n} \psi_b
=
(-1)^{l+1}\psi_\b \partial_{\mu_1}\dots \partial_{\mu_l} \partial^{2n} \psi_\a + \text{ total derivatives}\;,
\ee 
hence  $\psi_{(\a} \partial_{\mu_1}\dots \partial_{\mu_p} \partial^{2n} \psi_{b)}$ is nonzero only if $p+1\in2\Z^+$. This then indicates that the tower $[\psi_\a\psi_\b]_{n}$ has operators of even spin only. By a similar analysis, we show that only odd/even spin operators contribute to each double twist tower: we summarize them in \tabref{\ref{table:DoubleTwistFamilies}}.

\begin{table}
	\caption[Double-twist families for Majorana fermions in 3d]{\label{table:DoubleTwistFamilies}We list the different double-twist families for Majorana fermions in 3d, with their twist accumulation points, parities, and spins.}
	\begin{tabularx}{\textwidth}{ XXXX }
		\hline\hline
		\textbf{Family} & \textbf{Twist} & \textbf{Parity} & \textbf{Spin}\\[.01in]
		\hline
		$[\psi_{(\alpha}\psi_{\beta)}]_{n,l}$&$2\Delta_\psi+2n-1$&Even& $l\ge 2$, Even
		\\[.01in]	
		$[\psi_\alpha\psi^\alpha]_{n,l}$&$2\Delta_\psi+2n$&Odd
		& $l\ge 0$, Even
		\\[.01in]
		$[\psi_{(\rho}\partial^\rho_{\;\a}\psi_{\beta)}]_{n,l}$&$2\Delta_\psi+2n$&Odd &
		$l\ge 1$, Odd
		\\[.01in]
		$[\psi_\alpha(\partial\psi)^\alpha]_{n,l}$&$2\Delta_\psi+2n+1$&Even
		&
		$l\ge 0$, Even
		\\\hline\hline
	\end{tabularx}
\end{table}

\section{Large spin expansion}
\subsection{Lightcone bootstrap}
\subsubsection{Short review} 
\label{sec:LightconeBootstrapReview}

In the following calculations, we review the work of \cite{Simmons-Duffin:2016wlq} which solves the lightcone bootstrap in a perturbative expansion. We start by considering the 4-point function of identical scalars, $\<\f\f\f\f\>$. The 4-point function is invariant under $1\leftrightarrow3$ (or $s\leftrightarrow t$) crossing, which implies
\begin{equation}
	\left(\frac{(1-z)(1-\bar{z})}{z\zb}\right)^{2h_{\f}}\sum_{ \cO}P_{\f\f \cO}g_{h_ \cO ,\hb_ \cO}(z,\zb)=\sum_{ \cO}P_{\f\f \cO}g_{h_ \cO ,\hb_ \cO}(1-\zb,1-z)\;, \label{eq:4sCross}
\end{equation}
where  $P_{\f\f  \cO}=f_{\f\f\O}^{2}$.

We can now consider this equation in the lightcone limit $z\ll1-\bar{z}\ll1$. In the limit $z\ll1$, the identity operator dominates on the left-hand side, while taking $1-\bar{z}\ll1$ allows us to use the $\SL(2,\mathbb{R})$ expansion on the right-hand side. In this limit, the crossing equation becomes
\begin{equation}
	\left(\frac{1-\bar{z}}{z}\right)^{2h_{\f}}\approx\sum_{ \cO}P_{\f\f \cO}(1-\bar{z})^{h_{ \cO}}k_{2\hb_ \cO}(1-z)\;.
\end{equation}

By the arguments of \cite{Fitzpatrick:2012yx,Komargodski:2012ek}, in order to match the $z\rightarrow0$ divergence on the left-hand side, which is not present in any individual $t$-channel block, we need to sum over operators with unbounded spin on the right-hand side. Specifically, we need a tower of ``double-twist'' operators in the $t$-channel, $[\f\f]_{0,\ell}$, such that $h_{0,\ell}\rightarrow 2h_{\f}$ as $\ell\rightarrow\infty$. 

At this point we could use a Bessel function approximation of the blocks to derive the large-$\ell$ asymptotics of the OPE coefficients, but it will be more useful to use our knowledge of $1d$ generalized free field theories to write down the exact sum \cite{Simmons-Duffin:2016wlq}
\begin{subequations}
	\label{eq:SL2R_Sum}
	\be 
	\sum_{\substack{\bar{h}=-a+l \\ \ell=0,1,...}}S_{a}(\bar{h})k_{2\bar{h}}(1-z)=\left(\frac{z}{1-z}\right)^{a}\;,
	\ee 
	where
	\be 
	\label{eq: definition of S_a}
	S_{a}(\bar{h})=\frac{\Gamma(\bar{h})^{2}\Gamma(\bar{h}-a-1)}{\Gamma(-a)^{2}\Gamma(2\bar{h}-1)\Gamma(\bar{h}+a+1)}\;.
	\ee 
\end{subequations}

From (\ref{eq:SL2R_Sum}), we now have at large $\ell$, where $\bar{h}\approx 2h_{\f}+\ell$, the following result for the OPE coefficients:
\begin{equation}
	P_{\f\f \cO}(\bar{h})\sim S_{-2h_{\f}}(\bar{h})\;.
\end{equation}
The ``$\sim$'' is because with this approach we can only find the asymptotic expansion for the OPE coefficients at large $\bar{h}$. Note that by expanding to higher orders in $1-\bar{z}$ one can also prove the existence of operators $[\f\f]_{n,\ell}$ which have $h_{n,\ell}\rightarrow 2h_{\f}+n$ as $\ell\rightarrow\infty$. 

To extend these calculations to higher orders in the large-$\bar{h}$ expansion, we can use the $\SL(2,\mathbb{R})$ expansion on both sides of the crossing equation (\ref{eq:4sCross}), expand in $z\ll1-\bar{z}\ll1$, and then use (\ref{eq:SL2R_Sum}) to unambiguously match generic powers of $z$ in the $s$-channel to the large-spin asymptotics of double-twist operators in the $t$-channel.

Of course, there are subtleties in this procedure which we have glossed over. First, for the arguments of \cite{Fitzpatrick:2012yx,Komargodski:2012ek} to work when matching a power of $z$ in the $s$-channel to an infinite sum of $t$-channel blocks, we need the $s$-channel term to be more divergent than any individual $t$-channel block. From (\ref{eq:4sCross}), we see this is only true if $h_{ \cO}<2h_{\f}$. However, as noted in \cite{Alday:2015ewa}, we can make any generic, individual power of $z^{a}$ on the left-hand side of (\ref{eq:4sCross}) as divergent as we like by repeatedly acting with a $\SL(2,\mathbb{R})$ Casimir differential operator:
\be
\mathcal{C}\equiv (1-z)^{2}z\partial_{z}^{2}+(1-z)^{2}\partial_{z}\;.
\ee
Since the $t$-channel $\SL(2,\mathbb{R})$ blocks are eigenfunctions of this Casimir, these differential operators leave the form of the $t$-channel expansion unchanged. Therefore, by acting with this differential operator sufficiently many times we can make the $s$-channel more divergent than the crossed channels.

In the terminology of \cite{Simmons-Duffin:2016wlq}, generic powers $z^{a}$ are ``Casimir-singular''. On the other hand, terms like $z^{n}$ and $z^{n}\log(z)$, with $n$ a non-negative integer, are called ``Casimir-regular''. If we repeatedly act with $\mathcal{C}$ on these terms, we eventually get $0$. Therefore, we cannot use the arguments of \cite{Fitzpatrick:2012yx,Komargodski:2012ek} to match these terms with large-spin asymptotics of double-twist operators and they are more sensitive to finite-spin effects.

In the study of the lightcone bootstrap, there are a few places where Casimir-regular terms can appear. The first is if we start the $\SL(2,\mathbb{R})$ sum (\ref{eq:SL2R_Sum}) at a generic point $\bar{h}_{0}$:
\begin{equation}
	\sum_{\substack{\bar{h}=\bar{h}_{0}+\ell \\ \ell=0,1,...}}S_{a}(\bar{h})k_{2\bar{h}}(1-z)=\left(\frac{z}{1-z}\right)^{a}+\mathcal{A}(\bar{h}_0)\;,
\end{equation}
where $\mathcal{A}(\bar{h}_0)$ is defined in \cite{Simmons-Duffin:2016wlq} (we will not need its explicit form). Since the choice of starting point only affects a finite number of blocks, changing the lower limit of the sum will not affect predictions for large-spin asymptotics. 

Another important issue is that our sums over blocks are not actually integer spaced. In general, the double-twist operators will get anomalous dimensions which also depend on the spin. Therefore, for a tower of double-twist operators $ \cO_{\ell}$ parametrized by the spin $\ell$ we have 
\begin{equation}
	\bar{h}_{ \cO_{\ell}}=2 h_{\phi}+\ell+\delta h_{ \cO_{\ell}}\;,
\end{equation}
where $\delta h_{ \cO_\ell}$ is half the anomalous dimension with respect to the generalized free field value.

To account for this effect, we can reparametrize our sum by inserting a Jacobian:
\begin{equation}
	\sum\limits_{\ell=0}^{\infty}\frac{\partial \bar{h}_{ \cO_\ell}}{\partial \ell} P(\bar{h}_{ \cO_\ell})k_{2\bar{h}_{ \cO_\ell}}(1-z)=\sum\limits_{\ell=0}^{\infty}P(2h_{\phi}+\ell)k_{4h_{\phi}+2\ell}(1-z)+\ldots\;,
\end{equation}
where the dropped terms are Casimir-regular in $z$. In the current discussion we will not be concerned with matching Casimir-regular terms, but will rather focus on how we can use the $\SL(2,\mathbb{R})$ expansion to match individual conformal blocks in the $s$-channel.

Thus, let us now consider the effect of single generic conformal block $g_{h_i,\bar{h}_i}(z,\bar{z})$ in the s-channel. In the limit $z\ll1-\bar{z}\ll1$ it has the general form:
\be
g_{h_i,\bar{h}_i}(z,\bar{z})=\left(\frac{z}{1-z}\right)^{h_i}\left(A_i \log(1-\bar{z})+B_i+\O(1-\bar{z})\right)+\ldots\;,
\ee
where we have only written the leading order terms. It is straightforward to include $\SL(2,\mathbb{R})$ descendants using the results of \cite{Simmons-Duffin:2016wlq,Hogervorst:2016hal}, and to expand to higher orders in $(1-\bar{z})$ using the explicit form of the hypergeometric functions, where such terms are needed to fix the corrections for higher twist towers, i.e. $[\phi\phi]_{n>0}$.\footnote{We will drop the label $\ell$ when referring to a given twist tower.} Here we will be primarily interested in the form of the correction for the leading twist $[\f\f]_{0}$ operators.

The crossing equation then becomes
\begin{multline}
	\left(\frac{z}{1-z}\right)^{-2h_\phi}+\sum_i \left(\frac{z}{1-z}\right)^{h_i-2h_\phi}(A_i\log(1-\bar{z})+B_i+\order{1-\bar{z}})\\=\sum\limits_{ \cO\in[\phi\phi]_0}P_{\phi\phi \cO}(1-\bar{z})^{h_ \cO-2h_\phi}k_{2\bar{h}_ \cO}(1-z)+\ldots\;.\label{eq:CrossingForFourScalars}
\end{multline}

To match the $\log$ terms we have to expand in the anomalous dimension, $h_{ \cO}=2h_{\f}+\delta h_ \cO$, and we find
\bea[5] 
P_{\phi\phi \cO}\sim{}&2\frac{\partial \bar{h}_ \cO}{\partial l}\left[S_{-2h_\phi}(\bar{h}_ \cO)+\sum_{i} B_iS_{h_i-2h_\phi}(\bar{h}_ \cO)\right]\;,\label{eq:asymptoticP}\\
\delta h_ \cO P_{\phi\phi \cO}\sim{}&2\frac{\partial \bar{h}_ \cO}{\partial l}\left[\sum_{i} A_iS_{h_i-2h_\phi}(\bar{h}_ \cO)\right]\;.
\eea 
The factors of $2$ are because we only sum over double-twist operators  of even spin in the $t$-channel. To find the asymptotic, large-spin behavior of the anomalous dimensions we then just need the ratio of these terms:
\begin{equation}
	\delta h_ \cO \sim \frac{\sum_{i} A_iS_{h_i-2h_\phi}(\bar{h}_ \cO)}{S_{-2h_\phi}(\bar{h}_ \cO)+\sum_{i} B_iS_{h_i-2h_\phi}(\bar{h}_ \cO)}\;.\label{eq:asymptoticdh}
\end{equation}
In order to compute $P_{\phi\phi \cO}$ we can then plug this into the relation $\frac{\partial \bar{h}_ \cO}{\partial l}=\left(1-\frac{\partial \delta h_ \cO}{\partial \bar{h}_ \cO}\right)^{-1}$.

\subsubsection{Coefficient Expansions}
\label{sec:ExpansionOfS}
In \equref{eq: definition of S_a}, we provided the MFT coefficient $S_a(\hb)$ for identical operators. Its generalization $S_a^{r,s}(\hb)$ are defined as
\be
\label{eq:ExpansionCoefficientSGeneralized}
S_a^{r,s}(\hb)\coloneqq\frac{1}{\Gamma(-a-r)\Gamma(-a-s)}\frac{\Gamma(\hb-r)\Gamma(\hb-s)}{\Gamma(2\hb-1)}\frac{\Gamma(\hb-a-1)}{\Gamma(\hb+a+1)}\;,
\ee 
which satisfy the one dimensional Mean Field Theory sum~\cite{Simmons-Duffin:2016wlq}:
\be 
\sum\limits_{\substack{\hb=l-a\\l=0,1,\dots}}S_a^{r,s}(\hb)(1-z)^{\hb} \pFq{2}{1}{\hb-r,\hb+s}{2\hb}{1-z}=\frac{z^{r+a}}{(1-z)^a}\;.
\ee	

$S_a^{r,s}(\hb)$ scales like $4^{-\hb}\hb^{-\frac{3}{2}-2a-r-s}$ at large $\hb$, meaning that we can expand $S_a^{r,s}(\hb)$ as
\be 
S_a^{r,s}(\hb)=\sum\limits_{k=0}^\infty c_{k,a}^{r,s,m,n} S_{a-\frac{m+n-k}{2}}^{r+m,s+n}(\hb)\;,
\ee
with $\hb$-independent coefficients $c_{k,a}^{r,s,m,n}$. Since we are working at next-to-next-to-leading order in $1/\hb$, we can truncate this expansion as
\be 
S_a^{r,s}(\hb)\approxeq c_{0,a}^{r,s,m,n} S_{a-\frac{m+n}{2}}^{r+m,s+n}(\hb)+c_{1,a}^{r,s,m,n} S_{a-\frac{m+n-1}{2}}^{r+m,s+n}(\hb)+c_{2,a}^{r,s,m,n} S_{a-\frac{m+n-2}{2}}^{r+m,s+n}(\hb)\;.
\ee

Armed with this, let us consider the following summation which appears repeatedly in \equref{eq:tChannelExpression1} after insertion of the ansatz in \equref{eq:OPEAnsatz}: 
\be
\sum_{l=0}^\infty\frac{\partial \hb}{\partial l}S_a^{0,0}(\hb)(1-z)^{\hb}\pFq{2}{1}{\hb-m,\hb+n}{2\hb}{1-z}\;.
\ee
For parity-even structures, $m,n=0$, hence we can immediately use \equref{eq:SomeOverNandL}. For parity-odd structures, we can expand $S^{0,0}$ in terms of $S^{m,n}$ and obtain
\begin{multline}
	\sum_{l=0}^\infty\frac{\partial \hb}{\partial l}S_a^{0,0}(\hb)(1-z)^{\hb}\pFq{2}{1}{\hb-m,\hb+n}{2\hb}{1-z}\\
	=z^{a+\frac{m-n}{2}} \left(c_{0,a}^{r,s}+c_{1,a}^{r,s}{\sqrt{z}}+\left(\frac{2a-m-n}{2}c_{0,a}^{r,s}+c_{2,a}^{r,s}\right)z+\order{z^{3/2}}\right)\;,
\end{multline}
where $c_{k,a}^{m,n}\equiv c_{k,a}^{0,0,m,n}$. This equation simply means that we replace factors of $\left(\frac{z}{1-z}\right)^a$ in \equref{eq:CrossedChannelFinalForm} with the expression above for parity-odd structures. 

We list below the coefficients $c_0$, $c_1$, and $c_2$ for the reader's convenience:
\be 
c_{k,a}^{r,s,m,n}=\kappa_k\frac{\Gamma \left(-a-\frac{m}{2}+\frac{n}{2}-r-\frac{k}{2}\right) \Gamma \left(-a+\frac{m}{2}-\frac{n}{2}-s-\frac{k}{2}\right)}{\Gamma (-a-r) \Gamma (-a-s)}\;,\quad k\in \{0,1,2\}\;,
\ee 
with
\bea
\kappa_0=&1\;,\\
\kappa_1=&-\frac{m(m+2r)+n(n+2s)}{2}\;,\\
\kappa_2=&\frac{1}{8}m^4+\frac{4r-1}{8}m^3+\frac{n(1+2n+4s)+4r(r-1)-2(a+1)}{8}m^2\nn\\
& +\frac{n^2(1+4r)+n(8rs-4)-4a(n-2)+4(a^2-r^2+1)}{8}m\nn\\
&
+\frac{n}{8}\left((n+2s)^2(n-1)-2n-2a(n-4)+4(a^2+1)\right)\;.
\eea

\subsection{Fermion conformal blocks and the crossing equations}
\label{sec:fermion}

Having reviewed lightcone bootstrap for scalars, we will now move on to 4-point functions of fermions $\<\psi\psi\psi\psi\>$ in 3d. Our focus for now will be to compute the first several perturbative $1/\bar{h}$ corrections to the $[\psi\psi]_0$ coefficients and anomalous dimensions, generalizing the method of section~\ref{sec:LightconeBootstrapReview} to fermions. As illustrated in~\cite{Simmons-Duffin:2016wlq}, such perturbative calculations are in many cases numerically sufficient and also quite useful in providing consistency checks on formulas obtained from the inversion formula approach. 

We will be using the method of \cite{Iliesiu:2015qra} to generate the fermion conformal blocks by hitting the four-scalar conformal block with differential operators in the embedding space.\footnote{A review of the embedding space formalism can be found in \secref{\ref{sec:EmbeddingReview}}.} Specifically, a contribution to $\<\psi\psi\psi\psi\>$ takes the form:
\\
\begin{multline}
	\left(\frac{X_{24}}{X_{14}}\right)^{\frac{\Delta_{12}}{2}}\left(\frac{X_{14}}{X_{13}}\right)^{\frac{\Delta_{34}}{2}}\frac{t_Ig^{I;a,b}_{h,\bar{h}}(z,\bar{z})}{X_{12}^{\frac{\Delta_1+\Delta_2+1}{2}}X_{34}^{\frac{\Delta_3+\Delta_4+1}{2}}}\\=\DD_a\widetilde{\DD}_b\left[
	\left(\frac{X_{24}}{X_{14}}\right)^{\frac{\Delta_{12}}{2}}\left(\frac{X_{14}}{X_{13}}\right)^{\frac{\Delta_{34}}{2}}\frac{g_{h,\bar{h}}(z,\bar{z})}{X_{12}^{\frac{\Delta_1+\Delta_2}{2}}X_{34}^{\frac{\Delta_3+\Delta_4}{2}}}
	\right]\;,\label{eq:crossingStandardForm}
\end{multline}
where $t_I$ are different 4-point structures in embedding space, $a,b$ are indices denoting possible 3-point structures, and we have suppressed the dependence of the blocks on the external dimensions. The operators $\DD_a$ are defined as follows:
\bea[eq:DifferentialOperators]
&\DD_1=\<S_1S_2\>\shift{1^+2^+}\;,\\
&\DD_2=-\<S_1\frac{\delta}{\delta X_1}\frac{\delta}{\delta X_2}S_2\>\shift{1^-2^-}\;,\\
&\DD_3=\<S_1\frac{\delta}{\delta X_1}S_2\>\shift{1^-2^+}-\<S_2\frac{\delta}{\delta X_2}S_1\>\shift{1^+2^-}\;,\\
&\DD_4=\<S_1\frac{\delta}{\delta X_1}S_2\>\shift{1^-2^+}+\<S_2\frac{\delta}{\delta X_2}S_1\>\shift{1^+2^-}\;,
\eea
where $\shift{a^i b^j}$ are operators which shift external dimensions such that
\be 
\shift{a^i b^j}: \{\Delta_a,\Delta_b\}\rightarrow \{\Delta_a+\frac{i}{2},\Delta_b+\frac{j}{2}\}\;.
\ee 
The corresponding operators acting on the points $(X_3,X_4)$ are found via the replacement
\be 
\widetilde{\DD}_a\coloneqq\DD_a\evaluated_{(1,2)\rightarrow(3,4)}\;.
\ee

An alternative basis denoted by $\mathcal{D}_{i}$ was used in \cite{Iliesiu:2015qra}, given by:\footnote{There is also an additional term in $\cD_4$ which vanishes for identical external operators.}
\bea[changebasis]
\cD_1=&\;\DD_1\;,\\
\cD_2=&\;\frac{1}{4(\hb-h)(\hb+h-1)}\left(\DD_2-(2h+2\Delta_{\psi}-4)(2h-2\Delta_{\psi}+1)\DD_1\right)\;,\\
\cD_3=&\;\frac{1}{2(\hb+h-1)}\DD_3\;,\\
\cD_4=&\;\frac{1}{2(\hb-h)}\DD_4\;.
\eea

These two bases have different merits. One nice feature of the $\DD_i$ basis is that the operators are independent of $(h,\bar{h})$, so there is a clean separation between the calculation of double-twist data and the differential operators, i.e. we do not want these operators to also depend on the anomalous dimensions. By comparison $\cD_i$ generates the most natural basis of embedding space, 3-point tensor structures. We find that it is most convenient to use the $\DD_i$ basis when performing the calculations and presenting the results.

The operators $\DD_1$ and $\DD_2$ generate parity-even structures, whereas $\DD_3$ and $\DD_4$ generate parity-odd ones. When the external fermions are identical, there are also selection rules on the spins of the exchanged operators: $\DD_{1}$, $\DD_{2}$, and $\DD_{3}$ are associated to operators of even spin while odd spins are associated with $\DD_4$.

Now let us write down the condition from crossing symmetry. For convenience we will define the prefactor $\mathsf{p}$ as
\be
\pre\left(\Delta_1,\Delta_2,\Delta_3,\Delta_4\right) \equiv 
\left(\frac{X_{24}}{X_{14}}\right)^{\frac{\Delta_{12}}{2}}\left(\frac{X_{14}}{X_{13}}\right)^{\frac{\Delta_{34}}{2}} X_{12}^{-\frac{\Delta_1+\Delta_2}{2}}X_{34}^{-\frac{\Delta_3+\Delta_4}{2}}\;,
\ee
and introduce the shorthand notations
\bea\pre_i &\equiv\pre\left(\Delta_1,\Delta_2,\Delta_3,\Delta_4\right)\;,\\
\pre_\psi &\equiv\pre\left(\Delta_\psi+\frac{1}{2},\Delta_\psi+\frac{1}{2},\Delta_\psi+\frac{1}{2},\Delta_\psi+\frac{1}{2}\right)\;.
\eea 
Then we can write a more compact form of the conformal block for identical fermions:
\be
t_Ig^{I;a,b}_{h,\bar{h}}(z,\bar{z})=\frac{\left(\DD_a\widetilde{\DD}_b\left[\pre_i\;
	g_{h,\bar{h}}(z,\bar{z})
	\right]\right)_{\Delta_i\rightarrow\Delta_\psi}}{\pre_\psi}\;,\label{eq:FermionConformalBlock}
\ee
and crossing symmetry implies
\be
\pre_\psi\sum\limits_ \cO t_I  P _ \cO^{a,b}g^{I;ab}_{h,\hb}(z,\zb)
=-\left(\pre_\psi t_I\right)\evaluated_{1\leftrightarrow 3}\sum\limits_ \cO   P _ \cO^{a,b}g^{I;ab}_{h,\hb}(1-\zb,1-z)\;,\label{crossing_fermion}
\ee 
where we use $P$ for the coefficients in the $\DD_{a}$ differential basis.

To simplify some expressions, we will first consider the contributions of double-twist operators in the $(12)\rightarrow (34)$ OPE. Then later we will take $1\leftrightarrow 3$ so that they appear in the $t$-channel, and match their sum to the contributions of individual $s$-channel blocks, as in section~\ref{sec:LightconeBootstrapReview}.

The 3-point structures will be unimportant in the following discussion so we will simplify the notation of the left-hand side of (\ref{crossing_fermion}) to
\be
\sum\limits_{\substack{\text{different}\\\text{ structures}}}\sum\limits_ \cO\left(\DD\widetilde{\DD}\left[\pre_i P _ \cO g_{h,\hb}(z,\zb)\right]\right)\evaluated_{\Delta_i\rightarrow\Delta_\psi}\;.
\ee 

We can also interchange the order of differential operators and summation over relevant operator families. Since $\pre$ is independent of the exchanged operator, the double-twist sum reduces to
\be
\sum\limits_{\substack{\text{different}\\\text{ structures}}}\left(\DD\widetilde{\DD}\left[\pre_i \sum\limits_ \cO P _ \cO g_{h,\hb}(z,\zb)\right]\right)_{\Delta_i\rightarrow\Delta_\psi}\;.
\ee 

To go any further, we need to specify which operators $ \cO$ must appear to reproduce the lightcone limit of the crossed channel. In particular, we know that an infinite sum of double-twist operators is needed to reproduce the identity operator in the crossed channel \mbox{\cite{Simmons-Duffin:2016wlq,Komargodski:2012ek,Fitzpatrick:2012yx}}. The required operators and their quantum numbers are schematically shown in Table~\ref{table:DoubleTwistFamilies}.

To remove clutter, we will denote different double-twist families generically as $[\psi\psi]_n$ below. Their contribution then reads as
\be 
\sum\limits_{\substack{\text{different}\\\text{ structures}}}\Bigg(\DD\widetilde{\DD}\bigg[\pre_i\sum\limits_{n=0}^\infty\sum\limits_{l=0}^\infty P _{[\psi\psi]_n}
g_{h_{[\psi\psi]_n},\,\bar{h}_{[\psi \psi]_n}}(z,\zb)
\bigg]\Bigg)_{\Delta_i\rightarrow\Delta_\psi}\;,
\ee 
where the summation over all relevant families $[\psi\psi]_n$ appearing in Table~\ref{table:DoubleTwistFamilies} is implicit.

Restoring possible dependence on the external dimension differences $r,s$ (which arise from the shift operators $\Pi$), we will now use Eq.~(\ref{eq:SL2R}) in a slightly modified form after applying $g(z,\zb)=g(\zb,z)$ symmetry:
\be
g_{h,\hb}^{r,s}(z,\zb)=\sum\limits_{n=0}^\infty\sum\limits_{j=-n}^n A_{n,j}^{r,s}(h,\hb)\bar{z}^{h+n} k^{r,s}_{2(\bar{h}+j)}(z)\;.
\ee
By expanding
\be 
h_{[\psi\psi]_n}=h_{[\psi\psi]_0}+n+\delta h_{[\psi\psi]_n}\;,
\ee
the leading part of the double-twist sum at small $\bar{z}$ can then be rewritten as
\be
\sum\limits_{\substack{\text{different}\\\text{ structures}}}\Bigg(\DD\widetilde{\DD}\bigg[\pre_i\sum\limits_{l=0}^\infty P_{[\psi\psi]_0}\zb^{h_{[\psi\psi]_0}+\delta h_{[\psi\psi]_0}} k^{r,s}_{2\hb_{[\psi\psi]_0}}(z)
\bigg]\Bigg)_{\Delta_i\rightarrow\Delta_\psi}\left(1+\order{{\zb}}\right)\;.
\ee

To reproduce $\log$ terms in the crossed channel, we will expand to linear order in the anomalous dimension:\footnote{Higher order terms in $\delta h$ are matched with multi-twist operators in the $s$-channel.}
\small 
\be
\sum\limits_{\substack{\text{different}\\\text{ structures}}}\Bigg( \DD\widetilde{\DD}\bigg[\pre_i\;\zb^{h_{[\psi\psi]_0}}\bigg\{
\sum\limits_{l=0}^\infty P _{[\psi\psi]_0} k^{r,s}_{2\hb_{[\psi\psi]_0}}(z)
+\log(\zb)\sum\limits_{l=0}^\infty P _{[\psi\psi]_0}\delta h_{[\psi\psi]_0}
k^{r,s}_{2\hb_{[\psi\psi]_0}}(z)
\bigg\}\bigg]\Bigg)_{\Delta_i\rightarrow\Delta_\psi}.
\label{eq:tChannelExpression1}
\ee
\normalsize

In \cite{Simmons-Duffin:2016wlq} it was shown how to sum over $ {}_2F_1$ hypergeometric functions to reproduce terms Casimir-singular in $z$:
\be 
\sum_{l=0}^\infty\frac{\partial \hb}{\partial l}S_a^{r,s}(\hb)k^{r,s}_{2\hb}(1-z)
=\left(\frac{z}{1-z}\right)^{a}+[\cdots]_z\;, \label{eq:SomeOverNandL}
\ee 
which is the generalization of \equref{eq:SL2R_Sum} to non-identical external scaling dimensions. The explicit form of $S_a^{r,s}(h)$ is given in \equref{eq:ExpansionCoefficientSGeneralized}, though it will not be necessary for the following calculations.

As explained in section~\ref{sec:LightconeBootstrapReview}, we make the ansatz
\bea[eq:OPEAnsatz]
P _{[\psi\psi]_0}^{a,b}(\bar{h})=&\left(\frac{\partial \hb_{[\psi\psi]_0}}{\partial l}\right)\sum\limits_{\{i\}}A_{a,b,i}S_i^{0,0}(\hb_{[\psi\psi]_0})\;,
\label{eq:Definition of coefficient A}
\\
(\delta hP )^{a,b}_{[\psi\psi]_0}(\bar{h})=&\left(\frac{\partial \hb_{[\psi\psi]_0}}{\partial l}\right)\sum\limits_{\{j\}}B_{a,b,j}S_j^{0,0}(\hb_{[\psi\psi]_0})\;,
\eea
and insert this into \equref{eq:tChannelExpression1} to obtain
\be
\frac{1}{2}\sum\limits_{\substack{\text{different}\\\text{ structures}}}\left( \DD\widetilde{\DD}\left[\pre_i\zb^{h_{[\psi\psi]_0}}\left(
\sum\limits_{\{i\}}A_{a,b,i}\left(\frac{1-z}{z}\right)^{i}
+\log(\zb) 
\sum\limits_{\{j\}}B_{a,b,j}\left(\frac{1-z}{z}\right)^{j}
\right)\right]\right)_{\Delta_i\rightarrow\Delta_\psi}\;. \label{eq:CrossedChannelFinalForm}
\ee
The $1/2$ in front accounts for the fact that we are summing over operator families with even-integer spacing~\cite{Simmons-Duffin:2016wlq}, since we are dealing with identical external fermions. In particular, $P^{4,4}=0$ for even $l$, and $P^{1,1}=P^{1,2}=P^{2,1}=P^{2,2}=P^{3,3}=0$ for odd $l$~\cite{Iliesiu:2015qra}.

After acting with the differential operators we take $1\leftrightarrow 3$ and $(z,\bar{z})\rightarrow (1-z,1-\bar{z})$ and match to individual $s$-channel blocks. We will find the sets $\{i\}$ and $\{j\}$, and the coefficients $A$ and $B$ by matching the $s$-channel. 

There is actually a subtle point we skipped while going from \equref{eq:SomeOverNandL} to \equref{eq:CrossedChannelFinalForm}. As evident from \equref{eq:DifferentialOperators}, we need $S^{\pm\frac{1}{2},\pm\frac{1}{2}}$ to be able to use \equref{eq:SomeOverNandL} for parity-odd structures even though we used $S^{0,0}$ in our ansatz above. We resolve this in appendix~\ref{sec:ExpansionOfS} by expanding $S_a^{r+n,r+m}$ in terms of $S_{a+k}^{r,s}$.

\subsection{Results}

In this section, we first discuss identity matching and find the MFT solutions. Then we consider the exchange of parity-even and parity-odd operators of arbitrary dimension and spin, and calculate their contribution to the OPE coefficients and anomalous dimensions of the double-twist families $[\psi_{(\alpha}\psi_{\beta)}]_{0,l}$, $[\psi_\alpha\psi^\alpha]_{0,l}$, and $[\psi_{(\rho}\partial^\rho_{\;\alpha}\psi_{\beta)}]_{0,l}$ at leading and sub-leading order in the small $z$ expansion. As special cases, we will present the contributions due to stress tensor exchange and scalar exchanges.

The reader is reminded that the contributions of all double-twist families are present, but we simply match the subset of terms relevant for the above families. For example, we match the $\order{z^{1-\Delta_\psi}(1-\bar{z})^0}$ contribution of the stress tensor without matching the $\order{z^{-\Delta_\psi}(1-\bar{z})^1}$ contribution from identity exchange, even though the latter is more dominant in the lightcone limit. However, these contributions come from different twist families in the crossed channel, and there is no mixing for the terms leading order in $(1-\bar{z})$. For instance, both $[\psi_{(\a} \psi_{\b)}]_{0,l}$ and $[\psi_{(\a} \psi_{\b)}]_{1,l}$ bring contributions of order $\order{z^{-\Delta_\psi}(1-\bar{z})^1}$, however only $[\psi_{(\a} \psi_{\b)}]_{0,l}$ brings $\order{z^{-\Delta_\psi}(1-\bar{z})^0}$ terms. So by requiring $[\psi_{(\a} \psi_{\b)}]_{0,l}$ to reproduce these in the crossed channel, we can extract its OPE coefficients and anomalous dimensions.

Below, we will suppress the label for the double twist families whenever there is no ambiguity. For example, we will simply write the OPE coefficient $f^1_{[\psi_{(\a}\psi_{\b)}]_{0,l}}$ as $f^1$. We can extract the relevant family due to the conditions listed in Table~\ref{table:DoubleTwistFamilies} and the fact that the four types of 3-point functions $f^a$ are associated with (parity, spin) as: ($+$, even), ($+$, even), ($-$, even), and ($-$, odd), respectively.

\subsubsection{Identity matching}
Let us first focus on the identity contribution alone. In the $s$-channel the relevant terms are  $\order{z^{-\frac{1}{2}-\Delta_\psi}(1-\bar{z})^0}$ and $\order{z^{\frac{1}{2}-\Delta_\psi}(1-\bar{z})^0}$. We reproduce these terms in the $t$-channel by tuning $A_{a,b,i}$ in \equref{eq:Definition of coefficient A}; for example, we need
\be 
\left\{A_{2,2,\frac{3}{2}-\Delta_\psi},A_{2,2,\frac{5}{2}-\Delta_\psi }\right\}
\ee 
for the double-twist family $[\psi_{(\a} \psi_{\b)} ]_{0,l}$. The fact that $A_{1,1,i}$, $A_{1,2,i}$, and $A_{2,1,i}$ are zero reflects the vanishing of the 3-point coefficient $f^1$ at all orders.

As there is no anomalous dimension for the identity exchange alone we have $\frac{\partial \hb}{\partial l}=1$, hence we can immediately get $P^{a,b}$ with \equref{eq:OPEAnsatz}, then solve for the physical 3-point coefficients\footnote{\label{footnote: conformal block normalization difference}The $(-1)^l$ term here would be absent in the notation of~\cite{Iliesiu:2015qra}, however we need it as our conformal block normalization in \equref{2.39 of Dolan:2011dv} differs from that paper by a factor of $(-1)^l$.}
\be 
\label{eq: definition of P in perturbative section}
P_\cO^{a,b}=(-1)^lf_{\psi_1\psi_2\cO}^a f_{\psi_3\psi_4\cO}^b\;,
\ee 
where the identity itself has the OPE coefficients $f^1_{\psi\psi\mathbb{1}}=i$ and $f^2_{\psi\psi\mathbb{1}}=0$.

Using the steps described above, we compute the OPE coefficients
\bea[eq:identity OPE]
f^1=\,&0\;, \label{eq: identity lambda 1}
\\
f^2=\,&\frac{f_0}{4\hb^2}\left(1-\frac{8 \Delta_{\psi }-17}{16 \bar{h}}-\frac{256 \Delta_\psi^3-2112 \Delta_\psi ^2+4208 \Delta_\psi-2787}{1536 \hb^2}+\order{\frac{1}{\hb}}^3\right)\;,\\
f^3=\,&\frac{f_0}{2\hb}\sqrt{\frac{\Delta _{\psi }-1}{2\hb}}\left(1-\frac{24\Delta_{\psi }-31}{16 \bar{h}}+\order{\frac{1}{\hb}}^2\right)\;
,\\
f^4=\,&\frac{f_0}{2\hb}\sqrt{\frac{\Delta _{\psi }-1}{2\hb}}\left(1+\frac{8\Delta_{\psi }-1}{16\bar{h}}+\order{\frac{1}{\hb}}^2\right)\;,
\eea
where for convenience we have defined the prefactor $f_0$ as 
\be 
f_0\equiv i \sqrt[4]{\pi } \frac{2^{\frac{3}{2}-\bar{h}} \bar{h}^{\Delta _{\psi }-\frac{1}{4}}}{\Gamma \left(\Delta _{\psi }+\frac{1}{2}\right)}\;.\label{eq: Definition of lambda zero}
\ee 

Note that the results take a slightly simpler form in the $\cD$ basis. E.g., at leading order
\be 
f^2\evaluated_{\cD}=f_0\;,\qquad 
f^{3,4}\evaluated_{\cD}=f_0\sqrt{\frac{\Delta _{\psi }-1}{2\hb}}\;,
\ee 
which follows from \equref{changebasis}.

\subsubsection{Matching the exchange of a generic parity-even operator}
Let us turn to the contribution of the exchange of a generic parity-even operator $\cO^+_{\tau,l}$ of twist $\tau$ and spin $l$ in the $s$-channel to the double-twist families $[\psi_{(\alpha}\psi_{\beta)}]_0$, $[\psi_\alpha\psi^\alpha]_0$, and $[\psi_{(\rho}\partial^\rho_{\a}\psi_{\b )}]_0$ in the $t$-channel.

When calculating corrections to the anomalous dimensions of double-twist families we need to recall that the contributions of multiple operators are not additive. Additionally, in general there can be multiple double-twist families that mix with each other. The full formula for their anomalous dimension matrix is\footnote{The anomalous dimension matrix~(\ref{eq:anomalous dimension calculation}) is diagonal if there is no mixing between the double-twist families.}
\be
\frac{\gamma_{[\psi\psi]}}{2}=\delta h_{[\psi\psi]}=\frac{\sum\limits_{\cO}\left(\delta h_{[\psi\psi]}P_{[\psi\psi]}^{a,b}J^{-1}_{[\psi\psi]}\right)_{\cO}}{\sum\limits_{\cO}\left(P_{[\psi\psi]}^{a,b}J^{-1}_{[\psi\psi]}\right)_{\cO}}\;,\label{eq:anomalous dimension calculation}
\ee 
where $J$ is the Jacobian
\be 
J_{[\psi\psi]}\equiv\frac{\partial \delta h_{[\psi\psi]}}{\partial \hb}\;.
\ee 
Here $\cO$ runs over all exchanged operators in the $s$-channel. Likewise, the OPE coefficients $f^a_{[\psi\psi]}$ are given as
\be
(-1)^{\ell}f_{[\psi\psi]}^af_{[\psi\psi]}^b=J_{[\psi\psi]}\sum\limits_{\cO}\left(P^{a,b}_{[\psi\psi]}J^{-1}_{[\psi\psi]}\right)_\cO\;.
\ee 

In the large $\bar{h}$ expansion, we can of course truncate the summation over operators in twist to extract the large $\bar{h}$ behavior. For example, in the first few orders, we see that
\bea[eq:LeadingOrder]
\frac{\gamma_{[\psi\psi]}}{2}=\;\delta h_{[\psi\psi]}
&=\frac{\left(\delta h_{[\psi\psi]}P_{[\psi\psi]}^{a,b}\right)_{\mathbbm{O}}}{\left(P_{[\psi\psi]}^{a,b}\right)_{\mathbb{1}}+\left(P_{[\psi\psi]}^{a,b}\right)_{\mathbbm{O}}}\;,\\
(-1)^lf_{[\psi\psi]}^af_{[\psi\psi]}^b&=\left(P^{a,b}_{[\psi\psi]}\right)_\mathbb{1}+\left(P^{a,b}_{[\psi\psi]}\right)_\mathbbm{O}\;,
\eea
for the identity operator $\mathbb{1}$ along with the operator with minimum twist $\mathbbm{O}$, which is usually either the stress tensor or a scalar of low dimension. 

At leading order in $1/\bar{h}$ one can easily isolate the contribution of any operator. Only the identity operator contributes in the denominator in \equref{eq:anomalous dimension calculation}, allowing one to write an isolated contribution to the anomalous dimension. Likewise, at leading order, one can immediately calculate an individual contribution to $f^a$.

Once we go beyond leading order, we can work with $\left(P_{[\psi\psi]}^{a,b}J^{-1}_{[\psi\psi]}\right)_\cO$ and\linebreak $\left(\delta h^{a,b}_{[\psi\psi]}P_{[\psi\psi]}^{a,b}J^{-1}_{[\psi\psi]}\right)_\cO$, make an ansatz for their large $\bar{h}$ behavior, and then calculate the corrections to the 3-point coefficients and anomalous dimensions. We find
\begin{subequations}
	\label{eq: generic parity-even exchange}
	\footnotesize
	\be
	\left(P^{1,1}J^{-1}\right)_{\cO_{\tau,l}^+}
	=&-
	\frac{f_+^2}{16 \bar{h}^4 (f_{\cO}'^1)^2 (H_{l+\frac{\tau }{2}-1})^2}\\&\x\left[
	\left\{\left(f_{\cO}'^1\right)^2 (5-4 \Delta_\psi )-4 \left(f_{\cO}^2\right)^2 l (l+\tau -1) (-2 \Delta _\psi+\tau +1)^2\right\}^2+\order{\frac{1}{\hb}}
	\right]\;,
	\ee
	\be 
	\hspace{-2in}\left(P^{2,2}J^{-1}\right)_{\cO_{\tau,l}^+}
	=&-\frac{f_+^2 \left(f_{\cO}'^1\right)^2  }{16 \bar{h}^4}\Bigg[
	1+\frac{-8 \Delta _{\psi }+4
		\tau +17}{8 \bar{h}}
	+\order{\frac{1}{\hb}}^2\Bigg]\;,
	\ee 
	\be 
	\hspace{.07in}\left((\delta h P)^{2,2}J^{-1}\right)_{\cO_{\tau,l}^+}=&-\frac{f_+^2}{32\hb^4 H_{l+\frac{\tau }{2}-1}}\Bigg[
	\left(f_{\cO}'^1\right)^2\left(1+\frac{17+4\tau-8\Delta_\psi}{8\hb}\right)\\&
	+\frac{1}{(2\ell-1)(2\ell+2\tau-1)\bar{h}^2} \Bigg( \frac{\left(f_{\cO}'^1\right)^2}{384 } - f_{\cO}'^1f_{\cO}^2 l (\tau -1) (l+\tau -1) \left(-2 \Delta _{\psi }+\tau +1\right)^2
	\\&
	+\left(f_{\cO}^2\right)^2 l (l+\tau -1) \left(4 l \tau +4 (l-1) l+2 \tau ^2-5 \tau +2\right) \left(2 \Delta _{\psi }-\tau -1\right)^2 \Bigg)\\&
	\x\bigg\{
	4 l^2 \left(-8 \tau ^3+72 \tau ^2+512 \tau +1851\right)+16 \tau ^4-128 \tau ^3-928 \tau ^2-3214 \tau\\ 
	&+1827-4 l \left(8 \tau ^4-80 \tau ^3-440 \tau ^2-1339 \tau +1851\right) -128 (2 l-1) \Delta _{\psi }^3 (2 l+2 \tau -1)\\
	&
	-16 \Delta _{\psi } \left(l^2 (96 \tau +652)+l \left(96 \tau ^2+556 \tau -652\right)-42 \tau ^2-302 \tau +157\right)\\
	&+192 \Delta _{\psi }^2 \left(2 l^2 (\tau +13)+2 l \left(\tau ^2+12 \tau -13\right)-\tau ^2-12 \tau +6\right)
	\bigg\}
	+\order{\frac{1}{\hb}}^3
	\Bigg]\;,
	\ee 
	\be 
	\left(P^{3,3}J^{-1}\right)_{\cO_{\tau,l}^+}=&-\left(P^{4,4}J^{-1}\right)_{\cO_{\tau,l}^+}
	\\=&\frac{f_+^2f_{\cO}'^1 }{16\hb^3}\Bigg[
	\left(1 + \frac{-8 \Delta _{\psi }+4 \tau +15}{8 \bar{h}}\right) \\
	&\times \bigg\{f_{\cO}'^1 (\tau +2)-2 \Delta _{\psi } (f_{\cO}'^1-4 f_{\cO}^2 l (l+\tau -1)) -4 f_{\cO}^2 l (\tau +1) (l+\tau -1)
	\bigg\}
	+
	\order{\frac{1}{\hb}}^2
	\Bigg]\;,
	\ee 
	\be 
	\left( (\delta h P)^{3,3}J^{-1}\right)_{\cO_{\tau,l}^+}
	=& -\frac{f_+^2}{32 \bar{h}^3 H_{l+\frac{\tau }{2}-1} }\Bigg[
	f_{\cO}'^1 \left( f_{\cO}'^1(2\Delta_{\psi}-\tau-2) -4 f_{\cO}^2 \ell(2\Delta_{\psi}-\tau-1)(\ell+\tau-1) \right)\\
	&- \frac{1}{8\bar{h}} \Bigg( (f_{\cO}'^1)^2 \left(48 \Delta_{\psi}^2 + \tau (4\tau+35) - 2 \Delta_{\psi}(16\tau+55)+62 \right) \\
	&-4 f_{\cO}'^1 f_{\cO}^2 \ell (24 \Delta_{\psi}-4\tau-23)(2\Delta_{\psi}-\tau-1)(\ell+\tau-1) \\
	&+16 (f_{\cO}^2)^2 \ell (\tau-2)(\ell+\tau-1) (2\Delta_{\psi}-\tau-1)^2
	\Bigg)
	+\order{\frac{1}{\hb}}^2
	\Bigg]\;,
	\ee
	\be
	\left( (\delta hP)^{4,4}J^{-1}\right)_{\cO_{\tau,l}^+}
	=& - \left( (\delta hP)^{3,3}J^{-1}\right)_{\cO_{\tau,l}^+} +\frac{f_+^2}{32 \bar{h}^4 H_{l+\frac{\tau }{2}-1} }\Bigg[(f_{\cO}'^1)^2 ( 4(\Delta_{\psi}-1)(2\Delta_{\psi}-\tau-2) -\tau)  \\&- 8 f_{\cO}'^1 f_{\cO}^2 \ell (2\Delta_{\psi}-1)(2\Delta_{\psi}-\tau-1)(\ell+\tau-1) \\&+ 4 (f_{\cO}^2)^2 \ell (\tau-2)(\ell+\tau-1)(2\Delta_{\psi}-\tau-1)^2 +\order{\frac{1}{\hb}}  \Bigg]\;,
	\ee
	\normalsize
\end{subequations}
where $H_a$ is the Harmonic number, and we defined
\be 
f_{\cO}'^1\equiv f_{\cO}^1+\left(2\Delta_{\psi}-\tau-1\right)\left(2\Delta_{\psi}+\tau-4\right)f_{\cO}^2
\ee 
and
\be 
f_+\equiv \frac{2^{\frac{3}{2} -\bar{h}+ l + \frac{\tau}{2}} \bar{h}^{\Delta _{\psi }-\frac{1}{4} (1+2 \tau )}}{\Gamma \left(\Delta _{\psi }+\frac{1-\tau }{2}\right)}\sqrt{\left(l+\frac{\tau }{2}\right)_{\frac{1}{2}}H_{l+\frac{\tau }{2}-1}}
\ee 
for convenience. Note that $f_+$ reduces back to \equref{eq: Definition of lambda zero} for identity exchange after setting $l=\tau=\delta h=0$. The appearance of $(f'^1_{\cO})^2$ in the denominator of $P^{1,1}$ is because we compute it by first matching $P^{1,2}$ and then using the relation $P^{1,1} = (P^{1,2})^2/ P^{2,2}$.

\paragraph{Reproducing identity matching:}
As a consistency check, let us use this general form to reproduce the identity exchange contribution to the $[\psi_{(\a}\psi_{\b)}]_{0,l}$ family. We can first check that the anomalous dimension due to sole identity exchange is indeed zero:\footnote{One sees a possible divergence if one does not take $f_{\cO}^2 \rightarrow 0$ first. This is natural because the corresponding 3-point structure does not exist for scalars so its coefficient should be taken to vanish before setting $l = 0$.}
\be
\gamma_{[\psi_{(\a}\psi_{\b)}]_{0,l}}=2\delta h_{[\psi_{(\a}\psi_{\b)}]_{0,l}}=2\frac{\left((\delta h P)^{2,2}\right)_{\cO_{\tau,l}^+}}{\left(P^{2,2}\right)_{\cO_{\tau,l}^+}}\evaluated_{\substack{f_{\cO}^1\rightarrow i,\;f_{\cO}^2\rightarrow 0,\;l\rightarrow 0,\;\tau\rightarrow0}}=0\;.
\ee 

With this, one can now straightforwardly calculate
\bea
\left(f^1\right)^2&=\left(P^{1,1}\right)_{\cO_{\tau,l}^+}\evaluated_{\substack{f_{\cO}^1\rightarrow i,\;f_{\cO}^2\rightarrow 0,\;l\rightarrow 0,\;\tau\rightarrow0}}\;,
\\
\left(f^2\right)^2&=\left(P^{2,2}\right)_{\cO_{\tau,l}^+}\evaluated_{\substack{f_{\cO}^1\rightarrow i,\;f_{\cO}^2\rightarrow 0,\;l\rightarrow 0,\;\tau\rightarrow0}}\;,
\eea
which match \equref{eq:identity OPE}.

\paragraph{Parity-even scalar exchange:} We can also consider the special case of exchange of a parity-even scalar of twist $\tau_s$. Let us write down the anomalous dimension to leading order in $1/\hb$ for convenience. We then only need to use $P^{i,j}$ for identity exchange and $(\delta h P)^{i,j}$ for scalar exchange.

We can immediately read the leading-order anomalous dimensions due to a parity-even scalar exchange as follows:
\begin{subequations}
	\begin{equation}
		\gamma_{[\psi_{(\a}\psi_{\b)}]_{0}}=2\frac{\left((\delta h P)^{2,2}\right)_{\cO_{\tau,l}^+}\evaluated_{\substack{f_{\cO}^1\rightarrow f_\phi,\;f_{\cO}^2\rightarrow 0,\;l\rightarrow 0,\;\tau\rightarrow\tau_s}}}{\left(P^{2,2}\right)_{\cO_{\tau,l}^+}\evaluated_{\substack{f_{\cO}^1\rightarrow f_\mathbb{1},\;f_{\cO}^2\rightarrow 0,\;l\rightarrow 0,\;\tau\rightarrow 0}}}=
		\frac{f ^2_\phi}{\hb^{\tau_s}}\frac{2^{{\tau_s} } \left(\frac{{\tau_s} }{2}\right)_{\frac{1}{2}}  \left(\left(-\frac{{\tau_s} }{2}+\Delta _{\psi }+\frac{1}{2}\right)_{\frac{{\tau_s} }{2}}\right)^2}{\sqrt{\pi }}\;,\label{eq: anomalous of parity-even scalar}
	\end{equation}
	and likewise
	\begin{equation}
		\gamma_{[\psi_\a\psi^\a]_0}=\gamma_{[\psi_{(\rho}\partial^\rho_{\a}\psi_{\b )}]_0}=\frac{f ^2_\phi}{\hb^{\tau_s}}
		\frac{2^{{\tau_s} } \left(\frac{{\tau_s} }{2}\right)_{\frac{1}{2}}  \left(\left(-\frac{{\tau_s} }{2}+\Delta _{\psi }+\frac{1}{2}\right)_{\frac{{\tau_s}
				}{2}}\right)^2\left(\Delta _{\psi }-1-\frac{{\tau_s}}{2}\right)}{\sqrt{\pi } \left(\Delta _{\psi }-1\right)}\;.
	\end{equation}
	\label{eq:AnomalousDimensionOfParityEvenScalars}
\end{subequations}

There are two comments in order. Firstly, $\gamma_{[\psi_\a\psi^\a]_0}$ and $\gamma_{[\psi_{(\rho}\partial^\rho_{\a}\psi_{\b )}]_0}$ na\"ively seem to be divergent in the MFT limit $\Delta_\psi\rightarrow 1$. However, what really matters when solving the analytic bootstrap is the weighted contribution $(f_{{[\psi \psi]}_{0}})^{2} \gamma_{{[\psi \psi]}_{0}}$. We can check from \equref{eq:identity OPE} that the squared OPE coefficients also vanish linearly as $\Delta_{\psi}\rightarrow1$. So, we need to check that this weighted contribution in fact vanishes in mean field theory, or that the factor
\be
\left(-\frac{{\tau_s} }{2}+\Delta _{\psi }+\frac{1}{2}\right)_{\frac{{\tau_s}}{2}} &=& \frac{\Gamma(\Delta _{\psi }+\frac{1}{2})}{\Gamma(-\frac{{\tau_s} }{2}+\Delta _{\psi }+\frac{1}{2})}
\ee 
is zero. In mean field theory we have the contributions from scalar operators $[\psi_{\alpha}\partial^{\alpha\beta}\psi_{\beta}]_{n}$ with twist $2\Delta_{\psi}+1+2n$ for $n\in \N_0$. Plugging in the mean field theory twist we obtain a factor of $\Gamma(-n)^{-1}$, so the result vanishes as expected.

Secondly, we recall that $f_\phi\equiv f_{\psi\psi\phi}$ is purely imaginary due to the Grassmann nature of fermions, hence $\gamma_{[\psi_{(\a}\psi_{\b)}]_0}<0$, as expected for the leading double-twist trajectory~\cite{Komargodski:2012ek}. 

\paragraph{Stress tensor exchange:} As a last example we consider stress tensor exchange. From Ward identities, we know that\footnote{See \cite{Iliesiu:2015qra} for their calculation in the $\cD$ basis.}
\be 
f_{\psi\psi T}^1= \frac{3 i \left(\Delta _{\psi }-1\right) \left(2 \Delta _{\psi }+1\right)}{16 \sqrt{C_T}}\;,\quad 
f_{\psi\psi T}^2= -\frac{3 i}{32 \sqrt{C_T}}\;,
\ee 
where $C_T$ is the central charge. Hence, we compute
\begin{subequations}
	\begin{equation}
		\gamma_{[\psi_{(\a}\psi_{\b)}]_{0}}=2\frac{\left((\delta h P)^{2,2}\right)_{\cO_{\tau,l}^+}\evaluated_{\substack{f_{\cO}^1\rightarrow f_{\psi\psi T}^1,\;f_{\cO}^2\rightarrow f_{\psi\psi T}^2,\;l\rightarrow 2,\;\tau\rightarrow 1}}}{\left(P^{2,2}\right)_{\cO_{\tau,l}^+}\evaluated_{\substack{f_{\cO}^1\rightarrow f_\mathbb{1},\;f_{\cO}^2\rightarrow 0,\;l\rightarrow 0,\;\tau\rightarrow 0}}}=
		-\frac{48 \Gamma \left(\Delta _{\psi }+\frac{1}{2}\right)^2}{\pi C_T \Gamma \left(\Delta _{\psi }-1\right)^2\hb}\;,
	\end{equation}
	and
	\begin{equation}
		\gamma_{[\psi_\a\psi^\a]_0}=\gamma_{[\psi_{(\rho}\partial^\rho_{\a}\psi_{\b )}]_0}= -\frac{24 \left(\Delta _{\psi }-1\right) \left(2 \Delta _{\psi }+1\right) \Gamma \left(\Delta _{\psi }+\frac{1}{2}\right)^2}{\pi C_T \Gamma \left(\Delta _{\psi }\right)^2\hb}\;.
	\end{equation}
\end{subequations}

\subsubsection{Matching the exchange of a generic parity-odd operator}
Similar to the previous computation, we can also consider the contribution of the exchange of a generic parity-odd operator $\cO^-_{\tau,\ell}$ in the $s$-channel to the double-twist operators $[\psi_{(\alpha}\psi_{\beta)}]_0$, $[\psi_\alpha\psi^\alpha]_0$, and $[\psi_{(\rho}\partial^\rho_{\a}\psi_{\b )}]_0$ in the $t$-channel.

The analogue of \equref{eq: generic parity-even exchange} now reads as
\begin{subequations}
	\label{eq: generic parity-odd exchange}
	\footnotesize
	\be 
	\left(P^{1,2}J^{-1}\right)_{\cO_{\tau,l}^-}
	=-f_-^2\Bigg[\left(\left(f_\cO'^3\right)^2-\left(f_\cO'^4\right)^2\right)\left(1+\frac{-8 \Delta _{\psi }+4 \tau +13}{8 \bar{h}}\right)+\order{\frac{1}{\hb}}^2\Bigg]\;,
	\ee 
	\begin{multline}
		\left((\delta h P)^{2,2}J^{-1}\right)_{\cO^-_{\tau,l}}
		=-\frac{f_-^2}{8\hb^2}\Bigg[\frac{(2 l+\tau -1) \left(\left(f_\cO'^3\right)^2l-\left(f_\cO'^4\right)^2 (\tau+l-1)\right)}{l (l+\tau -1)}\\\x\left(1+\frac{-8 \Delta _{\psi }+4 \tau +21}{8 \bar{h}}\right)+\order{\frac{1}{\hb}}^2\Bigg],
	\end{multline}
	\be 
	\left((\delta h P)^{3,3}J^{-1}\right)_{\cO^-_{\tau,l}}
	=
	-\frac{f_-^2}{4}&\Bigg[\left(\left(f_\cO'^4\right)^2+\left(f_\cO'^3\right)^2\right) (2 l+\tau -1)+\frac{2l+\tau-1}{8\hb}\\&\x\bigg\{\frac{\left(f_\cO'^4\right)^2 \left(8 (1-3l) \Delta _{\psi }+4 (l-1) \tau +23 l-4\right)}{l}
	\\&+\frac{\left(f_\cO'^3\right)^2 \left(-8 \Delta _{\psi } (3l+3\tau-2)+(l+\tau)(4 \tau +23)-19\right)}{l+\tau -1}
	\bigg\}+\order{\frac{1}{\hb}}^2\Bigg],
	\ee 
	\be 
	\left((\delta h P)^{4,4}J^{-1}\right)_{\cO^-_{\tau,l}}
	=
	-\frac{f_-^2}{4}&\Bigg[\left(\left(f_\cO'^4\right)^2+\left(f_\cO'^3\right)^2\right) (2 l+\tau -1)+\frac{2l+\tau-1}{8\hb}\\&\x\bigg\{\frac{\left(f_\cO'^4\right)^2  \left(8 (l-1) \Delta _{\psi }+4 (l+1) \tau +7 l+4\right)}{l}
	\\&+\frac{\left(f_\cO'^3\right)^2 \left(8 \Delta _{\psi } (l+\tau)+l(4 \tau +7)+\tau(4\tau-1)-11\right)}{l+\tau -1}
	\bigg\}+\order{\frac{1}{\hb}}^2\Bigg],
	\ee 
	\normalsize
\end{subequations}
where for convenience we have defined
\bea 
f_\cO'^3\equiv{}&2(\tau+l-1)f_\cO^3\;,\\
f_\cO'^4\equiv{}&2lf_\cO^4\;,\\
f_-\equiv{}& \frac{2^{\frac{\tau }{2}-\bar{h}+l-1} \bar{h}^{\Delta _{\psi }-\frac{1}{4} (7+2 \tau )}}{\Gamma \left(\Delta _{\psi }-\frac{\tau }{2}\right)\sqrt{\left(l+\frac{\tau }{2}\right)_{\frac{1}{2}}}}\;.
\eea 
Note that $f_\cO'^3$ and $f_\cO'^4$ are actually the 3-point coefficients in the $\cD$ basis as one can see by comparing \equref{eq:DifferentialOperators} and \equref{changebasis}. Quantities not shown, e.g. $P^{1,1}$ and $P^{2,2}$, do not appear in the matching conditions at this order, hence we do not learn about any new contributions to them. Let us also comment that since $P^{1,2}=0$ in the free theory limit, $P^{1,2}$ being required to be nonzero serves as a probe of the effect of interactions.

\paragraph{Parity-odd scalar exchange:} At leading order, the exchange of a parity-odd scalar contributes to the anomalous dimension of the double-twist families as
\begin{subequations}
	\label{eq:AnomalousDimensionOfParityOddScalars}
	\be 
	\gamma_{[\psi_{(\a} \psi_{\b)}]_0}={}2\frac{\left((\delta h P)^{2,2}\right)_{\cO_{\tau,l}^-}\evaluated_{\substack{f_{\cO}'^3\rightarrow f_\phi,\;f_{\cO}^4\rightarrow 0,\;l\rightarrow 0,\tau\rightarrow\tau_s}}}{\left(P^{2,2}\right)_{\cO_{\tau,l}^+}\evaluated_{\substack{f_{\cO}^1\rightarrow f_\mathbb{1},\;f_{\cO}^2\rightarrow 0,\;l\rightarrow 0,\tau\rightarrow 0}}}=\frac{f_\phi^2}{\hb^{\tau_s+1}}
	\frac{ 2^{\tau _s-1} \left(\left(\frac{2 \Delta _{\psi }-\tau
			_s}{2}\right)_{\frac{\tau _s+1}{2} }\right)^2}{\sqrt{\pi }\left(\frac{\tau _s}{2}\right)_{\frac{1}{2}}}\;,
	\ee 
	where $f_\phi$ is given in the more standard $\cD$ basis. Similarly,
	\begin{align}
		\gamma_{[\psi_\a\psi^\a]_0}= - \gamma_{[\psi_{(\rho}\partial^\rho_{\a}\psi_{\b)}]_0} ={}&\frac{f ^2_\phi}{\hb^{\tau_s}}
		\frac{2^{{\tau_s} } \left(\left(\frac{2 \Delta _{\psi }-\tau
				_s}{2}\right)_{\frac{\tau _s+1}{2} }\right)^2}{\sqrt{\pi } \left(\frac{{\tau_s} }{2}\right)_{-\frac{1}{2}}\left(\Delta _{\psi }-1\right)}	\;.
	\end{align}
\end{subequations}
Comparing with \equref{eq:AnomalousDimensionOfParityEvenScalars}, we see that the contribution of parity-odd scalar exchange to the parity-even double-twist family $[\psi_{(\a} \psi_{\b)}]_0$ comes at the higher order $\hb^{-\tau_s-1}$ instead of $\hb^{-\tau_s}$.

\chapter{Fermions in Conformal Field Theories: Non-Perturbative Corrections}
\label{sec:nonperturbative}

\section{Perturbative vs Non-perturbative}
\label{sec:inversion}

In \secref{\ref{chapter: perturbative}}, we computed CFT data for fermions using large spin expansion. While large spin expansions has been historically useful and are numerically shown to provide surprisingly good results even at finite values of spin, there exists an alternative method which makes far more conceptual and practical sense to utilize and we will turn to that in this section.

This alternative elegant way to calculate OPE data is by making use of the Lorentzian inversion formula \cite{Caron-Huot:2017vep,Simmons-Duffin:2017nub}. In addition to providing a resummation of the $1/\ell$ expansion, this formalism also allows one to compute nontrivial nonperturbative effects which are exponentially suppressed at large $\ell$ but may be important at smaller values of $\ell$~\cite{Sleight:2018epi, Cardona:2018dov, Sleight:2018ryu, Liu:2018jhs, Cardona:2018qrt}. Such effects are in fact needed in order to obtain a resummation which is analytic in $\ell$. We would like to take the opportunity to review a derivation of these effects, generalizing previous computations to different external dimensions and arbitrary $\SL(2,\mathbb{R})$ blocks, and also to illustrate their importance in 3d CFTs.\footnote{
	\label{footnote: inversion of y monomials}	
	The Lorentzian inversion formula yields an analytic complete result only as long as it is used properly. As we will review, Lorentzian inversion formula relies on \emph{inversion} of the conformal blocks --- which will be made clearer below --- and in practice, this translates into inversion of some other function with which the conformal block is written as an expansion over since closed form expressions for $3d$ conformal blocks are not available. Now, if the chosen function provides a convergent expansion for the conformal block for the whole range $z\in [0,1]$, then the inversion formula works properly. For instance, $k_{2\hb}(z)$ defined in \equref{eq:SL2R - definition of k} is such a function hence the expansion in \equref{eq:SL2R - expansion} with $k_{2\hb}$ inverted does provide the full analytic results with non-perturbative corrections. On the other hand, one can also expand conformal block as functions of $y\coloneqq\frac{z}{1-z}$ and invert the monomial $y^a$. As this expansion is only convergent for $z\in[0,\half]$, the result of Lorentzian inversion formula only provides the perturbative terms, same terms that would be obtained by large spin computations as large spin expansion is an asymptotic expansion determined by the region $z\sim 0$ (same region $y^a$ expansion is convergent). In practice, this reflects on the result of the inversion formula: one observes non-analytic behavior at finite spin \cite{Cardona:2018dov} --- for further details see footnote 28 of \cite{Liu:2018jhs}. 
}

For a 4-point function of scalars $\<\f_1\f_2\f_3\f_4\>$, the CFT inversion formula gives the OPE data for the $s$-channel in terms of two integrals of the function $g(z,\bar{z})$:
\begin{equation}
	c(h,\bar{h})=c^{t}(h,\bar{h})+(-1)^{\hb-h}c^{u}(h,\bar{h})\;,
\end{equation}
where
\bea[eq: Lorentzian inversion formula for scalars]
c^{t}(h,\bar{h})=&\frac{\kappa_{2\bar{h}}}{4}\int\limits_{0}^{1}dzd\bar{z}\mu(z,\bar{z})g^{r,s}_{d-1-h,\bar{h}}(z,\bar{z})\text{dDisc}_{t}[g(z,\bar{z})]\;,\label{eq:ctInv}
\\
\kappa_{2\hb}\equiv&\frac{\Gamma(\hb+r)\Gamma(\hb-r)\Gamma(\hb+s)\Gamma(\hb-s)}{2\pi^2\Gamma(2\hb-1)\Gamma(2\hb)}\;,
\\
\mu(z,\bar{z})=&\bigg|\frac{z-\bar{z}}{z\bar{z}}\bigg|^{d-2}\frac{\left((1-z)(1-\bar{z})\right)^{s-r}}{(z\bar{z})^{2}}\;,
\eea
and we recall that $r=h_{1}-h_2$ and $s=h_3-h_4$. The double discontinuity around $z=0$, which we call the $s$-channel dDisc, is defined by
\be
\text{dDisc}_{s}[g(z,\bar{z})]=\cos\left(\pi(s-r)\right)g(z,\bar{z})-\frac{1}{2}e^{i\pi(s-r)}g(ze^{2\pi i},\bar{z})-\frac{1}{2}e^{i\pi(r-s)}g(ze^{-2\pi i},\bar{z})\;. \label{eq:dDisc}
\ee
The $t$ and $u$-channel double discontinuities are defined in the same way, except around $z=1$ and $z=\infty$ respectively. The term $c^{u}(h,\bar{h})$ is also defined in the same way as (\ref{eq:ctInv}), but with the integration being taken from $-\infty$ to $0$.

In \cite{Caron-Huot:2017vep} the inversion formula was written in terms of $(\Delta,\ell)$, in which case the OPE coefficients for generic $\Delta$ are given by
\begin{equation}
	\label{eq: OPE coefficients from Simon's form of Lorentzian inversion formula}
	f_{12\cO}f_{34\cO}=-\Res_{\Delta'=\Delta}c(\Delta',\ell)\;, \quad \text{for fixed $\ell$}.
\end{equation}
Here we need to take residues of $c(h,\bar{h})$ with respect to $h$ at fixed $\bar{h}-h=\ell$, which will introduce some extra Jacobians as in the lightcone bootstrap. We will focus on $c^{t}$ since the $u$-channel can always be found by taking $1\leftrightarrow3$ and multiplying by $(-1)^{\ell}$.

It is convenient to define a generating function for the poles of $c^{t}(h,\bar{h})$:
\be
c^{t}(h,\bar{h})\bigg|_{\text{poles}}=\int\limits_{0}^{1}\frac{dz}{2z}z^{-h}C^{t}(z,\bar{h})\;.
\ee
The outer integral turns powers of $z$ inside $C^{t}(z,\bar{h})$ into poles for $h$. Since we are interested in the low-twist data, and in particular the $n=0$ double-twist operators, we study the small $z$ limit of $C(z,\bar{h})$:
\be
C^{t}(z,\bar{h})\approx \int\limits_{0}^{1}d\bar{z}\frac{(1-\bar{z})^{s-r}}{\zb^{2}}\kappa_{2\bar{h}}k^{r,s}_{2\bar{h}}(\bar{z})\text{dDisc}_{t}\left[\frac{(z\bar{z})^{h_1+h_2}g(1-z,1-\bar{z})}{[(1-z)(1-\bar{z})]^{h_{2}+h_{3}}}\right]\;,
\ee
where we have used crossing symmetry inside the dDisc.
This is a generating function for the $\SL(2,\mathbb{R})$ primaries with respect to $\bar{z}$ and to subtract descendants along $z$ we have to expand the inverted block in (\ref{eq:ctInv}) in powers of $z$. 

One nice example is to consider a 4-point function of identical scalars, $\<\f\f\f\f\>$. As in the lightcone bootstrap, terms regular and logarithmic in $z$ in $C^{t}(z,\bar{h})$ will correspond to corrections of OPE coefficients and scaling dimensions of the double-twist towers $[\f\f]_{n}$, respectively. To see this we will assume the anomalous dimensions of double-twist operators are small and expand $C^{t}(z,\bar{h})$ both in $z$ and the anomalous dimension:
\be
C^{t}(z,\bar{h})\approx  z^{2h_\f} P_{[\f\f]_0}(\bar{h})(1+\delta h_{[\f\f]_0}(\bar{h})\log(z)) + \ldots\;,
\ee
where the $\log(z)$ comes from a single $t-$channel conformal block. Integrating over $z$ we see the term regular in $z$ becomes a single pole while the term logarithmic in $z$ becomes a double pole. Some of these corrections at finite spin were recently derived in the works~\cite{Sleight:2018epi, Cardona:2018dov, Sleight:2018ryu, Liu:2018jhs, Cardona:2018qrt}. We will review these results and present some generalizations.

As an example, we can consider the exchange of a scalar operator $\mathcal{O}$ of twist $\tau_\cO=2h_\cO$ in the $t$-channel and use the inversion formula to extract the anomalous dimension of the $[\f\f]_0$ tower. In the limit $z\rightarrow 0$, the $\log(z)$ piece of the scalar block $g_{h_\cO,h_\cO}(1-z,1-\bar{z})$ is known in closed form. In a general 4-point function $\<\f_{1}\f_{2}\f_{3}\f_{4}\>$ the $t$-channel blocks develop logs when $h_{1}+h_{2}=h_{3}+h_{4}$, with the coefficient given by\footnote{The conformal block normalization is the same as in Eq.~(\ref{2.39 of Dolan:2011dv}).}
\begin{multline}
	g_{h_\cO,h_\cO}(1-z,1-\bar{z})\bigg|_{\log(z)}=-\log(z)\frac{\Gamma (2 h_{\cO})}{\Gamma (h_{1}-h_{4}+h_{\cO}) \Gamma (-h_{1}+h_{4}+h_{\cO})}  \\ 
	\times (1-\bar{z})^{h_{\cO}} \pFq{2}{1}{h_{1}-h_{4}+h_{\cO},-h_{1}+h_{4}+h_{\cO}}{2 h_{\cO}-\frac{d-2}{2}}{1-\bar{z}}\;.
\end{multline}

Focusing for now on the case where the external scalars are the same, we can match the $\log(z)$ term in the generating function, yielding the correction
\begin{multline}
	\hspace{-.5cm}(\delta h P)_{[\phi\phi]_0}(\hb)=- f_{\f\f\cO}^{2}\frac{\Gamma(2h_\cO)}{\Gamma(h_\cO)^{2}} \kappa_{2\bar{h}}\int\limits_{0}^{1}\frac{d\bar{z}}{\bar{z}^{2}} k_{2\bar{h}}(\bar{z})\\\times\dDisct{\left(\frac{\bar{z}}{1-\bar{z}}\right)^{2h_{\f}}(1-\bar{z})^{h_\cO}\pFq{2}{1}{h_\cO,h_\cO}{2h_\cO-\frac{d-2}{2}}{1-\bar{z}}}\;.
\end{multline}
Notice that this formula gives the product $\delta h \times P $ and one still needs to compute the corrected OPE coefficients, as well as add the $u$-channel contribution (identical up to a factor $(-1)^{\bar{h}-h}$), in order to find the anomalous dimension.

Using a hypergeometric transformation, we can rewrite this as
\begin{multline}\label{eq:zbintegral}
	(\delta h P)_{[\phi\phi]_0}(\hb)= - f_{\f\f\cO}^{2} \frac{\Gamma(2h_\cO)}{\Gamma(h_\cO)^{2}}\kappa_{2\bar{h}}  \int_0^1 \frac{d\bar{z}}{\bar{z}^2} \left(\frac{1-\bar{z}}{\bar{z}}\right)^{-\hb}
	\pFq{2}{1}{\hb,\hb}{2\hb}{-\frac{\bar{z}}{1-\bar{z}}}
	\\
	\times \bar{z}^{-h_\cO}
	\pFq{2}{1}{h_\cO,h_\cO-\frac{d-2}{2}}{2h_\cO-\frac{d-2}{2}}{-\frac{1-\bar{z}}{\bar{z}}}  \dDisct{(1-\bar{z})^{h_\cO} \left(\frac{\bar{z}}{1-\bar{z}}\right)^{\Delta_{\phi}}}\;.
\end{multline}
Finally, we can write the hypergeometric functions as a Mellin-Barnes integral and perform the $\bar{z}$ integral using the identity
\be
\int_0^1 \frac{d\bar{z}}{\bar{z}(1-\bar{z})} \left(\frac{\bar{z}}{1-\bar{z}}\right)^{\alpha} = 2\pi \delta(i\alpha) \label{eq:zIntegration}
\ee
to find:
\be
(\delta h P)_{[\phi\phi]_0}(\hb)=&-f_{\f\f\cO}^2    \sin^2 \left(\pi(h_\cO-2h_{\phi})\right) \frac{\Gamma(2h_\cO)\Gamma\left(2h_\cO-\frac{d-2}{2}\right)}{\pi^2\Gamma(h_\cO)^3\Gamma\left(h_\cO-\frac{d-2}{2}\right)}   \frac{\Gamma(\hb)^2}{\Gamma(2\hb-1)}
\\&
\times
\frac{1}{2\pi i} \int_{-i\infty}^{i\infty} ds'
\Bigg( \frac{\Gamma\left(h_\cO-2h_{\phi}+1+s'\right)^2\Gamma\left(\hb-h_\cO+2h_{\phi}-1-s'\right)}{\Gamma\left(\hb+h_\cO-2h_{\phi}+1+s'\right)}
\\&\hspace{1.25in} \frac{\Gamma\left(h_\cO+s'\right)\Gamma\left(h_\cO-\frac{d-2}{2}+s'\right)\Gamma(-s')}{\Gamma\left(2h_\cO-\frac{d-2}{2}+s'\right)}\Bigg)\;.
\ee
Finally, summing over the poles of $\Gamma(-s')$ gives
\small 
\begin{multline}
	(\delta h P)_{[\phi\phi]_0}(\hb)\big|_{\text{pert}}
	=-f_{\f\f\cO}^{2} \sin ^2(\pi  (h_{\cO}-2 h_{\f}))  \frac{\Gamma (2 h_{\cO})   \Gamma (h_{\cO}-2 h_{\f}+1)^2}{ \pi^2  \Gamma (h_{\cO})^2}
	\\  \times  \frac{\Gamma (\bar{h})^2 \Gamma(\bar{h}-h_{\cO}+2h_{\phi}-1)}{\Gamma (2 \bar{h}-1) \Gamma(\bar{h}+h_{\cO}-2h_{\phi}+1)} \pFq{4}{3}{h_{\cO},h_{\cO}-\frac{d-2}{2},h_{\cO}-2 h_{\f}+1,h_{\cO}-2 h_{\f}+1}{-\bar{h}+h_{\cO}-2 h_{\f}+2,\bar{h}+h_{\cO}-2 h_{\f}+1,2 h_{\cO}-\frac{d-2}{2}}{1}\;,
\end{multline}
\normalsize
which when expanded at large $\bar{h}$ reproduces the perturbative $1/\bar{h}$ expansion from the lightcone bootstrap. It can also be obtained by expanding the hypergeometric function inside Eq.~(\ref{eq:zbintegral}) at small $\bar{z}$, performing the integrals term-by-term, and resumming the result. However, this resummation contains spurious poles in $\bar{h}$, leading to a non-analytic function, connected to the fact that performing the small $\bar{z}$ expansion inside the integral fails to correctly capture its behavior near $\bar{z} \sim 1$ as noted in footnote~\ref{footnote: inversion of y monomials}.  

Additionally summing over the poles of $\Gamma\left(\bar{h} -h_\cO+2h_\phi-1-s'\right)$ gives the correction
\be
(\delta h P)_{[\phi\phi]_0}(\bar{h})\big|_{\text{nonpert}}
=&- f_{\f\f\cO}^2    \sin^2 \left(\pi(h_\cO-2h_{\phi})\right) \frac{\Gamma(2h_\cO)\Gamma\left(2h_\cO-\frac{d-2}{2}\right)}{\pi^2\Gamma(h_\cO)^3\Gamma\left(h_\cO-\frac{d-2}{2}\right)}   \frac{\Gamma(\bar{h})^4 }{\Gamma(2\bar{h}-1)\Gamma(2\bar{h})}
\\
&\times \frac{ \Gamma(-\bar{h}+h_\cO-2h_{\phi}+1)\Gamma(\bar{h}+2h_{\phi}-1)\Gamma(\bar{h}+2h_{\phi}-\frac{d}{2})}{\Gamma(\bar{h}+h_\cO+2h_{\phi}-\frac{d}{2})}  \\&\x\pFq{4}{3}{\bar{h}, \bar{h}, \bar{h} + 2 h_{\f}-1,\bar{h}+2 h_{\f}-\frac{d}{2}}{2 \bar{h},\bar{h}-h_{\cO}+2 h_{\f},\bar{h}+h_{\cO}+2 h_{\f}-\frac{d}{2}}{1}\;.
\ee

The full sum $(\delta h P)_{[\phi\phi]_0}(\bar{h})=(\delta h P)_{[\phi\phi]_0}(\bar{h})\big|_{\text{pert}}+(\delta h P)_{[\phi\phi]_0}(\bar{h})\big|_{\text{nonpert}}$ has no spurious poles, and the same asymptotics as $(\delta h P)_{[\phi\phi]_0}(\bar{h}) \big|_{\text{pert}}$, since 
\be
\left(\delta h_{[\phi\phi]_0}(\bar{h})\right)_{\text{nonpert}} \sim 4^{-\bar{h}}\bar{h}^{1/2-4h_{\phi}}
\ee
is exponentially damped at asymptotically large $\bar{h}$. Such exponentially damped contributions can be understood as arising from the region of integration near $\bar{z} \sim 1$, while the perturbative contributions come from expanding the integrand near $\bar{z} \sim 0$. 

We will now generalize the log matching to different external dimensions by considering the 4-point function $\<\f_1\f_2\f_2\f_1\>$. The $s$-channel OPE data is given by integrating over the $t$- and $u$-channel double discontinuities, so the anomalous dimensions are given by
\begin{equation}
	\left(\delta h P\right)_{[\phi_1\phi_2]_0}(\hb)= f_{11\cO}f_{22\cO} \left(\delta h P\right)_{1221}(\hb)+ f_{12\cO}^{2}\left(\delta h P\right)_{1212}(\hb)\;,
\end{equation}
where
\be 
{}&\left(\delta h  P\right)_{1234}(\hb)\bigg|_{\text{pert}}=  -\sin (\pi  (h_{1}+h_{4}-h_{\cO})) \sin (\pi  (h_{2}+h_{3}-h_{\cO})) \\ 
&\;\times \frac{\Gamma (2 h_{\cO}) \Gamma(h_{\cO}-h_2-h_3+1) \Gamma(h_{\cO}-h_1-h_4+1)}{\pi^2 \Gamma (h_{\cO}+h_{2}-h_{3}) \Gamma (h_{\cO}-h_{2}+h_{3})}\\
&\;\times \frac{ \Gamma (\bar{h}+h_{1}-h_{2}) \Gamma(\bar{h}+h_3-h_4) \Gamma(\bar{h}-h_\cO+h_2+h_4-1) }{\Gamma (2 \bar{h}-1) \Gamma(\bar{h}+h_{\cO}-h_2-h_4+1)} \\
&\;\times \pFq{4}{3}{h_{\cO}-h_{2}+h_{3},h_{\cO}-h_{2}+h_{3}-\frac{d-2}{2},h_{\cO}-h_{2}-h_{3}+1,h_{\cO}-h_{1}-h_{4}+1}{-\bar{h}+h_{\cO}-h_{2}-h_{4}+2,\bar{h}+h_{\cO}-h_{2}-h_{4}+1,2 h_{\cO}-\frac{d-2}{2}}{1}\;,
\ee 
\be 
\left(\delta h P\right)_{1234}(\hb)&\bigg|_{\text{nonpert}}=- \sin (\pi  (h_{1}+h_{4}-h_{\cO})) \sin (\pi  (h_{2}+h_{3}-h_{\cO}))\\
&\times \frac{\Gamma (2 h_{\cO}) \Gamma \left(2 h_{\cO}-\frac{d-2}{2}\right) }{\pi^2  \Gamma (h_{\cO}+h_{2}-h_{3}) \Gamma (h_{\cO}-h_{2}+h_{3})^2 \Gamma \left(h_{\cO}-h_{2}+h_{3}-\frac{d-2}{2}\right)}
\\
&\times \frac{\Gamma (\bar{h}+h_{1}-h_{2}) \Gamma (\bar{h}-h_{1}+h_{2})  \Gamma (\bar{h}+h_{3}-h_{4}) \Gamma (\bar{h}-h_{3}+h_{4}) }{ \Gamma(2\bar{h}) \Gamma (2 \bar{h}-1) }
\\
&\times \frac{\Gamma(-\bar{h}+h_{\cO}-h_{2}-h_{4}+1)\Gamma (\bar{h}+h_{1}+h_{2}-1) \Gamma \left(\bar{h}+h_{1}+h_{2}-\frac{d}{2}\right)}{\Gamma(\bar{h}+h_{\cO}+h_2+h_4-\frac{d}{2})} \\
&\times \pFq{4}{3}{\bar{h}-h_{1}+h_{2},\bar{h}-h_{3}+h_{4},\bar{h}+h_{1}+h_{2}-1,\bar{h}+h_{1}+h_{2}-\frac{d}{2}}{2 \bar{h},\bar{h}-h_{\cO}+h_{2}+h_{4},\bar{h}+h_{\cO}+h_{2}+h_{4}-\frac{d}{2}}{1}\;.
\ee 

To derive these expressions from the inversion formula, we had to set $h_{1}+h_{2}=h_{3}+h_{4}$ so the $u$- and $t$-channel blocks have $\log(z)$ terms, but we left this equality implicit in the above expression.

Corrections to OPE coefficients can be derived in a similar way, by matching regular terms in $t$-channel conformal blocks. Somewhat cumbersome formulas for such corrections in general dimension and for general spin exchange were given in~\cite{Cardona:2018qrt}. In \secref{\ref{sec: scalar nonperturbative}} we will describe an alternate and perhaps simpler approach to obtaining anomalous dimensions and OPE coefficient corrections in 3d CFTs, via dimensional reduction.

\section{A detour: effect of non-perturbative terms in scalar CFTs}
\label{sec: scalar nonperturbative}
\subsubsection{Ising CFT}
To demonstrate why nonperturbative corrections can be important, we would like to see how they affect analytic predictions for the 3d Ising CFT. We will restrict ourselves to the 4-point function $\<\s\s\s\s\>$ and extract predictions for the $[\s\s]_{0}$ scaling dimensions and OPE coefficients. We will improve the results found in \cite{Alday:2015ewa,Simmons-Duffin:2016wlq}.

For the Ising CFT we will focus on the effects of three operators, the identity operator $\mathbb{1}$, the lightest parity-even scalar $\epsilon$, and the stress-tensor $T^{\mu\nu}$. We will also use the following results from the numerical bootstrap \cite{ElShowk:2012ht,Simmons-Duffin:2016wlq} as inputs:
\be
h_{\s}{}&={}0.25907445(50)\;, \qquad
h_{\epsilon}{}={}0.7063125(50)\;, \qquad
h_{T}{}={}0.5\;,
\\
&\hspace{.25in}f_{\s\s\epsilon}{}={}1.0518537(41)\;, \qquad
f_{\s\s T}{}={}0.32613776(45)\;.
\ee

To use the inversion formula, we will use dimensional reduction to write the 3d blocks as sums of 2d blocks \cite{Hogervorst:2016hal}. Specifically, we use the expansion\footnote{Our parametrization differs from \cite{Hogervorst:2016hal}, so $\mathcal{A}^{\text{here}}_{n,j}=\mathcal{A}^{\text{there}}_{\frac{n+j}{2},\bar{h}-h+j-n}$, and our normalization is such that $c^{(d)}_{\ell}=\frac{(d-2)_\ell}{\left(\frac{d-2}{2}\right)_\ell}$ in Eq. (2.35) of \cite{Hogervorst:2016hal}.}
\bea
g^{r,s,(3d)}_{h,\bar{h}}(z,\bar{z})&=\sum\limits_{n=0}^{\infty} \ \sum\limits_{\begin{tiny}j=\max(-n,n-\ell)\end{tiny}}^{n}\mathcal{A}^{r,s}_{n,j}(h,\bar{h})g^{r,s,(2d)}_{h+n,\bar{h}+j}(z,\bar{z})\;,
\\
g^{r,s,(2d)}_{h,\bar{h}}(z,\bar{z})&=\frac{1}{1+\delta_{\bar{h}-h,0}}\left(z^h {}_{2}F_{1}(h+r,h+s,2h,z)\bar{z}^{\bar{h}} {}_{2}F_{1}(\bar{h}+r,\bar{h}+s,2\bar{h},\bar{z})+(z\leftrightarrow\bar{z})\right)\;.
\eea
In \cite{Hogervorst:2016hal} this expansion was derived in closed form for $r=s=0$, which will be sufficient for our calculations.

Since each 2d block is a sum of hypergeometrics, we can use the same techniques as when inverting a single scalar block in the previous section. Specifically, after setting $r=s=0$ and extracting the leading $z \rightarrow 0$ behavior of the hypergeometrics in the $t$-channel, we have contributions from $\SL(2,\mathbb{R})$ blocks of the form
\be
g^{(2d)}_{h_\cO,\bar{h}_\cO}(1-z,1-\bar{z})\bigg|_{z\rightarrow 0} =&\; \frac{1}{1+\delta_{\bar{h}_\cO-h_\cO,0}} \left[ - \frac{\Gamma(2 h_\cO)}{\Gamma(h_\cO)^2} (\log(z) + 2 \psi^{(0)}(h_\cO) + 2 \gamma) + \ldots \right] \\
&\times (1-z)^{h_{\cO}} k_{2\bar{h}_{\cO}}(1-\bar{z}) + (h_{\cO} \leftrightarrow \bar{h}_{\cO})\;.
\ee
Here $\psi^{(0)}(z) = \Gamma'(z)/\Gamma(z)$ is the digamma function and $\gamma$ is the Euler constant. We then we find the following corrections to the OPE coefficients and anomalous dimensions after inverting the $t$-channel block: 
\newpage
\bea 
\left(\delta h P \right)^{h_\cO,\bar{h}_\cO}_{[\s\s]_{0}} (\bar{h})&\bigg|_{\text{pert}}=- f_{\s\s\cO}^{2} \sin ^2(\pi  (2 h_{\s}-\bar{h}_{\cO}))   \nn\\
&\times \frac{\Gamma(2 h_{\cO}) \Gamma (\bar{h}_{\cO}-2 h_{\s}+1)^2}{\pi^2 \Gamma (h_{\cO})^2 } \frac{\Gamma (\bar{h})^2}{\Gamma(2\bar{h}-1) } \frac{   \Gamma(\bar{h}-\bar{h}_{\cO}+2 h_{\s} -1)}{\Gamma(\bar{h}+\bar{h}_{\cO}-2h_\s+1)} \nn\\
&\times \pFq{4}{3}{\bar{h}_{\cO},\bar{h}_{\cO},\bar{h}_{\cO}-2 h_{\s}+1,\bar{h}_{\cO}-2 h_{\s}+1}{2 \bar{h}_{\cO},-\bar{h}+\bar{h}_{\cO}-2 h_{\s}+2,\bar{h}+\bar{h}_{\cO}-2 h_{\s}+1}{1}\;,
\\ \nn
\\
\left(\delta h P\right)^{h_\cO,\bar{h}_\cO}_{[\s\s]_{0}}(\bar{h})&\bigg|_{\text{nonpert}}=- f_{\s\s\cO}^{2} \sin ^2(\pi  (2 h_{\s}-\bar{h}_{\cO})) \nn\\
&\times \frac{\Gamma(2 h_{\cO})\Gamma(2 \bar{h}_{\cO})}{\pi^2 \Gamma (h_{\cO})^2 \Gamma (\bar{h}_{\cO})^2 }  \frac{ \Gamma (\bar{h})^4  }{ \Gamma(2\bar{h}-1) \Gamma(2\bar{h}) } \frac{\Gamma(-\bar{h}+\bar{h}_\cO-2h_\s+1) \Gamma (\bar{h}+2 h_{\s}-1)^2}{\Gamma (\bar{h}+\bar{h}_{\cO}+2 h_{\s} -1) } \nn\\
& \x\pFq{4}{3}{\bar{h},\bar{h},\bar{h}+2 h_{\s}-1,\bar{h}+2 h_{\s}-1}{2 \bar{h},\bar{h}-\bar{h}_{\cO}+2 h_{\s},\bar{h}+\bar{h}_{\cO}+2 h_{\s}-1}{1}\;,
\eea 
with the net contribution from a given 2d block in both the $t-$ and $u$-channels given by
\be
(\delta h P)_{[\s\s]_{0}} =&\; \frac{1+(-1)^{\bar{h}-h}}{1+\delta_{\bar{h}_\cO-h_\cO,0}} \left[\left(\delta h P \right)^{h_\cO,\bar{h}_\cO}_{[\s\s]_{0}} (\bar{h})\bigg|_{\text{pert}} + \left(\delta h P \right)^{h_\cO,\bar{h}_\cO}_{[\s\s]_{0}} (\bar{h})\bigg|_{\text{nonpert}} \right] \\
& + (h_\cO \leftrightarrow \bar{h}_\cO)\;.
\ee

\begin{figure}
	\centering
	\includegraphics[width=\textwidth]{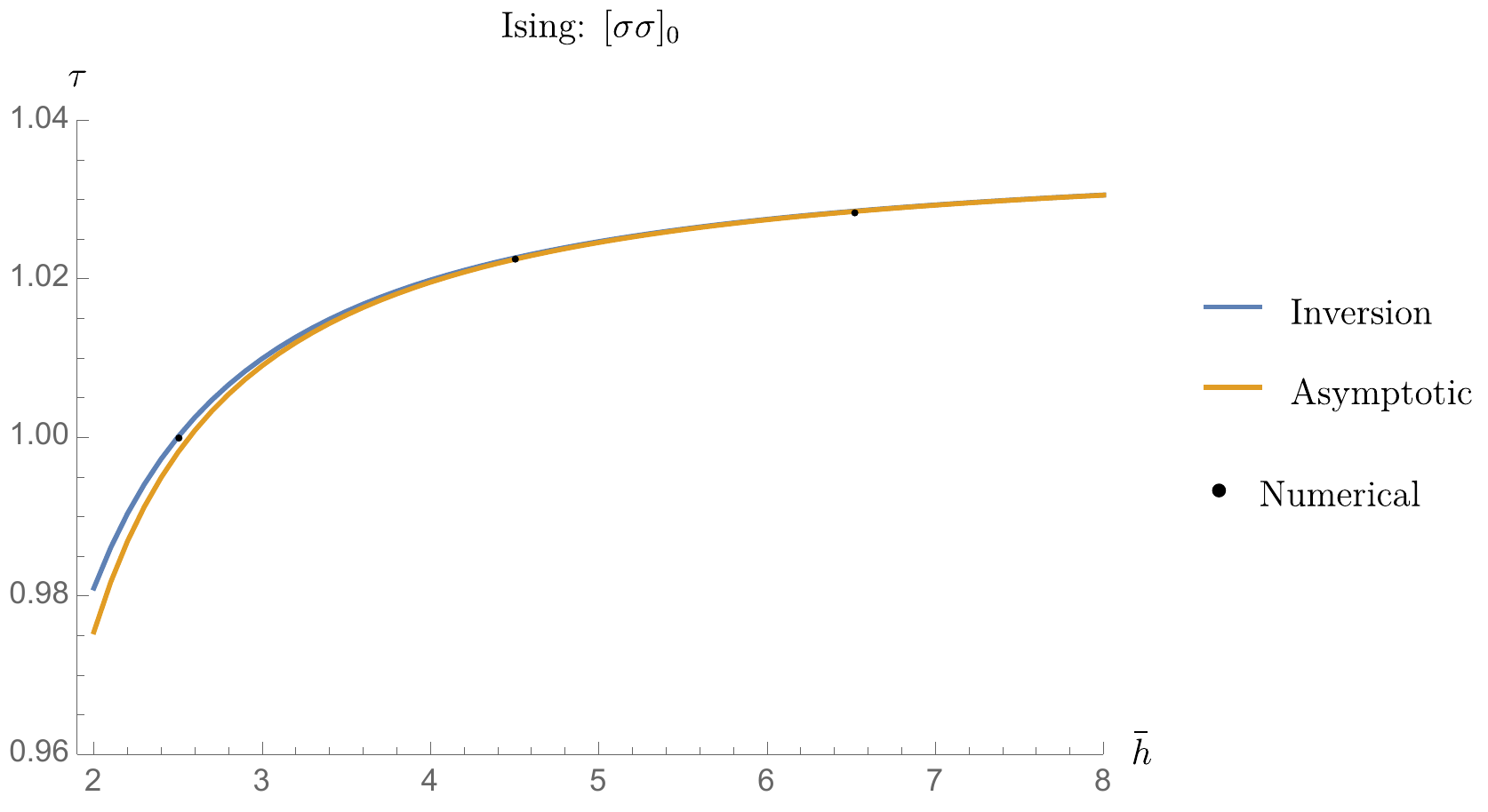}
	\caption[Spectrum  in the Ising model]{Spectrum for $[\sigma\sigma]_{0}$ in the Ising CFT derived using the inversion formula, asymptotic lightcone expansion, and numerical bootstrap. Numerical data is taken from \cite{Simmons-Duffin:2016wlq}. The curves in this and later plots are obtained by matching at $z=.1$.}
\end{figure}

Similarly, by matching regular terms we obtain the corrections to the OPE coefficients
\be
\delta P_{[\s\s]_{0}} =&\;  \frac{1+(-1)^{\bar{h}-h}}{1+\delta_{\bar{h}_\cO-h_\cO,0}} \left(2 \psi^{(0)}(h_\cO) + 2 \gamma\right)  \left[ \left(\delta h P \right)^{h_\cO,\bar{h}_\cO}_{[\s\s]_{0}} (\bar{h})\bigg|_{\text{pert}} + \left(\delta h P \right)^{h_\cO,\bar{h}_\cO}_{[\s\s]_{0}} (\bar{h})\bigg|_{\text{nonpert}} \right]\\&+ (h_\cO \leftrightarrow \bar{h}_\cO)\;.
\ee

Note that if we take $\cO = \mathbb{1}$ to be the identity operator and take the limit \mbox{$h_\cO = \bar{h}_\cO \rightarrow 0$} (as well as set $f_{\s\s\mathbb{1}} = 1$), then we reproduce the expected identity contribution $P_{[\s\s]_{0}} = (1+ (-1)^{\bar{h}-h}) S_{-2h_\s}(\bar{h})$.

At finite spin and finite anomalous dimensions one does not expect that it is sufficient to match the terms logarithmic and regular in $z$ to obtain the precise OPE data. Although inverting individual operators produces factors of $z^{h_{[\s\s]_{n,\ell}}}$ and $z^{h_{[\s\s]_{n,\ell}}}\log z$, we know the exact generating function $C^t(z,\bar{h})$ at small $z$ is~\cite{Caron-Huot:2017vep}:
\begin{align}
	C^t(z,\bar{h})=C_{[\s\s]_{0}}(\bar{h})z^{2h_{\s}+\delta h_{[\s\s]_{0}}(\bar{h})}+... \text{ ,}
\end{align}
where we ignore terms subleading in $z$.

We can then extract the anomalous dimension via:
\begin{align}
	\delta h_{[\s\s]_{0}}(\bar{h})= \lim\limits_{z\rightarrow 0}\frac{(z\partial_{z}-2h_{\s})C^t(z,\bar{h})}{C^t(z,\bar{h})}\;,
\end{align}
which we in practice evaluate by evaluating the generating function at small but finite $z$. We find the OPE coefficients in a similar way by taking our value for $\delta h_{[\s\s]_{0}}(\bar{h})$ and using:
\begin{align}
	C_{[\s\s]_{0}}(\bar{h})=\lim\limits_{z\rightarrow 0}\frac{C^t(z,\bar{h})}{z^{2h_{\s}+\delta h_{[\s\s]_{0}}(\bar{h})}}\;,
\end{align}
where we once again evaluate the right-hand side at small but finite $z$.

In evaluating these expressions one wishes to take $z$ small, but not too small so as to avoid neglected terms with higher powers of $\log z$ from becoming important. In~\cite{Simmons-Duffin:2016wlq} it was found that $z=.1$ is a good choice for the Ising model (there called $\bar{y}_0$), so we will present results at this value in our analysis. It may be helpful to further optimize the matching value of $z$. As more operators are included one should also see that the results become less and less sensitive to this choice.

Now the procedure should be clear: we can expand the 3d blocks as sums of 2d blocks and invert each block term by term. This procedure is sufficient to extract finite-spin data from the Lorentzian inversion formula. In practice we find that we need to expand to at most 10 to 15 orders in the 2d expansion such that the errors introduced by truncating this expansion are smaller than the errors from the numerical input. 

\begin{figure}
	\centering
	\includegraphics[width=\textwidth]{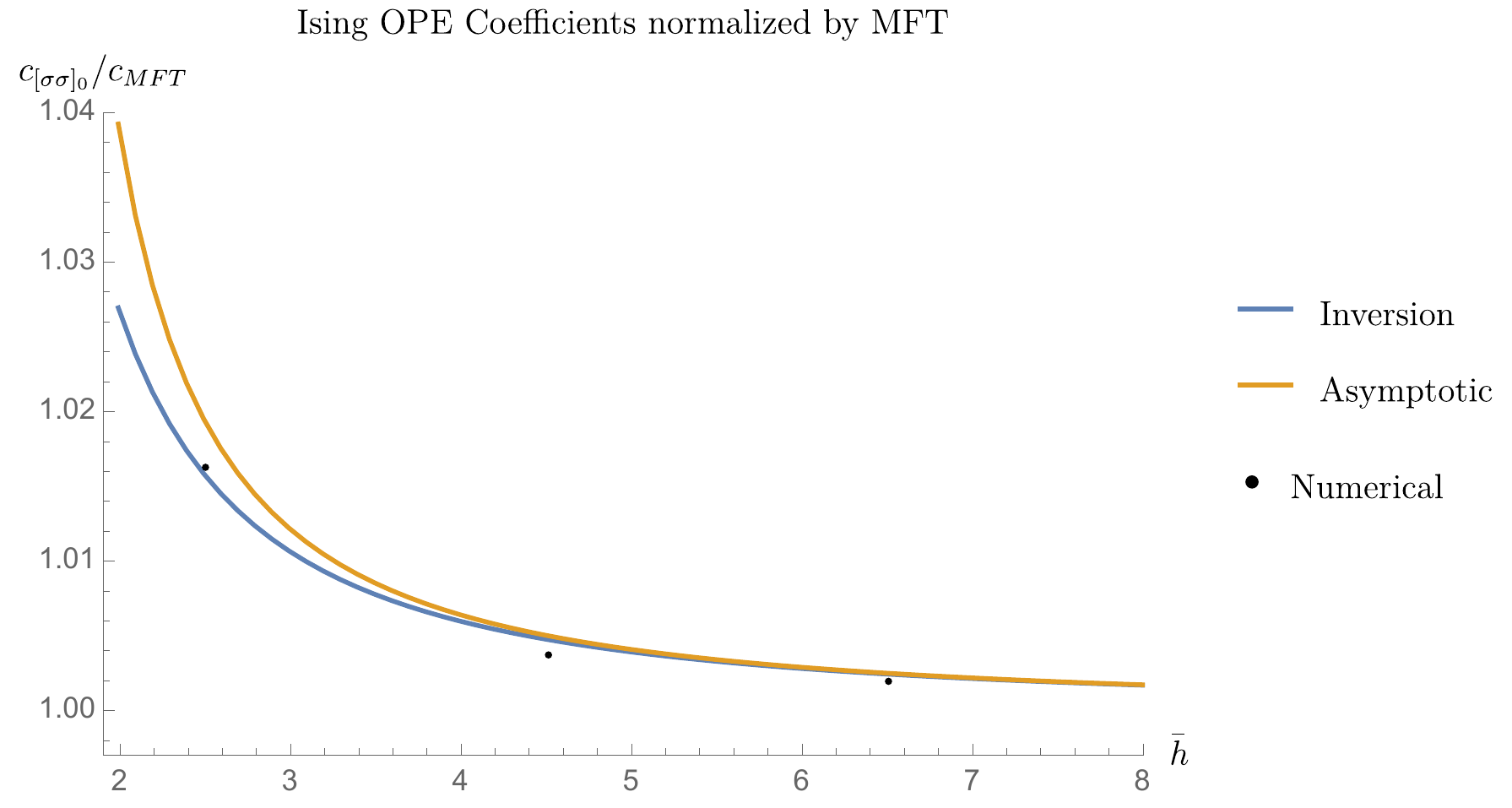}
	\caption[OPE coefficients in the Ising model]{OPE coefficients  $f_{\s\s[\s\s]_{0}}$ in the Ising CFT. Numerical data is taken from \cite{Simmons-Duffin:2016wlq} and the OPE coefficients are normalized by dividing by the mean field theory OPE coefficients.}
\end{figure}

With this data and the above expressions, we can extract $P_{[\s\s]_{0}}$ and $\delta h_{[\s\s]_{0}}$, but we have to do a little more work to extract the physical OPE coefficients and scaling dimensions. To find the scaling dimensions, we need to solve the equation
\be
\bar {h}-2h_{\s}-\delta h_{[\s\s]_{0}}(\bar{h})=\ell\;,
\ee
where $\ell$ is the spin of the local, double-twist operator. As the anomalous dimensions are expressed in terms of ${}_{4}F_{3}$ hypergeometric functions, we will solve this equation numerically. We then calculate the physical OPE coefficients $f_{\s\s[\s\s]_{0}}$ using the relation
\be
f^{2}_{\s\s[\s\s]_{0}}\approx\left(1-\frac{\partial \delta h_{[\s\s]_{0}}(\bar{h})}{\partial\bar{h}}\right)^{-1}P_{\s\s[\s\s]_{0}}\;,
\ee
where the Jacobian appears because we need to take residues of the OPE function $c(\Delta,\ell)$ in terms of $\Delta$ at fixed spin $\ell$.

A comparison of results from the numerical bootstrap~\cite{Simmons-Duffin:2016wlq}, the leading asymptotic lightcone bootstrap (\ref{eq:asymptoticP},~\ref{eq:asymptoticdh}), and the inversion formula result, can be found in Table \ref{table:IsingInversion}. We focus here on how accurately we can reproduce the low-spin data. We see that in all cases, including the nonperturbative effects from the inversion formula leads to more accurate results. This is clearest for the scaling dimensions, where we have at least an extra digit of precision for the lightest spin-$2$ and spin-$4$ operators. 

This improvement is especially marked for the stress-tensor and gives additional evidence that the stress-tensor should be thought of as a double-twist operator composed of two $\sigma$ operators. We see a similar improvement for the OPE coefficients, although it is smaller in comparison to the dimensions. The errors listed come from the errors in the numerical input and do not include errors from truncating the operator product expansion to include only a few light operators. As we include more operators beyond $\epsilon$ and $T^{\mu\nu}$ we expect the results to improve even further.

\begin{table}
	\caption[Results for the twists and OPE coefficients in the 3d Ising model]{\label{table:IsingInversion}We list results for the twists and OPE coefficients for the double-twist family $[\s\s]_{0,\ell}$ in the 3d Ising model by either matching at $z=.1$ or using the na\"ive $\log z$ matching valid for perturbative anomalous dimensions.  Approximate errors come from numerical input.}
	\begin{tabularx}{\textwidth}{%
			>{\hsize=.18\hsize\linewidth=\hsize}X%
			>{\hsize=.3\hsize\linewidth=\hsize}X%
			>{\hsize=.3\hsize\linewidth=\hsize}X%
			>{\hsize=.3\hsize\linewidth=\hsize}X%
			>{\hsize=.3\hsize\linewidth=\hsize}X%
			>{\hsize=.3\hsize\linewidth=\hsize}X%
		}
		\hline\hline
		{} & \textbf{Numerics}&  \textbf{Inversion${}_{z=.1}$}& \textbf{Inversion${}_{\log z}$}& \textbf{Lightcone${}_{z=.1}$}  & \textbf{Lightcone${}_{\log z}$}\\\hline
		$\tau_{[\s\s]_{0,2}}$& 1 & 1.000060(2) & 0.998459(4) &0.998082(4) &0.9962944(46) \\ 
		$\tau_{[\s\s]_{0,4}}$&1.022665(28)&1.0226890(7)&1.022472(3)&1.022510(3)&1.0222880(28)\\ 
		$f_{\s\s[\s\s]_{0,2}}$&0.32613776(45) &0.325981(1) &0.3262377(9)&0.327398(1) &0.3277057(10)\\ 
		$f_{\s\s[\s\s]_{0,4}}$&0.069076(43)&0.0691405(2)&0.0691445(2) &0.0691630(2)&0.0691671(2)\\ \hline \hline
	\end{tabularx}
\end{table}\

%
%
%

\subsubsection{O(2) model}

\begin{figure}
	\centering
	\includegraphics[width=\textwidth]{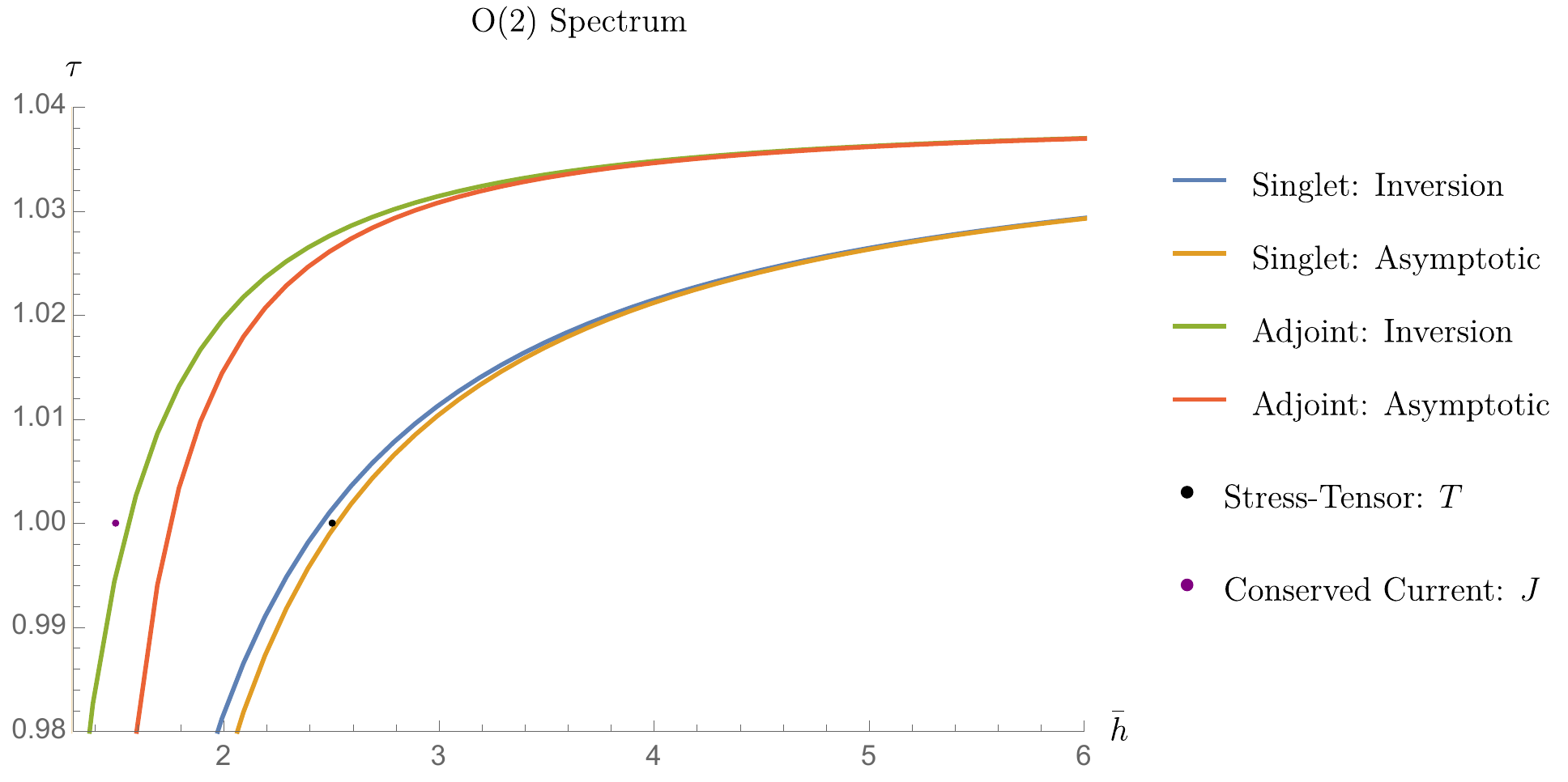}
	\caption[Spectrum in the O(2) model]{Spectrum of $[\f\f]_{0}^{(I)}$ and $[\f\f]_{0}^{(A)}$ in the O(2) model. The black dots corresponds to the stress tensor and conserved current which have twist one.}
\end{figure}

We can repeat the above analysis, but now for the O(2) vector model. We will study the 4-point function of fundamental scalars $\<\f^{i}\f^{j}\f^{k}\f^{\ell}\>$, which we can decompose in terms of the exchanged global symmetry representations as
\begin{multline}
	x_{12}^{2\Delta_{\f}}x_{34}^{2\Delta_{\f}}\<\f_{i}\f_{j}\f_{k}\f_{l}\>=\delta_{ij}\delta_{kl}I(u,v)+(\delta_{il}\delta_{jk}-\delta_{ik}\delta_{jl})A(u,v)
	\\ + \left(\delta_{il}\delta_{jk}+\delta_{ik}\delta_{jl}-\delta_{ij}\delta_{kl}\right)S(u,v)\;,
\end{multline}
where $I$, $A$, and $S$ correspond to contributions from exchanged operators that transform in the singlet, antisymmetric, and symmetric traceless representation of O(2), respectively. 

If we collect them into a vector $\vec{Z}(u,v)=\{I(u,v),A(u,v),S(u,v)\}$, then $(1,i)\leftrightarrow (3,k)$ crossing implies
\begin{subequations}
	\be 
	\left(\frac{u}{v}\right)^{\Delta_{\f}}\vec{Z}(u,v)=M\cdot \vec{Z}(u,v)\;,
	\ee 
	for
	\be 
	M=\left( \begin{array}{ccc}
		\frac{1}{2} & \frac{1}{2} &  1 \vspace{.085cm} \\
		\frac{1}{2} & \frac{1}{2} & -1  \vspace{.085cm} \\
		\frac{1}{2} & -\frac{1}{2} & 0 
	\end{array}
	\right). \label{eqn:matO2}
	\ee 
\end{subequations}

We will use the following results from the numerical bootstrap \cite{Kos:2013tga,Kos:2015mba,Kos:2016ysd}:
\begin{subequations}
	\begin{alignat}{2}
		h_{\f}&=0.25963(16)\;,
		& h_{\f^{2}}&=0.7559(13)\;,
		\\
		h_{t}&=0.6179(16)\;, 
		& f_{\f\f\f^{2}}&=0.68726(65)\;,
		\\
		f_{\f\f J}&=0.52558(46)\;,\quad  &
		f_{\f\f T}&=0.23146(16)\;.
	\end{alignat}
\end{subequations}

Here $t$ refers to the lightest symmetric, traceless scalar in the $\f_{i} \times \f_{j}$ OPE. There is one crucial piece of OPE data missing, the OPE coefficient $f_{\f\f t}$, although there are estimates from the $\epsilon$-expansion \cite{Dey:2016mcs}, which yield
\begin{align}
	f_{\f\f t} &\approx \{0.8944,\, 0.8246,\, 0.8850\} \quad \text{at} \quad \{\cO(\epsilon),\, \cO(\epsilon^2),\, \cO(\epsilon^3)\}.\label{eq:O2opeRange}
\end{align}

\begin{figure}
	\centering
	\includegraphics[width=\textwidth]{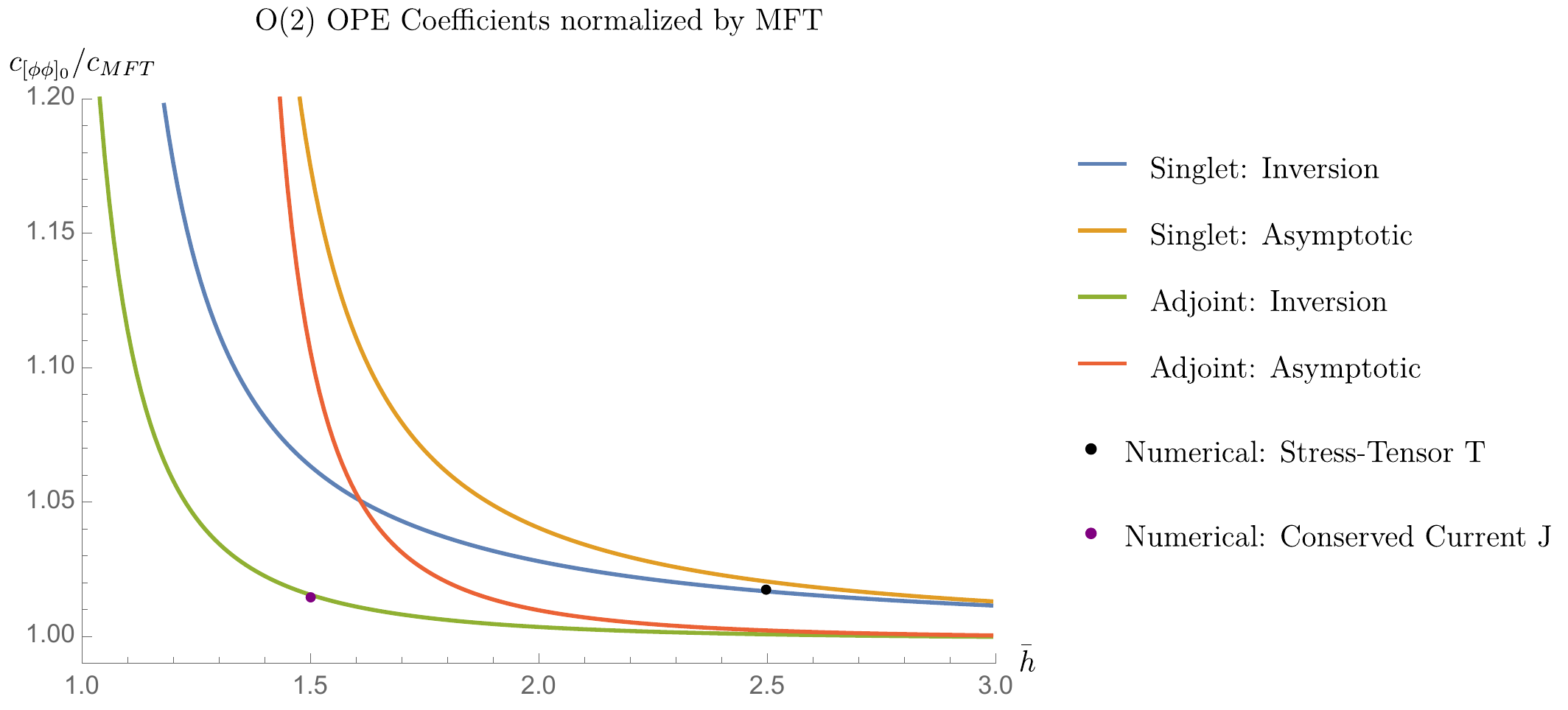}
	\caption[OPE coefficients in the O(2) model]{OPE coefficients for $f_{\f\f[\f\f]^{(I)}_{0}}$ and $f_{\f\f[\f\f]^{(A)}_{0}}$ in the O(2) model. Numerical data is taken from \cite{Kos:2013tga} and the OPE coefficients are normalized by dividing by the mean field theory OPE coefficients.}
\end{figure}

Using this data as input, we can calculate the low-spin spectrum for the O(2) vector model using either the asymptotic lightcone bootstrap or the inversion formula. In our calculations we expanded to $12^{th}$ and $20^{th}$ order in the 2d conformal blocks to obtain converged results for the stress-tensor and conserved current OPE data, respectively. The results are shown in Table \ref{table:O2Inversion}.

We see that the inversion formula in general gives more accurate results for both the conserved current $J^{\mu}$ and the stress-tensor $T^{\mu\nu}$. The improvement is particularly large for $c_{\f\f[\f\f]^{A}_{0,1}}$, or the coupling between two scalars and $J^{\mu}$. One reason the inversion formula gives an improved estimate for this OPE coefficient is because it also gives a much more accurate result for $\tau_{[\f\f]^{(A)}_{0,1}}$, which is used as input in the calculation of the OPE coefficient. Finally, we should note that the inversion formula is only guaranteed to hold for spin $J>1$, but we see for the O(2) vector model it likely holds down to at least $J=1$. 

The one exception appears to be the twist of the stress-tensor itself, for which the lightcone analysis gives a result which is slightly closer to the exact answer. We expect this is an artifact of truncating the $t$-channel expansion: as we include more operators the results for the twist will decrease which will push the lightcone result further from the exact result.

\begin{table}
	\caption[Results for the twists and OPE coefficients in the O(2) model]{\label{table:O2Inversion}We list results for the twists and OPE coefficients for the double-twist family $[\f\f]_{0,\ell}$ in the 3d O(2) model by either matching at $z=.1$ or using the na\"ive $\log z$ matching valid for perturbative anomalous dimensions. The errors are approximate and come from both numerical input and from using the lower and upper values in (\ref{eq:O2opeRange}).}
	\begin{tabularx}{\textwidth}{%
			>{\hsize=.18\hsize\linewidth=\hsize}X%
			>{\hsize=.3\hsize\linewidth=\hsize}X%
			>{\hsize=.3\hsize\linewidth=\hsize}X%
			>{\hsize=.3\hsize\linewidth=\hsize}X%
			>{\hsize=.3\hsize\linewidth=\hsize}X%
			>{\hsize=.3\hsize\linewidth=\hsize}X%
		}
		\hline\hline
		{} & \textbf{Numerics}& \textbf{Inversion${}_{z=.1}$} & \textbf{Inversion${}_{\log z}$} & \textbf{Lightcone${}_{z=.1}$} & \textbf{Lightcone${}_{\log z}$}\\
		\hline
		$\tau_{[\f\f]^{(I)}_{0,2}}$  &  1&1.0012(24) &0.9996(26) & 0.9992(25) & 0.9973(27)\\ 
		$\tau_{[\f\f]^{(A)}_{0,1}}$ & 1 & 0.9958(60)& 0.9933(66) & 0.9480(90)& 0.933(13) \\ 
		$f_{\f\f[\f\f]^{(I)}_{0,2}}$  &0.231462(16) &0.23128(46)&0.23147(48)&0.23231(50)&0.23254(52)\\
		$f_{\f\f[\f\f]^{(A)}_{0,1}}$  & 0.52558(46) & 0.5270(32)& 0.5286(36)&0.6005(97) &0.630(17)\\ \hline\hline 
	\end{tabularx}
\end{table}

%

We can also take a different point of view and use the inversion formula to make a prediction for $f_{\f\f t}$. For example, if we require that the inversion formula reproduces the exact twist of $T^{\mu\nu}$ then we find the following range:
\be
\qquad f_{\f\f t}\in(0.857,0.951)\,,
\ee
with a central value of approximately $f_{\f\f t}=0.9038$. Using results from Monte Carlo~\cite{Campostrini:2006ms} as input, setting $h_{\f}=0.259525(50)$, $h_{\f^{2}}=0.75562(11)$, and $h_{t}=0.6180(5),$\footnote{This range comes from comparing the bootstrap data in Figure 9 of~\cite{Kos:2015mba} with the Monte Carlo allowed region.} and repeating the above analysis, the window shrinks to
\be
\qquad f_{\f\f t}\in (0.883,0.901)\,.
\ee

By including more operators in the inversion formula or the effects of operator mixing it should be possible to improve the above results further. It will be interesting to understand which operators need to be included in order to reproduce the current beyond the $10^{-3}$ level. It would also be interesting to extend this work to higher orders in the small-$z$ expansion to understand the higher-twist families.

\section{Review: harmonic analysis on the conformal group}

In \secref{\ref{sec:inversion}} we compared large spin expansions and the inversion formula and stated that the inversion formula is far more elegant both conceptually and in practical computations; furthermore, it allows us to reach non-perturbative corrections with which the analytic results behave far better at finite values of spin. We demonstrated this by carrying out the computations via inversion formula for 3d Ising and $\mathrm{O}(2)$ models and compared those results with numerics and large spin expansions in \secref{\ref{sec: scalar nonperturbative}}. In the rest of the thesis, we will extend the application of inversion formula to CFTs with fermionic operators.

Our strategy to study the inversion formula for spinning operators involves a combination of Euclidean and Lorentzian ingredients. Our starting point for relating the fermionic $6j$ symbol to the scalar one involves their Euclidean definition as an overlap of partial waves. Then we use weight-shifting operators \cite{Karateev:2017jgd}, which transform in a finite-dimensional representation of the conformal group, to expand the fermionic $6j$ symbol as a sum over scalar symbols. We can plug in the explicit form of the scalar $6j$ symbol as calculated via the Lorentzian inversion formula to obtain the fermionic $6j$ symbol in closed form.\footnote{The full $6j$ symbol is only known in $d=1, 2, 4$ but the poles and residues are computable in general dimensions.} Finally, since a partial wave for general external operators can be split as a sum over two blocks, there exists a similar split for the $6j$ symbol in terms of the inversion of two blocks. By splitting the scalar and fermionic $6j$ symbols, we then find the inversion of a single block when we have external fermions. 

In this section, we will review the harmonic analysis.

\subsection{Shadow transform and shadow coefficients}
\label{sec: shadow}
In \equref{eq: shadow definition} we introduced the shadow transformation $\cO\rightarrow \mathbf{S}[\cO]$ as an application of the conformally-invariant pairing given in \equref{eq: conformally invariant pairing}. We remind the reader that in our conventions it reads as 
\be 
\mathbf{S}[\cO](x)\equiv \int dy \cO(y)\<\tl\cO(y)\tl\cO(x)\>.
\ee 
As $\mathbf{S}[\cO]$ is in the same conformal representation as $\tl\cO$, we expect that the tree point structures of one can be expanded about those of the other one: we define the \emph{shadow matrix} as the transformation matrix between these two basis, i.e.
\be 
\label{eq: shadow coefficients}
\<\cO_1\cO_2\mathbf{S}[\cO_3]\>^a=S^{a}_{c}(\cO_1\cO_2[\cO_3])\<\cO_1\cO_2\widetilde{\cO}_{3}\>^c.
\ee 

One nice way to understand such relations is through the diagrammatic notation introduced in \cite{Karateev:2018oml}. In this language, one denotes two point functions as 
\be 
\includegraphics[scale=1.2]{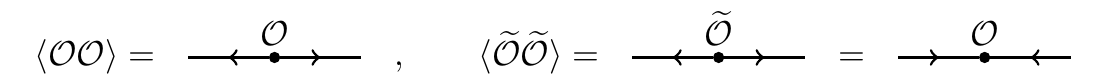},
\ee
where we see that taking the shadow is equivalent to changing the direction of arrow. Likewise, pairing operators is gluing the arrows; for example the diagrammatic equation 
\be 
\includegraphics[scale=1.6]{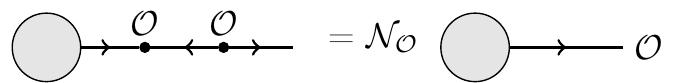}
\ee
stands for 
\be 
\<\cdots \mathbf{S}^{2}[\cO](x)\cdots\>=\int dx dy \<\cdots \cO(x)\cdots\>\<\tl\cO(x)\tl\cO(y)\>\<\cO(y)\cO(z)\>=\cN_{\cO}\<\cdots\cO(z)\cdots\>,
\ee 
or $\mathbf{S}^2[\cO]=\cN_{\cO}\cO $. This follows from the definition of the shadow transformation in \equref{eq: shadow definition} and the irreducibility of the representations. The factor $\cN$ in our conventions is
\be 
\cN_{\Delta,l} =\frac{\pi ^3 \tan (\pi  \left(\Delta+l\right) )}{(\Delta -\frac{3}{2}) (-\Delta +l+2) (\Delta +l-1)}\;.
\ee 

\begin{figure}
	\centering
	\includegraphics[scale=1]{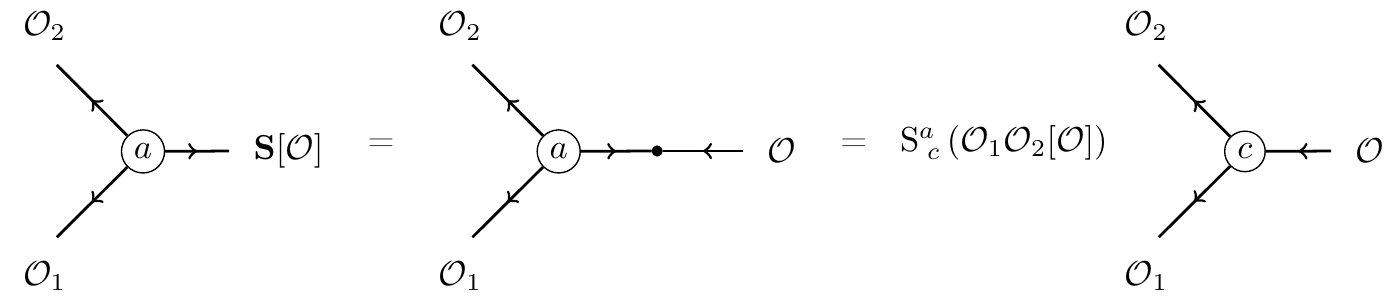}
	\caption[Definition of the shadow coefficients in diagrammatic notation]{\label{fig:shadow} Diagrammatic definition of shadow coefficients. Note that $a$ and $c$ label the three-point structures, and arrows allow us to keep track of scaling dimensions. We use the standard convention where an operator $\cO_{\Delta,J}$ with an outgoing arrow from a three-point structure enters that structure as itself. On the contrary, changing the direction of arrow is equivalent to changing $\cO_{\Delta,J}$ to $\tl\cO_{\Delta,J}\equiv\cO_{\tl\Delta,J}\equiv \cO_{3-\Delta,J}$.}
\end{figure}

In the diagrammatic language, three point structures are denoted by three arrows connected by a node with an additional label for the basis index. For example, one can see \equref{eq: shadow coefficients} in \figref{\ref{fig:shadow}}.

One can compute shadow matrices by either in position space through the action of the weight shifting operators, or in Fourier space where the shadow transformation turns into a simple convolution \cite{Karateev:2018oml}. Either way, we will not go into the details and simply provide the final form of the shadow matrices in our conventions for the relevant three point structures in this thesis:
\scriptsize
\be 
S^1_2([\psi_{\De_\psi}]\phi_{\De_\phi}\cO_{\De,l})=\,& -\frac{i \pi ^{3/2} \Gamma \left(\Delta _{\psi }-1\right) \Gamma \left(\frac{1}{2} \left(l+\Delta -\Delta _{\phi }-\Delta _{\psi
	}+\frac{5}{2}\right)\right) \Gamma \left(\frac{1}{2} \left(l-\Delta +\Delta _{\phi }-\Delta _{\psi }+\frac{7}{2}\right)\right)}{\Gamma
	\left(\frac{7}{2}-\Delta _{\psi }\right) \Gamma \left(\frac{1}{2} \left(l+\Delta -\Delta _{\phi }+\Delta _{\psi }+\frac{1}{2}\right)\right)
	\Gamma \left(\frac{1}{2} \left(l-\Delta +\Delta _{\phi }+\Delta _{\psi }-\frac{1}{2}\right)\right)},
\\
S^1_1(\psi_{\De_\psi}[\phi_{\De_\phi}]\cO_{\De,l})=\,& \frac{\pi  \sin \left(\pi  \Delta _{\phi }\right) \Gamma \left(2 \left(\Delta _{\phi }-2\right)\right) \Gamma
	\left(\frac{1}{2} \left(l+\Delta -\Delta _{\phi }-\Delta _{\psi }+\frac{5}{2}\right)\right) \Gamma \left(\frac{1}{2} \left(l-\Delta -\Delta
	_{\phi }+\Delta _{\psi }+\frac{5}{2}\right)\right)}{2^{2 \Delta _{\phi }-5}\Gamma \left(\frac{1}{2} \left(l+\Delta +\Delta _{\phi }-\Delta _{\psi
	}-\frac{1}{2}\right)\right) \Gamma \left(\frac{1}{2} \left(l-\Delta +\Delta _{\phi }+\Delta _{\psi }-\frac{1}{2}\right)\right)},
\\
S^1_2(\psi_{\De_\psi}\phi_{\De_\phi}[\cO_{\De,l}])=\,&
\frac{(-1)^{l+1} \Gamma (\Delta -1) \Gamma (l+\Delta -1) \Gamma \left(\frac{1}{2} \left(l-\Delta +\Delta _{\phi }-\Delta _{\psi
	}+\frac{7}{2}\right)\right) \Gamma \left(\frac{1}{2} \left(l-\Delta -\Delta _{\phi }+\Delta _{\psi }+\frac{5}{2}\right)\right)}{\pi ^{-3/2} \Gamma
	\left(\Delta -\frac{1}{2}\right) \Gamma (l-\Delta +3) \Gamma \left(\frac{1}{2} \left(l+\Delta +\Delta _{\phi }-\Delta _{\psi
	}-\frac{1}{2}\right)\right) \Gamma \left(\frac{1}{2} \left(l+\Delta -\Delta _{\phi }+\Delta _{\psi }+\frac{1}{2}\right)\right)},
\\
S^1_3([\psi_{\De_1}]\psi_{\De_2}\cO_{\De,l})=\,&
\frac{i \left(-\Delta +\Delta _1+\Delta _2-2\right) \Gamma \left(\Delta _1-1\right) \Gamma \left(\frac{1}{2} \left(l+\Delta -\Delta
	_1-\Delta _2+2\right)\right) \Gamma \left(\frac{1}{2} \left(l-\Delta -\Delta _1+\Delta _2+3\right)\right)}{2 \pi ^{-3/2} \Gamma \left(\frac{7}{2}-\Delta
	_1\right) \Gamma \left(\frac{1}{2} \left(l+\Delta +\Delta _1-\Delta _2\right)\right) \Gamma \left(\frac{1}{2} \left(l-\Delta +\Delta _1+\Delta
	_2+1\right)\right)},
\\
S^1_3(\psi_{\De_1}[\psi_{\De_2}]\cO_{\De,l})=\,&
S^1_3([\psi_{\De_2}]\psi_{\De_1}\cO_{\De,l}),
\\
S^1_1(\psi_{\De_1}\psi_{\De_2}[\cO_{\De,l}])=\,&
\frac{\pi ^{3/2} (-1)^l \Gamma \left(\Delta -\frac{3}{2}\right) \Gamma (l+\Delta -1) \Gamma \left(\frac{1}{2} \left(l-\Delta +\Delta _1-\Delta
	_2+3\right)\right) \Gamma \left(\frac{1}{2} \left(l-\Delta -\Delta _1+\Delta _2+3\right)\right)}{\Gamma (\Delta -1) \Gamma (l-\Delta +3) \Gamma
	\left(\frac{1}{2} \left(l+\Delta +\Delta _1-\Delta _2\right)\right) \Gamma \left(\frac{1}{2} \left(l+\Delta -\Delta _1+\Delta _2\right)\right)},
\ee 
\normalsize
where we can get all other nonzero components from the relations
\scriptsize
\be 
S^1_2([\psi_{\De_\psi}]\phi_{\De_\phi}\cO_{\De,l})=-S^2_1([\psi_{\De_\psi}]\phi_{-\De_\phi}\cO_{-\De,l}),
\\
S^1_1(\psi_{\De_\psi}[\phi_{\De_\phi}]\cO_{\De,l})=S^2_2(\psi_{\De_\psi}[\phi_{\De_\phi}]\cO_{\De,l-1}),
\\
S^1_2(\psi_{\De_\psi}\phi_{\De_\phi}[\cO_{\De,l}])=-S^2_1(\psi_{-\De_\psi}\phi_{-\De_\phi}[\cO_{\De,l}]),
\\
\frac{S^1_4([\psi_{\De_1}]\psi_{\De_2}\cO_{\De,l})}{\frac{l}{\Delta -\Delta _1-\Delta _2+2}}=\frac{S^2_3([\psi_{\De_1}]\psi_{\De_2}\cO_{\De,l})}{\frac{-\Delta +\Delta _1+\Delta _2+l-1}{2 \left(\Delta -\Delta _1-\Delta _2+2\right)}}=\frac{S^2_4([\psi_{\De_1}]\psi_{\De_2}\cO_{\De,l})}{-\frac{-\Delta +\Delta _1+\Delta _2+l-1}{2 \left(\Delta -\Delta _1-\Delta _2+2\right)}}=S^1_3([\psi_{\De_1}]\psi_{\De_2}\cO_{\De,l}),
\\
\frac{S^3_1([\psi_{\De_1}]\psi_{\De_2}\cO_{\De,l})}{\frac{-\Delta -\Delta _1+\Delta _2+l+2}{\Delta +\Delta _1-\Delta _2-2}}=\frac{S^3_2([\psi_{\De_1}]\psi_{\De_2}\cO_{\De,l})}{\frac{2 l}{\Delta +\Delta _1-\Delta _2-2}}=\frac{S^4_1([\psi_{\De_1}]\psi_{\De_2}\cO_{\De,l})}{\frac{-\Delta -\Delta _1+\Delta _2+l+2}{\Delta +\Delta _1-\Delta _2-2}}=\frac{S^4_2([\psi_{\De_1}]\psi_{\De_2}\cO_{\De,l})}{-\frac{2 \left(\Delta +\Delta _1-\Delta _2-1\right)}{\Delta +\Delta _1-\Delta _2-2}}=S^1_3([\psi_{\De_1}]\psi_{-\De_2}\cO_{-\De,l}),
\\
\frac{S^1_4(\psi_{\De_1}[\psi_{\De_2}]\cO_{\De,l})}{-\frac{l}{\Delta -\Delta _1-\Delta _2+2}}=\frac{S^2_3(\psi_{\De_1}[\psi_{\De_2}]\cO_{\De,l})}{\frac{-\Delta +\Delta _1+\Delta _2+l-1}{2 \left(\Delta -\Delta _1-\Delta _2+2\right)}}=\frac{S^2_4(\psi_{\De_1}[\psi_{\De_2}]\cO_{\De,l})}{\frac{-\Delta +\Delta _1+\Delta _2+l-1}{2 \left(\Delta -\Delta _1-\Delta _2+2\right)}}=S^1_3(\psi_{\De_1}[\psi_{\De_2}]\cO_{\De,l}),
\\
\frac{S^3_1(\psi_{\De_1}[\psi_{\De_2}]\cO_{\De,l})}{\frac{-\Delta +\Delta _1-\Delta _2+l+2}{\Delta -\Delta _1+\Delta _2-2}}=\frac{S^3_2(\psi_{\De_1}[\psi_{\De_2}]\cO_{\De,l})}{\frac{2 l}{\Delta -\Delta _1+\Delta _2-2}}=\frac{S^4_1(\psi_{\De_1}[\psi_{\De_2}]\cO_{\De,l})}{\frac{\Delta -\Delta _1+\Delta _2-l-2}{\Delta -\Delta _1+\Delta _2-2}}=\frac{S^4_2(\psi_{\De_1}[\psi_{\De_2}]\cO_{\De,l})}{\frac{2 \left(\Delta -\Delta _1+\Delta _2-1\right)}{\Delta -\Delta _1+\Delta _2-2}}=S^1_3(\psi_{-\De_1}[\psi_{\De_2}]\cO_{-\De,l}),
\\
\frac{S^2_1(\psi_{\De_1}\psi_{\De_2}[\cO_{\De,l}])}{-\frac{2 \Delta -3}{2 (\Delta -1)}}=\frac{S^2_2(\psi_{\De_1}\psi_{\De_2}[\cO_{\De,l}])}{-\frac{\Delta -2}{\Delta -1}}=S^1_1(\psi_{\De_1}\psi_{\De_2}[\cO_{\De,l}]),
\\
\frac{S^3_3(\psi_{\De_1}\psi_{\De_2}[\cO_{\De,l}])}{(\Delta -1) \left(\Delta +\Delta _1-\Delta _2-2\right) \left(\Delta -\Delta _1+\Delta _2-2\right)-(\Delta -2) l^2-(\Delta -2) l}=\frac{S^1_1(\psi_{-\De_1}\psi_{-\De_2}[\cO_{\De+1,l}])}{2 (2 \Delta -3) (-\Delta +l+2) (\Delta +l-1)},
\\
\frac{S^4_4(\psi_{\De_1}\psi_{\De_2}[\cO_{\De,l}])}{-(\Delta -2) \left((\Delta -1)^2-\Delta _1^2-\Delta _2^2+2 \Delta _1 \Delta _2\right)+(\Delta -1) l^2+\Delta  l-l}=\frac{S^1_1(\psi_{-\De_1}\psi_{-\De_2}[\cO_{\De+1,l}])}{2 (2 \Delta -3) (-\Delta +l+2) (\Delta +l-1)},
\\
\frac{S^3_4(\psi_{\De_1}\psi_{\De_2}[\cO_{\De,l}])}{\left(2\Delta -3\right) \left(\Delta _2-\Delta _1\right) l}=\frac{S^4_3(\psi_{\De_1}\psi_{\De_2}[\cO_{\De,l}])}{(2 \Delta -3) \left(\Delta _1-\Delta _2\right) (l+1)}=\frac{S^1_1(\psi_{-\De_1}\psi_{-\De_2}[\cO_{\De+1,l}])}{2 (2 \Delta -3) (-\Delta +l+2) (\Delta +l-1)}.
\ee 
\normalsize

The block (anti-)diagonal form of shadow matrices reflects the property that shadow transformation is parity-definite and that we have chosen our three-point structures with definite parity. As the two point function in \equref{eq: shadow definition} carries a definite parity, the shadow matrix relates the same (opposite) parity structures if the shadowed operator is of integer (half-integer) spin; this is why, say, $S^a_b([\psi]\psi\cO)$ is block anti-diagonal whereas $S^a_b(\psi\psi[\cO])$ is block diagonal.\footnote{We remind the reader that what we refer to here as parity is simply the inversion $X_i\rightarrow -X_i$ in embedding space.}

\subsection{Euclidean pairings}

In this section and the rest of the thesis, we will denote conformally invariant pairings of correlation functions as $\left(\<\dots\>,\<\dots\>\right)$ which is defined as
\begin{multline}
	\Big(\<\cO_1(x_1)\cdots\cO_n(x_n)\>\;,\;\<\tl\cO_{\pi_1}(x_{\pi_1})\cdots \tl\cO_{\pi_n}(x_{\pi_n})\>\Big)
	\\
	\coloneqq \int\frac{d^dx_1\cdots d^dx_n}{\vol(\SO(d+1,1))}\<\cO_1(x_1)\cdots\cO_n(x_n)\>\<\tl\cO_{\pi_1}(x_{\pi_1})\cdots \tl\cO_{\pi_n}(x_{\pi_n})\>
\end{multline}
for any permutation $\pi$. We remind the reader that suppressed indices of any operator and its dual are contracted from southwest to northeast; for instance,
\begin{multline}
	\left(\<\f_1(x_1)\f_2(x_2)\cO_3(x_3)\>,\<\tl\f_1(x_1)\tl\f_2(x_2)\tl\cO_3(x_3)\>\right)
	\\
	=\int\frac{d^3x_1d^3x_2 d^3x_3}{\vol(\SO(4,1))}\<\f_1(x_1)\f_2(x_2)(\cO_3)_{\a_1\dots\a_{2l_3}}(x_3)\>\<\tl\f_1(x_1)\tl\f_2(x_2)(\tl\cO_3)^{\a_1\dots\a_{2l_3}}(x_3)\>\;.
\end{multline}

\subsubsection{Two Point Pairings and Plancherel Measure}
\label{sec: two point pairing}

Let us first consider the pairing of scalar two-point functions.We denote it as 
\be 
{}\left(\<\phi(x_1)\phi(x_2)\>,\<\tl\phi(x_1)\tl\phi(x_2)\>\right)
=\,&\int \frac{d^dx_1d^dx_2}{\vol(\SO(d+1,1))}\<\phi(x_1)\phi(x_2)\>\<\tl\phi(x_1)\tl\phi(x_2)\>
\\
=\,& \frac{1}{2^d\vol(\SO(1,1))\x\vol(\SO(d))}\<\phi(0)\phi(\infty)\>\<\tl\phi(0)\tl\phi(\infty)\>,
\ee 
where $\vol(\SO(1,1))\x\vol(\SO(d))$ is the stabilizer group for two points and the factor $2^d$ is the Fadeev-Popov determinant.

We define an operator at infinity as 
\be 
\cO(\infty)\equiv \lim\limits_{L\rightarrow\infty} L^{2\De}\cO(\hat{e}L)
\ee 
for a unit vector $\hat e$ hence $\<\tl\phi(0)\tl\phi(\infty)\>=1$ in our conventions, meaning
\be 
\left(\<\phi(x_1)\phi(x_2)\>,\<\tl\phi(x_1)\tl\phi(x_2)\>\right)
=\frac{1}{64\pi^2\vol(\SO(1,1))}\;.
\label{eq: scalar two point pairing}
\ee 

We can use this result to compute the pairing of spinning two point functions. To do this, we first rewrite the two-point function $\<\cO\cO\>^{\Delta,J}$ in terms of weight-shifting operators $\cD^{a,b}$ acting on $\<\cO\cO\>^{\Delta-a,J-b}$,\footnote{We review these operators in \secref{\ref{}}.} then integrate it by parts, and finally act with the adjoint weight-shifting operators $\left(\cD^{a,b}\right)^*\propto \cD^{a,-b}$ on the other two-point function. Diagrammatically,
\begin{equation*}
	\begin{aligned}
		\includegraphics[scale=1]{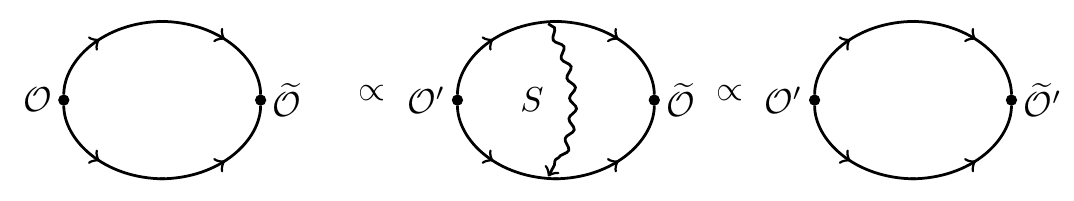}
	\end{aligned}\;.
\end{equation*}
We can find the coefficient between the first two diagrams above by direct calculation. For example, if we choose $a=b=\half$, we have
\be 
\<\cO_1\cO_2\>^{\Delta,J}=
-\frac{i \cD^{(+,+)}_{1A}\cD^{(+,+)A}_{2}}{16 \left(\Delta -2\right) \left(\Delta -\frac{3}{2}\right) \left(\Delta
	+J-2\right) \left(\Delta +J-1\right)}
\<\cO_1\cO_2\>^{\De-\half,J-\half},
\ee 
where we can integrate by parts and carry these differential operators to the other two-point function using the relation
\begin{multline}
	\left(\cD^{+,+}_\a \cO^{\De-\half,J-\half}, \cO^{3-\De,J}\right)
	= \left(\cO^{\De-\half,J-\half}, \left(\cD^{+,+}_\a\right)^* \cO^{3-\De,J}\right)
	\\=-\frac{1}{2J} \left(\cO^{\De-\half,J-\half}, \cD^{+,-}_\a \cO^{3-\De,J}\right).
\end{multline}
Carrying out the computation, we find that
\be 
\left(\<\cO_1\cO_2\>^{\Delta,J}, \<\cO_1\cO_2\>^{3-\Delta,J}\right)
= \frac{2J+1}{2J}
\left(\<\cO_1\cO_2\>^{\Delta,J-\half},\<\cO_1\cO_2\>^{3-\Delta,J-\half}\right).
\ee 
Note that this recursion relation is independent of which weight-shifting operator we choose: we get exactly the same relation for all $a,b=\pm\half$ choices.

Using \equref{eq: scalar two point pairing}, we get the final result
\be 
\left(\<\cO_1\cO_2\>^{\Delta,J},\<\cO_1\cO_2\>^{3-\Delta,J}\right)
=\frac{2J+1}{64\pi^2\vol(\SO(1,1))}\;.
\label{eq: spinning two point pairing}
\ee 

We can use this expression to compute the Plancherel measure\footnote{Plancherel measure of a locally compact group $G$ describes the decomposition of the irreducible unitary representations (IUR) into regular representations and are defined on the set of  IUR.} for the conformal group. It is easy to see this diagrammatically:
\begin{equation*}
	\begin{aligned}
		\includegraphics[scale=1.7]{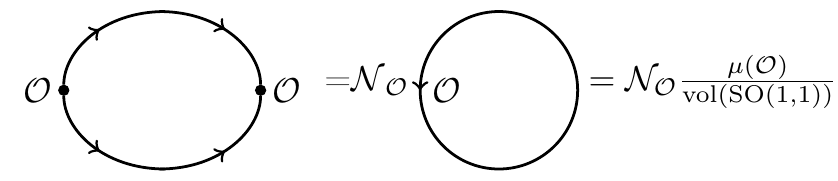}
	\end{aligned}\;,
\end{equation*}
where we first make use of $\mathbf{S}^2=\cN$ to convert the pairing into a circle and then identify the circle as the Plancherel measure up to the volume factor; see \cite{Karateev:2018oml} for details. Therefore, we conclude that
\be 
\mu(\cO_{\Delta,l})=\frac{\vol(\SO(1,1))}{\cN_{\Delta,l}}
\<\cO_1\cO_2\>^{\Delta,l}\.\<\cO_1\cO_2\>^{3-\Delta,l}\;,
\ee 
and we compute it as
\be 
\mu(\cO_{\Delta,l})=\frac{(2 \Delta -3) (-\Delta +l+2) (\Delta +l-1) \Gamma (2 l+2) \cot (\pi  (\Delta +l))}{128\pi ^5}\;.
\ee 

\subsubsection{Three Point Pairings}
In our conventions we have
\begin{multline}
	\left(\<\phi_1\phi_2\cO\>,\<\tl\phi_1\tl\phi_2\tl\cO\>\right)={}(-2)^J\int\frac{d^dx_1d^dx_2d^dx_3}{\vol(\SO(d+1,1))}\<\phi_1(x_1)\phi_2(x_2)\cO_{\mu_1\dots\mu_J}(x_3)\>\\\x\<\tl\phi_1(x_1)\tl\phi_2(x_2)\tl\cO^{\mu_1\dots\mu_J}(x_3)\>\;,
\end{multline}
which can be calculated by gauge fixing
\be 
\left(\<\phi_1\phi_2\cO_J\>,\<\tl\phi_1\tl\phi_2\tl\cO_J\>\right)=\frac{(-1)^J\widehat C_J(1)}{2^{d-J}\;\vol(\SO(d-1))}\;,
\ee
where $2^d$ is the appropriate Fadeev-Popov determinant. In $3d$ this reads as\footnote{We have the convention $\vol(\SO(n))=\vol(\SO(n-1))\vol(S^{n-1})$ with \mbox{$\vol(\SO(2))=2\pi$}. As what really matters is only the ratios of group volumes, this choice does not affect any physical result.}
\be 
\left(\<\phi_1\phi_2\cO_J\>,\<\tl\phi_1\tl\phi_2\tl\cO_J\>\right)=\frac{(-1)^J\G\left(J+1\right)}{16\sqrt{\pi}\G\left(J+\frac{1}{2}\right)}\;.
\label{eq: scalar scalar three point pairing}
\ee 

The pairing of spinning three-point functions can be calculated by reducing them via the weight-shifting operators and using the scalar pairing above. Schematically,
\begin{equation*}
	\begin{aligned}
		\includegraphics[scale=1.8]{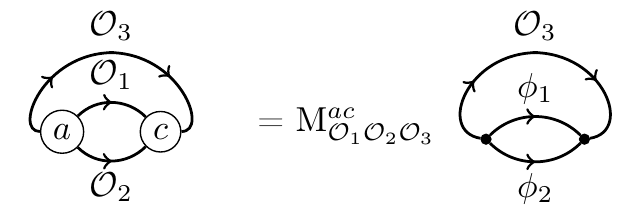}
	\end{aligned}\;.
\end{equation*}

The procedure to calculate the matrix $M^{ac}$ is as follows. We first expand the $\<\psi\psi\cO\>^a$ and $\<\psi\phi\cO\>^a$ three-point functions in terms of $\<\phi\phi\cO\>$, schematically
\begin{equation*}
	\begin{aligned}
		\includegraphics[scale=.75]{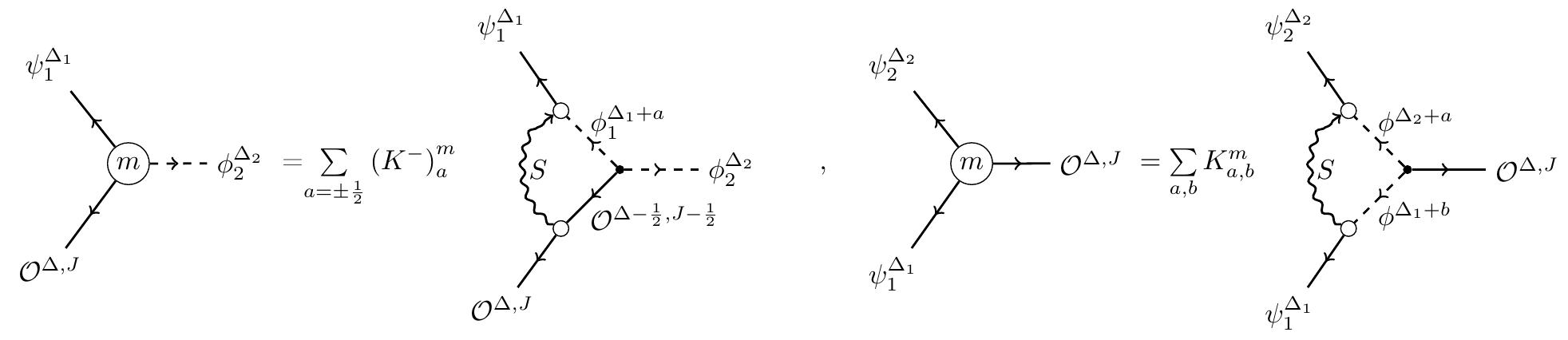}
	\end{aligned}\;.
\end{equation*}
We then integrate by parts and act with the adjoint of these weight-shifting operators on the other spinning three-point function, which produces $\<\phi\phi\cO\>$ up to overall coefficients.\footnote{In $d=3$, there actually exists a compact expression for three point pairings of operators of any spin in \cite{Karateev:2018oml}, thus one can bypass this computation.} By this procedure, we find that
\begin{footnotesize}
	\bea 
	\left(\<\psi_1^{\Delta_1}\phi_2^{\Delta_2}\cO^{\Delta,J}\>^m,\<\psi_1^{3-\Delta_1}\phi_2^{3-\Delta_2}\cO^{3-\Delta,J}\>^n\right)={}&{}
	\frac{(-1)^{J-\half} \Gamma \left(J+\frac{3}{2}\right)}{16 \sqrt{\pi } \Gamma (J+1)}
	\begin{pmatrix}
		-1&0\\0&1
	\end{pmatrix},
	\\
	\left(\<\psi_1^{\Delta_1}\psi_2^{\Delta_2}\cO^{\Delta,J}\>^m,\<\psi_1^{3-\Delta_1}\psi_2^{3-\Delta_2}\cO^{3-\Delta,J}\>^n\right)={}&{}
	\left\{
	\begin{aligned}
		&\frac{(-1)^J \Gamma (J+1)}{8 \sqrt{\pi } \Gamma \left(J+\frac{1}{2}\right)}
		\left(
		\begin{array}{cccc}
			-1 & \frac{1}{2} & 0 & 0 \\
			\frac{1}{2} & -\frac{2 J+1}{4 J} & 0 & 0 \\
			0 & 0 & 1 & 0 \\
			0 & 0 & 0 & -\frac{J+1}{J} \\
		\end{array}
		\right)
		&& J>0
		\\
		&\frac{1}{8\pi}\begin{pmatrix}
			-1&0\\0&1
		\end{pmatrix}
		&& J=0
	\end{aligned}
	\right.
	\label{eq: fermion fermion three point pairing}
	\eea
\end{footnotesize}

\subsection{Euclidean inversion formula}
An $n-$point correlator can be expanded as a tensor product of two irreducible representations of the Euclidean conformal group, which basically provides us with an integral representation of a higher-point correlator in terms of lower point ones. This has been known for almost half a century since the early work of Dobrev et. al. \cite{Dobrev:1977qv} and was revived in recent years~\cite{Gadde:2017sjg,Caron-Huot:2017vep,Karateev:2018oml,Kravchuk:2018htv}. In the notation of \cite{Kravchuk:2018htv}, we can schematically write
\be 
\<\cO_1\cdots \cO_n\>=\int d\cO \int d^dx 
\<\cO_1(x_1)\cO_2(x_2)\cO(x)\>^aP^a_\cO(x_3,\dots x_n;x)
\label{eq: old syle partial wave expansion}
\ee 
for a generic $n-$point correlator.  This corresponds to the following diagram\footnote{For notational brevity, we denote the matrix inverse of the pairings of three point structures as pairings in denominator, i.e.
	\be 
	\frac{\left(\<ABC\>^a,\<\tl A\tl B\tl C\>^b\right)}{\left(\<ABC\>^c,\<\tl A\tl B\tl C\>^a\right)}=
	\frac{\left(\<ABC\>^b,\<\tl A\tl B\tl C\>^a\right)}{\left(\<ABC\>^a,\<\tl A\tl B\tl C\>^c\right)}=
	\delta_{bc}\;.
	\ee 
}

\begin{equation}
	\label{eq: old style partial wave expansion}
	\begin{aligned}
		\includegraphics[scale=1.18]{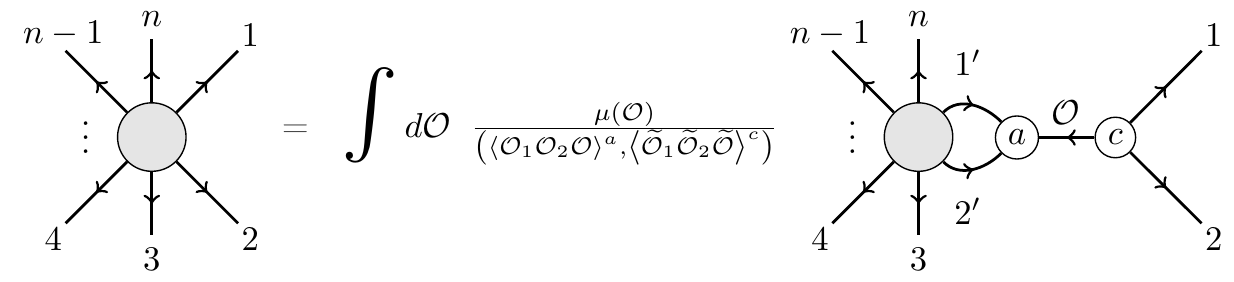}
	\end{aligned}\;,
\end{equation}
where we identify
\begin{equation}
	\begin{aligned}
		\includegraphics[scale=1.23]{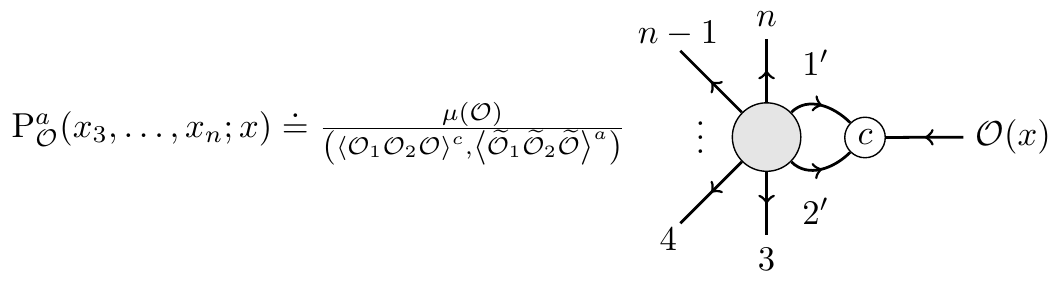}
	\end{aligned}\;.
\end{equation}
Here the integration measure is $d\cO=2\pi\Delta_{ll'}\delta\left(\nu-\nu'\right)$ and it is defined over the Euclidean principal series 
\be 
\Delta=\frac{d}{2}+i\nu, \quad \nu\geq0, \qquad l\in\mathbb{Z}.
\ee
We are glossing over the details in this quick review and refer the reader to \cite{Karateev:2018oml,Kravchuk:2018htv} for more details.

Let us consider this general expression in the case of $\<\cO_1\cO_2\cO_3\cO_4\>$. For four-point functions, we can decompose $P^a_\cO(x_3,x_4;x)$ in terms of three-point structures:
\be 
P^a_\cO(x_3,x_4;x)=\rho^{(s)}_{ab}(\cO)\<\cO_3(x_3)\cO_4(x_4)\tl\cO(x)\>^b.
\label{eq: definition of rho}
\ee 
Here $\rho^{(s)}_{ab}(\cO)$ are partial wave expansion coefficients and are related to OPE coefficients via \equref{eqn:OPEfunc_To_Coefs_schannel} as we will see below. With \equref{eq: partial wave convention}, we can use the equation above to obtain the partial wave expansion\footnote{In some papers $P^a_\cO(x_3,\dots,x_n;x)$ is referred to as conformal partial wave as well. We will not be using these objects in this paper and will reserve this term for $\Psi^{ab}_\cO$ defined in \equref{eq: partial wave convention}.} of four-point function:
\be 
\label{eq: partial wave expansion}
\<\cO_1\cO_2\cO_3\cO_4\>=
\<\cO_1\cO_2\>\<\cO_3\cO_4\>+\int_{\cC}d\cO \rho^{(s)}_{ab}(\cO)\Psi^{(s)ab}_{\cO}(x_i)\;,
\ee 
where we define the s-channel conformal partial wave as the gluing of two three-point functions
\be 
\label{eq: partial wave convention}
\Psi^{(s)ac}_{\cO}(x_i)=\int d^dx\<\cO_1\cO_2\cO(x)\>^a\<\cO_3\cO_4\tl\cO(x)\>^c=
\begin{aligned}
	\includegraphics[scale=1.2]{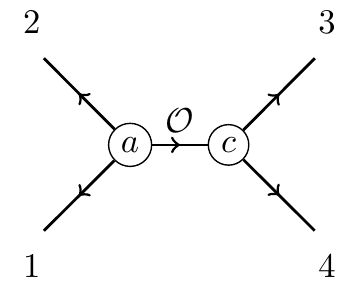}.
\end{aligned}\;.
\ee

\begin{figure}
	\centering
	\includegraphics[scale=1.25]{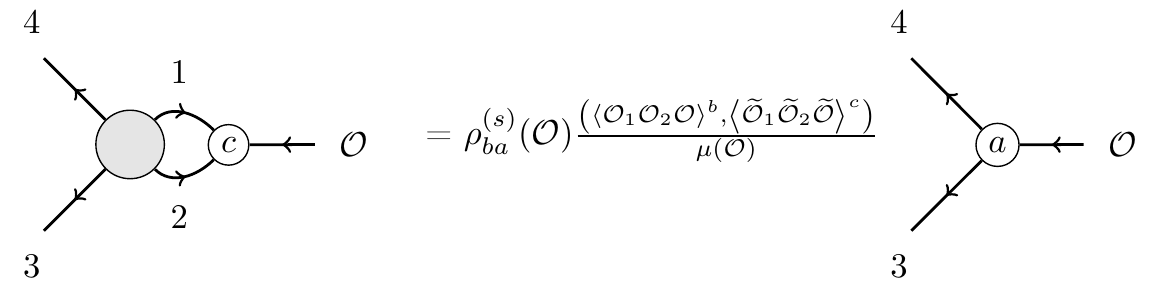}
	\caption[Definition of the OPE function in diagrammatic notation]{\label{fig:ope definition} We can take the definition of the OPE function $\rho$ to be the coefficients
		of the pairing between 	a four-point correlator $\<\cO_1\cO_2\cO_3\cO_4\>$ and a three-point structure $\<\cO_3\cO_4\cO\>^c$ in the basis of the three point structures $\<\cO_1\cO_2\cO\>^a$. Note that the overall coefficient also depends on the bubble coefficient $\cB$ which is a calculable kinematic term. By pairing both sides with $\<\cO_1\cO_2\tl\cO\>^e$, we can reduce this relation to the more standard definition generally used in the literature, such as (2.33) of \cite{Liu:2018jhs}, (2.40) of \cite{Karateev:2018oml}, or (1.6) of \cite{Simmons-Duffin:2017nub}. Note that these references use different conventions so the formulas are not entirely the same.}
\end{figure}

We would like to note two points about \equref{eq: partial wave expansion}. The first point is the fact that we are explicitly writing the identity contribution because the identity block is actually orthogonal to the partial waves, hence it cannot be expanded in terms of them \cite{Caron-Huot:2017vep}. It is further argued in \cite{Simmons-Duffin:2017nub} that there may be other non-normalizable contributions to the four-point function that need to be written out explicitly. In particular, any scalar operator with $\Delta<\frac{d}{2}$ gives such a contribution. We will assume that either there is no scalar with $\Delta<\frac{d}{2}$ in the spectrum of the theory or that their contributions can be obtained by analytic continuation from the principal series.

The second point we would like to draw attention is the integration in \equref{eq: partial wave expansion}: we  specified that the integration is over the contour $C$ which we define as
\be 
\label{eq: contours}
\int_{\cC}d\cO\equiv \sum\limits_{J_\cO}\int\limits_{\frac{d}{2}}^{\frac{d}{2}+i\infty}\frac{d\Delta_{\cO}}{2\pi i}\;,\quad \int_{\cC'}d\cO\equiv \sum\limits_{J_\cO}\int\limits_{\frac{d}{2}-i\infty}^{\frac{d}{2}+i\infty}\frac{d\Delta_{\cO}}{2\pi i}
\ee 
for convenience.\footnote{To be precise, in the measure $d\cO$ we now have a sum over either integer or half-integer spin, depending on the four-point function --- in general dimensions $d$ we have to sum over all allowed $\SO(d)$ representations. In odd dimensions we can also have the discrete series of $\De$, but they will be canceled by poles in $\rho^{(s)}_{ab}(\mathcal{O})$, so we will not include them explicitly.}  Also, note that we give the expansion in terms of s-channel partial waves. This is indicated by the explicit $(s)$ superscript on $\rho$ and $\Psi$. Additionally, we leave the dependence of $\rho$ and $\psi$ on external operators implicit.

\begin{figure}
	\centering
	\includegraphics[scale=.21]{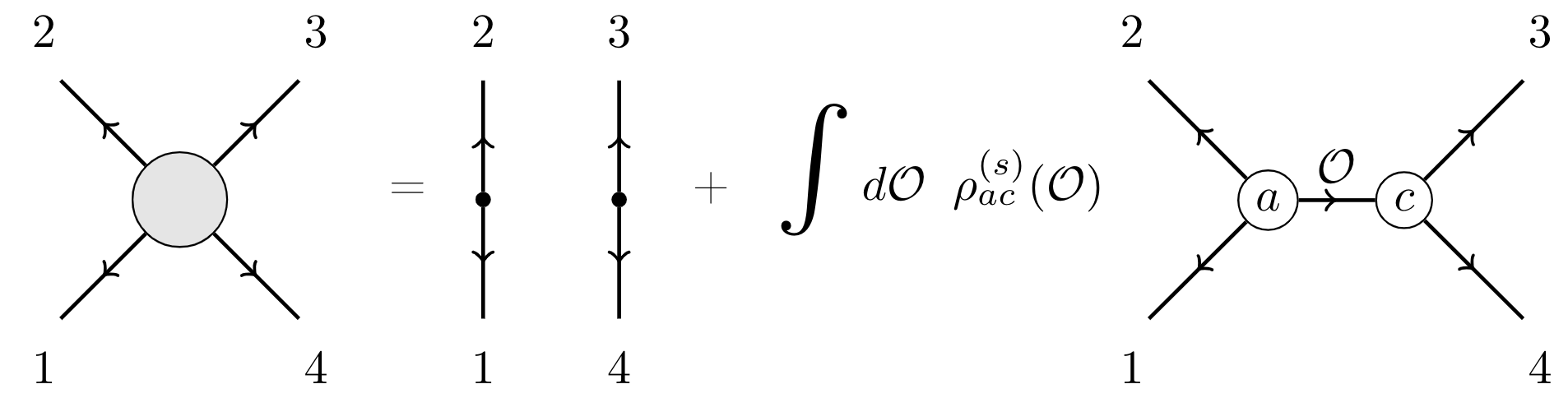}
	\caption[Partial wave expansion in diagrammatic notation]{\label{fig: partial wave expansion} Diagrammatic illustration of the s-channel partial wave expansion of the four-point function, assuming that the identity contribution is the only non-normalizable contribution. Instead of separating it, we can deform it onto the principal series and deform back after the analytic continuation from principal series to physical poles.
	}
\end{figure}

The definition of $\rho$ in \equref{eq: definition of rho} is diagrammatically shown in figure \ref{fig:ope definition}. We can pair both sides with a three-point function and obtain the Euclidean inversion formula:
\be 
\rho_{ab}^{(s)}(\cO_5)
=
\frac{\int \frac{d^{d}x_{1}...d^{d}x_5}{\vol(\SO(d+1,1))}\mu(\cO_5)\<\tl\cO_1\tl\cO_2\tl\cO_5\>^c\<\cO_1\cO_2\cO_3\cO_4\>\<\tl\cO_3\tl\cO_4\cO_5\>^d}{\left(\<\tl\cO_1\tl\cO_2\tl\cO_5\>^a,\<\cO_1\cO_2\cO_5\>^c\right)\left(\<\cO_3\cO_4\tl\cO_5\>^d,\<\tl\cO_3\tl\cO_4\cO_5\>^b\right)}\;.
\label{eq: Euclidean inversion formula long form}
\ee 
Note that we can rewrite this as\footnote{Remember that suppressed indices are contracted and they always go from southwest to northeast, hence the order of correlators are important in our conventions, i.e. $\<\dots\>\<\cdots\>= (-1)^{\#}\<\cdots\>\<\dots\>$.}
\be 
\label{eq: Euclidean inversion formula 2}
\rho_{ab}^{(s)}(\cO)
=
\frac{\int \frac{d^{d}x_{1}...d^{d}x_4}{\vol(\SO(d+1,1))}\left(\int d^dx_5\<\tl\cO_1\tl\cO_2\tl\cO_5\>^c\<\tl\cO_3\tl\cO_4\cO_5\>^d\right)\<\cO_1\cO_2\cO_3\cO_4\>}{\frac{(-1)^{\Sigma_{55}}}{\mu(\cO_5)}\left(\<\tl\cO_1\tl\cO_2\tl\cO_5\>^a,\<\cO_1\cO_2\cO_5\>^c\right)\left(\<\tl\cO_3\tl\cO_4\cO\>^b,\<\cO_3\cO_4\tl\cO_5\>^d\right)}\;,
\ee 
for the shorthand notation
\be 
\Sigma_{ij}=\left\{\begin{aligned}
	-1 && \text{if }\cO_i\text{ and }\cO_j\text{ are both fermions}\\
	1 &&\text{otherwise}
\end{aligned}\right.
\ee 

As we will show below, the ``denominator'' above is actually the inverse of the \emph{partial wave normalization} hence we conclude
\be 
\label{eq: Euclidean inversion formula}
\rho_{ab}^{(s)}(\cO)
= \eta_{(ac)(bd)}^\cO
\left(\Psi^{\tl{(s)}cd}_{\tl\cO}(x_i), \<\cO_1\cO_2\cO_3\cO_4\>\right).
\ee 
where $\eta_{(ac)(bd)}^\cO \eta^{(ce)(df)}_\cO=\delta^{e}_{a}\delta^{f}_{b}$ for  $\eta^{(ce)(df)}_\cO$ given in \equref{eq: partial wave normalization}.\footnote{In the rest of the thesis, we use the shorthand notation $\Psi^{\tl{(\chi)}ab}_{\cO}(x_i)$ which is defined as  \mbox{$\Psi^{\tl{(\chi)}ab}_{\cO}(x_i)=\Psi^{(\chi)ab}_{\cO}(x_i)\lvert_{\cO_1\rightarrow\tl\cO_1,\dots,\cO_4\rightarrow\tl\cO_4}$} for any channel $\chi=s,t,u$.}

We could have derived this result by starting from the partial wave expansion of figure \ref{fig: partial wave expansion}, pairing it with $\Psi^{\tl{(s)}cd}_{\tl\cO}$, and utilizing the orthogonality of the partial waves directly.\footnote{\label{footnote: tadpole}When we pair the partial wave expansion \equref{eq: partial wave expansion} with a partial wave of the same channel, there is actually another term coming from the pairing of identity exchange with the partial wave. However the pairing of identity exchange with the partial wave of the same channel is proportional to a tadpole diagram:
	\begin{equation}
		\label{fig:tadpole}
		\begin{aligned}		\includegraphics[scale=1]{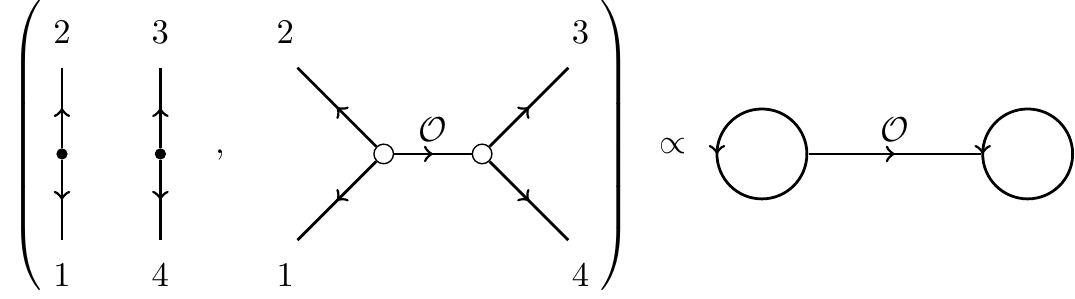}
		\end{aligned}\;.
	\end{equation}
	Such diagrams are zero by the irreducibility of the representations unless $\cO=\mathbb{1}$, which is never the case for the partial waves on the principal series.
}

We would like to  remind the reader that the diagrams, albeit useful, are to be considered as schematic expressions only. In particular, they are agnostic to possible signs associated to the orderings of fermions. As an example, consider \equref{eq: old style partial wave expansion}: in our conventions it stands for
\begin{subequations}
	\begin{equation}
		\hspace{-.2in}	\<\cO_1\cO_2\cO_3\cdots\cO_n\>=\int d\cO \mu(\cO) \int d^dx d^dx_{1}'d^dx_{2}' \frac{\<\cO_1\cO_2\cO(x)\>^a\<\tl\cO(x)\tl\cO'_{2}\tl\cO'_{1}\>^c\<\cO'_{1}\cO'_{2}\cO_3\cdots\cO_n\>}{\left(\<\tl\cO\tl\cO_{2}\tl\cO_{1}\>^a,\<\cO_{1}\cO_{2}\cO\>^c\right)}\;,
		\label{eq: correct equation}
	\end{equation}
	but not
	\begin{equation}
		\hspace{-.2in}	\<\cO_1\cO_2\cO_3\cdots\cO_n\>\neq\int d\cO \mu(\cO) \int d^dx d^dx_{1}'d^dx_{2}' \frac{\<\cO_1\cO_2\cO(x)\>^a\<\cO'_{1}\cO'_{2}\cO_3\cdots\cO_n\>\<\tl\cO(x)\tl\cO'_{2}\tl\cO'_{1}\>^c}{\left(\<\tl\cO\tl\cO_{2}\tl\cO_{1}\>^a,\<\cO_{1}\cO_{2}\cO\>^c\right)}\;.
	\end{equation}
	However, one cannot deduce this from the diagram alone.
\end{subequations}

\subsubsection{Bubble Coefficients and Partial Wave Normalization}

\begin{figure}
	\centering
	\includegraphics[scale=1.4]{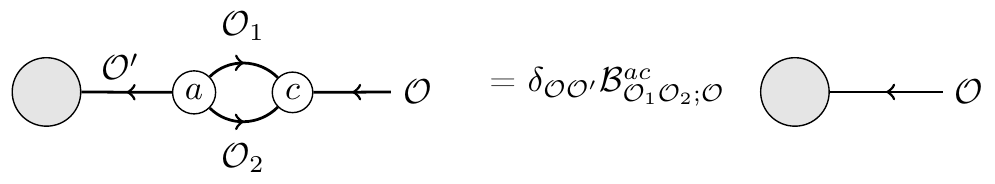}
	\caption[Bubble annihilation coefficients in diagrammatic notation]{\label{fig:bubble} Definition of the bubble annihilation matrix $\cB$. One can explicitly calculate $\cB$ by removing the gray blob above and connecting both ends: this relates $\cB$ times the Plancherel measure to the pairing of two three-point functions, which can then be computed by going to a fixed conformal frame and carrying out the explicit calculations. A similar calculation is carried out in $4d$ in \cite{Karateev:2018oml}, see appendix C there. Here we repeated it for $3d$.}
\end{figure}

One of the interesting pairings that we can consider is
the so-called bubble integral\footnote{This follows from the irreducibility of representations.} 
\begin{multline}
	\label{eq: bubble coefficient}
	\<\cdots\tl\cO'(x)\cdots\>
	\int d^{d}x_1d^{d}x_2
	\<\cO_{1}(x_1)\cO_{2}(x_2)\cO'(x)\>^a
	\<\tl\cO_{1}(x_1)\tl\cO_{2}(x_2)\tl\cO(y)\>^b
	\\= \delta_{\cO\cO'}
	\delta^d(x-y)\cB_{\cO_1\cO_2;\cO}^{ab}\<\cdots\tl\cO(y)\cdots\>,
\end{multline}
which we can see in figure \ref{fig:bubble} in diagrammatic language. By imposing $\cO'=\cO$ and taking the trace of both sides without acting on $\<\cdots\tl\cO'(x)\cdots\>$, we can relate the bubble coefficient $\cB_{\cO_1\cO_2;\cO}^{ab}$ to the three point pairing and the Plancherel measure:
\be
\label{eq: bubble coefficient in explicit form}
\cB_{\cO_1\cO_2;\cO}^{ab}=&
\frac{\left(\<\cO_{1}\cO_{2}\cO\>^a,\<\tl\cO_{1}\tl\cO_{2}\tl\cO\>^b\right)}{\ac{mu}}.
\ee

One straightforward usage of these bubble matrices is the calculation of the normalization of the conformal partial wave defined in \equref{eq: partial wave convention}:
\be 
\left(\Psi^{\tl{(s)}ab}_{\cO_5},\Psi^{(s)cd}_{\cO_6}\right)=
\int \frac{d^{d}x_1...d^dx_6}{\SO(d+1,1)}\<\tl\cO_1\tl\cO_2\cO_5\>^a\.\<\tl\cO_3\tl\cO_4\widetilde{\cO}_5\>^b\.\<\cO_1\cO_2\cO_6\>^c\.\<\cO_3\cO_4\widetilde{\cO}_6\>^d\;,
\ee 
If we perform the $x_{3,4}$ integrals, we obtain
\be 
\left(\Psi^{\tl{(s)}ab}_{\cO_5},\Psi^{(s)cd}_{\cO_6}\right)=\delta_{\cO_5\tl\cO_6}
(-1)^{\Sigma_{66}}
\cB_{\tl\cO_3\tl\cO_4;\cO_6}^{bd}
\left(\<\tl\cO_1\tl\cO_2\tl\cO_6\>^a, \<\cO_1\cO_2\cO_6\>^c\right),
\ee 
where $\de_{\cO_5\tl\cO_6}=\de_{l_5l_6}2\pi\de(\nu_5-\nu_6)$ for $\De=\frac{d}{2}+i\nu$  and where $(-1)^{\Sigma_{66}}$ follows from the change of the order of the three-point functions. Next we use \equref{eq: bubble coefficient in explicit form} to find the more symmetric form:
\be
\label{eq: definition of eta}
\left(\Psi^{\tl{(s)}ab}_{\cO_5},\Psi^{(s)cd}_{\cO_6}\right)=\delta_{\cO_5\tl\cO_6}\eta_{\cO_5}^{(ac)(bd)}
\ee 
for
\be 
\label{eq: partial wave normalization}
\eta_{\cO_5}^{(ac)(bd)}\coloneqq \frac{(-1)^{\Sigma_{55}}}{\mu(\cO_5)}\left(\<\tl\cO_1\tl\cO_2\tl\cO_{5}\>^a\.\<\cO_1\cO_2\cO_{5}\>^c\right)\left(\<\tl\cO_3\tl\cO_4\cO_{5}\>^b\.\<\cO_3\cO_4\tl\cO_{5}\>^d\right)\;.
\ee 

Note that changing the order of the three-point functions brings an overall sign\linebreak \mbox{$(-1)^{2(l_1+l_2+l_3+l_4)}=1$}.\footnote{That this term is 1 follows from Lorentz invariance as we need an even number of fermions in a non-zero vacuum expectation value.} Since we also have $\mu\left(\cO\right)=\mu(\tl\cO)$, the partial wave normalization satisfies the following symmetries:
\bea[eq: symmetry of partial wave normalization]
\left(\Psi^{\tl{(s)}ab}_{\tl\cO},\Psi^{(s)cd}_{\cO}\right)=
\left(\Psi^{(s)cd}_{\cO},\Psi^{\tl{(s)}ab}_{\tl\cO}\right),
\\
\left(\Psi^{\tl{(s)}ab}_{\tl\cO},\Psi^{(s)cd}_{\cO}\right)=
\left(\Psi^{\tl{(s)}cd}_{\tl\cO},\Psi^{(s)ab}_{\cO}\right).
\eea

\subsection{Partial Waves and Conformal Blocks}
\label{app:Scalar_Conventions}
In this section we will briefly review the relation between the conformal partial wave expansion and the conformal block expansion. The goal is to establish the general dictionary between the two for general four-point functions. The method we use is not new, but it will be useful to present the results in our conventions, taking care of signs with fermionic operators. 

First we recall that our definition for the partial wave and shadow transform are:
\begin{align}
	\Psi^{(s)ab}_{\cO_5}&=\int d^{d}x_{5}\<\cO_1\cO_2\cO_5\>^a\<\cO_3\cO_4\widetilde{\cO}_{5}\>^b\;, \label{eqn:DefCPWApp}
	\\
	\mathbf{S}[\cO](x)&=\int d^dy \cO(y)\<\cO(y)\cO(x)\>\;.
\end{align}

We will also find it useful to define the kinematic functions $C$:
\begin{align}
	\lim\limits_{x_1\rightarrow x_2}\<\cO_1(x_1)\cO_2(x_2)\cO_5(x_5)\>^a\sim C^{a}_{\cO_1\cO_2\cO_5}(x_{12})\<\cO_2(x_2)\cO_2(x_5)\>\;.
\end{align}

We can then define s-channel conformal blocks for $\<\cO_1\cO_2\cO_3\cO_4\>$ as solutions to the conformal Casimir equation with the following behavior in the limit $x_3\rightarrow x_4$ and $x_1\rightarrow x_2$:
\begin{align}
	G^{(s)ab}_{\cO_5}(x_i)\approx C^{p}_{\cO_1\cO_2\cO_5}(x_{12})C^{q}_{\cO_3\cO_4\cO_5}(x_{34})\<\cO_{5}(x_2)\cO_{5}(x_4)\>\;.
\end{align}
Here we work in Euclidean space and the order of limits does not matter. With this definition the four-point function has the following conformal block expansion:
\be 
\label{eq: conformal block expansion}
\<\cO_1\cO_2\cO_3\cO_4\>=\sum\limits_{\cO}\lambda_{\cO_1\cO_2\cO}^a\lambda_{\cO_3\cO_4\cO}^bG^{(s)ab}_{\cO}(x_i)\;,
\ee 
for the OPE coefficients $\lambda$ are defined as in \equref{eq: definition of OPE coefficients}.

Now we have to expand the conformal partial wave as a sum of two conformal blocks. To extract their coefficients, we just need to study the integrand in certain limits. We start by taking the limit $x_1\rightarrow x_2$ under the integrand in (\ref{eqn:DefCPWApp}) and then performing the $x_5$ integral. In this limit we have:
\begin{align}
	\Psi^{(s)ab}_{\cO_5}(x_i)&\supset \int d^{d}x_{5}C^a_{\cO_1\cO_2\cO_5}(x_{12})\<\cO_5(x_2)\cO_5(x_5)\>\<\cO_3\cO_5\widetilde{\cO}_{5}(x_5)\>^b \nonumber
	\\ &\supset C^a_{\cO_1\cO_2\cO_5}(x_{12})S^{b}_{c}(\cO_3\cO_4[\widetilde{\cO}_{5}])\<\cO_3\cO_4\cO_5(x_2)\>^c\;.
\end{align}
To get the second line we have to reorder the operators in the two-point function and implicitly raise and lower the spinor indices, so the two possible signs cancel. Taking the $x_3\rightarrow x_4$ limit we find:
\begin{align}
	\Psi^{(s)ab}_{\cO_5}(x_i)&\supset C^a_{\cO_1\cO_2\cO_5}(x_{12})S^{b}_{c}(\cO_3\cO_4[\widetilde{\cO}_{5}])C^{c}_{\cO_3\cO_4\cO_5}(x_{34})(-1)^{\Sigma_{55}}\<\cO_{5}(x_2)\cO_5(x_4)\>\;.
\end{align}

To get the coefficient for the other block we take the limit $x_3\rightarrow x_4$ under the integrand, perform the $x_5$ integral and then take the limit $x_1\rightarrow x_2$:
\begin{align}
	\Psi^{(s)ab}_{\cO_5}(x_I)&\supset S^{a}_{c}(\cO_3\cO_4[\cO_{5}])C^{c}_{\cO_1\cO_2\cO_5}(x_{12})C^b_{\cO_3\cO_4\cO_5}(x_{34})(-1)^{\Sigma_{55}}\<\cO_{5}(x_2)\cO_5(x_4)\>\;.
\end{align}

We therefore find that the full partial wave is:
\begin{align}
	\Psi^{(s)ab}_{\cO_5}(x_i)=(-1)^{\Sigma_{55}}\left[S^{b}_{c}(\cO_3\cO_4[\widetilde{\cO}_{5}])G^{(s)ac}_{\cO_5}(x_i)+S^{a}_{c}(\cO_1\cO_2[\cO_5])G^{(s)cb}_{\widetilde{\cO}_{5}}(x_i)\right]\;.
\end{align}

We can now use this equation inside \equref{eq: partial wave expansion} to obtain\footnote{There are also non-renormalizable contributions such as identity contribution, but we are dropping them here with the assumption that they can be obtained later by analytic continuation.}
\be 
\<\cO_1\cO_2\cO_3\cO_4\>=&\int_{\cC}d\cO \rho^{(s)}_{ab}(\cO)(-1)^{\Sigma_{55}}\left[S^{b}_{c}(\cO_3\cO_4[\widetilde{\cO}_{5}])G^{(s)ac}_{\cO_5}(x_i)+S^{a}_{c}(\cO_1\cO_2[\cO_5])G^{(s)cb}_{\widetilde{\cO}_{5}}(x_i)\right]
\\
=&\int_{\cC'}d\cO \rho^{(s)}_{ab}(\cO)(-1)^{\Sigma_{55}}S^{b}_{c}(\cO_3\cO_4[\widetilde{\cO}_{5}])G^{(s)ac}_{\cO_5}(x_i)
\ee 
where in the second line we used the shadow symmetry of the integrand and the definitions of the contours, i.e. \equref{eq: contours}. By closing the contour to the right and comparing with \equref{eq: conformal block expansion}, we find the following relation between the OPE function and the OPE coefficients:
\be 
\label{eq: OPE function to OPE coefficients}
\lambda_{\cO_1\cO_2\cO_5}^a\lambda_{\cO_3\cO_4\cO_5}^b=(-1)^{1+\Sigma_{55}}\text{Res}_{\Delta=\Delta_{5}}\rho^{(s)}_{ac}S^{c}_{b}(\cO_3\cO_4[\widetilde{\cO}_{5}])\bigg|_{J=J_5}\;,
\ee
where we first set $J=J_5$ and then evaluate the residue.

\subsection{$6j$ Symbols}
\label{sec:6jreview}

In \equref{eq: OPE function to OPE coefficients}, we found the relation to compute the OPE coefficients from the Euclidean inversion formula. We can now ask the general question: what does the contribution of a single operator $\cO$ in the t-channel map to in the s-channel under crossing? As realized in \cite{Komargodski:2012ek,Fitzpatrick:2012yx}, by studying the lightcone limit, an isolated operator $\cO$ maps to double-twist operators in the crossed channel. To review this result in the current language, let us first introduce the $6j$ symbol of the conformal group.

The $6j$ symbol is defined as the overlap of a t- and s-channel partial wave, which for external scalars is
\begin{align}
	\label{eq:6jDefScalar}
	\sixj{\f_1}{\f_2}{\f_3}{\f_4}{\cO_5}{\cO_6}&=\left(\widetilde{\Psi}^{(s)}_{\tl\cO_5},\Psi^{(t)}_{\cO_6}\right)
	\nonumber \\ &=\int dx_1... dx_6 \<\widetilde{\f}_1\widetilde{\f}_2\widetilde{\cO}_5\>\<\widetilde{\f}_{3}\widetilde{\f}_{4}\cO_5\>\<\f_3\f_2\cO_6\>\<\f_1\f_4\widetilde{\cO}_{6}\>\;.
\end{align}
Using the $6j$ symbol it is possible to write a single t-channel partial wave as a spectral integral over s-channel partial waves:
\begin{align}
	\Psi^{(t)}_{\cO_6}(x_i)=\int\limits_{\mathcal{C}}d\cO_{5}\frac{1}{\eta_{\cO_5}}\sixj{\f_1}{\f_2}{\f_3}{\f_4}{\cO_5}{\cO_6}\Psi^{(s)}_{\cO_5}(x_i)\;,
\end{align}
where $\eta_{\cO_5}$ is the normalization of partial waves of external scalars, special case of the general formula in \equref{eq: partial wave normalization}.

In practice the $6j$ symbol (\ref{eq:6jDefScalar}) has been calculated using the Lorentzian inversion formula in $d=2$ and $d=4$. The Lorentzian inversion formula gives another integral representation of the OPE function, but now with the correlator integrated over a causal diamond in Minkowski space:\footnote{Our convention is $\vol(\SO(n))=\vol(\SO(n-1))\vol(S^{n-1})$.}
\begin{align}
	\rho^{(s)}(\Delta,J)&=\alpha_{\Delta,J}\int\limits_{0}^{1}\int\limits_{0}^{1}dzd\bar{z}\mu(z,\bar{z})G^{(s)}_{J+d-1,\Delta+1-d}(z,\bar{z})\frac{\<[\f_3,\f_2][\f_1,\f_4]\>}{T^{(s)}}+(\text{u-channel})\;,
	\\
	\alpha_{\Delta,J}&=-\frac{\pi^{d-2}\Gamma(\frac{d-2}{2})\Gamma(J+d-2)}{2^{d+J+3}\text{vol}(\text{SO}(d-1))\Gamma(d-2)\Gamma(J+\frac{d-2}{2})}\frac{\Gamma(J+1)\Gamma(\Delta-\frac{d}{2})}{\Gamma(J+\frac{d}{2})\Gamma(\Delta-1)}
	\nonumber \\ & \qquad \frac{\Gamma(\frac{\Delta_{12}+J+\Delta}{2})\Gamma(\frac{\Delta_{21}+J+\Delta}{2})\Gamma(\frac{\Delta_{34}+J+\widetilde{\Delta}}{2})\Gamma(\frac{\Delta_{43}+J+\widetilde{\Delta}}{2})}{\Gamma(J+\Delta)\Gamma(J+d-\Delta)}\;.
\end{align}
Here $T^{(s)}$ is a kinematic s-channel prefactor:
\begin{align} 
	T^{(s)}&=\frac{1}{|x_{12}|^{\Delta_1+\Delta_2}|x_{34}|^{\Delta_3+\Delta_4}}\left(\frac{|x_{24}|}{|x_{14}|}\right)^{\Delta_1-\Delta_2}\left(\frac{|x_{14}|}{|x_{13}|}\right)^{\Delta_{3}-\Delta_4}\;.
\end{align}
The u-channel term is the same but with $3\leftrightarrow 4$. One can show that this form of the Lorentzian inversion formula is equivalent to the form we used above in. \equref{eq: Lorentzian inversion formula for scalars}, where we extract by comparing  \equref{eq: OPE function to OPE coefficients} and \equref{eq: OPE coefficients from Simon's form of Lorentzian inversion formula} that $c(\De,l)=\rho^{(s)}(\De,l)S(\cO_3\cO_4[\tl\cO_{\De,l}])$.

By first inverting individual blocks we can find the $6j$ symbol:
\begin{align}
	\sixj{\f_1}{\f_2}{\f_3}{\f_4}{\cO_5}{\cO_6}&=S(\f_3\f_4[\widetilde{\cO}_{6}])\sixjBlock{\f_1}{\f_2}{\f_3}{\f_4}{\cO_5}{\cO_{6}}+S(\f_1\f_2[\cO_{6}])\sixjBlock{\f_1}{\f_2}{\f_3}{\f_4}{\cO_5}{\widetilde{\cO}_{6}}, \label{eq:6j_symbol_split_Scalars}
	\\
	\sixjBlock{\f_1}{\f_2}{\f_3}{\f_4}{\cO_5}{\cO_{6}}&=\left(\Psi^{\tilde{1}\tilde{2}\tilde{3}\tilde{4}}_{\Delta_5,J_6},G^{3214}_{\Delta_6,J_6}\right)_{L}, \label{eq:scalar_Inv_Block}
\end{align}
where the subscript $L$ in (\ref{eq:scalar_Inv_Block}) is to emphasize that we use the Lorentzian inversion formula \cite{Liu:2018jhs}.

We do not have a closed form expression for the $6j$ symbol in generic dimensions, but for $d=3$ it is straightforward to calculate its poles and residues by using dimensional reduction and the explicit $d=2$ expressions \cite{Hogervorst:2016hal,Albayrak:2019gnz}. Let us now focus on the problem of inverting a single operator. The contribution of a single block for $\cO_{6}$ exchange gives
\begin{align}
	\<\f_1\f_2\f_3\f_4\>\supset \lambda_{\f_3\f_2\cO_{6}}\lambda_{\f_1\f_4\cO_6}\int\limits_{\mathcal{C}'}d\cO_{5}\frac{1}{\eta_{\cO_{5}}}\sixjBlock{\f_1}{\f_2}{\f_3}{\f_4}{\cO_5}{\cO_{6}}S(\f_3\f_4[\widetilde{\cO}_{5}])G^{1234}_{\cO_{5}}(x_i)\;. \label{eq:ind_Block_Inv}
\end{align}
As a function of $\Delta_{5}$ the integrand of (\ref{eq:ind_Block_Inv}) has poles at the following locations:
\begin{eqnarray}
	\Delta_{5}&=\Delta_{1}+\Delta_{2}+2n+J_5\;,
	\\
	\Delta_{5}&=\Delta_{3}+\Delta_{4}+2n+J_5\;.
\end{eqnarray}
These are the dimensions and spins of the double-twist operators $[\phi_1\phi_2]_{n,J}$ and $[\phi_3\phi_4]_{n,J}$ in mean field theory (MFT) \cite{Komargodski:2012ek,Fitzpatrick:2012yx}. A special case is when $\Delta_1+\Delta_2=\Delta_3+\Delta_4$ in which case we get single and double poles  corresponding to corrections to the MFT spectrum and OPE coefficients of $[\f_1\f_2]_{n,J}$. An important exception is when we are inverting the identity block, $\cO_{6}=\mathbb{1}$, in which case we get single poles and find the MFT OPE coefficients themselves. We will work out the MFT OPE coefficients for fermions in section \ref{sec:MFTOPE}.

We will now extend this discussion to non-scalar external operators. To keep track of possible signs associated with fermionic operators, let us proceed step by step. For spacelike seperated operators, we have the relation $\<\cO_1\cO_2\cO_3\cO_4\>=(-1)^{\Sigma_{23}+\Sigma_{12}+\Sigma_{13}}\<\cO_3\cO_2\cO_1\cO_4\> $, with which \equref{eq: Euclidean inversion formula} reads
\be 
\rho_{ab}^{(s)}(\cO_5)
= (-1)^{\Sigma_{23}+\Sigma_{12}+\Sigma_{13}}\eta_{(ac)(bd)}^{\cO_5}
\left(\Psi^{\tl{(s)}cd}_{\tl\cO_5}(x_i), \<\cO_3\cO_2\cO_1\cO_4\>\right).
\ee 
If we now apply the partial wave expansion in \equref{eq: partial wave expansion} we obtain
\be 
\label{eq: general 6j symbol relation middle step 1}
\rho_{ab}^{(s)}(\cO_5)
= (-1)^{\Sigma_{23}+\Sigma_{12}+\Sigma_{13}}\eta_{(ac)(bd)}^{\cO_5}
\int_{\cC}d\cO_6  \rho^{(t)}_{ef}(\cO_6)
\left(\Psi^{\tl{(s)}cd}_{\tl\cO_5}(x_i), \Psi^{(t)ef}_{\cO_6}(x_i)\right).
\ee 
We identify the conformally invariant pairing above as the generalization of $6j$ symbol in 	\equref{eq:6jDefScalar}:
\begin{multline}
	\sixj{\cO_1}{\cO_2}{\cO_3}{\cO_4}{\cO_5}{\cO_6}^{abcd}=\left(\widetilde{\Psi}^{(s),ab}_{\tl\cO_5},\Psi^{(t),cd}_{\cO_6}\right)=\int d^{d}x_{1}...d^{d}x_{6}\<\widetilde{\cO}_1\widetilde{\cO}_2\widetilde{\cO}_{5}\>\<\widetilde{\cO}_3\widetilde{\cO}_4\cO_5\>\\\x\<\cO_3\cO_2\cO_6\>\<\cO_1\cO_4\widetilde{\cO}_{6}\>. \label{eqn:Gen6jSymbol}
\end{multline}
As in the scalar case we can split this into two pieces, corresponding to the inversion of individual blocks:
\begin{align}
	\sixj{\cO_1}{\cO_2}{\cO_3}{\cO_4}{\cO_5}{\cO_6}^{abcd}&=(-1)^{\Sigma_{66}}\bigg[S^{d}_{e}\left(\cO_1\cO_4[\widetilde{\cO}_{6}]\right)\sixjBlock{\cO_1}{\cO_2}{\cO_3}{\cO_4}{\cO_5}{\cO_6}^{abce} \nonumber
	\\ &\hspace{.75in}+S^{c}_{e}\left(\cO_3\cO_2[\cO_{6}]\right)\sixjBlock{\cO_1}{\cO_2}{\cO_3}{\cO_4}{\cO_5}{\widetilde{\cO}_6}^{abed}\bigg]\;,\label{eqn:GenSplit6j}
	\\
	\sixjBlock{\cO_1}{\cO_2}{\cO_3}{\cO_4}{\cO_5}{\cO_6}^{abcd}&=\left(\widetilde{\Psi}^{(s),ab}_{\cO_{5}},G^{(t),cd}_{\cO_{6}}\right)_{L}\;.&\label{eqn:InvSingBlockGenSpin}
\end{align}
By using the shadow symmetry of the integrand, we can then rewrite \equref{eq: general 6j symbol relation middle step 1} as
\begin{multline}
	\rho_{ab}^{(s)}(\cO_5)
	= (-1)^{\Sigma_{66}+\Sigma_{23}+\Sigma_{12}+\Sigma_{13}}\eta_{(ac)(bd)}^{\cO_5}
	\\\x\int_{\cC'}d\cO_6  \rho^{(t)}_{ef}(\cO_6)
	S^{f}_{g}\left(\cO_1\cO_4[\widetilde{\cO}_{6}]\right)\sixjBlock{\cO_1}{\cO_2}{\cO_3}{\cO_4}{\cO_5}{\cO_6}^{cdeg}\;.
\end{multline}
If we now close the contour $\cC'$ to the right and pick up the residues at $\De=\De_6$, multiply both sides with $S^{c}_{b}(\cO_3\cO_4[\widetilde{\cO}_{5}])$, and then pick up the residues of both sides at $\De=\De_5$, we obtain the formula for the correction due to the inversion of  a single block $G^{(t),fg}_{\cO_6}$ in the crossed channel:
\begin{multline}
	\lambda_{125,a}\lambda_{345,b}\bigg|_{G^{(t),fg}_{\cO_6}}=(-1)^{1+\Sigma_{55}+\Sigma_{12}+\Sigma_{13}+\Sigma_{23}}\lambda_{326,f}\lambda_{146,g}\\\x\Res_{\Delta=\Delta_{5}}\eta_{(ad)(ce)}^{\De,J}\sixjBlock{\cO_1}{\cO_2}{\cO_3}{\cO_4}{\cO_{\De,J}}{\cO_6}^{defg}S^c_b(\cO_{3}\cO_{4}[\widetilde{\cO}])\bigg|_{J=J_{5}}\;, \label{eqn:InvSingBlock_to_OPEdata}
\end{multline}
where we use \equref{eq: OPE function to OPE coefficients} both in $s$ and $t$ channels.

The problem now is how to reduce the full $6j$ symbol, (\ref{eqn:Gen6jSymbol}), to a sum of scalar $6j$ symbols, (\ref{eq:6jDefScalar}). To do this we will need to use weight-shifting operators, which allow us to reduce general conformal integrals involving fermions to those involving scalars. We will review these operators in the next section.

\subsection{Weight-Shifting Operators}
\label{sec: shadow matrices}

The work \cite{Karateev:2017jgd} introduced differential operators, $\mathcal{D}^{A}$, which transform in a finite-dimensional representation of the conformal group $\SO(d+1,1)$ given by the index $A$. By acting with weight-shifting operators we can transform a conformally-invariant tensor structure involving an operator $\cO$ to a conformally covariant structure involving a new operator $\cO'$. Here, we will use weight-shifting operators which change the spin by half-integers.

In $d=3$ the double cover of the conformal group $\SO(3,2)$ is $\Sp(4,\mathbb{R})$, so we will use weight-shifting operators which transform in the fundamental representation of $\Sp(4,\mathbb{R})$. We will use the notation of  \cite{Karateev:2017jgd} and write the four possible operators as
\begin{align}
	\mathcal{D}^{\pm\pm}: [\Delta,\ell]\rightarrow [\Delta\pm \frac{1}{2},\ell\pm\frac{1}{2}]\;,
\end{align}
which in embedding space are given by
\bea[eq: differential operators in embedding space]
\left(\cD^{-+}\right)_a=& S_a\;,\\
\left(\cD^{--}\right)_a=& X_{ab}\pdr{}{S_b}\;,\\
\left(\cD^{++}\right)_a=& -2(\Delta-1)S_b(\partial_X)^b_{\;\;a}
-S_aS_b(\partial_X)^b_{\;\;c}\pdr{}{S_c}\;,
\\
\left(\cD^{+-}\right)_a=&
4(1+l-\De)(\Delta-1)\Omega_{ab}\pdr{}{S_b}
+2(1+l-\De)X_{ab}(\partial_X)^b_{\;c}\pdr{}{S_c}
\nonumber
\\
&-S_a\pdr{}{S_b}X_{bc}(\partial_X)^c_{\;d}\pdr{}{S_d}\;.
\eea
In the rest of the paper, we will suppress spinor index of the weight-shifting operators.

We will use these operators to relate conformal integrals involving fermionic tensor structures to known integrals involving bosonic structures. For this, we need to define the adjoint of a weight-shifting operator with respect to our bilinear pairing:
\begin{align}
	&\left(\mathcal{D}\cO,\widetilde{\cO}'\right)=\left(\cO,\mathcal{D}^{*}\widetilde{\cO}'\right)\;,
	\\
	&\left(\cO,\widetilde{\cO}\right)=\int d^dx\cO(x)\widetilde{\cO}(x)\;.
\end{align}
We should stress that here $\cO$ is shorthand for some representation of the conformal group and does not need to obey the spin-statistics theorem. We will define the adjoint by moving from the left to the right, where we recall there are suppressed spinor indices.

It is not hard to see that
\begin{align}
	\label{eq: adjoints of weight-shifting operators}
	\left(\mathcal{D}^{pq}\right)^{*}\bigg|_{\Delta,\ell}=\zeta^{pq}_{\ell}\mathcal{D}^{p,-q}\bigg|_{\widetilde{\Delta}\mp \frac{p}{2},\ell\pm \frac{q}{2}}\;,
\end{align}
where $p,q=\pm1$ and we have emphasized that the adjoint of a weight-shifting operator depends explicitly on the representation it acts on, although the coefficient $\zeta$ only depends on the spin. 

To calculate the adjoints in practice we go to the Poincar\'e section, or work in physical space. The result is summarized as:
\begin{eqnarray}
	&\zeta^{--}_{\ell}=-2\ell\;, \qquad &\zeta^{-+}_{\ell}=\frac{1}{2\ell+1}\;,
	\\
	&\zeta^{+-}_{\ell}=2\ell\;, \qquad  &\zeta^{++}_{\ell}=-\frac{1}{2\ell+1}\;.
\end{eqnarray}

The next ingredient we need is the crossing equation for covariant two-point functions:\footnote{To avoid clutter, we will use $\cD_n^{\pm\pm}$ to denote a weight-shifting operator acting on the $n^{\text{th}}$ operator in the correlator that follows the weight-shifting operator. For example, $\cD_2^{ab}\<\cO_{\De_1,l_1}(x_1)\cO_{\De_2,l_2}(x_2)\cO_{\De_3,l_3}(x_3)\>$ stands for $\cD^{ab}(X_2,S_2)\<\cO_{\De_1,l_1}(X_1,S_1)\cO_{\De_2,l_2}(X_2,S_2)\cO_{\De_3,l_3}(X_3,S_3)\>$ in the embedding space.}
\begin{align}
	\mathcal{D}_{1}^{pq}\<\cO_{\Delta,\ell}(x_1)\cO_{\Delta +\frac{p}{2},\ell+\frac{q}{2}}(x_2)\>=\alpha^{pq}_{\Delta,\ell}\mathcal{D}_{2}^{-p,-q}\<\cO_{\Delta,\ell}(x_1)\cO_{\Delta+\frac{p}{2},\ell+\frac{q}{2}}(x_2)\>\;,
\end{align}
and we find
\bea[eq: Definition of alpha]
\a^{--}_{\De,l}=&{} \frac{i l}{(2 \Delta -3) (\Delta +l-1)}\;,\\
\a^{-+}_{\De,l}=&{} -\frac{i}{2 (2 \Delta -3) (2 l+1)(-\Delta +l+2)}\;,\\
\a^{+-}_{\De,l}=&{} -8 i (\Delta -1) l (-\Delta +l+1)\;,\\
\a^{++}_{\De,l}=&{} \frac{4 i (\Delta -1) (\Delta +l)}{2 l+1}\;.
\eea 

We can now find the shadow transform of weight-shifted operators:
\begin{align}
	\mathbf{S}[\mathcal{D}^{pq}\cO_{\Delta,\ell}](x)&=\int d^{d}y\left(\mathcal{D}^{pq}\cO_{\Delta,\ell}(y)\right)\<\widetilde{\cO}_{\Delta+\frac{p}{2},\ell+\frac{q}{2}}(y)\widetilde{\cO}_{\Delta+\frac{p}{2},\ell+\frac{q}{2}}(x)\>
	\nonumber \\ &=\int d^{d}y \zeta^{pq}_{\ell}\alpha^{p,-q}_{\widetilde{\Delta}-\frac{p}{2},\ell+\frac{q}{2}}\cO_{\Delta,\ell}(y)\mathcal{D}^{-a,b}_{2}\<\widetilde{\cO}_{\Delta,\ell}(y)\widetilde{\cO}_{\Delta,\ell}(x)\>
	\nonumber \\ &= \zeta^{pq}_{\ell}\alpha^{p,-q}_{\widetilde{\Delta}-\frac{p}{2},\ell+\frac{q}{2}}\mathcal{D}^{-pq}\mathbf{S}[\cO_{\Delta,\ell}](x)\;,
\end{align}
or
\bea[eq:SDtoDS_Coef]
\mathbf{S}[\mathcal{D}^{pq}\cO_{\Delta,\ell}](x)&=\chi^{pq}_{\Delta,\ell}\mathcal{D}^{-p,q}\mathbf{S}[\cO_{\Delta,\ell}]\;,\label{eq:SDtoDS}
\\
\chi^{pq}_{\Delta,\ell}&=\zeta^{pq}_{\ell}\alpha^{p,-q}_{\widetilde{\Delta}-\frac{p}{2},\ell+\frac{q}{2}}\;.
\eea 
This gives a way to push the shadow transform past the weight-shifting operators. Then to calculate the shadow transform of a three-point tensor structure we will relate the fermionic structures to the bosonic ones. For simplicity we focus on three-point function structures involving one fermion and one scalar:
\begin{align}
	\<\psi_{\Delta_1}\phi_{\Delta_2}\cO_{\Delta_3,\ell_3}\>\;,
\end{align}
where $\cO$ has half-integer spin. We want to use weight-shifting operators to write such a three-point function in terms of a structure involving two scalars:
\begin{align}
	\<\phi_{\Delta_{1}'}\phi_{\Delta_{2}}\cO_{\Delta_{3}',\ell_{3}'}\>\;,
\end{align}
where the third operator $\cO$ now has integer spin. There are two equivalent ways to do this:
\bea
\<\psi_{\Delta_1}\phi_{\Delta_2}\cO_{\Delta_3,\ell_3}\>^a=\sum\limits_{p}\kappa^{a}_{1,p}(\psi_{\Delta_1}\phi_{\Delta_2}\cO_{\Delta_3,\ell_3})\mathcal{D}_{1}^{-p,+}\mathcal{D}_{3}^{-+}\<\phi_{\Delta_1+\frac{p}{2}}\phi_{\Delta_2}\cO_{\Delta_{3}+\frac{1}{2},\ell-\frac{1}{2}}\>,\label{eq: fsO decomposition 1}
\\
\<\psi_{\Delta_1}\phi_{\Delta_2}\cO_{\Delta_3,\ell_3}\>^a=\sum\limits_{p}\kappa^{a}_{2,p}(\psi_{\Delta_1}\phi_{\Delta_2}\cO_{\Delta_3,\ell_3})\mathcal{D}_{1}^{-p,+}\mathcal{D}_{3}^{++}\<\phi_{\Delta_1+\frac{p}{2}}\phi_{\Delta_2}\cO_{\Delta_{3}-\frac{1}{2},\ell-\frac{1}{2}}\>, \label{eq: fsO decomposition 2}
\eea
where each matrix, $\kappa_{1,2}$, is invertible. As the next simplest case, we can consider the three point structure involving two fermions
\begin{align}
	\label{eq: expansion of psipsiO}
	\<\psi_{\Delta_1}\psi_{\Delta_2}\cO_{\Delta_3,\ell_3}\>^a=\sum\limits_{p,q}\kappa^{a}_{3,pq}(\psi_{\Delta_1}\psi_{\Delta_2}\cO_{\Delta_3,\ell_3})\mathcal{D}_{1}^{-p,+}\mathcal{D}_{2}^{-q,+}\<\phi_{\Delta_1+\frac{p}{2}}\phi_{\Delta_2+\frac{q}{2}}\cO_{\Delta_{3}}\>\;.
\end{align}
where now the index $a$ runs from $1$ to $4$ hence $\ka_3$ is again invertible. The explicit forms of these $\ka$ matrices are easily computable though we will not present them as they are rather lengthy and do not provide any particular insight.

As we stated in \secref{\ref{sec: shadow}}, one can use weight shifting operators to find shadow matrices. To illustrate this, let us consider a few simple examples. The first one is the calculation of the matrix $S^{a}_{b}(\psi_1 [\phi_2]\cO_3)$:
\small 
\begin{align}
	{}&\<\psi_{\Delta_1}\mathbf{S}[\phi_{\Delta_2}]\cO_{\Delta_3,\ell_3}\>^{a}\\
	&=\mathbf{S}_{2}\sum\limits_{p}\kappa^{a}_{1,p}(\psi_{\Delta_1}\phi_{\Delta_2}\cO_{\Delta_3,\ell_3})\mathcal{D}_{1}^{-p,+}\mathcal{D}_{3}^{-+}\<\phi_{\Delta_1+\frac{p}{2}}\phi_{\Delta_2}\cO_{\Delta_{3}+\frac{1}{2},\ell-\frac{1}{2}}\>,
	\nonumber \\ &= \sum\limits_{p}\kappa^{a}_{1,p}(\psi_{\Delta_1}\phi_{\Delta_2}\cO_{\Delta_3,\ell_3})S(\phi_{\Delta_1+\frac{p}{2}}[\phi_{\Delta_2}]\cO_{\Delta_{3}+\frac{1}{2},\ell-\frac{1}{2}})\mathcal{D}_{1}^{-p,+}\mathcal{D}_{3}^{-+}\<\phi_{\Delta_1+\frac{p}{2}}\phi_{\widetilde{\Delta}_2}\cO_{\Delta_{3}+\frac{1}{2},\ell-\frac{1}{2}}\>,
	\nonumber \\ &= \sum\limits_{p,b}\kappa^{a}_{1,p}(\psi_{\Delta_1}\phi_{\Delta_2}\cO_{\Delta_3,\ell_3})S(\phi_{\Delta_1+\frac{p}{2}}[\phi_{\Delta_2}]\cO_{\Delta_{3}+\frac{1}{2},\ell-\frac{1}{2}})\left(\kappa^{-1}_{1}(\psi_{\Delta_1}\phi_{\widetilde{\Delta}_{2}}\cO_{\Delta_3,\ell_3})\right)^{p}_{b}\<\psi_{\Delta_1}\phi_{\widetilde{\Delta}_{2}}\cO_{\Delta_{3},\ell_{3}}\>^b.
\end{align}
\normalsize
Here we first wrote the fermionic structure in terms of the bosonic one, then acted with the shadow transform on the simple three point function involving two scalars, and then finally acted with the weight-shifting operators. After acting with the weight-shifting operators we expressed the answer in the original basis.  Therefore, the relevant shadow matrix is
\small 
\begin{align}
	S^{a}_{b}(\psi_{\Delta_1}[\phi_{\Delta_2}]\cO_{\Delta_3,\ell_3})=\sum\limits_{p}\kappa^{a}_{1,p}(\psi_{\Delta_1}\phi_{\Delta_2}\cO_{\Delta_3,\ell_3})S(\phi_{\Delta_1+\frac{p}{2}}[\phi_{\Delta_2}]\cO_{\Delta_{3}+\frac{1}{2},\ell-\frac{1}{2}})\left(\kappa^{-1}_{1}(\psi_{\Delta_1}\phi_{\widetilde{\Delta}_{2}}\cO_{\Delta_3,\ell_3})\right)^{p}_{b}.
\end{align}
\normalsize
As the next example, let us turn to $S^{a}_{b}([\psi_{\Delta_1}]\phi_{\Delta_2}\cO_{\Delta_3,\ell_3})$. To find the shadow transform of $\psi_{\Delta_1}$, we now have to pass the shadow transform past the weight-shifting operators using (\ref{eq:SDtoDS}) and (\ref{eq:SDtoDS_Coef}):
\begin{align}
	S^{a}_{b}(&[\psi_{\Delta_1}]\phi_{\Delta_2}\cO_{\Delta_3,\ell_3}) \\
	&=\sum\limits_{p}\kappa^{a}_{1,p}(\psi_{\Delta_1}\phi_2\cO_{\Delta_3,\ell_3})\chi^{-p,+}_{\Delta+\frac{p}{2},0}S\left([\phi_{\Delta_1+\frac{p}{2}}]\phi_{\Delta_2}\cO_{\Delta_3+\frac{1}{2},\ell-\frac{1}{2}}\right)\left(\kappa^{-1}_{1}(\psi_{\widetilde{\Delta}_{1}}\phi_{\Delta_2}\cO_{\Delta_3,\ell_3})\right)^{-p}_{b}. \nonumber
\end{align}
Lastly, one needs to pass shadow transform past the weight shifting operator for the shadow transform of the $\cO_{\Delta_3,\ell_3}$ in a similar manner:
\begin{align}
	S^{a}_{b}(\psi_{\Delta_1}\phi_{\Delta_2}[\cO_{\Delta_3,\ell_3}])=\sum\limits_{p}\kappa^{a}_{1,p}(\psi_{\Delta_1}\phi_2\cO_{\Delta_3,\ell_3})\chi^{-,+}_{\Delta_3+\frac{1}{2},\ell_3-\frac{1}{2}}\left(\kappa^{-1}_{2}(\psi_{\Delta_1}\phi_{\Delta_2}\cO_{\widetilde{\Delta}_3,\ell_3})\right)^{p}_{b}.
\end{align}

The computation of shadow matrices via weight shifting operators are analogus for three point structures of two fermions. For the sake of completeness, the shadow matrix for the shadow transform of spin$-l$ operator reads as
\small 
\begin{align}
	S^{a}_{b}(\psi_1\psi_2[\cO_{\Delta_3,\ell_3}])=\sum\limits_{p,q}\kappa^{a}_{3,pq}(\psi_{\Delta_1}\psi_{\Delta_2}\cO_{\Delta_3,\ell_{3}})S(\phi_{\Delta_1+\frac{p}{2}}\phi_{\Delta_2+\frac{q}{2}}[\cO_{\Delta_3,\ell_3}])\left(\kappa^{-1}_{3}(\psi_{\Delta_1}\psi_{\Delta_2}\cO_{\widetilde{\Delta}_3,\ell_3})\right)^{pq}_{b},
\end{align}
\normalsize
whereas the expression for the shadow transform of one of the $\psi$ operators is
\begin{multline}
	S^{a}_{b}([\psi_1]\psi_2\cO_3])
	=\sum\limits_{p,q}\kappa^{a}_{3,pq}(\psi_{\Delta_1}\psi_{\Delta_2}\cO_{\Delta_3,\ell_{3}})S([\phi_{\Delta_1+\frac{p}{2}}]\phi_{\Delta_2+\frac{q}{2}}\cO_{\Delta_3,\ell_3})\\\x\chi^{-p,+}_{\Delta_1+\frac{p}{2},0}\left(\kappa^{-1}_{3}(\psi_{\widetilde{\Delta}_1}\psi_{\Delta_2}\cO_{\Delta_3,\ell_3})\right)^{-p,q}_{b}.
\end{multline}

\section{Mean Field Theory OPE Coefficients}

\label{sec:MFTOPE}
In mean field theory we have the factorized correlator
\be 
\<\cO_1\cO_2\cO_3\cO_4\>=\<\cO_1\cO_2\>\<\cO_3\cO_4\>
+(-1)^{\Sigma_{23}}\<\cO_1\cO_3\>\<\cO_2\cO_4\>+\<\cO_1\cO_4\>\<\cO_2\cO_3\>\;.\ee 
The two-point function $\<\cO_i(x_i)\cO_j(x_j)\>$ is only non-zero when $\cO_i=\cO_j$, but we will leave this restriction implicit in this section.

We will now expand the MFT four-point function in s-channel partial waves and extract the partial wave expansion coefficient. The first term is automatically separated in the s-channel partial wave expansion, so we can focus on the latter two Wick contractions.

The $\<\cO_1\cO_4\>\<\cO_2\cO_3\>$ contraction gives
\small 
\be 
\rho_{ab}^{(s)}(\cO)\bigg|_{\<14\>\<23\>} &= \eta^{\cO_5}_{(ac)(bd)}\left(\widetilde{\Psi}^{(s),cd}_{\cO_5},\<\cO_1\cO_4\>\<\cO_2\cO_3\>\right)
\\ &= \eta^{\cO_5}_{(ac)(bd)}\int \frac{d^{d}x_{1}...d^{d}x_{5}}{\vol(\SO(d+1,1))}\<\widetilde{\cO}_{1}\widetilde{\cO}_{2}\widetilde{\cO}_{5}\>^c\<\widetilde{\cO}_3\widetilde{\cO}_4\cO_5\>^d\<\cO_1\cO_4\>\<\cO_2\cO_3\>
\\ &= \eta^{\cO_5}_{(ac)(bd)} (-1)^{\Sigma_{11}+\Sigma_{22}}S^{d}_{e}(\widetilde{\cO}_2[\widetilde{\cO}_1]\cO_5)S^{e}_{f}([\widetilde{\cO}_2]\cO_{1}\cO_5)\left(\<\widetilde{\cO}_{1}\widetilde{\cO}_{2}\widetilde{\cO}_{5}\>^c,\<\cO_2\cO_1\cO_5\>^f\right)\;,
\ee 
\normalsize
where we have made the replacements $\cO_{3,4}\rightarrow \cO_{2,1}$, respectively.  

The other contraction gives a similar result:
\small 
\be
\rho_{ab}^{(s)}(\cO)\bigg|_{\<13\>\<24\>}&=(-1)^{\Sigma_{23}} \eta^{\cO_5}_{(ac)(bd)}\left(\widetilde{\Psi}^{(s),cd}_{\cO_5},\<\cO_1\cO_3\>\<\cO_2\cO_4\>\right)
\\
&=(-1)^{\Sigma_{23}+\Sigma_{11}+\Sigma_{22}}\eta^{\cO_5}_{(ac)(bd)}S^{d}_{e}([\widetilde{\cO}_{1}]\widetilde{\cO}_{2}\cO_5)S^{e}_{f}(\cO_1[\widetilde{\cO}_{2}]\cO_5)\left(\langle\widetilde{\cO}_{1}\widetilde{\cO}_{2}\widetilde{\cO}_{5}\rangle ^{c},\<\cO_1\cO_2\cO_5\>^f\right)\;.
\ee
\normalsize

The full result in MFT is then
\be
\rho^{(s),\text{MFT}}_{ab}(\cO)=\hat{\delta}_{\cO_1\cO_4}\hat{\delta}_{\cO_2\cO_3}\rho_{ab}^{(s)}(\cO)\bigg|_{\<14\>\<23\>}+\hat{\delta}_{\cO_2\cO_4}\hat{\delta}_{\cO_1\cO_3}\rho_{ab}^{(s)}(\cO)\bigg|_{\<13\>\<24\>}\;, \label{eq: generic MFT result}
\ee
where $\hat{\delta}_{\cO_1\cO_2}=\delta_{\Delta_1\Delta_2}\delta_{l_1l_2}$ and should not be confused with the delta function on the principal series. Next we will apply this explicitly to various correlators containing fermions in order to calculate their MFT coefficients.

\subsection{$\<\psi\phi\phi\psi\>$}
For the correlator $\<\psi\phi\phi\psi\>$, the identity operator only appears in the t-channel, hence \equref{eq: generic MFT result} becomes
\be 
\label{eq: 2f2s MFT rho}
\rho_{ab}^{(s)}(\cO)=
& -\eta^{\cO}_{(ac)(bd)}S^{d}_{e}(\widetilde{\phi}[\widetilde{\psi}]\cO)S^{e}_{f}([\widetilde{\phi}]\psi\cO)\left(\<\widetilde{\psi}\widetilde{\phi}\widetilde{\cO}\>^c,\<\phi\psi\cO\>^f\right)\;.
\ee 
By explicit calculation, we find
\begin{multline}
	\rho^{(s)}_{ac}(\cO)
	S^{c}_{b}(\phi\psi[\tl\cO])
	=
	\frac{\pi ^{5/2} (2 \Delta -3) (2 J+1) 4^{\Delta _{\phi }-2} \Gamma \left(\Delta -\frac{3}{2}\right) (-\Delta +J+2)
		(\Delta +J-1)
	}{\Gamma (\Delta -1) \Gamma \left(J+\frac{3}{2}\right)
		\Gamma \left(\Delta _{\psi }-1\right) \Gamma \left(\Delta _{\psi }+\frac{1}{2}\right) }
	\\
	\x \frac{ \Gamma (J+1)\csc \left(\pi  \Delta _{\psi }\right) \sin \left(\pi  \Delta _{\phi }\right)\csc \left(2 \pi  \Delta _{\phi }\right) \csc (\pi  (J-\Delta ))}{\Gamma \left(2 \Delta _{\phi
		}-1\right) \Gamma (-J+\Delta -1) \Gamma (J+\Delta )}
	\left(
	\begin{array}{cc}
		c_1 & 0 \\
		0 & c_2 \\
	\end{array}
	\right)\;,
\end{multline}
with coefficients
\begin{subequations}
	\begin{multline}
		c_1=-\frac{\Gamma \left(\frac{1}{4} \left(2 J+2 \Delta _{\mathcal{O}\phi \psi }-1\right)\right) \Gamma \left(\frac{1}{4}
			\left(2 J+2 \Delta _{\mathcal{O}\psi \phi }+1\right)\right)}{\Gamma \left(\frac{1}{4} \left(2 J-2 \Delta _{\mathcal{O}\phi \psi }+5\right)\right) \Gamma
			\left(\frac{1}{4} \left(2 J-2 \Delta _{\mathcal{O}\psi \phi }+7\right)\right)}
		\\\x\frac{\Gamma \left(\frac{1}{4} \left(2 J+2 \Delta _{\phi \psi
				\mathcal{O}}-1\right)\right) \Gamma \left(\frac{1}{4} \left(2 J+2 \Delta +2 \Delta _{\phi }+2 \Delta _{\psi
			}-5\right)\right)}{\Gamma \left(\frac{1}{4} \left(2 J-2
			\Delta _{\phi \psi \mathcal{O}}+5\right)\right) \Gamma \left(\frac{1}{4} \left(2 J-2 \Delta -2 \Delta _{\phi }-2 \Delta
			_{\psi }+13\right)\right)}
	\end{multline}
	and
	\begin{multline}
		c_2=\frac{\Gamma \left(\frac{1}{4} \left(2 J+2 \Delta _{\mathcal{O}\phi \psi }+1\right)\right) \Gamma \left(\frac{1}{4}
			\left(2 J+2 \Delta _{\mathcal{O}\psi \phi }-1\right)\right)}{\Gamma \left(\frac{1}{4} \left(2 J-2 \Delta _{\mathcal{O}\phi \psi }+7\right)\right) \Gamma
			\left(\frac{1}{4} \left(2 J-2 \Delta _{\mathcal{O}\psi \phi }+5\right)\right)}
		\\\x\frac{\Gamma \left(\frac{1}{4} \left(2 J+2 \Delta _{\phi \psi
				\mathcal{O}}+1\right)\right) \Gamma \left(\frac{1}{4} \left(2 J+2 \Delta +2 \Delta _{\phi }+2 \Delta _{\psi
			}-7\right)\right)}{\Gamma \left(\frac{1}{4} \left(2 J-2
			\Delta _{\phi \psi \mathcal{O}}+7\right)\right) \Gamma \left(\frac{1}{4} \left(2 J-2 \Delta -2 \Delta _{\phi }-2 \Delta
			_{\psi }+11\right)\right)}\;,	
	\end{multline}
\end{subequations}
where $\De_{abc}\coloneqq\De_a+\De_b-\De_c$ as in \equref{eq: definition of delta with three indices}.

We see that the first component has residues at $\Delta=\Delta_\psi+\Delta_\phi+J-\half+2n$, which corresponds to the double twist families $[\phi\psi_\a]_{n,J}$. In contrast, the last component has residues at $\Delta=\Delta_\psi+\Delta_\phi+J+\half+2n$, which corresponds to the double twist families $[\phi\partial_{\a\b}\psi^\b]_{n,J}$.

By taking their respective residues, we can find the OPE coefficients. For the leading ($n=0$) tower,\footnote{Results for higher-twist towers are given in \secref{\ref{app:MFT_higher_twist}}.} they read as
\footnotesize
\bea 
P_{11}^{(s)}\left([\phi\psi]_{0,J}\right)=&
-\frac{\Gamma \left(J+\Delta _{\psi }\right) \Gamma \left(J+\Delta _{\phi }-\frac{1}{2}\right) \Gamma
	\left(J+\Delta _{\phi }+\Delta _{\psi }-1\right)}{\Gamma \left(J+\frac{1}{2}\right) \Gamma \left(\Delta _{\psi
	}+\frac{1}{2}\right) \Gamma \left(\Delta _{\phi }\right) \Gamma \left(2 J+\Delta _{\phi }+\Delta _{\psi
	}-\frac{3}{2}\right)}\;,
\\
P_{22}^{(s)}\left([\phi\partial\psi]_{0,J}\right)=&
-\frac{ (2 J+1) 2^{3-2 \Delta _{\phi }} \Gamma (J+1) \Gamma \left(\Delta _{\psi }\right) \Gamma
	\left(J+\Delta _{\psi }\right) \left(\Delta _{\psi }+\Delta _{\phi }+J-1\right)}{\sqrt{\pi } \left(2 \Delta _{\psi
	}-1\right) \Gamma (2 J+3) \Gamma \left(2 \Delta _{\psi }-2\right) \Gamma \left(\Delta _{\phi }\right) \Gamma
	\left(J+\Delta _{\phi }+\Delta _{\psi }-\frac{1}{2}\right) }
\nn\\&\x\frac{\Gamma \left(J+\Delta _{\phi
	}+\frac{1}{2}\right) \Gamma \left(2 J+2 \Delta _{\phi }+2 \Delta _{\psi }-3\right)}{\Gamma \left(2 J+\Delta _{\phi }+\Delta _{\psi
	}-\frac{1}{2}\right)}\;.
\eea 
\normalsize
where we define shorthand notation\footnote{The reason for the difference between \equref{eq: definition of opeP} and $P_\cO^{a,b}=(-1)^lf_{\psi_1\psi_2\cO}^a f_{\psi_3\psi_4\cO}^b$ as given in \equref{eq: definition of P in perturbative section} is the difference in the conformal block normalization used in this chapter and chapter \ref{chapter: perturbative} --- also see footnote~\ref{footnote: conformal block normalization difference}.}
\be 
\label{eq: definition of opeP}
P^{(s)}_{ab}(\cO)\coloneqq\lambda^a_{\cO_1\cO_2\cO}\lambda^b_{\cO_3\cO_4\cO}\;.
\ee 

At large $J$, the leading behavior is
\bea[eq: MFT coefficients at large spin for two fermion and two scalars]
P_{11}^{(s)}\left([\phi\psi]_{0,J}\right)\sim&
-\frac{\sqrt{\pi } 2^{-\Delta_ \psi -\Delta_ \phi -2 J+\frac{5}{2}} J^{\Delta_ \psi +\Delta_ \phi
		-1}}{\Gamma \left(\Delta_ \psi +\frac{1}{2}\right) \Gamma (\Delta _\phi )}\;,
\\
P_{22}^{(s)}\left([\phi\partial\psi]_{0,J}\right)\sim&
-\frac{\sqrt{\pi }  \left(\Delta _{\psi }-1\right) 2^{-\Delta _{\psi }-\Delta _{\phi }-2 J+\frac{3}{2}}
	J^{\Delta _{\psi }+\Delta _{\phi }-2}}{\Gamma \left(\Delta _{\psi }+\frac{1}{2}\right) \Gamma \left(\Delta _{\phi
	}\right)}\;.
\eea 
As we are working with three-point structures such that $\lambda_{\phi\psi\cO,a}=(-1)^{J-\half}\lambda_{\psi\phi\cO,a}$, we see that
\bea 
\left(\lambda_{\psi\phi\cO,1}\right)^2\sim& -(-1)^{J-\half}
\frac{\sqrt{\pi }2^{-\Delta_ \psi -\Delta_ \phi -2 J+\frac{5}{2}} J^{\Delta_ \psi +\Delta_ \phi
		-1}}{\Gamma \left(\Delta_ \psi +\frac{1}{2}\right) \Gamma (\Delta _\phi )}\;,
\label{eq: OPE coefficient squared for fermion scalar parity even double twist family}
\\
\left(\lambda_{\psi\phi\cO,2}\right)^2\sim& -(-1)^{J-\half}
\frac{\sqrt{\pi } \left(\Delta _{\psi }-1\right) 2^{-\Delta _{\psi }-\Delta _{\phi }-2 J+\frac{3}{2}}
	J^{\Delta _{\psi }+\Delta _{\phi }-2}}{\Gamma \left(\Delta _{\psi }+\frac{1}{2}\right) \Gamma \left(\Delta _{\phi
	}\right)}\;.
\eea 
In the free theory limit with $\{\De_\phi,\De_\psi\}\rightarrow\{\half,1\}$, these become
\be 
\left(\lambda_{\psi\phi\cO,1}\right)^2\sim -(-1)^{J-\half}4^{1-J}\sqrt{\frac{J}{\pi}}\;,\quad \left(\lambda_{\psi\phi\cO,2}\right)^2\sim 0\;.
\ee 
We see that this matches the results of \cite{Iliesiu:2015akf} once we take the normalizations into account. 
Specifically, they normalize their operators as
\begin{align} 
	\<\cO_{\Delta,\ell}(X_1,S_1)\cO_{\Delta,\ell}(X_2,S_2)\> &= c^{\text{there}}_{\cO}i^{2l}\frac{\<S_1S_2\>^{2l}}{X_{12}^{2\De+2l}}\;, 
	\\
	c^{\text{there}}_{\cO} &=\frac{4^{\Delta -1} (-1)^{\frac{1}{2}-J} \Gamma \left(J+\frac{3}{2}\right)}{\left(\frac{1}{2}\right)_{J+\frac{1}{2}}}\;,
\end{align}
while we take $c^{\text{here}}_{\cO}=1$.
Thus we have
\be 
c_\cO^{\text{there}}\left(\lambda_{\psi\phi\cO,1}\right)^2\sim -4 J\;,
\ee 
which can be compared with Eq. (4.4) of \cite{Iliesiu:2015akf}.

\subsection{$\<\psi_1\psi_2\psi_2\psi_1\>$}
Now we turn to correlators containing four fermions. For a correlator of the form $\<\psi_1\psi_2\psi_2\psi_1\>$, the identity operator only appears in the $14\rightarrow23$ channel, hence \equref{eq: generic MFT result} becomes
\be 
\rho_{ab}^{(s)}(\cO)
=
\eta^{\cO}_{(ac)(bd)} S^{d}_{e}(\widetilde{\psi}_2[\widetilde{\psi}_1]\cO)S^{e}_{f}([\widetilde{\psi}_2]\psi_{1}\cO)\left(\<\widetilde{\psi}_{1}\widetilde{\psi}_{2}\widetilde{\cO}\>^c,\<\psi_2\psi_1\cO\>^f\right)\;.
\ee 

When we calculate $\rho$, which we will not reproduce here, we see that it is block-diagonal, with the upper $2\x2$ block having poles at $\Delta=\Delta_{\psi_1}+\Delta_{\psi_2}+J-1+2n$, and with the lower $2\x2$ block having poles at $\Delta=\Delta_{\psi_1}+\Delta_{\psi_2}+J+2n$. These correspond to the double-twist families $[\psi_{1\a}\psi_{2\b}]_{n,J}$ and $[\psi_{1\a}\psi_2^\a]_{n,J}$ respectively. We then can read off the explicit results for OPE coefficients. 

For the leading tower ($n=0$) we obtain:
\footnotesize
\be 
{}&P^{(s)}_{[\psi_{1\a}\psi_{2\b}]_{0,J}}
=
-\frac{\Gamma \left(J+\Delta _1-\frac{1}{2}\right) \Gamma \left(J+\Delta _2-\frac{1}{2}\right) \Gamma \left(J+\Delta
	_1+\Delta _2-1\right)}{\Gamma \left(\Delta _1+\frac{1}{2}\right) \Gamma \left(\Delta _2+\frac{1}{2}\right) \Gamma (J)
	\Gamma \left(2 J+\Delta _1+\Delta _2-2\right)}\begin{pmatrix}
	0&0\\0&1
\end{pmatrix},
\\
&P^{(s)}_{[\psi_{1\a}\psi_2^\a]_{0,J}}
=
-\frac{\left(\Delta _1+\Delta _2+2 J-1\right) \Gamma \left(J+\Delta _1-\frac{1}{2}\right) \Gamma \left(J+\Delta
	_2-\frac{1}{2}\right) \Gamma \left(J+\Delta _1+\Delta _2-2\right)}{4 \Gamma \left(\Delta _1+\frac{1}{2}\right) \Gamma
	\left(\Delta _2+\frac{1}{2}\right) \Gamma (J+2) \Gamma \left(2 J+\Delta _1+\Delta _2\right)}
\\
&\x \begin{pmatrix}
	\left(\Delta _1+\Delta _2-2\right) \left(\Delta _1+J-\frac{1}{2}\right) \left(\Delta _2+J-\frac{1}{2}\right) +c_1& \left(\Delta _1-\Delta _2\right) J (J+1) \left(\Delta _1+\Delta _2+J-\frac{3}{2}\right)\\
	-\left(\Delta _1-\Delta _2\right) J (J+1) \left(\Delta _1+\Delta _2+J-\frac{3}{2}\right)& \left(\Delta _1+\Delta _2-2\right) \left(\Delta _1+J-\frac{1}{2}\right) \left(\Delta _2+J-\frac{1}{2}\right)-c_1
\end{pmatrix}\;,
\ee 
\normalsize
with
\begin{multline}
	c_1= \frac{1}{4} \left(2 \Delta _1-1\right) \left(\Delta _1+\Delta _2-2\right) \left(2 \Delta _2-1\right)\\
	+\left(\Delta
	_1+\Delta _2-2\right) J^3+\left(\Delta _1^2+\left(4 \Delta _2-\frac{9}{2}\right) \Delta _1+\Delta _2^2-\frac{9 \Delta
		_2}{2}+3\right) J^2\\+\left(\Delta _1+\Delta _2-1\right) \left(\Delta _1 \left(2 \Delta _2-1\right)-\Delta _2\right) J\;.
\end{multline}
Results for higher-twist towers are given in section~\ref{app:MFT_higher_twist}.

\subsection{$\<\psi\psi\psi\psi\>$}
For identical fermions $\<\psi\psi\psi\psi\>$, both the t- and u-channels contribute, hence we have
\be 
\rho_{ab}^{(s)}(\cO)
=\, &\eta^{\cO}_{(ac)(bd)} S^{d}_{e}(\widetilde{\psi}[\widetilde{\psi}]\cO)S^{e}_{f}([\widetilde{\psi}]\psi\cO)\left(\<\widetilde{\psi}\widetilde{\psi}\widetilde{\cO}\>^c,\<\psi\psi\cO\>^f\right)
\nonumber 
\\
&-\eta^{\cO}_{(ac)(bd)}S^{d}_{e}([\widetilde{\psi}]\widetilde{\psi}\cO)S^{e}_{f}(\psi[\widetilde{\psi}]\cO)\left(\<\widetilde{\psi}\widetilde{\psi}\widetilde{\cO}\>^{c},\<\psi\psi\cO\>^f\right).
\ee 

For the leading ($n=0$) tower, the explicit expressions for the OPE coefficients are as follows:
\footnotesize
\be
\label{eq: MFT coefficients for four fermions}
P^{(s)}_{[\psi_\a\psi_\b]_{n,J}}
=&\frac{2 (-1)^{J+1} \left(\Delta _{\psi }+J-1\right) \Gamma \left(J+\Delta _{\psi
	}-\frac{1}{2}\right){}^2 \Gamma \left(J+2 \Delta _{\psi }-1\right)}{\Gamma (J) \Gamma \left(\Delta _{\psi
	}+\frac{1}{2}\right){}^2 \Gamma \left(2 J+2 \Delta _{\psi }-1\right)}\begin{pmatrix}
	0&0\\0&1+(-1)^J
\end{pmatrix},
\\
P^{(s)}_{[\psi_\a\psi^\a]_{n,J}}
=&
\frac{(-1)^{J+1} 2^{-2 J} J \Gamma \left(\Delta _{\psi }\right) \Gamma \left(J+\Delta _{\psi }+\frac{1}{2}\right) \Gamma
	\left(J+2 \Delta _{\psi }\right)}{\left(1-2 \Delta _{\psi }\right){}^2 \Gamma (J+1) \Gamma \left(\Delta _{\psi
	}-\frac{1}{2}\right) \Gamma \left(2 \Delta _{\psi }-2\right) \Gamma \left(J+\Delta _{\psi }\right)}\left(
\begin{array}{cc}
	\frac{1+(-1)^J}{J \left(J+2 \Delta _{\psi }-2\right)} & 0 \\
	0 & \frac{1-(-1)^{J}}{(J+1) \left(J+2 \Delta _{\psi }-1\right)} \\
\end{array}
\right).
\ee
\normalsize
After an appropriate change of basis, these results match perfectly to those calculated using the lightcone bootstrap at large $J$~\cite{Albayrak:2019gnz}.

\subsection{MFT Coefficients for Higher Twist Towers}
\label{app:MFT_higher_twist}

In the subsections above we presented the MFT coefficients for the leading twist towers; here we will present the results for higher twist towers.

\paragraph{For $\<\f\psi\psi\f\>:$}	
\footnotesize
\bea 
P_{11}^{(s)}\left([\phi\psi]_{n,J}\right)=&\,
\frac{(2 J+1) \Gamma (J+1) \cos \left(\pi  \left(\Delta _{\psi }+\Delta
	_{\phi }\right)\right) \Gamma \left(n+\Delta _{\psi }-1\right) \Gamma \left(n+\Delta _{\phi }-\frac{1}{2}\right) \Gamma
	\left(J+n+\Delta _{\psi }\right)}{2^{4-2 \Delta _{\phi }}\pi ^{3/2} n! \Gamma \left(J+\frac{3}{2}\right) \Gamma
	\left(\Delta _{\psi }-1\right) \Gamma \left(\Delta _{\psi }+\frac{1}{2}\right) \Gamma \left(2 \Delta _{\phi }-1\right)
	\Gamma (J+n+1)}
\nn\\
&\x \frac{\left(2 \Delta _{\psi }+2 \Delta _{\phi }+4 J+4 n-3\right) \Gamma \left(J+n+\Delta
	_{\phi }-\frac{1}{2}\right) \Gamma \left(-2 n-\Delta _{\phi }-\Delta _{\psi }+\frac{7}{2}\right) }{\Gamma \left(2 J+2 n+\Delta _{\phi
	}+\Delta _{\psi }-\frac{1}{2}\right)}
\nn\\
&\x\frac{\Gamma \left(n+\Delta
	_{\phi }+\Delta _{\psi }-\frac{5}{2}\right) \Gamma \left(J+n+\Delta _{\phi }+\Delta _{\psi }-\frac{3}{2}\right) \Gamma
	\left(J+2 n+\Delta _{\phi }+\Delta _{\psi }-1\right)}{\Gamma \left(J+2 n+\Delta _{\phi }+\Delta _{\psi }-\frac{3}{2}\right)},
\\
P_{22}^{(s)}\left([\phi\partial\psi]_{n,J}\right)=&\,
P_{11}^{(s)}\left([\phi\psi]_{n,J}\right) \frac{\left(\Delta _{\psi }+n-1\right) \left(\Delta _{\phi }+J+n-\frac{1}{2}\right) \left(\Delta _{\psi }+\Delta _{\phi
	}+n-\frac{5}{2}\right)}{2 (J+n+1)
	\left(\frac{1}{4} \left(2 \Delta _{\psi }+2 \Delta _{\phi }-5\right)+n\right) \left(\Delta _{\psi }+\Delta _{\phi }+J+2
	n-\frac{3}{2}\right)}
\nn\\&\x\frac{\left(\frac{1}{2} \left(\Delta _{\psi }+\Delta _{\phi }+J-1\right)+n\right)}{ \left(J+\frac{1}{4} \left(2 \Delta _{\psi }+2 \Delta _{\phi }+4 n-3\right)\right)}\;.
\eea 
\normalsize

\paragraph{For $\<\psi_1\psi_2\psi_2\psi_1\>:$}
\small 
\begin{multline}
	P^{(s)}_{[\psi_{1\a}\psi_{2\b}]_{n,J}}= -\begin{pmatrix}
		\frac{n \left(\Delta _1+\Delta _2+2 J+2 n-2\right) }{4 J}c_1 & c_2\\ c_2& \left(\Delta _1+\Delta _2+J+2 n-2\right) \left(\Delta _1+\Delta _2+2 J+2 n-2\right) c_3
	\end{pmatrix}
	\\\x \frac{\Gamma \left(J+n+\Delta
		_2-\frac{1}{2}\right) \Gamma \left(J+n+\Delta _1+\Delta _2-\frac{5}{2}\right) \Gamma \left(J+2 n+\Delta _1+\Delta
		_2-3\right)}{\Gamma \left(2
		n+\Delta _1+\Delta _2-3\right) \Gamma \left(J+2 n+\Delta _1+\Delta _2-\frac{5}{2}\right) \Gamma \left(2 J+2 n+\Delta
		_1+\Delta _2-1\right)}
	\\\x  \frac{\Gamma \left(J+\frac{3}{2}\right) \Gamma \left(n+\Delta _1-1\right) \Gamma \left(n+\Delta _2-1\right) \Gamma
		\left(n+\Delta _1+\Delta _2-3\right) \Gamma \left(J+n+\Delta _1-\frac{1}{2}\right) }{(J+1) n! \Gamma \left(\Delta _1-1\right) \Gamma \left(\Delta _1+\frac{1}{2}\right) \Gamma \left(\Delta
		_2-1\right) \Gamma \left(\Delta _2+\frac{1}{2}\right) \Gamma (J) \Gamma \left(J+n+\frac{3}{2}\right) },
\end{multline} 
\normalsize
and
\small 
\begin{multline}
	P^{(s)}_{[\psi_{1\a}\psi_2^\a]_{n,J}}=-2^{2 \Delta _1+2 \Delta _2-6}\begin{pmatrix}
		\frac{1+J}{2}d_1&d_2\\-d_2&-\frac{J}{2}d_3
	\end{pmatrix}
	\\\x
	\frac{\Gamma \left(J+\frac{3}{2}\right) \left(\Delta _1+\Delta _2+2 J+2 n-1\right) \Gamma
		\left(n+\Delta _1-1\right) \Gamma \left(n+\Delta _2-1\right) \Gamma \left(n+\Delta _1+\Delta _2-2\right) }{\pi  \left(2 \Delta _1-1\right) \left(2 \Delta
		_2-1\right) (J+1) n! \Gamma \left(2 \Delta _1-2\right) \Gamma \left(2 \Delta _2-2\right) \Gamma (J+1) \Gamma
		\left(J+n+\frac{3}{2}\right)}
	\\\x \frac{\Gamma
		\left(J+n+\Delta _1-\frac{1}{2}\right) \Gamma \left(J+n+\Delta _2-\frac{1}{2}\right) \Gamma \left(J+n+\Delta _1+\Delta
		_2-\frac{3}{2}\right) \Gamma \left(J+2 n+\Delta _1+\Delta _2-2\right)}{\Gamma \left(2 n+\Delta _1+\Delta _2-2\right)\Gamma \left(J+2 n+\Delta _1+\Delta
		_2-\frac{3}{2}\right) \Gamma \left(2 J+2 n+\Delta _1+\Delta _2\right)},
\end{multline}
\normalsize
where
\be 
c_1=&
4 J^2 \left(\Delta _1+\Delta _2+n-3\right)+4 J \left(\Delta _1+\Delta _2+n-3\right) \left(\Delta _1+\Delta _2+2
n-2\right)\\&+\left(\Delta _1+\Delta _2+2 n-3\right) \left(2 \Delta _1+2 \Delta _2+2 n-5\right),
\\
c_2=&4 n \left(\Delta _1+\Delta _2+n-\frac{7}{2}\right) \left(\frac{1}{2} \left(\Delta _1+\Delta _2+J-2\right)+n\right)
\left(J+\frac{1}{2} \left(\Delta _1+\Delta _2+2 n-2\right)\right),
\\
c_3=&  J^2+J \left(\Delta _1+\Delta
_2+2 n-2\right)+(2 n+1) \left(\Delta _1+\Delta _2+n-3\right),
\\
d_1=& 2 J^2 \left(\Delta _1+\Delta _2+2 n-2\right)+\left(2 \Delta _1+2 n-1\right) \left(\Delta
_1+\Delta _2+2 n-2\right) \left(2 \Delta _2+2 n-1\right)
\\&+J \left(-9 \Delta _1-9 \Delta _2+2 \left(\Delta _1^2+\Delta _2^2+4 \Delta _1
\Delta _2+6 n^2+\left(6 \Delta _1+6 \Delta _2-9\right) n\right)+6\right),
\\
d_2=& \left(\Delta _1-\Delta _2\right) J (J+1) \left(\Delta _1+\Delta _2+J+2 n-\frac{3}{2}\right),
\\
d_3=&J \left(-11 \Delta _1-11 \Delta _2+2 \left(\Delta _1^2+\Delta _2^2+4 \Delta
_1 \Delta _2+6 n^2+\left(6 \Delta _1+6 \Delta _2-11\right) n\right)+10\right)
\\&+4 \left(\Delta _1+n-1\right) \left(\Delta
_2+n-1\right) \left(\Delta _1+\Delta _2+2 n-1\right)
+ 2 J^2 \left(\Delta _1+\Delta _2+2 n-2\right)\;.
\ee 

\paragraph{For $\<\psi\psi\psi\psi\>:$}
\begin{multline}
	P^{(s)}_{[\psi_\a\psi_\b]_{n,J}}
	=
	n\frac{(-1)^{J+1} \left((-1)^J+1\right) 2^{-2 J-4 n+1} \Gamma \left(J+\frac{3}{2}\right) \Gamma \left(n+\Delta _{\psi
		}-1\right) \Gamma \left(n+2 \Delta _{\psi }-3\right) }{\left(1-2 \Delta _{\psi
		}\right)^2 \Gamma (J+2) \Gamma (n+1) \Gamma \left(2 \left(\Delta _{\psi }-1\right)\right){}^2 \Gamma
		\left(J+n+\frac{3}{2}\right)}
	\\\x
	\frac{\Gamma \left(J+n+\Delta _{\psi }-\frac{1}{2}\right) \Gamma
		\left(J+n+2 \Delta _{\psi }-\frac{5}{2}\right) \Gamma \left(J+2 n+2 \Delta _{\psi }-3\right)}{\Gamma \left(n+\Delta _{\psi }-\frac{3}{2}\right) \Gamma \left(J+n+\Delta _{\psi
		}-1\right) \Gamma \left(J+2 n+2 \Delta _{\psi }-\frac{5}{2}\right)}
	\begin{pmatrix}
		c_1&  c_2\\  c_2&\frac{1}{n}c_3
	\end{pmatrix},
\end{multline}
and
\small 
\begin{multline}
	P^{(s)}_{[\psi_\a\psi^\a]_{n,J}}
	=
	\frac{(-1)^J 2^{-2 J-4 n} \Gamma \left(J+\frac{3}{2}\right) \Gamma \left(n+\Delta _{\psi }\right) \Gamma \left(n+2
		\Delta _{\psi }-2\right)\Gamma \left(J+2 n+2 \Delta _{\psi }\right)}{\left(1-2 \Delta _{\psi }\right){}^2 \Gamma (J+2)
		\Gamma (n+1) \Gamma \left(2 \Delta _{\psi }-2\right){}^2 \Gamma \left(J+n+\frac{3}{2}\right)\Gamma \left(J+2 n+2 \Delta _{\psi
		}-\frac{3}{2}\right)}
	\\\x \frac{\Gamma \left(J+n+\Delta _{\psi }+\frac{1}{2}\right) \Gamma \left(J+n+2 \Delta _{\psi
		}-\frac{3}{2}\right)}{\Gamma \left(n+\Delta
		_{\psi }-\frac{1}{2}\right) \Gamma \left(J+n+\Delta _{\psi }\right)}\left(
	\begin{array}{cc}
		-\frac{\left(1+(-1)^J\right) (J+1)}{J+2 \left(n+\Delta _{\psi }-1\right)} & 0 \\
		0 & \frac{\left(-1+(-1)^J\right) J}{J+2 n+2 \Delta _{\psi }-1} \\
	\end{array}
	\right),
\end{multline}
\normalsize
where
\small 
\be 
c_1=&4 J^2 \left(2 \Delta _{\psi }+n-3\right)+8 J \left(\Delta _{\psi }+n-1\right) \left(2 \Delta _{\psi
}+n-3\right)+\left(2 \Delta _{\psi }+2 n-3\right) \left(4 \Delta _{\psi }+2 n-5\right),\\
c_2=&2J \left(4 \Delta _{\psi }+2 n-7\right) \left(J+2 \left(\Delta _{\psi }+n-1\right)\right),\\
c_3=&4J \left(J+2 \left(\Delta _{\psi }+n-1\right)\right) \left(J^2+2 J \left(\Delta _{\psi }+n-1\right)+(2 n+1) \left(2
\Delta _{\psi }+n-3\right)\right).
\ee 
\normalsize

\section{CFT data of a general spectrum}

In this section, we will use the techniques we developed in previous section to calculate the anomalous dimensions of the double-twist operators exchanged in the $\<\psi\f\f\psi\>$ and $\<\psi\psi\psi\psi\>$ correlators. For this, we will primarily focus on using the coefficient of the double poles in \equref{eq: OPE function in terms of scalar 6j symbol} to obtain $\delta h P$, and we will divide it by $P^{\text{MFT}}$ to obtain $\delta h=\gamma/2$.\footnote{One can further improve these results by considering the residues of \equref{eq: OPE function in terms of scalar 6j symbol} to obtain $\delta P$ as in (\ref{eq: OPE data in terms of scalar 6j symbol}) or the example in~\ref{app:OPE}, with which one schematically has $\delta h=\frac{\delta h P}{P^{\text{MFT}}+\delta P}$ up to possible mixing between different twist towers. Necessary $\cK$ coefficients needed for these computations are provided in \cite{Albayrak:2020rxh}.} We also give an example of calculating OPE coefficient corrections in section~\ref{app:OPE}.

\subsection{Applications in the computation of anomalous dimensions}
\label{sec:applications}

As we reviewed in section~\ref{sec:6jreview}, $6j$ symbols develop double poles when certain relations are satisfied; in the case of \equref{eq: OPE function in terms of scalar 6j symbol}, we have double poles in $\De'$ at \mbox{$\De'=\De_1+\De_2+J'$} if $\Delta_1+\Delta_2=\Delta_3+\Delta_4$. Here these correspond to double-twist operators $[\f_1\f_2]$. We will stick to decomposition coefficients $\cK$ with $\cO_6'=\cO_6$, so taking the coefficient of double poles of \equref{eq: OPE function in terms of scalar 6j symbol} roughly translates into relating $\delta h P_{[\cO_1\cO_2]}\evaluated_{\cO_6}$ to $\delta h P_{[\f_1\f_2]}\evaluated_{\cO_6}$, where here $\f_i$ are some fictitious scalar operators. For later convenience, we define
\small
\bea 
\mathfrak{dp}_1^{J,n}(\f_1,\f_2,\cO_6)\equiv &
\lim\limits_{\De\rightarrow\De_1+\De_2+J+2n} \left(\De -\De_1-\De_2-J-2n\right)^2 \frac{S(\f_3 \f_4 [\widetilde{\cO}_{5}])}{\eta^{(s)}_{\cO_{5}}} \sixjBlock{\f_1}{\f_2}{\f_2}{\f_1}{\cO_{\De,J}}{\cO_{6}},
\\
\mathfrak{dp}_2^{J,n}(\f_1,\f_2,\cO_6)\equiv &
\lim\limits_{\De\rightarrow\De_1+\De_2+J+2n} \left(\De -\De_1-\De_2-J-2n\right)^2  \frac{S(\f_3 \f_4 [\widetilde{\cO}_{5}])}{\eta^{(s)}_{\cO_{5}}}  \sixjBlock{\f_1}{\f_2}{\f_1}{\f_2}{\cO_{\De,J}}{\cO_{6}},
\eea
\normalsize
which describe the corrections $\delta h P$ with external scalars after the OPE coefficients are stripped off:
\be 
\delta h P_{[\f_1\f_2]}\evaluated_{\cO_6\in \phi_1\x \phi_1}=&\lambda_{\f_1\f_1\cO_6}\lambda_{\f_2\f_2\cO_6}\mathfrak{dp}_{1}^{J,n}(\f_1,\f_2,\cO_6)\;,
\\
\delta h P_{[\f_1\f_2]}\evaluated_{\cO_6\in \phi_1\x \phi_2}=&\lambda_{\f_1\f_2\cO_6}^2\mathfrak{dp}_{2}^{J,n}(\f_1,\f_2,\cO_6)\;.
\ee 
As a reminder, in these formulas $\eta$ is the normalization for the scalar partial wave.

The particular decomposition we will be using in \equref{eq: OPE function in terms of scalar 6j symbol} will  be the one with $\cO_6$ kept constant, hence by taking the coefficients of double poles on both sides, we obtain
\begin{multline}
	\label{eq: general form for deltahP}
	(\delta h P)_{ab}([\cO_1\cO_2]_{n,J})\evaluated_{G^{(t),fg}_{\cO_6}}=
	\\
	-\lambda_{326,f}\lambda_{146,g}\sum\limits_{\f_i,J',n'}
	\opeFuncDecomp{fg}{ab}{\cO_1}{\cO_2}{\cO_3}{\cO_4}{[\cO_1\cO_2]_{n,J}}{\cO_6}{\f_1}{\f_2}{\f_2}{\f_1}{[\f_1\f_2]_{n',J'}}{\cO_{6}}
	\mathfrak{dp}_1^{J',n'}(\f_1,\f_2,\cO_6)
	\\
	-\lambda_{326,f}\lambda_{146,g}\sum\limits_{\f_i,J',n'}
	\opeFuncDecomp{fg}{ab}{\cO_1}{\cO_2}{\cO_3}{\cO_4}{[\cO_1\cO_2]_{n,J}}{\cO_6}{\f_1}{\f_2}{\f_1}{\f_2}{[\f_1\f_2]_{n',J'}}{\cO_{6}}
	\mathfrak{dp}_2^{J',n'}(\f_1,\f_2,\cO_6)\;.
\end{multline}

For example, for the leading parity-even tower in the s-channel of $\<\psi\f\f\psi\>$, we find\footnote{In this section, we use the following shorthand notation for brevity: $\f^a_i\equiv\f_{\De_i+a}$, $\tl\f^a_i\equiv\f_{3-\De_i+a}$, \mbox{$\cO_i^{a,b}\equiv\cO_{\De_i+a,l_i+b}$}, and $\tl\cO_i^{a,b}\equiv\cO_{3-\De_i+a,l_i+b}$.}  
\begin{multline}
	\label{eq: parity even tower of two fermions and two scalars}
	(\delta h P)_{11}([\f\psi]^+_{0,J})\evaluated_{G^{(t)}_{\cO_6}}=-i \lambda_{\f\f\cO_6}\lambda_{\psi\psi\cO_6}^1\mathfrak{dp}_1^{J-\half,0}(\psi^{\half},\f,\cO_6)
	\\
	-i\lambda_{\f\f\cO_6}\lambda_{\psi\psi\cO_6}^2(1-\delta_{0,l_6})\bigg(\frac{\left(J+\frac{1}{2}\right) (J+1)}{\left(\Delta _6-1\right) l_6}
	\mathfrak{dp}_1^{J+\half,0}(\psi^{-\half},\f,\cO_6)\\
	+\frac{2 \left(\Delta _{\psi }+J-1\right){}^2 \left(\Delta _{\psi }+\Delta _{\phi
		}+J-2\right) \left(2 \Delta _{\psi }+2 \Delta _{\phi }+2 J-5\right)}{\left(\Delta
		_6-1\right) l_6 \left(2 \Delta _{\psi }+2 \Delta _{\phi }+4 J-5\right) \left(2 \Delta
		_{\psi }+2 \Delta _{\phi }+4 J-3\right)}\mathfrak{dp}_1^{J-\half,0}(\psi^{-\half},\f,\cO_6)
	\\
	+\frac{\left(-2 \Delta _{\psi }-\Delta _6+l_6+4\right) \left(2 \Delta _{\psi }-\Delta
		_6+l_6-1\right)}{4 \left(\Delta _6-1\right) l_6} \mathfrak{dp}_1^{J-\half,0}(\psi^{\half},\f,\cO_6)
	\bigg)\;.
\end{multline}
For the parity-odd tower, we instead have
\footnotesize
\begin{multline}
	\label{eq: parity odd tower of two fermions and two scalars}
	(\delta h P)_{22}\left([\f\psi]^-_{0,J}\right)\evaluated_{G^{(t)}_{\cO_6}}=i \lambda_{\f\f\cO_6}\lambda_{\psi\psi\cO_6}^1\bigg(
	\frac{J+\frac{1}{2}}{J+1}
	\mathfrak{dp}_1^{J+\half,0}(\psi^{\half},\f,\cO_6)\\
	+\frac{2 \left(2 \Delta _{\phi }+2 J-1\right)^2 \left(\Delta _{\psi }+\Delta _{\phi
		}+J-1\right)}{\left(2 \Delta _{\psi }+2 \Delta _{\phi }+2 J-3\right) \left(2 \Delta
		_{\psi }+2 \Delta _{\phi }+4 J-3\right) \left(2 \Delta _{\psi }+2 \Delta _{\phi }+4
		J-1\right)}\mathfrak{dp}_1^{J-\half,0}(\psi^{\half},\f,\cO_6)\bigg)
	\\
	+i \lambda_{\f\f\cO_6}\lambda_{\psi\psi\cO_6}^2(1-\delta_{0,l_6})\bigg(
	\frac{1}{\left(\Delta _6-1\right) l_6}\mathfrak{dp}_1^{J-\half,1}(\psi^{-\half},\f,\cO_6)
	\\
	+\frac{(2 J+1) \left(-2 \Delta _{\psi }-\Delta _6+l_6+4\right) \left(2 \Delta _{\psi
		}-\Delta _6+l_6-1\right)}{8 \left(\Delta _6-1\right) (J+1) l_6}\mathfrak{dp}_1^{J+\half,0}(\psi^{\half},\f,\cO_6)
	\\
	+\frac{\left(J+\frac{1}{2}\right) \left(\Delta _{\psi }-1\right){}^2 \left(\Delta _{\psi
		}+\Delta _{\phi }-\frac{5}{2}\right) \left(\Delta _{\psi }+\Delta _{\phi
		}+J-1\right)}{\left(\Delta _6-1\right) (J+1) l_6 \left(\Delta _{\psi }+\Delta _{\phi
		}-\frac{3}{2}\right) \left(\Delta _{\psi }+\Delta _{\phi }+J-\frac{3}{2}\right)}\mathfrak{dp}_1^{J-\half,0}(\psi^{\half},\f,\cO_6)
	\\
	+
	\frac{\left(2 \Delta _{\phi }+2 J-1\right){}^2 \left(-2 \Delta _{\psi }-\Delta
		_6+l_6+4\right) \left(2 \Delta _{\psi }-\Delta _6+l_6-1\right) \left(\Delta _{\psi
		}+\Delta _{\phi }+J-1\right)}{2 \left(\Delta _6-1\right) l_6 \left(2 \Delta _{\psi }+2
		\Delta _{\phi }+2 J-3\right) \left(2 \Delta _{\psi }+2 \Delta _{\phi }+4 J-3\right)
		\left(2 \Delta _{\psi }+2 \Delta _{\phi }+4 J-1\right)}
	\mathfrak{dp}_1^{J-\half,0}(\psi^{\half},\f,\cO_6)
	\bigg)\;.
\end{multline}
\normalsize
We would like to point out the appearance of  $\mathfrak{dp}_1^{J-\half,1}(\psi^{-\half},\f,\cO_6)$ which indicates that we may need to extract the data of \emph{non-leading twist} scalar towers from the scalar $6j$ symbol in order to obtain the \emph{leading twist} spinning towers. This happens for the cases where the scaling dimensions of $\cO_{1,2}$ and $l_5$ are shifted downwards while the scaling dimension of $\cO_5$ is shifted upwards. We also note that the absence of $\lambda_{\psi\psi\cO_6}^{3,4}$ follows from our parity-definite choice of three-point structures.\footnote{The OPE coefficients $\lambda_{\psi\psi\cO_6}^{3,4}$ can only appear through the product of $3-$point structures $\<\f\f\cO_6\>\<\psi\psi\cO_6\>^{3,4}$, which are parity-odd under $X\rightarrow -X$ in embedding space (see footnote \ref{footnote: parity of three-point structures} as well). In the s-channel this can only match to the non-diagonal pieces $\<\psi\f\cO_5\>^{1,2}\<\psi\f\cO_5\>^{2,1}$ for which no double pole appears: \mbox{$(\delta h P)_{12}\left([\f\psi]_{0,J}\right)=(\delta h P)_{21}\left([\f\psi]_{0,J}\right)=0$}.
}

Using the $\cK$ coefficients given explicitly in the attached \texttt{Mathematica} notebook of \cite{Albayrak:2020rxh}, one can similarly obtain all cases for four fermions as well. For brevity, we reproduce here a few cases for $l_6=0$:
\begin{multline}
	\label{eq: parity even tower of four fermions}
	(\delta h P)_{22}\left([\psi\psi]^+_{0,J}\right)\evaluated_{G^{(t)}_{\f_6}}= \lambda_{\psi\psi\cO_6}^1\lambda_{\psi\psi\cO_6}^1\mathfrak{dp}_1^{J-1,0}(\psi^{\half},\psi^{\half},\f_6)
	\\+\lambda_{\psi\psi\cO_6}^3\lambda_{\psi\psi\f_6}^3
	\bigg(
	\frac{2 J}{\left(\Delta _6-1\right)^2}\mathfrak{dp}_2^{J,0}(\psi^{-\half},\psi^{\half},\f_6)
	\\+\frac{\left(\Delta _{\psi }+\frac{J-2}{2}\right) \left(\Delta _{\psi
		}+J-\frac{3}{2}\right)}{\left(\Delta _6-1\right){}^2 \left(\Delta _{\psi }+J-1\right)}\mathfrak{dp}_2^{J-1,0}(\psi^{-\half},\psi^{\half},\f_6)
	\bigg)\;,
\end{multline}
and 
\be 
(\delta h P)_{11}\left([\psi\psi]^+_{0,J}\right)=(\delta h P)_{12}\left([\psi\psi]^+_{0,J}\right)=(\delta h P)_{21}\left([\psi\psi]^+_{0,J}\right)=0\; .
\ee 
Unlike the parity-even case, parity-odd families have non-zero off-diagonal components:
\footnotesize
\be 
{}&(\delta h P)_{34}\left([\psi\psi]^+_{0,J}\right)\evaluated_{G^{(t)}_{\f_6}}=\frac{\lambda_{\psi\psi\f_6}^3\lambda_{\psi\psi\f_6}^3}{(\De_6-1)^2}\bigg(
-\mathfrak{dp}_2^{J-1,1}(\psi^{-\half},\psi^{\half},\f_6) 
+\frac{J\left(J+1\right) \left(2 J+3\right)}{2 J+1}
\mathfrak{dp}_2^{J+1,0}(\psi^{-\half},\psi^{\half},\f_6) 
\\
{}&\qquad
+\frac{\left(2 \Delta _{\psi }+J-2\right) \left(2 \Delta _{\psi }+J-1\right) \left(2 \Delta
	_{\psi }+2 J-3\right) \left(2 \Delta _{\psi }+2 J-1\right) \left(4 \Delta _{\psi }+2
	J-5\right)}{64 \left(\Delta _{\psi }+J-1\right){}^2 \left(4 \Delta _{\psi }+2
	J-3\right)}\mathfrak{dp}_2^{J-1,0}(\psi^{-\half},\psi^{\half},\f_6) 
\\
{}&\qquad
+\frac{J \left(2 \Delta _{\psi }+J-1\right) \left(8 \Delta _{\psi }^3+2 \left(4 J^2-8
	J+3\right) \Delta _{\psi }-4 J^2+4 (4 J-3) \Delta _{\psi }^2+4 J-3\right)}{8 \left(2
	\Delta _{\psi }-1\right) \left(\Delta _{\psi }+J-1\right) \left(\Delta _{\psi
	}+J\right)}\mathfrak{dp}_2^{J,0}(\psi^{-\half},\psi^{\half},\f_6)
\bigg),
\\
{}&(\delta h P)_{43}\left([\psi\psi]^+_{0,J}\right)\evaluated_{G^{(t)}_{\f_6}}=(\delta h P)_{34}\left([\psi\psi]^+_{0,J}\right)\evaluated_{G^{(t)}_{\f_6}}\,,
\ee  
\normalsize
where one can write down $(\delta h P)_{33}\left([\psi\psi]^+_{0,J}\right)\ne 0$ and $(\delta h P)_{44}\left([\psi\psi]^+_{0,J}\right)\ne 0$ as well. However, all non-block-diagonal terms such as $(\delta h P)_{13}\left([\psi\psi]^+_{0,J}\right)$ are zero as the corresponding structures do not develop double poles. This is why we do not have terms with the mixed coefficient $\lambda_{\psi\psi\cO_6}^1\lambda_{\psi\psi\cO_6}^3$.

To calculate the scalar coefficients $\mathfrak{dp}_{i}^{J,0}$ we will use the Lorentzian inversion formula~\cite{Caron-Huot:2017vep,Liu:2018jhs,Albayrak:2019gnz}, combined with either dimensional reduction of the 3d block~\cite{Hogervorst:2016hal,Albayrak:2019gnz} or resummations of the lightcone expansion~\cite{Li:2019dix}. Using dimensional reduction we find for the $n=0$ double-twist operators:
\begin{align}
	\label{eq: doublePoles1}
	\mathfrak{dp}_1^{J,0}(\phi_1,\phi_2,\mathcal{O}_{6})&=-\sum\limits_{p=0}^{\infty}\sum\limits_{q=\text{max}(-p,p-2J_6)}^{p}2\kappa^{0,0}_{2\bar{h}}\sin(\pi(h_6-2h_1))\sin(\pi(h_6-2h_2))
	\nonumber \\ &\hspace{4cm}  \times \mathcal{A}^{0,0}_{p,q}\Omega^{h_1h_2h_2h_1}_{\bar{h},h_{6}+p,2h_2}\frac{\Gamma(2(\bar{h}+q))}{\Gamma^{2}(\bar{h}+q)}\bigg|_{\bar{h}=h_1+h_2+J},
	\\
	\label{eq: doublePoles2}
	\mathfrak{dp}_2^{J,0}(\phi_1,\phi_2,\mathcal{O}_{6})&=-\sum\limits_{p=0}^{\infty}\sum\limits_{q=\text{max}(-p,p-2J_6)}^{p}2\kappa^{h_{21},h_{12}}_{2\bar{h}}\sin^{2}(\pi(h_6-h_1-h_2))
	\nonumber \\ &  \times \mathcal{A}^{h_{21},h_{12}}_{p,q}\Omega^{h_1h_2h_1h_2}_{\bar{h},h_{6}+p,h_1+h_2}\frac{\Gamma(2(\bar{h}+q))}{\Gamma(h_{12}+\bar{h}+q)\Gamma(h_{21}+\bar{h}+q)}\bigg|_{\bar{h}=h_1+h_2+J}.
\end{align} 
Here we have defined $h=\frac{1}{2}(\Delta-J)$, $\bar{h}=\frac{1}{2}(\Delta+J)$, $h_{ij}=h_i-h_j$, and 
\begin{align}
	\kappa^{a,b}_{2\bar{h}}=\frac{\Gamma(\bar{h}+a)\Gamma(\bar{h}-a)\Gamma(\bar{h}+b)\Gamma(\bar{h}-b)}{2\pi^{2}\Gamma(2\bar{h})\Gamma(2\bar{h}-1)}\,.
\end{align}

The coefficients $\mathcal{A}^{a,b}_{p,q}$ come from performing dimensional reduction for 3d blocks in terms of the chiral, 2d blocks and were found for $a=b=0$ in  \cite{Hogervorst:2016hal}.\footnote{In comparison to \cite{Hogervorst:2016hal} we use $\mathcal{A}^{\text{here}}_{p,q}=\mathcal{A}^{\text{there}}_{\frac{p_q}{2},\bar{h}-h+q-p}$, and for Eq. (2.35) there we use $c^{(d)}_{\ell}=\frac{(d-2)_{\ell}}{\left(\frac{d-2}{2}\right)_{\ell}}$.} For general $a$ and $b$ we can compute $\mathcal{A}^{a,b}_{h,\bar{h}}$ recursively using the Casimir equation. For explicit results, we will mainly be interested in the large spin asymptotics, in which case we can restrict to $p=q=0$ and use $\mathcal{A}^{a,b}_{0,0}=1$. Finally, the function $\Omega$ is given by \cite{Liu:2018jhs}:
\begin{align}
	\label{eq: definition of omega}
	\Omega^{h_1,h_2,h_3,h_4}_{h_5,h_6,p} &= \frac{\Gamma (2 h_5) \Gamma (h_6-p+1) \Gamma (h_5+h_{12}-h_6+p-1) \Gamma (-h_{12}+h_{34}+h_6-p+1)}{\Gamma (h_5+h_{12}) \Gamma (h_5+h_{34}) \Gamma (h_5-h_{12}+h_6-p+1)}
	\nonumber \\ &
	\hspace{1cm}   \pFq{4}{3}{h_{23}+h_6,h_6-h_{14},-h_{12}+h_{34}+h_6-p+1,h_6-p+1}{2 h_6,h_5-h_{12}+h_6-p+1,-h_5-h_{12}+h_6-p+2}{1}
	\nonumber\\ \nonumber &
	+\frac{\Gamma (2 h_6) \Gamma (h_5+h_{13}+p-1) \Gamma (h_5+h_{42}+p-1) \Gamma (-h_5-h_{12}+h_6-p+1)}{\Gamma (h_6-h_{14}) \Gamma (h_{23}+h_6) \Gamma (h_5+h_{12}+h_6+p-1)} 
	\nonumber \\ &
	\hspace{1cm} \pFq{4}{3}{h_5+h_{13}+p-1,h_5+h_{42}+p-1,h_5+h_{34},h_5+h_{12}}{h_5+h_{12}+h_6+p-1,2 h_5,h_5+h_{12}-h_6+p}{1}\,.
\end{align}
When we study double-twist operators with large spin, or equivalently large $\bar{h}$, the first $_4F_3$ hypergeometric yields the asymptotic, large-spin prediction while the second $_4F_3$ gives effects which are exponentially suppressed.  

By inserting \equref{eq: doublePoles1} and \equref{eq: doublePoles2} into  \equref{eq: general form for deltahP}, we can obtain $\delta h P$ for various double-twist operators of fermions. Below we will consider some examples.

\subsubsection*{$\delta h P$ of double-twist towers $[\phi\psi_\a]^\pm_{0,l_5}$ due to scalar exchange:}
From \equref{eq: parity even tower of two fermions and two scalars} and \equref{eq: parity odd tower of two fermions and two scalars}, we find that
\begin{subequations}
	\begin{multline}
		(\delta h P)_{11}^{(p)}=
		-\lambda_{\f\f\f_6}\lambda_{\psi\psi\f_6}^1
		\frac{ (-1)^{l_5+1} \Gamma \left(\Delta _6\right) \left(\sin \left(\pi  \left(-\Delta _{\psi }-\Delta _{\phi }+\Delta
			_6\right)\right)+\sin \left(\pi  \left(\Delta _{\psi }-\Delta _{\phi }\right)\right)\right)}{\Gamma \left(\frac{\Delta _6}{2}\right)}
		\\ \x \Omega^{\frac{2 \Delta _{\psi}+1}{4} ,\frac{\Delta _{\phi }}{2},\frac{\Delta _{\phi }}{2},\frac{2 \Delta _{\psi }+1}{4}} _{\frac{2\Delta _{\psi }+2 \Delta _{\phi }+4l_5-1}{4},\frac{\Delta _6}{2},\Delta _{\phi }}
	\end{multline}
	and 
	\small 
	\begin{multline}
		(\delta h P)_{22}^{(p)}=
		-\lambda_{\f\f\f_6}\lambda_{\psi\psi\f_6}^1
		\frac{ (-1)^{l_5} \Gamma \left(\Delta _6\right) \cos \left(\pi  \left(\Delta _{\psi }-\frac{\Delta _6}{2}\right)\right) \sin
			\left(\pi  \left(\Delta _{\phi }-\frac{\Delta _6}{2}\right)\right)}{\Gamma \left(\frac{\Delta _6}{2}\right)^2}
		\\
		\x\Bigg(
		\frac{\left(2 l_5+1\right)}{l_5+1}
		\Omega^{\frac{2 \Delta_{\psi }+1}{4},\frac{\Delta _{\phi }}{2},\frac{\Delta _{\phi }}{2},\frac{2 \Delta _{\psi }+1}{4}}
		_{\frac{2\Delta _{\psi }+2\Delta _{\phi }+4l_5+3}{4},\frac{\Delta _6}{2},\Delta _{\phi }}
		\\	
		-\frac{4 \left(2 \Delta _{\phi
			}+2 l_5-1\right){}^2 \left(\Delta _{\psi }+\Delta _{\phi }+l_5-1\right)}{\left(2 \Delta _{\psi }+2 \Delta
			_{\phi }+2 l_5-3\right) \left(2 \Delta _{\psi }+2 \Delta _{\phi }+4 l_5-3\right) \left(2 \Delta _{\psi }+2 \Delta _{\phi }+4
			l_5-1\right)}
		\Omega^{\frac{2 \Delta _{\psi}+1}{4},\frac{\Delta _{\phi }}{2},\frac{\Delta _{\phi }}{2},\frac{2 \Delta _{\psi }+1}{4}}_{\frac{2\Delta _{\psi }+2\Delta _{\phi }+4l_5-1}{4},\frac{\Delta _6}{2},\Delta _{\phi }} 
		\Bigg)\,.
	\end{multline}
	\normalsize
\end{subequations}

By expanding the perturbative terms at large $l$, we can obtain the leading order behavior
\bea
(\delta h P)_{11}([\f\psi]^+_{0,l_5})\evaluated_{G^{(t)}_{\f_6}}\sim{}&{} \lambda_{\f\f\f_6}\lambda_{\psi\psi\f_6}^1 \frac{\sqrt{\pi } (-1)^{l_5+1} \Gamma \left(\Delta _6\right) 2^{-\Delta _{\psi }-\Delta _{\phi
		}-2 l_5+\frac{5}{2}} l_5^{\Delta _{\psi }+\Delta _{\phi }-\Delta _6-1}}{\Gamma \left(\frac{\Delta
		_6}{2}\right){}^2 \Gamma \left(-\frac{\Delta _6}{2}+\Delta _{\psi }+\frac{1}{2}\right) \Gamma \left(\Delta
	_{\phi }-\frac{\Delta _6}{2}\right)},
\\
(\delta h P)_{22}([\f\psi]^-_{0,l_5})\evaluated_{G^{(t)}_{\f_6}}\sim{}&{}
\frac{\lambda_{\f\f\f_6}\lambda_{\psi\psi\f_6}^1(-1)^{l_5} \left(-2 \Delta _{\psi }+\Delta _6+2\right) \Gamma \left(\frac{1}{2}
	\left(\Delta _6+1\right)\right) l_5^{\Delta
		_{\psi }+\Delta _{\phi }-\Delta _6-2}}{2^{\Delta _{\psi }+\Delta _{\phi }-\Delta _6+2 l_5+\frac{1}{2}}\Gamma \left(\frac{\Delta _6}{2}\right) \Gamma \left(-\frac{\Delta
		_6}{2}+\Delta _{\psi }+\frac{1}{2}\right) \Gamma \left(\Delta _{\phi }-\frac{\Delta _6}{2}\right)}.
\eea
By dividing these by the MFT coefficients at large $l$ give in \equref{eq: MFT coefficients at large spin for two fermion and two scalars}, we obtain the anomalous dimensions at leading order:
\footnotesize 
\bea 
\gamma_{[\f\psi]^+_{0,l_5}}\evaluated_{G^{(t)}_{\f_6}}={}&{}\frac{2}{P_{11}^{(s)}} (\delta h P)_{11}\evaluated_{G^{(t)}_{\f_6}}=
\frac{i\lambda_{\f\f\f_6}\lambda_{\psi\psi\f_6}^1}{l_5^{\Delta _6}}\frac{2^{\Delta _6} \Gamma \left(\frac{1}{2} \left(\Delta _6+1\right)\right) \Gamma
	\left(\Delta _{\psi }+\frac{1}{2}\right) \Gamma \left(\Delta _{\phi }\right)}{\sqrt{\pi } \Gamma
	\left(\frac{\Delta _6}{2}\right) \Gamma \left(\frac{1}{2} \left(-\Delta _6+2 \Delta _{\psi }+1\right)\right)
	\Gamma \left(\frac{1}{2} \left(2 \Delta _{\phi }-\Delta _6\right)\right)},
\\
\gamma_{[\f\psi]^-_{0,l_5}}\evaluated_{G^{(t)}_{\f_6}}={}&{}\frac{2}{P_{22}^{(s)}}(\delta h P)_{22}\evaluated_{G^{(t)}_{\f_6}} = -\frac{i\lambda_{\f\f\f_6}\lambda_{\psi\psi\f_6}^1}{l_5^{\Delta _6}}\frac{2^{\Delta _6-1} \left(\frac{\Delta _6}{2}-\Delta _{\psi }+1\right) \Gamma \left(\frac{\Delta
		_6+1}{2}\right)  \Gamma \left(\Delta _{\psi }+\frac{1}{2}\right) \Gamma \left(\Delta _{\phi
	}\right)}{\sqrt{\pi } \left(\Delta _{\psi }-1\right) \Gamma \left(\frac{\Delta _6}{2}\right) \Gamma
	\left(\frac{-\Delta _6+2 \Delta _{\psi }+1}{2}\right) \Gamma \left(\frac{2 \Delta
		_{\phi }-\Delta _6}{2}\right)}.
\eea 
\normalsize
\subsubsection*{$\delta h P$ of parity-even double-twist tower $[\phi\psi_\a]^+_{0,l_5}$ due to stress tensor exchange:}
By inserting \equref{eq: doublePoles1} and \equref{eq: doublePoles2} into \equref{eq: parity even tower of two fermions and two scalars}, we obtain
\begin{multline}
	(\delta h P)_{11}([\f\psi]^+_{0,\ell_5})\evaluated_{G^{(t)}_{T}}=
	\frac{3 \Delta _{\phi } \sin \left(\pi  l_5\right) \Gamma \left(l_5+\Delta _{\psi }\right) \Gamma
		\left(l_5+\Delta _{\phi }-\frac{1}{2}\right) \Gamma \left(l_5+\Delta _{\phi }+\Delta _{\psi
		}-\frac{3}{2}\right)}{2 \sqrt{2} \pi ^2 c_{T} \Gamma \left(l_5+1\right) \Gamma \left(\Delta _{\psi
		}-1\right) \Gamma \left(\Delta _{\phi }-\frac{1}{2}\right) \Gamma \left(2 l_5+\Delta _{\phi }+\Delta _{\psi
		}-\frac{1}{2}\right)}
	\\\x\Bigg(
	\left(2 \Delta _{\psi }+1\right) \left(2 \Delta _{\psi }+2 \Delta _{\phi }+4 l_5-3\right)\pFq{4}{3}{\frac{1}{2},\frac{1}{2},\frac{3}{2}-\Delta _{\phi },1-\Delta _{\psi }}{1,l_5+1,-l_5-\Delta _{\phi
		}-\Delta _{\psi }+\frac{5}{2}}{1}
	\\
	-4 \left(\Delta _{\psi }+l_5-1\right) \left(\Delta _{\psi }+\Delta _{\phi }+l_5-2\right)\pFq{4}{3}{\frac{1}{2},\frac{1}{2},\frac{3}{2}-\Delta _{\phi },2-\Delta _{\psi }}{1,l_5+1,-l_5-\Delta _{\phi
		}-\Delta _{\psi }+\frac{7}{2}}{1}
	\\
	+4 \left(l_5+\frac{1}{2}\right) \left(\Delta _{\phi }+l_5-\frac{1}{2}\right)
	\pFq{4}{3}{\frac{1}{2},\frac{1}{2},\frac{3}{2}-\Delta _{\phi },2-\Delta _{\psi }}{1,l_5+2,-l_5-\Delta _{\phi
		}-\Delta _{\psi }+\frac{5}{2}}{1}
	\Bigg)
	\\+\text{ (non-perturbative terms)},
\end{multline}
where we set
\be 
\label{eq: conditions for stress tensor}
l_6=2\;,\quad \De_6=3\;,\quad \lambda_{\f\f T}=-\frac{3 \Delta_\phi  \Gamma \left(\frac{3}{2}\right)}{2 (2 \pi )^{3/2} \sqrt{c_T}}\;,\quad \lambda_{\psi\psi T}^1=\frac{3 i (\Delta_\psi -1)}{4 \sqrt{c_T}}\;,\quad  \lambda_{\psi\psi T}^2=-\frac{3 i}{2 \sqrt{c_T}},
\ee 
where $c_{T}$ is the central charge, which here is defined as the normalization of the stress tensor two-point function. At large spin, the leading order term is then
\be
(\delta h P)_{11}([\f\psi]^+_{0,l_5})\evaluated_{G^{(t)}_{T}}\sim \frac{3 i (-1)^{l_5} \Delta _{\phi } 2^{-\Delta _{\psi }-\Delta _{\phi }-2 l_5+4} l_5^{\Delta _{\psi }+\Delta
		_{\phi }-2}}{\pi ^{3/2} c_T\Gamma \left(\Delta _{\psi }-1\right) \Gamma \left(\Delta _{\phi
	}-\frac{1}{2}\right)},
\ee
and we obtain the anomalous dimension via \equref{eq: MFT coefficients at large spin for two fermion and two scalars}:
\be 
\gamma_{[\f\psi]^+_{0,l_5}}\evaluated_{G^{(t)}_{T}}=\frac{2}{P_{11}^{(s)}} (\delta h P)_{11}\evaluated_{G^{(t)}_{T}}= \frac{1}{ l_5 }\frac{3 \sqrt{2} \Gamma \left(\Delta _{\psi }+\frac{1}{2}\right) \Gamma \left(\Delta _{\phi }+1\right)}{\pi ^2
	c_T\Gamma \left(\Delta _{\psi }-1\right) \Gamma \left(\Delta _{\phi }-\frac{1}{2}\right)}.
\ee 

\subsubsection*{$\delta h P$ of parity-even double-twist tower $[\psi\psi]^+_{0,l_5}$ due to the exchange of a generic parity-even operator or parity-odd scalar:}
Similar to the previous examples, we obtain the relevant $\delta h P$ by inserting \equref{eq: doublePoles1} and \equref{eq: doublePoles2} into \equref{eq: parity even tower of four fermions}. As the expressions are quite lengthy, we will not reproduce the full results but instead present their asymptotic forms at large spin. We find that 
\small 
\be
\label{eq: delta hP for four fermion}
(\delta h P)_{11}([\psi\psi]^+_{0,l_5})\evaluated_{G^{(t)}_{\cO_{6}}}={}&{}(\delta h P)_{12}([\psi\psi]^+_{0,l_5})\evaluated_{G^{(t)}_{\cO_6}}=(\delta h P)_{21}([\psi\psi]^+_{0,l_5})\evaluated_{G^{(t)}_{\cO_6}}=0,
\\
(\delta h P)_{22}([\psi\psi]^+_{0,l_5})\evaluated_{G^{(t)}_{\cO_6}}\sim{}&{}\left(\lambda_{\psi\psi\cO_6}^1\right)^2\frac{(-1)^{l_5+1} 2^{-2 \Delta _{\psi }+\Delta _6-2 l_5+l_6+2} l_5^{2 \Delta _{\psi }-\Delta _6+l_6-\frac{1}{2}}
	\Gamma \left(\frac{1}{2} \left(l_6+\Delta _6+1\right)\right)}{\Gamma \left(\frac{1}{2} \left(l_6+\Delta
	_6\right)\right) \Gamma \left(\frac{1}{2} \left(l_6-\Delta _6+1\right)+\Delta _{\psi }\right)^2},
\ee
\normalsize
where contributions due to $\lambda_{\psi\psi\cO_6}^{2,3,4}$ come at subleading order.\footnote{We see that this result matches the one calculated using lightcone bootstap methods in \cite{Albayrak:2019gnz}, once the change of basis and the difference in conformal block normalization is taken into account; compare \equref{eq: delta hP for four fermion} here with (3.30c) there.} By dividing by the MFT coefficients given in \equref{eq: MFT coefficients for four fermions}, we obtain the anomalous dimension at large spin:
\small 
\be 
\gamma_{[\psi\psi]^+_{0,l_5}}\evaluated_{G^{(t)}_{\cO_6}}=\frac{2}{P_{22}^{(s)}}(\delta h P)_{22} \evaluated_{G^{(t)}_{\cO_6}}=
\frac{\left(\lambda_{\psi\psi\cO_6}^1\right)^2}{l_5^{\Delta _6-l_6}}\frac{ 2^{\Delta _6+l_6} \Gamma \left(\Delta _{\psi
	}+\frac{1}{2}\right){}^2 \Gamma \left(\frac{1}{2} \left(l_6+\Delta _6+1\right)\right)}{\sqrt{\pi } \Gamma
	\left(\frac{1}{2} \left(l_6+\Delta _6\right)\right) \Gamma \left(\frac{1}{2} \left(l_6-\Delta
	_6+1\right)+\Delta _{\psi }\right)^2}\;.
\ee
\normalsize 
For example, for stress tensor exchange we can impose \equref{eq: conditions for stress tensor} which yields
\be 
\label{eq: anomalous dimension of parity even fermion fermion double-twist operator due to stress tensor}
\gamma_{[\psi\psi]^+_{0,l_5}}\evaluated_{G^{(t)}_{T}}=
-\frac{1}{l_5}\frac{48 \Gamma \left(\Delta _{\psi }+\frac{1}{2}\right)^2}{\pi c_T \Gamma \left(\Delta _{\psi
	}-1\right)^2}\;,
\ee 
whereas for parity-even scalar exchange it becomes
\be 
\label{eq: anomalous dimension of parity even fermion fermion double-twist operator due to parity even scalar}
\gamma_{[\psi\psi]^+_{0,l_5}}\evaluated_{G^{(t)}_{\f_6}}=
\frac{\left(\lambda_{\psi\psi\cO_6}^1\right)^2}{l_5^{\De_6}}\frac{2^{\Delta _6} \Gamma \left(\frac{\Delta _6+1}{2}\right) \Gamma \left(\Delta _{\psi
	}+\frac{1}{2}\right)^2}{\sqrt{\pi } \Gamma \left(\frac{\Delta _6}{2}\right) \Gamma \left(-\frac{\Delta
		_6}{2}+\Delta _{\psi }+\frac{1}{2}\right)^2}\;.
\ee 

For the exchange of a parity-odd scalar in the crossed channel, we still have $(\delta h P)_{ij}=0$ unless $i=j=2$, which now becomes
\be
\label{eq: delta hP for four fermion 2}
(\delta h P)_{22}([\psi\psi]^+_{0,l_5})\evaluated_{G^{(t)}_{\f_6}}\sim{}&{}\left(\lambda_{\psi\psi\f_6}^3\right)^2\frac{(-1)^{l_5} \Gamma \left(\frac{\Delta _6}{2}\right) 2^{-2 \Delta _{\psi }+\Delta _6-2 l_5+1} l_5^{2 \Delta
		_{\psi }-\Delta _6-\frac{3}{2}}}{\Gamma \left(\frac{1}{2} \left(\Delta _6+1\right)\right) \Gamma \left(\Delta
	_{\psi }-\frac{\Delta _6}{2}\right){}^2}\;,
\ee
from which we can extract the anomalous dimension as
\be 
\label{eq: anomalous dimension of parity even fermion fermion double-twist operator due to parity odd scalar}
\gamma_{[\psi\psi]^+_{0,l_5}}\evaluated_{G^{(t)}_{\f_6}}=\frac{2}{P_{22}^{(s)}}(\delta h P)_{22} \evaluated_{G^{(t)}_{\f_6}}= \frac{\left(\lambda_{\psi\psi\f_6}^3\right)^2}{l_5^{\De_6+1}} \frac{2^{\Delta _6-1} \Gamma \left(\frac{\Delta _6}{2}\right) \Gamma \left(\Delta _{\psi
	}+\frac{1}{2}\right){}^2}{\sqrt{\pi } \Gamma \left(\frac{\Delta _6+1}{2} \right) \Gamma
	\left(\Delta _{\psi }-\frac{\Delta _6}{2}\right)^2}\;.
\ee 
We see that the anomalous dimensions in \equref{eq: anomalous dimension of parity even fermion fermion double-twist operator due to stress tensor}, \equref{eq: anomalous dimension of parity even fermion fermion double-twist operator due to parity even scalar}, and \equref{eq: anomalous dimension of parity even fermion fermion double-twist operator due to parity odd scalar} match precisely to the results computed using large-spin expansions in \cite{Albayrak:2019gnz}.

\subsection{Computation of corrections to OPE Coefficients: a working example}
\label{app:OPE}
In \equref{eq: OPE function in terms of scalar 6j symbol} we relate the OPE function of spinning operators to the scalar $6j$ symbols, which we reproduce for reader's convenience:
\begin{multline}
	\rho_{ac}^{(s)}(\cO)S^c_b(\cO_3\cO_4[\tl\cO])\evaluated_{G^{(t),fg}_{\cO_6}}=\lambda_{326,f}\lambda_{146,g}\sum\limits_{\f_i,\cO',\cO'_{6}}
	\opeFuncDecomp{fg}{ab}{\cO_1}{\cO_2}{\cO_3}{\cO_4}{\cO}{\cO_6}{\f_1}{\f_2}{\f_3}{\f_4}{\cO'}{\cO'_{6}} \\\x \qquad\frac{S(\phi_3\phi_4[\widetilde{\cO'}])}{\eta^{(s)}_{\cO'}}\sixjBlock{\f_1}{\f_2}{\f_3}{\f_4}{\cO'}{\cO'_{6}}.
\end{multline}
By taking the double poles in $\De$ on both sides, we can extract $\delta hP$ for spinning operators in terms of scalar data, which we detailed and illustrated in section~\ref{sec:applications}. In this section, we will use this equation to extract correction to OPE coefficients for double twist operators $[\phi\psi]^+_0$ due to an exchange of a scalar in the crossed channel.

We see in \equref{eq: parity even tower of two fermions and two scalars} that the $\delta h P$ for $[\f\psi]^+_{0,J}$ reads as
\be
(\delta h P)_{11}([\f\psi]^+_{0,J})\evaluated_{G^{(t)}_{\phi_6}}=&
-i \lambda_{\f\f\f_6}\lambda_{\psi\psi\f_6}^1\mathfrak{dp}_1^{J-\half,0}(\psi^{\half},\f,\f_6)\;,
\\
(\delta h P)_{12}([\f\psi]^+_{0,J})\evaluated_{G^{(t)}_{\phi_6}}=&(\delta h P)_{21}([\f\psi]^+_{0,J})\evaluated_{G^{(t)}_{\phi_6}}=(\delta h P)_{22}([\f\psi]^+_{0,J})\evaluated_{G^{(t)}_{\phi_6}}=0\;,
\ee
as only the first term in \equref{eq: K coefficients for 2f2s with an exchange of a scalar t channel} contributes. For $(\delta P)$ on the other hand, we do not need double poles (single poles are sufficient) and there are also cross terms, hence we have
\be
(\delta P)_{11}([\f\psi]^+_{0,J})\evaluated_{G^{(t)}_{\phi_6}}=&	-i \lambda_{\f\f\f_6}\lambda_{\psi\psi\f_6}^1\mathfrak{p}_{1,+}^{J-\half,0}\;,
\\
(\delta P)_{12}([\f\psi]^+_{0,J})\evaluated_{G^{(t)}_{\phi_6}}=&-\lambda_{\f\f\f_6}\lambda_{\psi\psi\f_6}^3\bigg(
\frac{i \left(2 l_5+1\right)}{\Delta _6-1}\mathfrak{p}_{1,-}^{J+\half,0}
\\&\;\;+
\frac{8 i \left(\Delta _{\psi }+l_5-1\right) \left(\Delta _{\phi }+l_5-1\right) \left(\Delta _{\psi }+\Delta
	_{\phi }+l_5-2\right)}{\left(\Delta _6-1\right) \left(2 \Delta _{\psi }+2 \Delta _{\phi }+4 l_5-5\right)
	\left(2 \Delta _{\psi }+2 \Delta _{\phi }+4 l_5-3\right)}\mathfrak{p}_{1,-}^{J-\half,0}
\bigg)\;,
\\
(\delta P)_{12}([\f\psi]^+_{0,J})\evaluated_{G^{(t)}_{\phi_6}}=&i\lambda_{\f\f\f_6}\lambda_{\psi\psi\f_6}^3\frac{1}{\Delta _6-1}\mathfrak{p}_{2,-}^{J-\half,0}\;,
\\
(\delta P)_{22}([\f\psi]^+_{0,J})\evaluated_{G^{(t)}_{\phi_6}}=&0\;,
\ee
where we define the shorthand notation
\be 
\mathfrak{p}_{i,\pm}^{J,n}\equiv \mathfrak{p}_i^{J,n}(\psi^{\pm},\f,\f,\psi^{\half},\f_6)
\ee 
for
\footnotesize
\bea 
\mathfrak{p}_1^{J,n}(\f_1,\f_2,\f_3,\f_4,\cO_6)\equiv &
\lim\limits_{\De\rightarrow\De_1+\De_2+J+2n} \left(\De -\De_1-\De_2-J-2n\right) \frac{S(\f_3 \f_4 [\widetilde{\cO}_{\De,J}])}{\eta^{(s)}_{\cO_{\De,J}}} \sixjBlock{\f_1}{\f_2}{\f_3}{\f_4}{\cO_{\De,J}}{\cO_{6}},
\\
\mathfrak{p}_2^{J,n}(\f_1,\f_2,\f_3,\f_4,\cO_6)\equiv &
\lim\limits_{\De\rightarrow\De_3+\De_4+J+2n} \left(\De -\De_3-\De_4-J-2n\right) \frac{S(\f_3 \f_4 [\widetilde{\cO}_{\De,J}])}{\eta^{(s)}_{\cO_{\De,J}}} \sixjBlock{\f_1}{\f_2}{\f_3}{\f_4}{\cO_{\De,J}}{\cO_{6}}.
\eea
\normalsize

We can compute $\mathfrak{p}$ similar to $\mathfrak{dp}$ and include both perturbative and nonperturbative corrections to OPE coefficients. For brevity, we only reproduce the leading piece of the perturbative correction at large $l$:
\begin{multline}
	(\delta P)([\f\psi]^+_{0,\ell_5})\evaluated_{G^{(t)}_{\phi_6}}= \frac{i \sqrt{\pi } (-1)^{l_5-\frac{1}{2}} 2^{-\Delta _{\psi }-\Delta _{\phi }+\Delta _6-2 l_5+\frac{5}{2}}
		l_5^{\Delta _{\psi }+\Delta _{\phi }-\Delta _6-1}}{\Gamma \left(-\frac{\Delta _6}{2}+\Delta _{\psi
		}+\frac{1}{2}\right) \Gamma \left(\Delta _{\phi }-\frac{\Delta _6}{2}\right)}
	\\\x 
	\begin{pmatrix}
		-\lambda_{\f\f\f_6}\lambda_{\psi\psi\f_6}^1\frac{\Gamma \left(\frac{\Delta _6+1}{2}\right) H_{\frac{\Delta
					_6-2}{2}}}{\sqrt{\pi } \Gamma \left(\frac{\Delta _6}{2}\right)}
		& \lambda_{\f\f\f_6}\lambda_{\psi\psi\f_6}^3\frac{\Gamma \left(-\frac{\Delta _6}{2}+\Delta _{\psi }+\frac{1}{2}\right)}{2 \sqrt{l_5} \Gamma \left(\Delta
			_{\psi }-\frac{\Delta _6}{2}\right)}
		\\
		\lambda_{\f\f\f_6}\lambda_{\psi\psi\f_6}^3\frac{\Gamma \left(-\frac{\Delta _6}{2}+\Delta _{\psi }+\frac{1}{2}\right)}{2 \sqrt{l_5} \Gamma \left(\Delta
			_{\psi }-\frac{\Delta _6}{2}\right)} 
		& 0
	\end{pmatrix},
\end{multline}
where $H_a$ is the Harmonic number. As a consistency check, we see that setting
\be 
\De_6\rightarrow 0\;,\quad \lambda_{\f\f\f_6}\rightarrow 1\;,\quad \lambda_{\psi\psi\f_6}^1\rightarrow i\;,\quad \lambda_{\psi\psi\f_6}^3 \rightarrow 0
\ee 
reduces the result to the MFT coefficient \equref{eq: OPE coefficient squared for fermion scalar parity even double twist family}.

\chapter{Discussion}
In this thesis we reviewed various analytic tools that have been developed in recent years and used them to extend the progress of the analytic conformal bootstrap program to fermionic conformal field theories in three dimensional spacetimes. 

In the first part of the thesis, we used the $\mathrm{SL}(2,\R)$ expansion of the conformal blocks and the known formulae for the summation of the \emph{Casimir-irregular} terms to derive the relation between the CFT data of different channels in the lightcone bootstrap. This old-style lightcone bootstrap has been used for scalars and is sufficient to derive the large spin behavior of the double twist families in the spectrum, although the results surprisingly match the numerics even at low spins for scalar theories such as the $3d$ Ising and $\mathrm{O}(N)$ models. We extended this technique to fermions and derived the OPE coefficients and the anomalous dimensions for the large spin spectrum of fermionic CFTs. In particular, we provided both leading and next-to-leading order terms in the large spin expansion.

In the second part of the thesis, we provided a different approach for the analytic analysis of CFTs with spinning operators (such as fermions). This approach relies on deriving relations between the $6j$ symbols of the conformal group and using the known $6j$ symbols for scalars. The relations between the $6j$ symbols are derived using the conformally invariant pairings of the three point structures and the differential operators that can be used to shift the conformal weights of a given operator. When combined, these technologies allow one to compute the CFT data of fermionic operators \emph{non-perturbatively}, that is  by including both the perturbative and exponentially suppressed terms in spin. The second piece is inaccessible with the traditional lightcone bootstrap that we used in the first part of the thesis, and it ensures that the result is analytic all the way down to low spins, yielding expressions far more compatible with the numerical results. We first demonstrated the importance of these non-perturbative pieces in the $3d$ Ising and $\mathrm{O}(N)$ models, then extended the calculations to generic unitary fermionic CFTs by deriving the analytic expressions for their double twist spectrum.

While extending the application of the Lorentzian inversion formula to spinning correlators, we derived relations between the $6j$ symbols of different representations of the conformal group.\footnote{The reader can refer to\equref{eq: generalized J coefficients} for this relation and the definition of $\mathcal{J}-$coefficients.} Aside from their employment in the construction of the OPE function (hence the derivation of the CFT data), the $\mathcal{J}-$coefficients that we have constructed for this relation may be useful on their own. As $6j$ symbols are kinematic objects of the conformal group without any dynamic data, the relations between these objects are valid in any system with the conformal symmetry, and we believe that our results for these coefficients may have pure mathematical value, considering our derivation (and the resultant $J-$coefficients) are completely exact without any sort of approximation. We hope that further mathematical insight can be obtained from these objects in the future.

There are several open questions left that can be considered by the tools and results of this thesis. For instance, our demonstration for the improvement of the analytic predictions for $3d$ Ising and $\mathrm{O}(N)$ models with the inclusion of non-perturbative effects can be extended to concrete fermionic CFTs, such as GNY models. Would we then expect new precise analytic predictions that match the results of the numerical bootstrap~\cite{Iliesiu:2015qra,Iliesiu:2017nrv}? It is also natural to ask if we can predict analytic trajectories that cannot be accessed using scalar correlators in $3d$ $\mathcal{N}=1$ SCFTs such as the supersymmetric Ising~\cite{Rong:2018okz,Atanasov:2018kqw} and Wess-Zumino~\cite{Rong:2019qer} models. This is an interesting question as the implications of imposing analytic bootstrap constraints for all external operators in the same supermultiplet has not yet been fully understood, so our results for external fermionic operators can be utilized to address this question in SCFTs.

Moreover, it is also straightforward to go to higher spin now that we understand how to spin down a fermionic $6j$ symbol in $3d$. As a simple example, our results could then also be used to study correlation functions of conserved currents $J_{\mu}$ in the $O(N)$ vector model. There are many physically relevant observables, such $\<JJT\>$, which are only accessible with spinning correlators. Based on \cite{Simmons-Duffin:2016wlq,Caron-Huot:2017vep,Albayrak:2019gnz} we now know that the current and stress tensor lie on the double-twist trajectories composed of the fundamental scalars, $\phi$, so these correlators are now within reach of analytic methods.

Finally, we note that our results are directly applicable to the study of Witten diagrams
with external fermionic operators. For example, by studying the contribution of the stress tensor $T^{\mu\nu}$ to a fermionic correlator, e.g. $\<\phi\psi\psi\phi\>$, we can derive the binding energy for a two-particle state dual to $[\phi\psi]_n$, due to tree-level graviton exchange. The anomalous dimension, or corresponding 6j symbol, can then be used to bootstrap a graviton loop in
AdS$_4$~\cite{Aharony:2016dwx,Liu:2018jhs}. Likewise, by extending our computations to external currents, we can derive similar results for gauge interactions in AdS$_4$, which would provide valuable data that can be cross-checked with perturbative bulk computations.\footnote{As part of his PhD work not included in this thesis \cite{Albayrak:2018tam,Albayrak:2019asr,Albayrak:2019yve,Albayrak:2020isk,Albayrak:2020bso,Albayrak:2020saa,Albayrak:2020fyp}, the author studied AdS$_4$ gluons in momentum space, showing that written in a particular gauge any tree-level gluon Witten diagram can be purely computed algebraically (without any bulk-integration) which is demonstrated by the computation of the explicit results for several higher point amplitudes \cite{Albayrak:2018tam,Albayrak:2019asr}. It would be interesting to connect to these bulk computations from a boundary perspective.} Lastly, if one wants to study an AdS theory with fermions, we need to understand the tree-level fermionic correlators to fully determine a one-loop scalar four point function. We therefore hope the results presented here are useful in the wider study of AdS$_4$ correlators.

	\appendix

\chapter{$6j$ Symbols}
\label{qppandix: 6j symbols}

In \secref{\ref{sec:6jreview}} we defined and reviewed $6j$ symbols. In this appendix, we will further discuss $6j$ symbols and detail our method of deriving relations between $6j$ symbols for operators of different spins.

\section{Spinning Down the $6j$ Symbol}
\label{sec:Inversion_Isolated_Ops}

We first draw the attention of the reader to the problem of inverting a single partial wave (or block) with external spinning operators. This allows us to compute corrections to the anomalous dimensions of double twist operators. As a reminder, the general form of the $6j$ symbol is given by
\begin{align}
	\sixj{\cO_1}{\cO_2}{\cO_3}{\cO_4}{\cO_5}{\cO_6}^{abcd} &=\left(\widetilde{\Psi}^{(s),ab}_{\cO_5},\Psi^{(t),cd}_{\cO_6}\right) \nonumber\\
	&=\int d^{d}x_{1}...d^{d}x_{6}\<\widetilde{\cO}_1\widetilde{\cO}_2\widetilde{\cO}_{5}\>\<\widetilde{\cO}_3\widetilde{\cO}_4\cO_5\>\<\cO_3\cO_2\cO_6\>\<\cO_1\cO_4\widetilde{\cO}_{6}\>. \label{eqn:Gen6jSymbol2}
\end{align}
Our strategy will in a way be the reverse of the scalar case \cite{Liu:2018jhs}. Instead of using the Lorentzian inversion formula, we will follow the strategy outlined in \cite{Karateev:2017jgd} and use weight-shifting operators to calculate the $6j$ symbol for external fermions in terms of the $6j$ symbol for external scalars. We then use the expression (\ref{eq:6j_symbol_split_Scalars}), which splits the scalar $6j$ symbol into two pieces from inverting the physical block and its shadow to find the corresponding split for the fermionic $6j$ symbol.

To start, we use the results of \cite{Karateev:2017jgd} to write the t-channel spinning partial wave as a differential operator acting on the partial wave for external scalars: 
\begin{align}
	\Psi^{(t),ab}_{\cO}(x_i)=\mathfrak{D}^{ab}_{t}\Psi^{(t),\text{scalar}}_{\cO'}(x_i)\;,
\end{align}
where $\Psi^{(t),\text{scalar}}_{\cO'}(x_i)$ is a partial wave for four external scalars, $\<\f_3\f_2\f_1\f_4\>$. We are being very schematic here and it should be understood that $\mathfrak{D}^{ab}_{t}$ is a sum of multiple weight-shifting operators. For each term in the sum we may have to choose a different shifted operator $\cO'$, with scaling dimension and spin shifted from $\cO$, as well as different external scaling dimensions $\Delta_i$ of the scalar partial wave.

Given this expression we can simplify the spinning $6j$ symbol by taking the adjoint of this operator:
\begin{align}
	\left(\widetilde{\Psi}^{(s),ab}_{\cO_5},\Psi^{(t),ab}_{\cO_{6}}\right)&=\left(\mathfrak{D}^{ab}_{s}\widetilde{\Psi}^{(s),\text{scalar}}_{\cO'_{5}},\mathfrak{D}^{ab}_{t}\Psi^{(t),\text{scalar}}_{\cO'_{6}}\right) \nonumber
	\\ &=\left(\widetilde{\Psi}^{(s),\text{scalar}}_{\cO'_{5}},\mathfrak{D}^{*,ab}_{s}\mathfrak{D}^{ab}_{t}\Psi^{(t),\text{scalar}}_{\cO'_{6}}\right).
\end{align}
The two weight-shifting operators acting on the t-channel conformal partial wave then give a linear combination of undifferentiated t-channel conformal partial waves, at the price of more shifts for the internal and external labels. In the end we are left with an equality of the form
\begin{align}
	\sixj{\cO_1}{\cO_2}{\cO_3}{\cO_4}{\cO_5}{\cO_6}^{abcd}=\sum\limits_{\f_i,\cO'_{5,6}}
	\sixjDecomp{abcd}{\cO_1}{\cO_2}{\cO_3}{\cO_4}{\cO_5}{\cO_6}{\f_1}{\f_2}{\f_3}{\f_4}{\cO'_{5}}{\cO'_{6}}
	\sixj{\f_1}{\f_2}{\f_3}{\f_4}{\cO'_{5}}{\cO'_{6}}, \label{eq:6jDecompSpintoScalars}
\end{align}
where the sum runs over some set of scaling dimensions for the fictitious external scalars $\f_i$, whose dimensions are related to $\cO_i$ by some (half-)integer shift, and over both scaling dimensions and spins for $\cO'_{5,6}$, which are again related to the $\cO_{5,6}$ by (half-)integer shifts in both labels. We now turn to how to compute these coefficients.

The general strategy to compute the decomposition factors $J^{abcd}$ in \equref{eq:6jDecompSpintoScalars} can be systematized by the following procedure:
\begin{itemize}
	\item Write a three-point structure in terms of weight-shifting operators acting on three-point structures of lower spins.
	\item Use integration by parts and crossing symmetry of covariant three-point structures to move the weight-shifting operators such that they act on the same operator.
	\item By using irreducibility of the representations, the weight-shifting operators become multiples of the identity.
	\item Repeat until all three-point structures are of the form $\<\f\f\cO\>$, i.e.~we are left with three-point functions involving two scalars.
\end{itemize}
Below, we will unpack this procedure further by detailing each step in the explicit decomposition of the $6j$ symbol of two external scalars and two external fermions.

In our conventions, the $6j$-symbol for the correlator $\<\phi\phi\psi\psi\>$ reads as
\be 
\label{eq: fermion-scalar 6j symbol}
\left\{
\begin{matrix}
	\f_1 & \f_2& \cO_6 \\
	\psi_3 & \psi_4 & \cO_5
\end{matrix}
\right\}^{\uniq s_2t_1t_2}
=\int d^dx_1...d^dx_6\<\tl\f_1\tl\f_2\tl\cO_5\>\< \tl\psi_3\tl\psi_4\cO_5\>^{s_2}\<\psi_3\phi_2\cO_6\>^{t_1}\< \f_1\psi_4\tl\cO_6\>^{t_2},
\ee 
where the first three-point structure is already in the appropriate form for a $6j$ symbol of four external scalars, so all we need to do is to massage the remaining structures. As the first step, we use \equref{eq: expansion of psipsiO} to rewrite $\<\tl\psi_3\tl\psi_4\cO_5\>^{s_2}$:\footnote{We use the same shorthand notation as is used in \secref{\ref{sec:applications}}: $\f^a_i\equiv\f_{\De_i+a}$, $\tl\f^a_i\equiv\f_{3-\De_i+a}$, \mbox{$\cO_i^{a,b}\equiv\cO_{\De_i+a,l_i+b}$}, and $\tl\cO_i^{a,b}\equiv\cO_{3-\De_i+a,l_i+b}$.}
\begin{multline}
	\label{eq: decomposition of fermionic 6j symbol, step 0}
	\int d^dx_3d^dx_4d^dx_6\<\tl\psi_3\tl\psi_4\cO_5\>^{s_2}\<\psi_3\f_2\cO_6\>^{t_1}\< \f_1\psi_4\tl\cO_6\>^{t_2}\\=\int d^dx_3d^dx_4d^dx_6\sum\limits_{a,b}\kappa^{s_2}_{3,ab}(\tl\psi_3\tl\psi_4\cO_5)
	\cD_1^{-a,+}\cD_2^{-b,+}\<\tl\f_3^a\tl\f_4^b\cO_5 \>
	\<\psi_3\f_2\cO_6\>^{t_1}
	\<\f_1\psi_4\tl\cO_6\>^{t_2}\;.
\end{multline}
After that, we integrate by parts to obtain
\begin{multline}
	\label{eq: decomposition of fermionic 6j symbol, step 1}
	\int d^dx_3d^dx_4d^dx_6\<\tl\psi_3\tl\psi_4\cO_5\>^{s_2}\<\psi_3\f_2\cO_6\>^{t_1}\< \f_1\psi_4\tl\cO_6\>^{t_2}=\int d^dx_3d^dx_4d^dx_6\\\sum\limits_{a,b}\kappa^{s_2}_{3,ab}(\tl\psi_3\tl\psi_4\cO_5)
	\<\tl\f_3^a\tl\f_4^b\cO_5 \>
	\left[\left(\cD_1^{-a,+}\right)^*_{A}\<\psi_3\f_2\cO_6\>^{t_1}\right]
	\left[\left(\cD_2^{-b,+}\right)^{A*}
	\<\f_1\psi_4\tl\cO_6\>^{t_2}\right]\;,
\end{multline}
where we are showing the spinor indices of weight-shifting operators explicitly. With\linebreak \equref{eq: adjoints of weight-shifting operators}, the equation further reduces to
\begin{align}
	\label{eq: decomposition of fermionic 6j symbol, step 2}
	\int d^dx_3&d^dx_4d^dx_6\<\tl\psi_3\tl\psi_4\cO_5\>^{s_2}\<\psi_3\f_2\cO_6\>^{t_1}\< \f_1\psi_4\tl\cO_6\>^{t_2}
	\nonumber
	\\=&\int d^dx_3d^dx_4d^dx_6\sum\limits_{a,b}\kappa^{s_2}_{3,ab}(\tl\psi_3\tl\psi_4\cO_5)
	\zeta_{0}^{-a,+}\zeta_{0}^{-b,+}
	\<\tl\f_3^a\tl\f_4^b\cO_5 \>
	\nonumber
	\\ &\x\left[\left(\cD_1^{-a,-}\right)_{A}\<\psi_3\f_2\cO_6\>^{t_1}\right]
	\left[E_{\phi\psi\cO\rightarrow \cO\phi\psi}^{t_2t_2'}\left(\cD_3^{-b,-}\right)^{A}
	\<\tl\cO_6\f_1\psi_4\>^{t_2'}\right],
\end{align}
where we defined the exchange matrix $E$
\be 
\<\cO_2\cO_1\cO\>^a
=E^{ab}_{21\cO\to 12\cO}\<\cO_1\cO_2\cO\>^b
\ee 
to reorder the last tensor structure so that the weight-shifting operator acts on the third operator.\footnote{We do this because we will use the convention of \cite{Karateev:2017jgd} for finite-dimensional representations, where the weight-shifting operator acts on the first (third) operator in the $s$ ($t$) channel.} We can then use the crossing for conformally covariant three-point structures as derived in \cite{Karateev:2017jgd}, which reads as
\be \label{eq:3ptcrossing}
\left(\cD^{-a,-b}_3\right)^A\<\cO_1\cO_2\cO_3^{a,b}\>^{m}=\sum_{c,d,n}
\left\{
\begin{matrix}
	\cO_1 & \cO_2 & \cO_1^{c,d}\\
	\cO_3 & S & \cO_3^{a,b}
\end{matrix}
\right\}^{mn}
\left(\cD^{-c,-d}_1\right)^A\<\cO_1^{c,d}\cO_2\cO_3\>^{n}
\ee 
for the fermionic representation of weight-shifting operators in $3d$. Thus we obtain
\small 
\begin{align}
	\label{eq: decomposition of fermionic 6j symbol, step 3}
	\int d^dx_3&d^dx_4d^dx_6\<\tl\psi_3\tl\psi_4\cO_5\>^{s_2}\<\psi_3\f_2\cO_6\>^{t_1}\< \f_1\psi_4\tl\cO_6\>^{t_2}
	\nonumber \\=& \int d^dx_3d^dx_4d^dx_6\sum\limits_{a,b}\kappa^{s_2}_{3,ab}(\tl\psi_3\tl\psi_4\cO_5)
	\zeta_{0}^{-a,+}\zeta_{0}^{-b,+}
	\<\tl\f_3^a\tl\f_4^b\cO_5 \>
	\nonumber \\\x &\left[\left(\cD_1^{-a,-}\right)_{A}\<\psi_3\f_2\cO_6\>^{t_1}\right]
	\left[E_{\phi\psi\cO\rightarrow \cO\phi\psi}^{t_2t_2'}\sum_{c,d}
	\left\{
	\begin{matrix}
		\tl\cO_6 & \f_1 & \tl\cO_6^{c,d}\\
		\f_4^{-b} &S& \psi_4 
	\end{matrix}
	\right\}^{t_2'\uniq}
	\left(\cD^{-c,-d}_1\right)^A
	\<\tl\cO_6\f_1\phi_4^{-b}\>\right].
\end{align}
\normalsize

We can integrate by parts and use the crossing again to get both weight-shifting operators to act on the same operator:
\small 
\begin{align}
	\int & d^dx_3d^dx_4d^dx_6\<\tl\psi_3\tl\psi_4\cO_5\>^{s_2}\<\psi_3\f_2\cO_6\>^{t_1}\< \f_1\psi_4\tl\cO_6\>^{t_2}
	\nonumber \\
	=&\int d^dx_3d^dx_4d^dx_6\sum\limits_{a,b,c,d}\kappa^{s_2}_{3,ab}(\tl\psi_3\tl\psi_4\cO_5)
	\zeta_{0}^{-a,+}\zeta_{0}^{-b,+}\zeta_{l_t+d}^{-c,-d}\left\{
	\begin{matrix}
		\tl\cO_6 & \f_1 & \tl\cO_6^{c,d}\\
		\f_4^{-b} &S& \psi_4 
	\end{matrix}
	\right\}^{t_2'\uniq} E_{\phi\psi\cO\rightarrow \cO\phi\psi}^{t_2t_2'}
	\nonumber \\ &\x \<\tl\f_3^a\tl\f_4^b\cO_5 \>\Bigg[
	\sum_{e,f,n}
	\left\{
	\begin{matrix}
		\psi_3 & \f_2 & \cO_3^{e,f}\\
		\cO_6^{-c,d} & S & \cO_6
	\end{matrix}
	\right\}^{t_1n}
	\left(\cD_1^{-a,-}\right)_{A}
	\left(\cD^{-e,-f}_1\right)^A\<\cO_3^{e,f}\f_2\cO_6^{-c,d}\>^{n}
	\Bigg]
	\<\tl\cO_6\f_1\phi_4^{-b}\>. \label{eq: decomposition of fermionic 6j symbol, step 4}
\end{align}
\normalsize

By the irreducibility of the representations, we have 
\be 
\left(\cD^{-a,-b}\right)_A\left(\cD^{c,d}\right)^A\cO_{\De,l}= \delta^{ac}\delta^{bd}\beta^{\De,l}_{ab}\cO_{\De,l}
\ee 
with which we finally obtain\footnote{One can derive $\beta$ by acting on the two point function with weight shifting operators in embedding space. In our conventions, we have 
	\be 
	\beta_{a,b}^{\De,l}\equiv b (a+2 \Delta -3) (b+2 l+1) (2 a b \Delta +a (2 b+1) (a+b-2)+2 l)\quad\text{ for } a,b=\pm 1 \.
	\ee 
}
\begin{align}
	\int & d^dx_3d^dx_4d^dx_6\<\tl\psi_3\tl\psi_4\cO_5\>^{s_2}\<\psi_3\f_2\cO_6\>^{t_1}\< \f_1\psi_4\tl\cO_6\>^{t_2}
	\nonumber \\ &=d^dx_3d^dx_4d^dx_6\sum\limits_{a,b,c,d}\beta_{a,+}^{\Delta_3-a,0}\kappa^{s_2}_{3,ab}(\tl\psi_3\tl\psi_4\cO_5)
	\zeta_{0}^{-a,+}\zeta_{0}^{-b,+}\zeta_{l_t+d}^{-c,-d}E_{\phi\psi\cO\rightarrow \cO\phi\psi}^{t_2t_2'}
	\nonumber \\ &\x  \left\{
	\begin{matrix}
		\tl\cO_6 & \f_1 & \tl\cO_6^{c,d}\\
		\f_4^{-b} &S& \psi_4 
	\end{matrix}
	\right\}^{t_2'\uniq}\left\{
	\begin{matrix}
		\psi_3 & \f_2 & \phi_3^{-a}\\
		\cO_6^{-c,d} & S & \cO_6
	\end{matrix}
	\right\}^{t_1\uniq} \<\tl\f_3^a\tl\f_4^b\cO_5 \>
	\<\phi_3^{-a}\f_2\cO_6^{-c,d}\>^{n}
	\<\f_1\phi_4^{-b}\tl\cO_6^{c,d}\>. \label{eq: decomposition of fermionic 6j symbol, step 5}
\end{align}

Inserting this into \equref{eq: fermion-scalar 6j symbol}, we get
\begin{multline}
	\label{eq: 6j decomposition of 2f2s}
	\left\{
	\begin{matrix}
		\f_1 & \f_2& \cO_6 \\
		\psi_3 & \psi_4 & \cO_5
	\end{matrix}
	\right\}^{\uniq s_2t_1t_2}
	=
	\sum\limits_{a,b,c,d}
	\sixjDecomp{\uniq s_2t_1t_2}{\f_1}{\f_2}{\psi_3}{\psi_4}{\cO_5}{\cO_6}{\f_1}{\f_2}{\f_3^{-a}}{\f_4^{-b}}{\cO_5}{\cO_6^{-c,d}}
	\\\x\left\{
	\begin{matrix}
		\f_1 &\f_2& \cO_6^{-c,d} \\
		\f_3^{-a} & \f_4^{-b} & \cO_5
	\end{matrix}
	\right\}\;,
\end{multline} 
where
\begin{multline}
	\label{eq: J for 2f2s}
	\sixjDecomp{\uniq s_2t_1t_2}{\f_1}{\f_2}{\psi_3}{\psi_4}{\cO_5}{\cO_6}{\f_1}{\f_2}{\f_3^{-a}}{\f_4^{-b}}{\cO_5}{\cO_6^{-c,d}}=\beta_{a,+}^{\Delta_3-a,0}\kappa^{s_2}_{3,ab}(\tl\psi_3\tl\psi_4\cO_5)
	\zeta_{0}^{-a,+}\zeta_{0}^{-b,+}\zeta_{l_t+d}^{-c,-d}\\\x E_{\phi\psi\cO\rightarrow \cO\phi\psi}^{t_2t_2'}
	\left\{
	\begin{matrix}
		\tl\cO_6 & \f_1 & \tl\cO_6^{c,d}\\
		\f_4^{-b} &S& \psi_4 
	\end{matrix}
	\right\}^{t_2'\uniq}\left\{
	\begin{matrix}
		\psi_3 & \f_2 & \phi_3^{-a}\\
		\cO_6^{-c,d} & S & \cO_6
	\end{matrix}
	\right\}^{t_1\uniq}.
\end{multline}

Note that if $l_5=0$, there are only 2 independent structures for $\<\psi_3\psi_4\phi_5\>^{s_2}$, hence we need to take a $2\times 2$ invertible submatrix of $\kappa_3$. We can do this by restricting to the structures $\<\psi_3\psi_4\phi_5\>^{1,3}$ in \equref{eq: 3pt structures}, and fixing $b=\half$ in \equref{eq: 6j decomposition of 2f2s} instead of summing over $b=\pm\half$.

\begin{figure}
	\centering
	$\begin{aligned}
		\includegraphics[scale=.55]{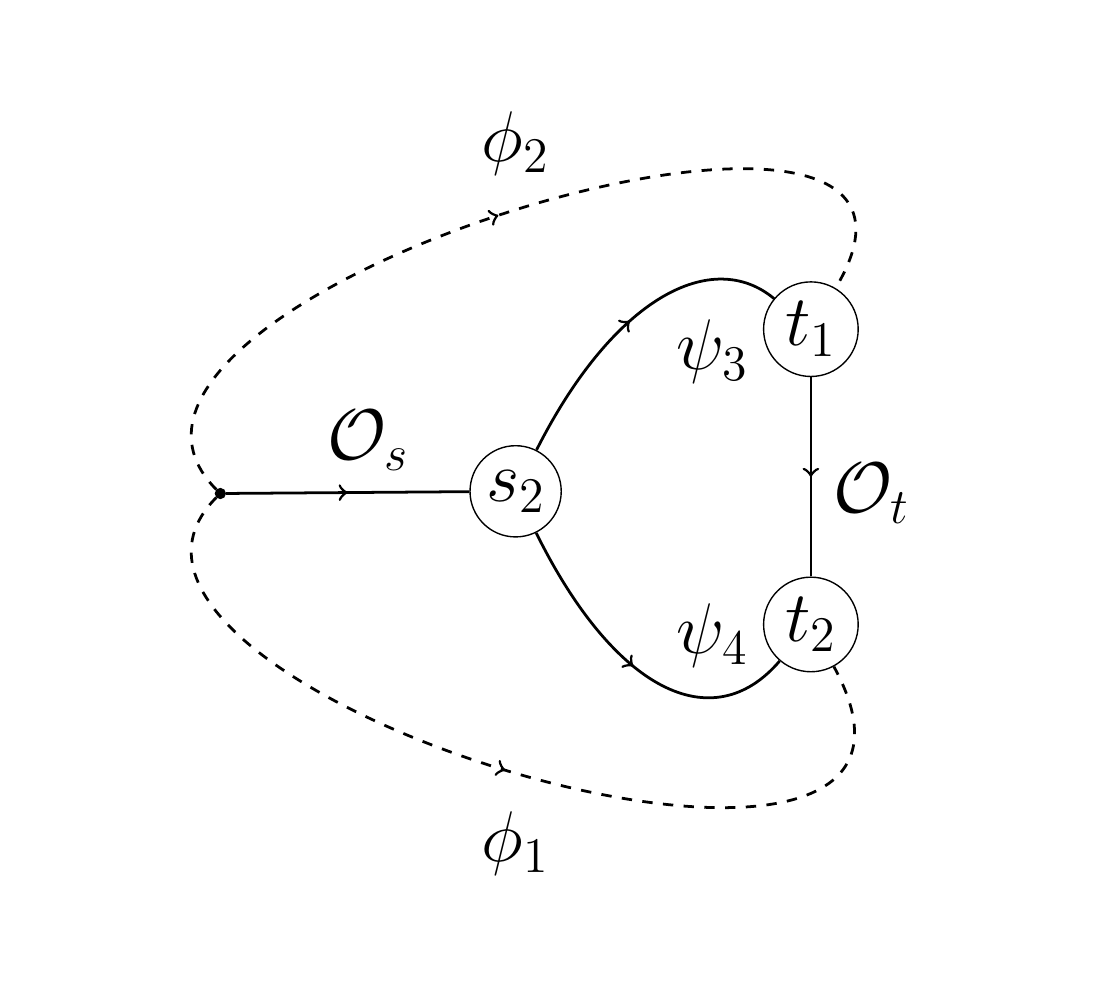}
	\end{aligned}$ $\Rightarrow$ $\begin{aligned}
		\includegraphics[scale=.55]{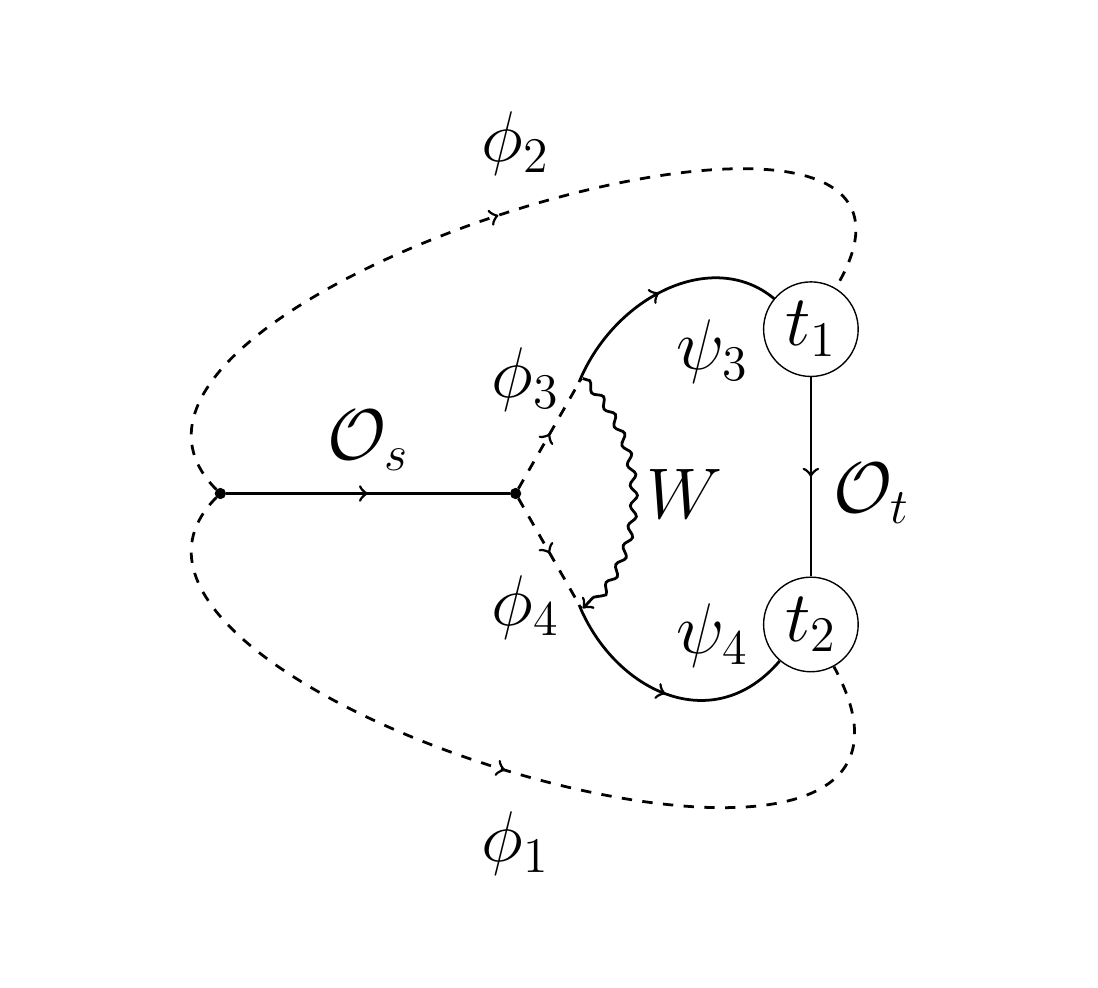}
	\end{aligned}$ $\Rightarrow$ $\begin{aligned}
		\includegraphics[scale=.55]{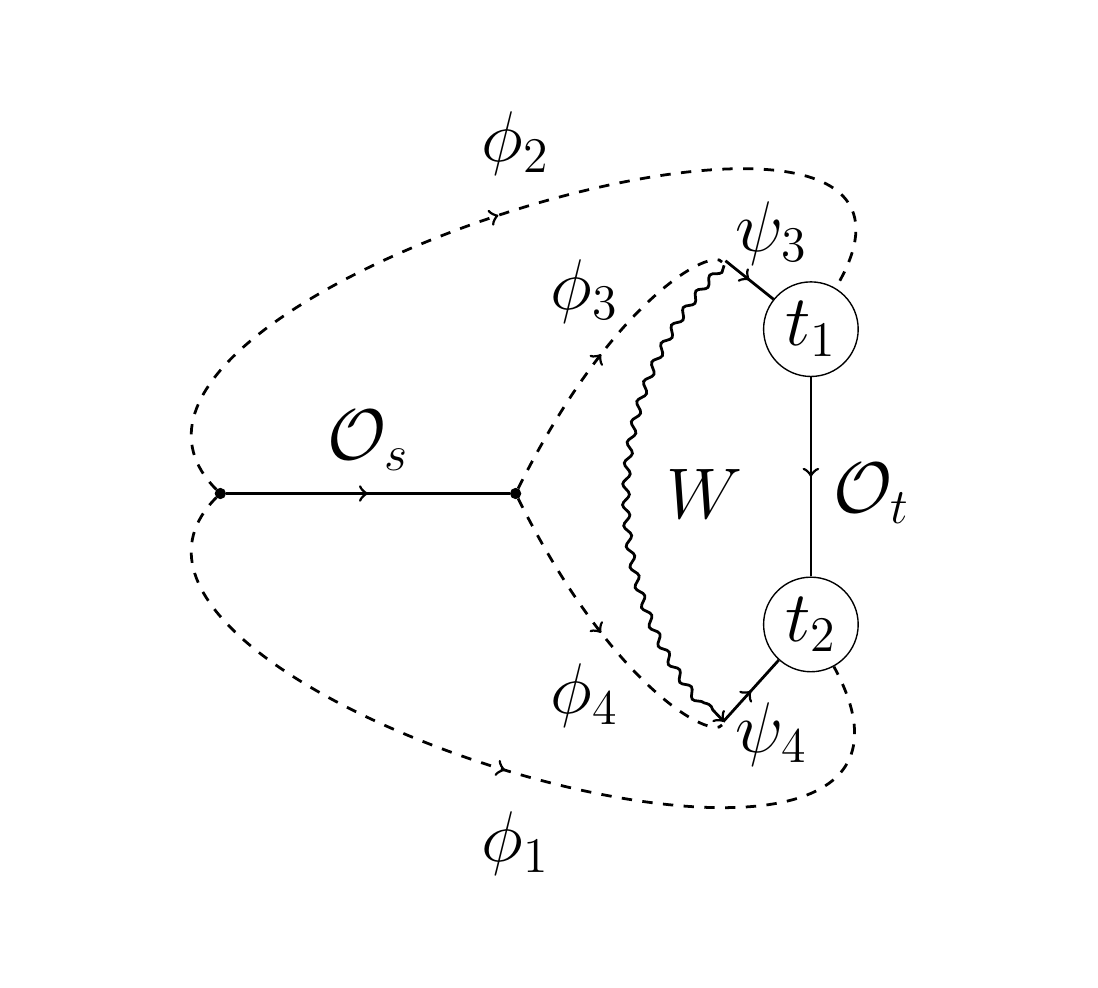}
	\end{aligned}$ $\Rightarrow$ $\begin{aligned}
		\includegraphics[scale=.55]{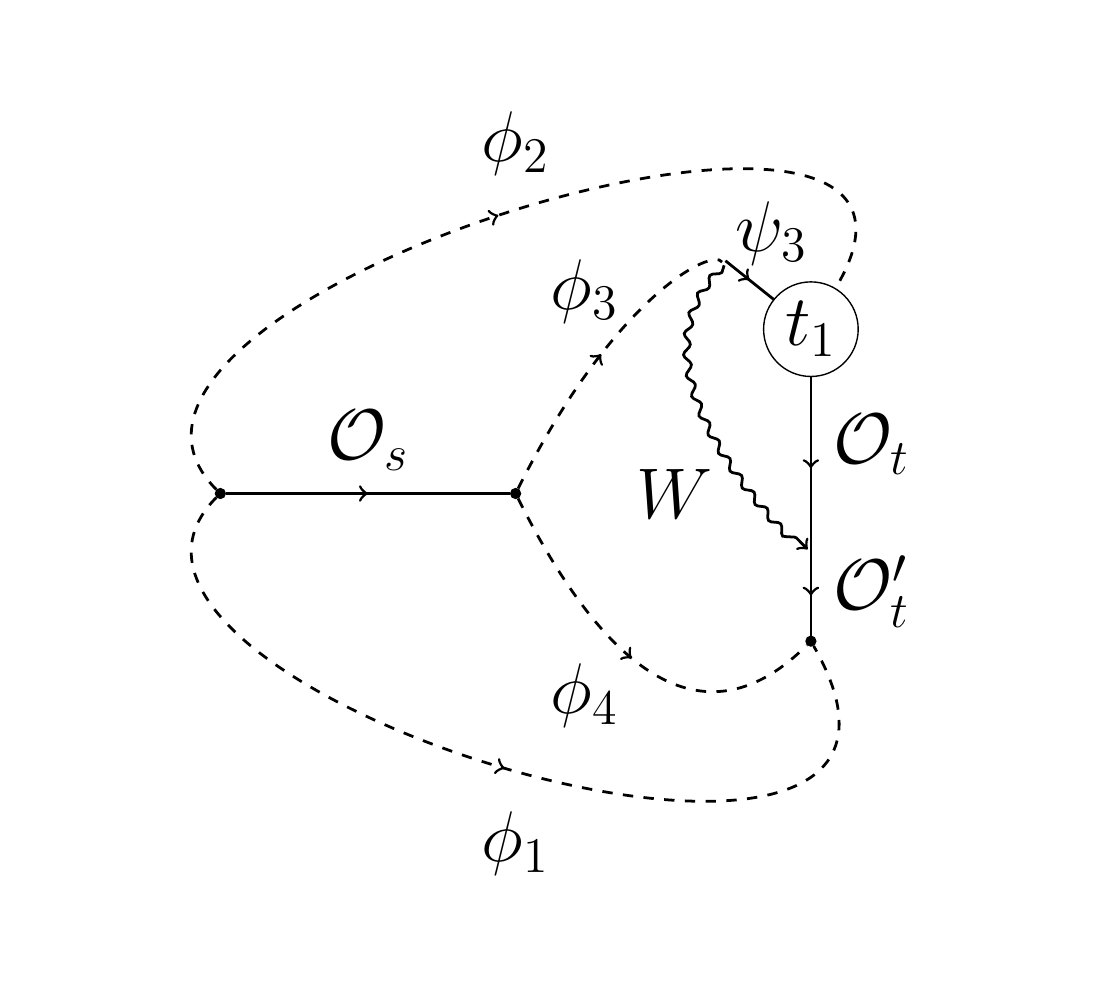}
	\end{aligned}$ $\Rightarrow$ $\begin{aligned}
		\includegraphics[scale=.55]{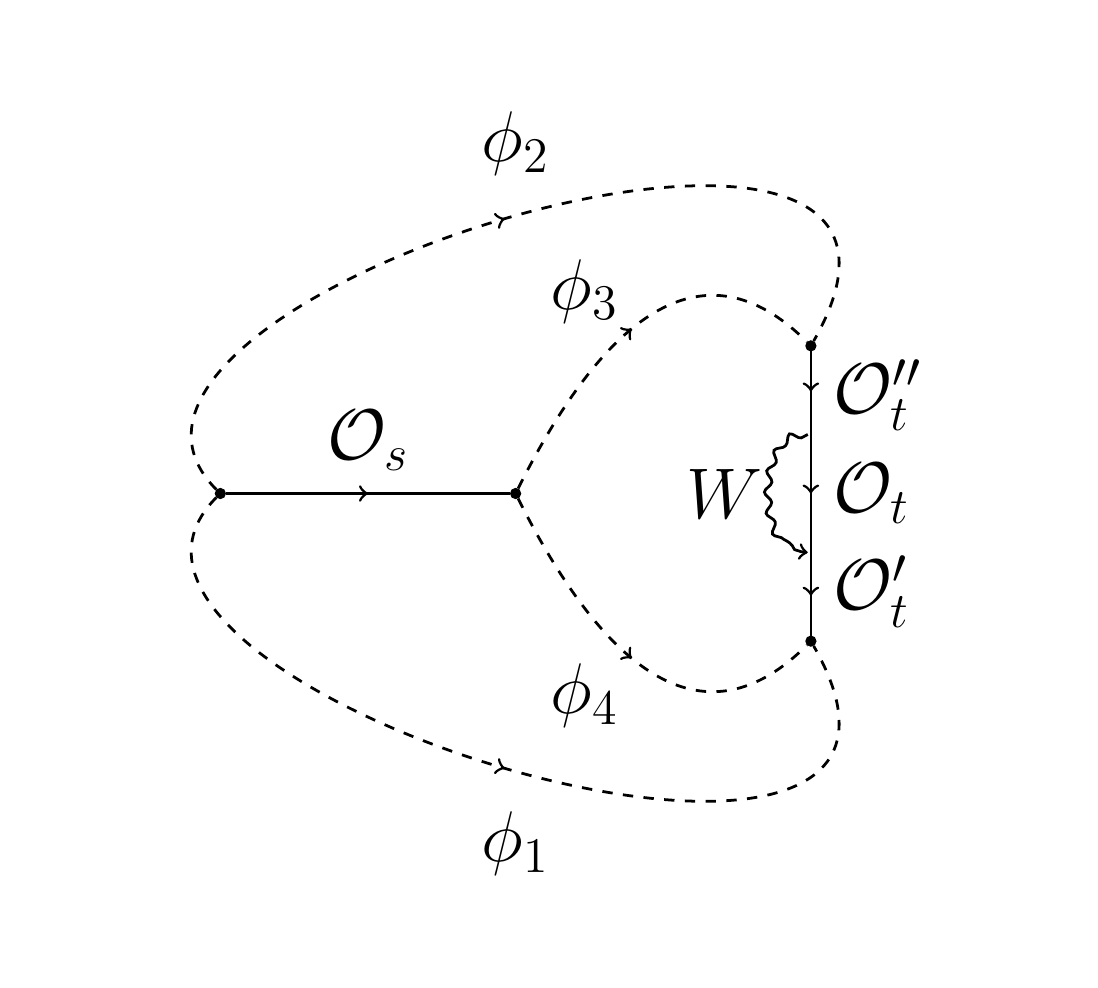}
	\end{aligned}$ $\Rightarrow$ $\begin{aligned}
		\includegraphics[scale=.55]{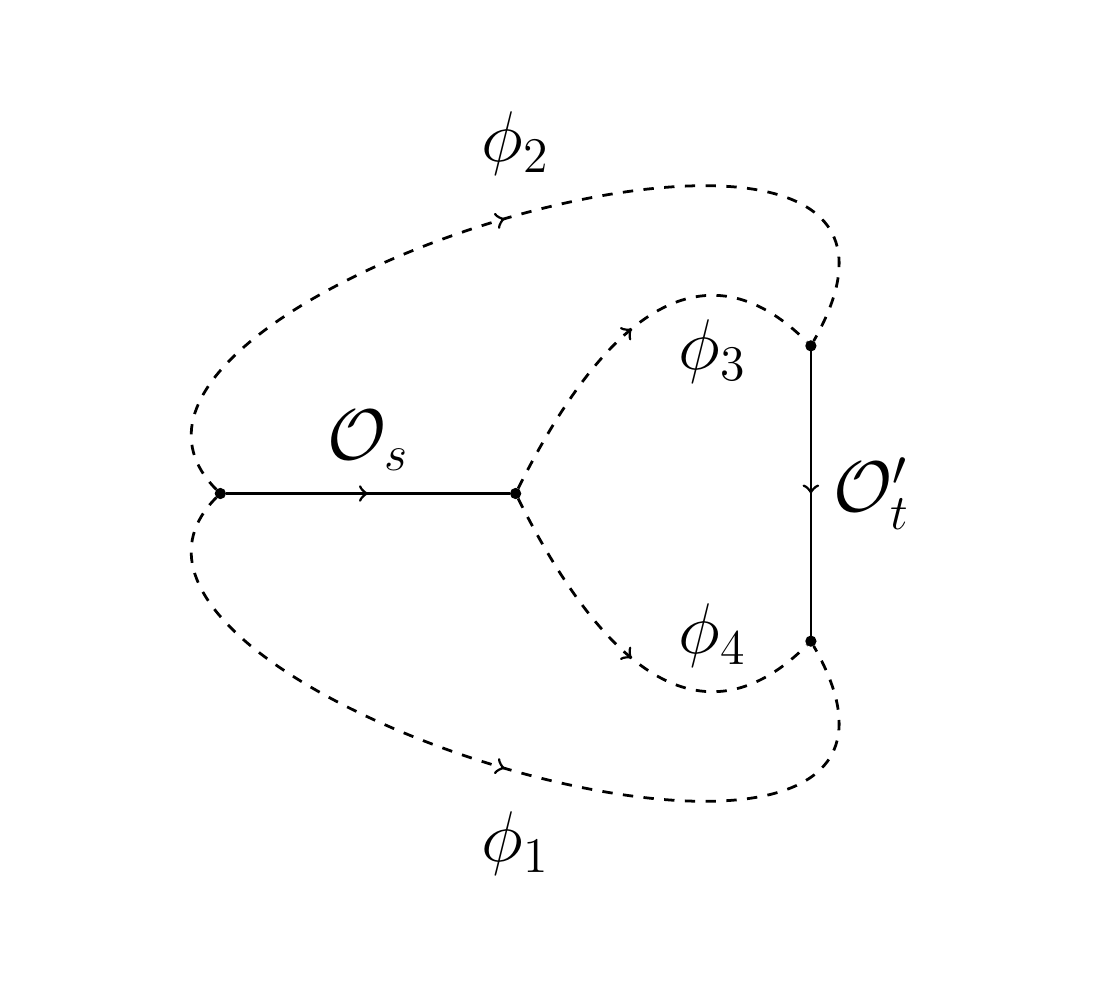}
	\end{aligned}$
	\caption[Diagrammatic illustration of the decomposition of two fermion $6j$ symbol]{\label{fig: ssff decomposition}
		Step by step diagrammatic illustration for the decomposition of the $6j$ symbol  for $\<\phi \psi \psi \phi\>$ into the  $6j$ symbol for $\<\f\f\f\f\>$. The idea is as follows: one re-expresses a fermionic three-point structure, $\<\psi\psi\cO\>$, in terms of weight-shifting operators acting on a scalar three-point structure, $\<\phi\phi\cO\>$. The weight-shifting operators are then moved inside the diagrammatic loop until they act on the same leg. By the irreducibility of the representations, i.e. $\left(\cD^{a,b}\right)_A\left(\cD^{-c,-d}\right)^A\sim \delta^{ac}\delta^{bd}$, the diagram reduces to that of a scalar $6j$ symbol. To be able to move around the weight-shifting operators, one either integrates by parts or uses the crossing relation for covariant three-point structures as explained in the main text. The diagrams above correspond to the equations \equref{eq: fermion-scalar 6j symbol}, \equref{eq: decomposition of fermionic 6j symbol, step 0}, \equref{eq: decomposition of fermionic 6j symbol, step 2}, \equref{eq: decomposition of fermionic 6j symbol, step 3}, \equref{eq: decomposition of fermionic 6j symbol, step 4}, and \equref{eq: decomposition of fermionic 6j symbol, step 5} respectively.
	}
\end{figure}

Despite the complicated and lengthy expressions, the procedure is actually quite straightforward and most easily tractable in the diagrammatic notation, see figure \ref{fig: ssff decomposition} for a summary of the decomposition above. However, one should only use the diagrammatic expressions as a guide, as there are ambiguities in their meaning, most notably sign ambiguities as they do not carry the information of the order of operators in the equations. 

For spinning correlators beyond $\<\f\f\psi\psi\>$ we can repeat the procedure above and recursively spin-down the $6j$ symbol. For this, we first define a generalized form of \equref{eq:6jDecompSpintoScalars}:
\small 
\be
\label{eq: generalized J coefficients}
\sixj{\cO_1}{\cO_2}{\cO_3}{\cO_4}{\cO_5}{\cO_6}^{abcd}=\sum\limits_{\f_i,\cO'_{5,6}}
\sixjDecompp{abcd}{efgh}{\cO_1}{\cO_2}{\cO_3}{\cO_4}{\cO_5}{\cO_6}{\cO'_{1}}{\cO'_{2}}{\cO'_{3}}{\cO'_{4}}{\cO'_{5}}{\cO'_{6}}
\sixj{\cO'_{1}}{\cO'_{2}}{\cO'_{3}}{\cO'_{4}}{\cO'_{5}}{\cO'_{6}}^{efgh},
\ee
\normalsize
where we obtain the ultimate result $J^{abcd}$ of \equref{eq:6jDecompSpintoScalars} by summing over these intermediate factors $J^{abcd}_{efgh}$.

For four external fermions, we only need to repeat this process twice, where in the first step we reduce from four external fermions to two external fermions and two external scalars, and then in the second step we reduce from two external fermions and two external scalars to four external scalars. The second step is already what we derived above, so the only new ingredient is the first step:
\begin{multline}
	\sixj{\psi_1}{\psi_2}{\psi_3}{\psi_4}{\cO_5}{\cO_6}^{s_1s_2t_1t_2}=\sum\limits_{a,b,c,d,t_1',t_2'}
	\sixjDecompp{s_1s_2t_1t_2}{\uniq s_2t_1't_2'}{\psi_1}{\psi_2}{\psi_3}{\psi_4}{\cO_5}{\cO_6}{\f_1^{-a}}{\f_2^{-b}}{\psi_3}{\psi_4}{\cO_5}{\cO_6^{-c,d}}\\\x
	\sixj{\f_1^{-a}}{\f_2^{-b}}{\psi_3}{\psi_4}{\cO_5}{\cO_6^{-c,d}}^{\uniq s_2t_1't_2'}.
\end{multline}
Combining this with \equref{eq: 6j decomposition of 2f2s}, we obtain the final result
\begin{multline}
	\label{eq: 6j decomposition of 4f}
	\sixj{\psi_1}{\psi_2}{\psi_3}{\psi_4}{\cO_5}{\cO_6}^{s_1s_2t_1t_2}=\sum\limits_{a,\dots,h}
	\sixjDecomp{s_1s_2t_1t_2}{\psi_1}{\psi_2}{\psi_3}{\psi_4}{\cO_5}{\cO_6}{\f_1^{-a}}{\f_2^{-b}}{\f_3^{-e}}{\f_4^{-f}}{\cO_5}{\cO_6^{-c-g,d+h}}\\\x
	\sixj{\f_1^{-a}}{\f_2^{-b}}{\f_3^{-e}}{\f_4^{-f}}{\cO_5}{\cO_6^{-c-g,d+h}},
\end{multline}
where
\begin{multline}
	\sixjDecomp{s_1s_2t_1t_2}{\psi_1}{\psi_2}{\psi_3}{\psi_4}{\cO_5}{\cO_6}{\f_1^{-a}}{\f_2^{-b}}{\f_3^{-e}}{\f_4^{-f}}{\cO_5}{\cO_6^{-c-g,d+h}} \\=\sum\limits_{t_1',t_2'} 
	\sixjDecompp{s_1s_2t_1t_2}{\uniq s_2t_1't_2'}{\psi_1}{\psi_2}{\psi_3}{\psi_4}{\cO_5}{\cO_6}{\f_1^{-a}}{\f_2^{-b}}{\psi_3}{\psi_4}{\cO_5}{\cO_6^{-c,d}}\\\x
	\sixjDecomp{\uniq s_2t_1't_2'}{\f_1^{-a}}{\f_2^{-b}}{\psi_3}{\psi_4}{\cO_5}{\cO_6^{-c,d}}{\f_1^{-a}}{\f_2^{-b}}{\f_3^{-e}}{\f_4^{-f}}{\cO_5}{\cO_6^{-c-g,d+h}}.
\end{multline}

\begin{figure}
	\centering
	$\begin{aligned}
		\includegraphics[scale=.55]{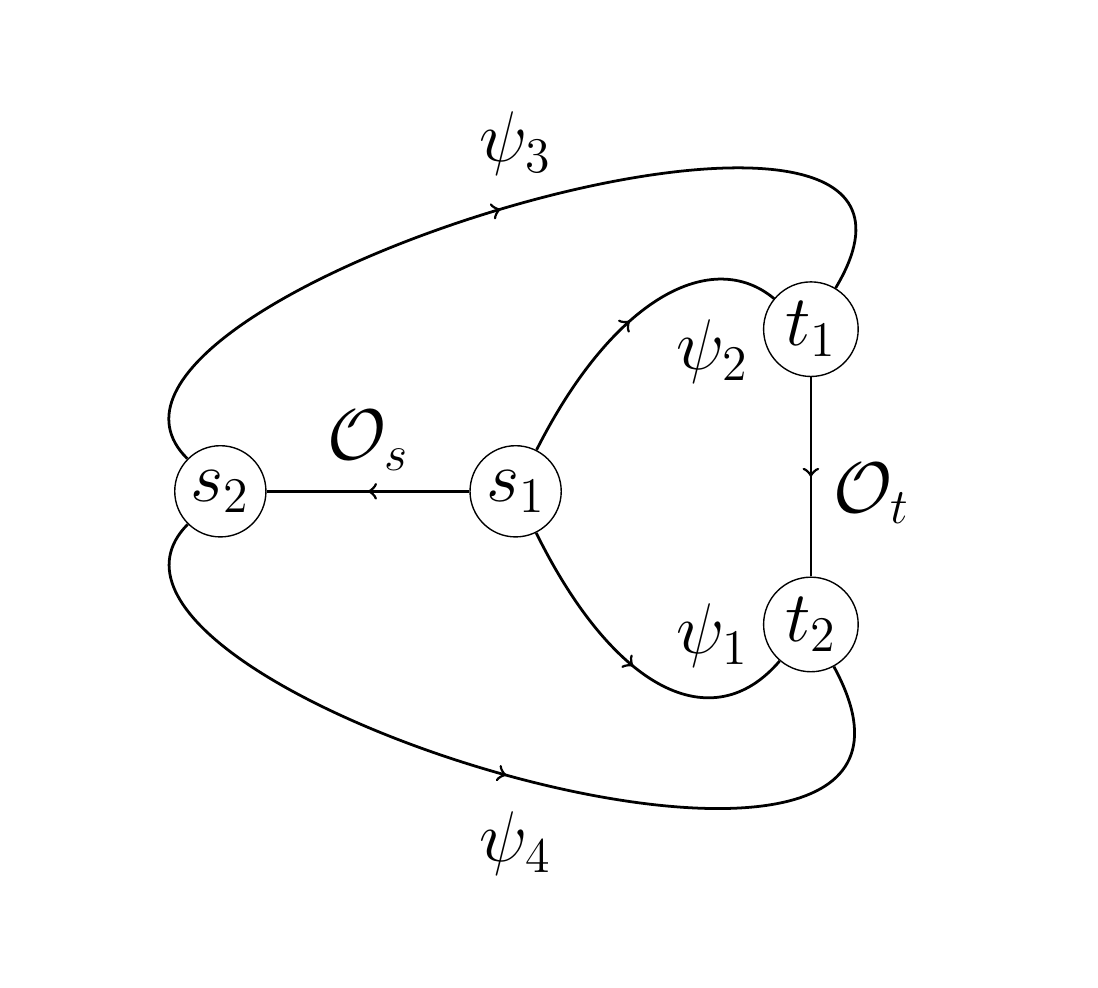}
	\end{aligned}$ $\Rightarrow$ $\begin{aligned}
		\includegraphics[scale=.55]{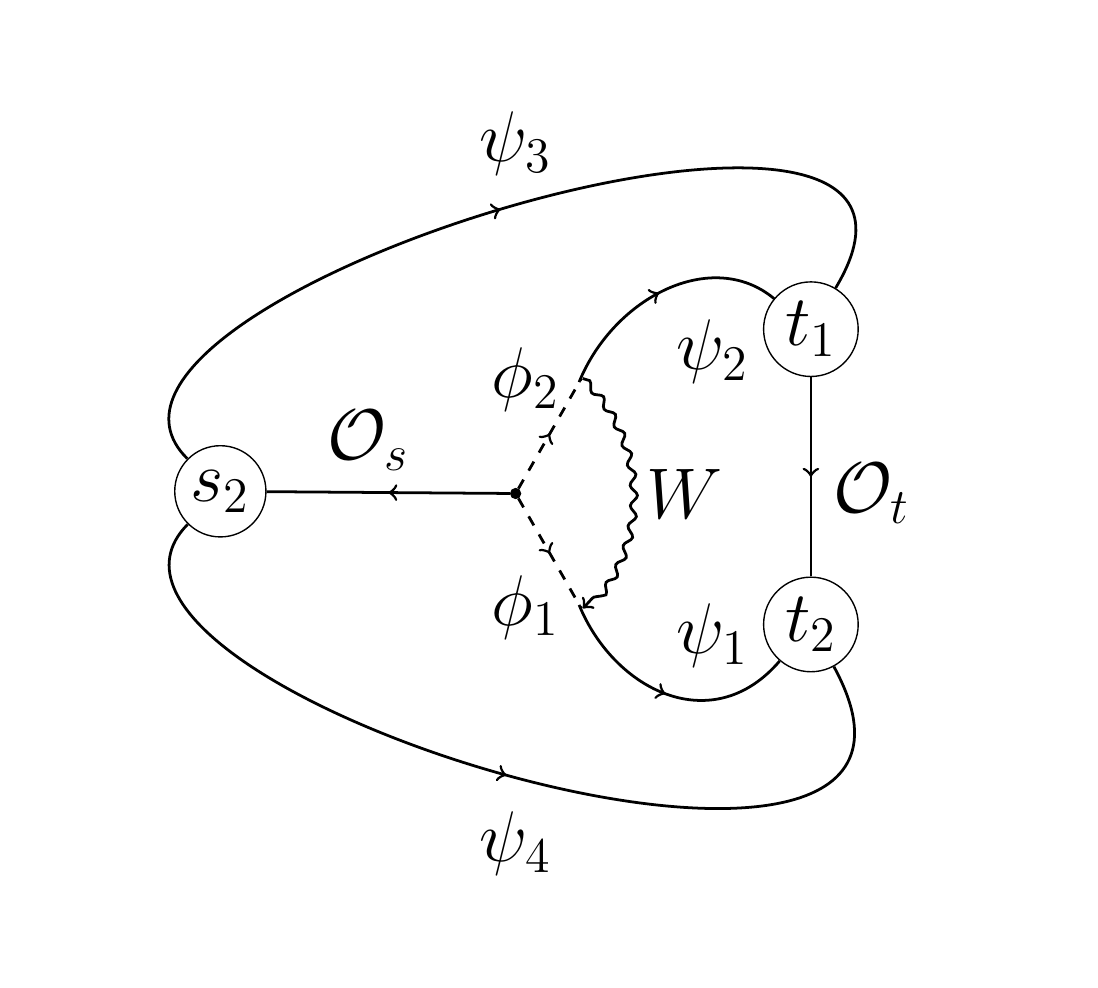}
	\end{aligned}$ $\Rightarrow$ $\begin{aligned}
		\includegraphics[scale=.55]{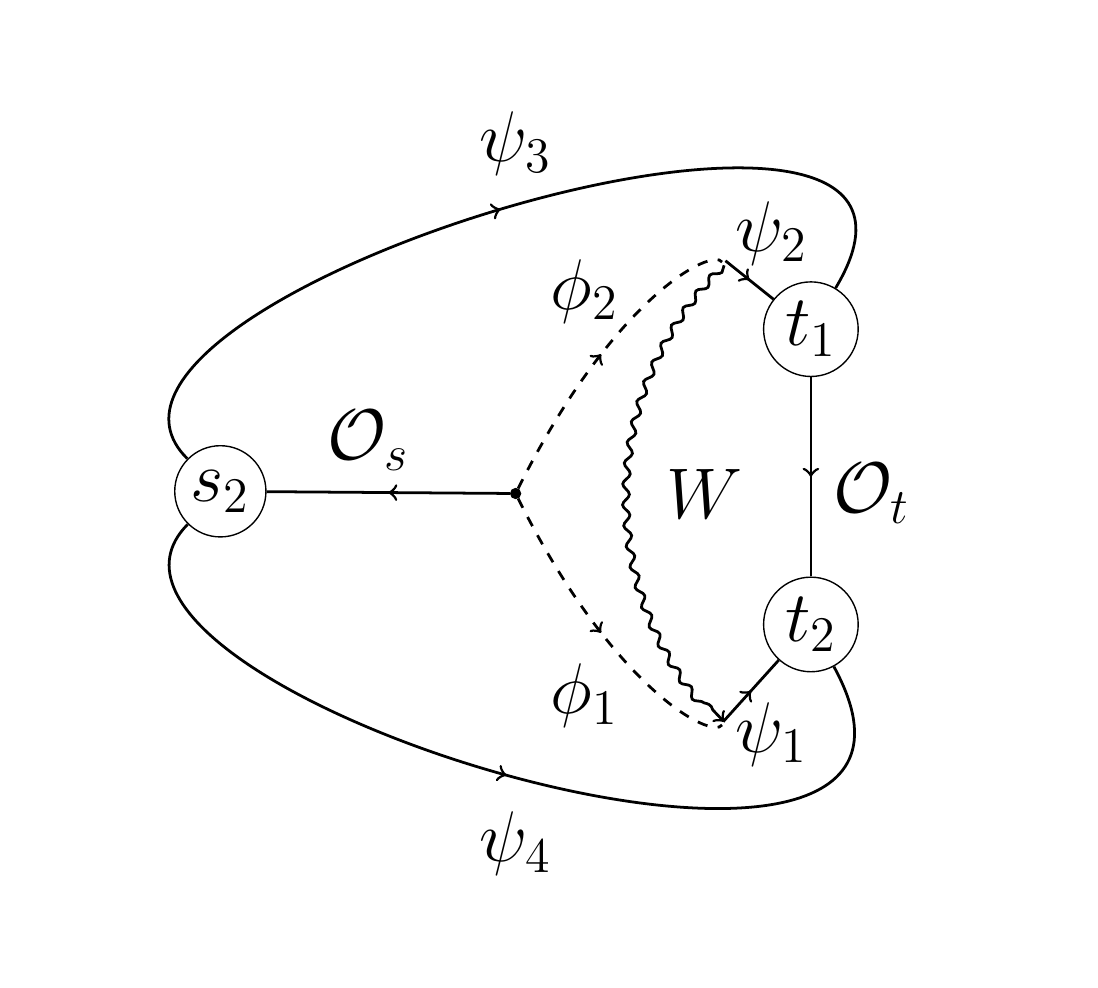}
	\end{aligned}$ $\Rightarrow$ $\begin{aligned}
		\includegraphics[scale=.55]{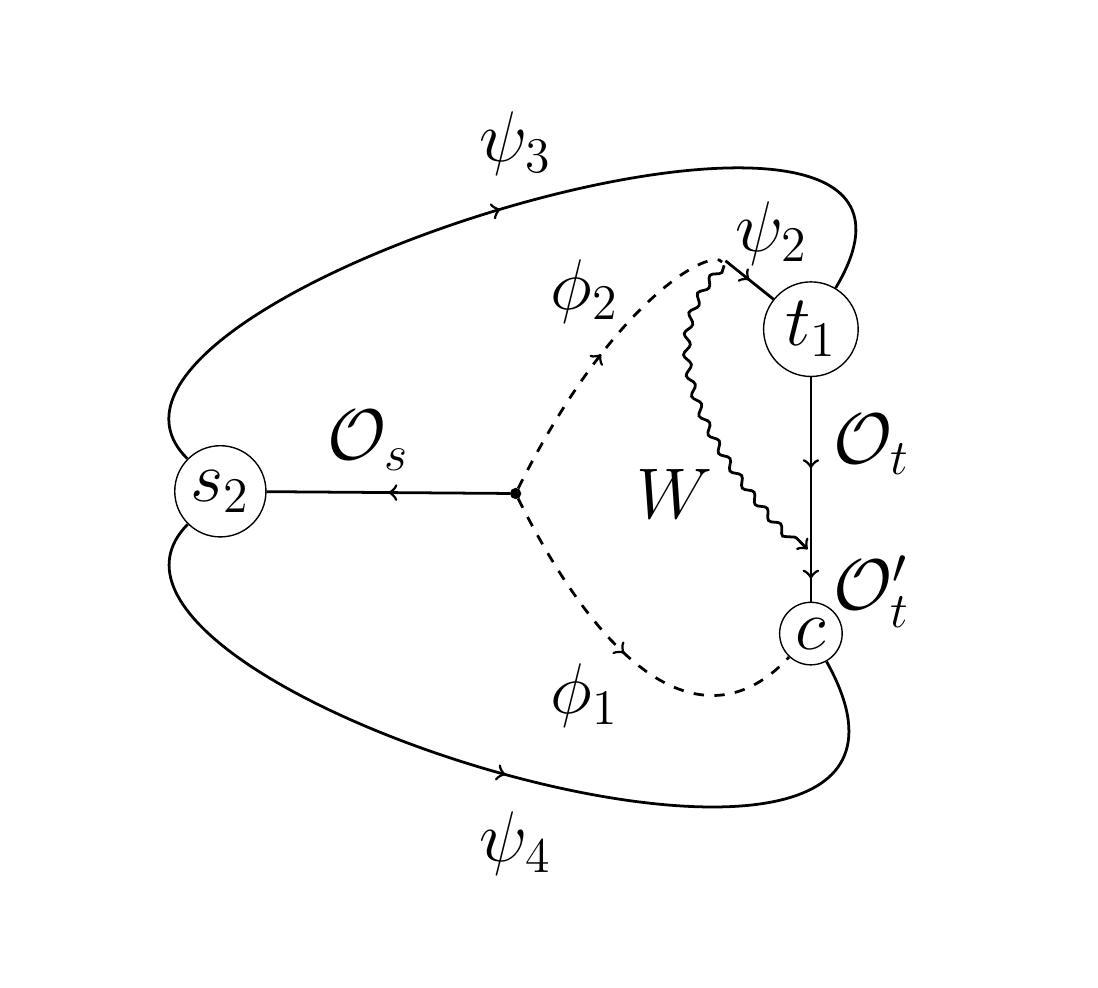}
	\end{aligned}$ $\Rightarrow$ $\begin{aligned}
		\includegraphics[scale=.55]{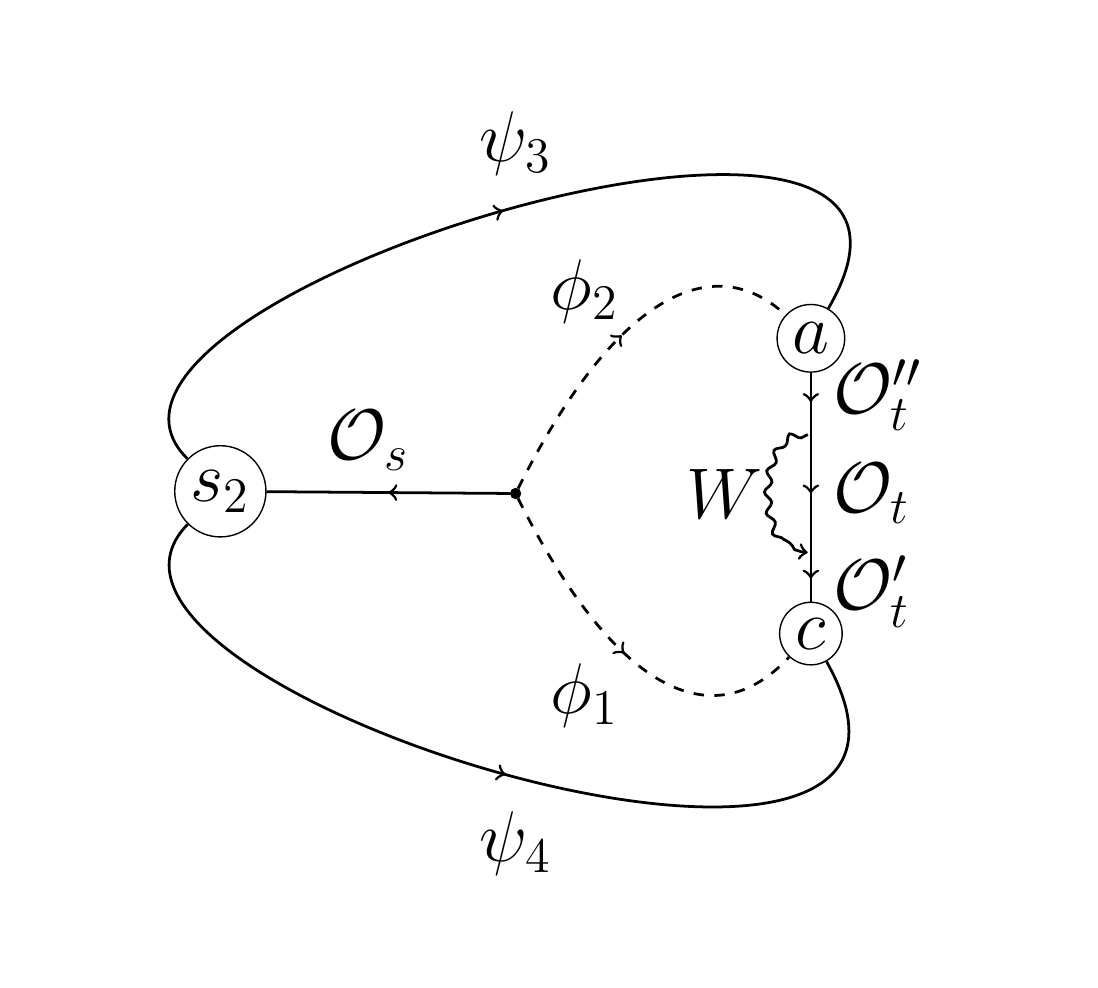}
	\end{aligned}$ $\Rightarrow$ $\begin{aligned}
		\includegraphics[scale=.55]{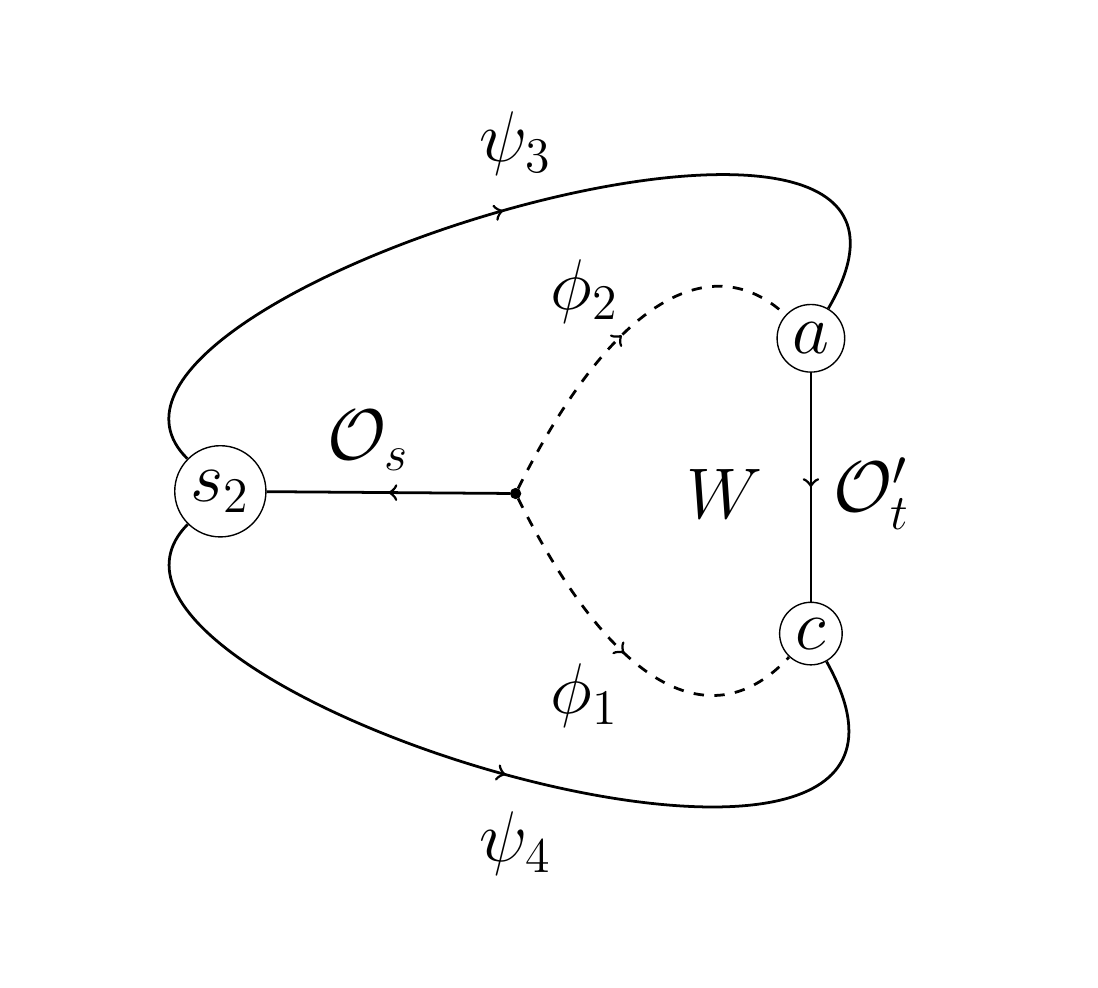}
	\end{aligned}$
	\caption[Diagrammatic illustration of the decomposition of four fermion $6j$ symbol]{\label{fig: ffff decomposition}
		Step by step diagrammatic illustration for the decomposition of $6j$ symbol of four external fermions, $\<\psi \psi \psi \psi\>$, in terms of $6j$ symbols of two external fermions and two external scalars.
	}
\end{figure}

We can derive $J^{s_1s_2t_1t_2}_{\uniq s_2t_1't_2'}$ in a similar manner to how we derived $J^{\uniq s_2t_1t_2}$. For brevity we skip the intermediate steps, illustrated diagrammatically in figure \ref{fig: ffff decomposition}, and only present the final result here:
\begin{multline}
	\sixjDecompp{s_1s_2t_1t_2}{\uniq s_2t_1't_2'}{\psi_1}{\psi_2}{\psi_3}{\psi_4}{\cO_5}{\cO_6}{\f_1^{-a}}{\f_2^{-b}}{\psi_3}{\psi_4}{\cO_5}{\cO_6^{-c,d}}=\sum\limits_{u_1,u_2,u_1',u_2'}
	\kappa^{s_1}_{3,ab}(\tl\psi_1\tl\psi_2\tl\cO_5)
	\zeta_{0}^{-b,+}\zeta_{0}^{-a,+}\zeta_{l_t+d}^{-c,-d}\\\x\beta_{b,+}^{\Delta_2-b,0}
	E_{\psi_1\psi_2\cO\rightarrow \psi_2\psi_1\cO}^{t_1u_1}
	\left\{
	\begin{matrix}
		\psi_2 & \psi_3 & \f_2^{-b}\\
		\cO_6^{-c,d} & S & \cO_6
	\end{matrix}
	\right\}^{u_1u_1'}
	E_{\phi\psi\cO\rightarrow \psi\phi\cO}^{u_1't_1'}
	E_{\psi_1\psi_2\cO\rightarrow \cO\psi_2\psi_1}^{t_2u_2}
	\\\x\left\{
	\begin{matrix}
		\tl\cO_6 & \psi_4 & \tl\cO_6^{c,d}\\
		\f_1^{-a} & S & \psi_1
	\end{matrix}
	\right\}^{u_2u_2'} 
	E_{\cO\psi\phi\rightarrow \phi\psi\cO}^{u_2't_2'}\;.
\end{multline}

\section{OPE Function and its Decomposition}
In \secref{\ref{sec:6jreview}} we discussed how $6j$ symbols are related to OPE coefficients. We reproduce \equref{eqn:InvSingBlock_to_OPEdata} for convenience:
\begin{align}
	\lambda_{125,a}\lambda_{345,b}\bigg|_{G^{(t),fg}_{\cO_6}}&=-\Res_{\Delta=\Delta_{5}}\rho_{ac}^{(s)}(\cO)S^c_b(\cO_3\cO_4[\tl\cO])\bigg|_{G^{(t),fg}_{\cO_6}}
	\nonumber \\ &=
	(-1)^{1+\Sigma_{55}+\Sigma_{12}+\Sigma_{13}+\Sigma_{23}}\lambda_{326,f}\lambda_{146,g}
	\nonumber \\
	& \quad \x\Res_{\Delta=\Delta_{5}}\eta_{(ad)(ce)}^{\De,J}\sixjBlock{\cO_1}{\cO_2}{\cO_3}{\cO_4}{\cO_{\De,J}}{\cO_6}^{defg}S^c_b(\cO_{3}\cO_{4}[\widetilde{\cO}])\bigg|_{J=J_{5}}.
\end{align}

We aim to relate the inversion of a single block for external fermions to the inversion of a single block for a scalar four-point function. For that, by comparing (\ref{eq:6j_symbol_split_Scalars}), (\ref{eqn:GenSplit6j}), and (\ref{eq:6jDecompSpintoScalars}) we write down
\begin{multline}
	\sixjBlock{\cO_1}{\cO_2}{\cO_3}{\cO_4}{\cO_5}{\cO_6}^{abcd}=(-1)^{\Sigma_{66}}\left(S^{-1}(\cO_1\cO_4[\widetilde{\cO_6}])\right)^{d}_{e}\\\x\sum\limits_{\f_i,\cO'_{5,6}}
	\sixjDecomp{abce}{\cO_1}{\cO_2}{\cO_3}{\cO_4}{\cO_5}{\cO_6}{\f_1}{\f_2}{\f_3}{\f_4}{\cO'_{5}}{\cO'_{6}} K^{\f_1\f_4}_{\widetilde{\cO}'_{6}}\sixjBlock{\f_1}{\f_2}{\f_3}{\f_4}{\cO'_{5}}{\cO'_{6}}.
\end{multline}
We then find the following expression for the OPE function:
\begin{align}
	\label{eq: OPE function in terms of scalar 6j symbol}
	\rho_{ac}^{(s)}(\cO)S^c_b(\cO_3\cO_4[\tl\cO])\evaluated_{G^{(t),fg}_{\cO_6}}=\lambda_{326,f}\lambda_{146,g}\sum\limits_{\f_i,\cO',\cO'_{6}}
	&\opeFuncDecomp{fg}{ab}{\cO_1}{\cO_2}{\cO_3}{\cO_4}{\cO}{\cO_6}{\f_1}{\f_2}{\f_3}{\f_4}{\cO'}{\cO'_{6}}
	\nonumber \\ & \qquad\frac{S(\phi_3\phi_4[\widetilde{\cO'}])}{\eta^{(s)}_{\cO'}}\sixjBlock{\f_1}{\f_2}{\f_3}{\f_4}{\cO'}{\cO'_{6}},
\end{align}
where $\eta^{(s)}_{\cO_5}$ is the normalization of the scalar partial wave for $\<\f_1\f_2\f_3\f_4\>$ and we have defined
\begin{multline}
	\label{eq: K coefficients}
	\opeFuncDecomp{fg}{ab}{\cO_1}{\cO_2}{\cO_3}{\cO_4}{\cO}{\cO_6}{\f_1}{\f_2}{\f_3}{\f_4}{\cO'}{\cO'_{6}}\equiv (-1)^{\Sigma_{55}+\Sigma_{66}+\Sigma_{12}+\Sigma_{13}+\Sigma_{23}} K^{\f_1\f_4}_{\widetilde{\cO}'_{6}}
	\frac{S^c_b(\cO_{3}\cO_{4}[\widetilde{\cO}])}{S(\f_3\f_4[\tl \cO'])}\\\x \left(S^{-1}(\cO_1\cO_4[\widetilde{\cO_6}])\right)^{g}_{h} \eta_{\cO'}^{(s)}\eta_{(ad)(ce)}^{(s)\cO}\sixjDecomp{defh}{\cO_1}{\cO_2}{\cO_3}{\cO_4}{\cO}{\cO_6}{\f_1}{\f_2}{\f_3}{\f_4}{\cO'}{\cO'_{6}}. 
\end{multline}

We expect the physical poles for inverting a fermionic block to come from the physical poles from inverting a scalar block. We then have for example
\begin{multline}
	\lambda_{125,a}\lambda_{345,b}\evaluated_{G^{(t),fg}_{\cO_6}}=-\lambda_{326,f}\lambda_{146,g}\sum\limits_{\f_i,\cO',\cO'_{6}}\opeFuncDecomp{fg}{ab}{\cO_1}{\cO_2}{\cO_3}{\cO_4}{\cO_5}{\cO_6}{\f_1}{\f_2}{\f_3}{\f_4}{\cO'_{5}}{\cO'_{6}}
	\\\x \Res_{\Delta=\Delta_{5}}\frac{S(\phi_3\phi_4[\widetilde{\cO'_{\De,J}}])}{\eta^{(s)}_{\cO'_{\De,J}}}\sixjBlock{\f_1}{\f_2}{\f_3}{\f_4}{\cO_{\De,J}'}{\cO'_{6}}\evaluated_{J=J_5}. \label{eq: OPE data in terms of scalar 6j symbol} 
\end{multline}

Here we have assumed the inversion of a single block just has single poles. In general when studying a correlator $\<\cO_1\cO_2\cO_3\cO_4\>$ with $\Delta_1+\Delta_2=\Delta_3+\Delta_4$ we find both single and double poles. The double poles give the OPE coefficients times the anomalous dimensions while the single poles gives the OPE coefficients themselves \cite{Caron-Huot:2017vep}.

Equation~\ref{eq: OPE function in terms of scalar 6j symbol} is the main result of this approach: by using weight-shifting operators successively, we can express CFT data of spinning operators in terms of $6j$ symbols of external scalars and the decomposition coefficients $\cK$. The former can be calculated efficiently using the Lorentzian inversion formula whereas the latter is given in terms of partial wave normalization factors, shadow matrices, and $6j$ symbol decomposition coefficients $\cJ$, each of which we have computed explicitly. 

We presented the most general form in \equref{eq: OPE function in terms of scalar 6j symbol}, however one can in fact choose either $\cO_5$ or $\cO_6$ to stay the same by moving the weight-shifting operators through the other leg only. Indeed, we kept $\cO_5$ the same in the calculation of the $\cJ$ coefficients both for $\<\f\f\psi\psi\>$ and for $\<\psi\psi\psi\psi\>$, as we can observe in \equref{eq: 6j decomposition of 2f2s} and \equref{eq: 6j decomposition of 4f}. One can similarly compute $\cJ$ while keeping $\cO_6$ constant, though a separate calculation is not necessary: there are several identities between various $\cJ$ coefficients, which follow from the symmetries of the $6j$ symbols that we have summarized in appendix \ref{sec: symmetries of 6j symbols}. In particular, via \equref{eq: 6j symmetry regarding exchange of 5 and 6}, we have
\begin{multline}
	\label{eq: relation between J coefficients for 5 and 6 exchange}
	\sixjDecompp{abcd}{efgh}{\cO_1}{\cO_2}{\cO_3}{\cO_4}{\cO_5}{\cO_6}{\cO_1^{\delta_1,\epsilon_1}}{\cO_2^{\delta_2,\epsilon_2}}{\cO_3^{\delta_3,\epsilon_3}}{\cO_4^{\delta_4,\epsilon_4}}{\cO_5^{\delta_5,\epsilon_5}}{\cO_6^{\delta_6,\epsilon_6}}=(-1)^{\Sigma_{55}+\Sigma_{5'5'}+\Sigma_{66}+\Sigma_{6'6'}}
	\\\x
	\sixjDecompp{dcba}{hgfe}{\tl\cO_1}{\tl\cO_4}{\tl\cO_3}{\tl\cO_2}{\cO_6}{\cO_5}{\tl\cO_1^{-\delta_1,\epsilon_1}}{\tl\cO_4^{-\delta_4,\epsilon_4}}{\tl\cO_3^{-\delta_3,\epsilon_3}}{\tl\cO_2^{-\delta_2,\epsilon_2}}{\cO_6^{\delta_6,\epsilon_6}}{\cO_5^{\delta_5,\epsilon_5}}.
\end{multline}

Let us now turn to the explicit results for the $\cK^{fg}_{ab}$ coefficients. Despite the complicated intermediate steps, the final form they take is quite simple as they are relatively short meromorphic functions in scaling dimensions and spins. For example, the only nonzero $\cK$ coefficients for $\<\psi\f\f\psi\>$ arising from the exchange of a scalar $\f_6$ in the t-channel are
\small
\bea[eq: K coefficients for 2f2s with an exchange of a scalar t channel]
\opeFuncDecomp{\uniq 1}{11}{\psi_1}{\f_2}{\f_2}{\psi_1}{\cO_5}{\f_6}{\f_1^{\half}}{\f_2}{\f_2}{\f_1^{\half}}{\cO_{5}^{\half,-\half}}{\f_6} ={}&{} i, \\
\opeFuncDecomp{\uniq 1}{11}{\psi_1}{\f_2}{\f_2}{\psi_1}{\cO_5}{\f_6}{\f_1^{\half}}{\f_2}{\f_2}{\f_1^{\half}}{\cO_{5}^{-\half,\half}}{\f_6} ={}&{}  \frac{i \left(\Delta _5-\frac{3}{2}\right) \left(l_5+\frac{1}{2}\right) \left(l_5-\De_{251}+\frac{5}{2}\right)^2}{4\left(\Delta _5-2\right) \left(l_5+1\right) \left(l_5-\Delta _5+2\right) \left(l_5-\Delta _5+3\right)},
\\
\opeFuncDecomp{\uniq 1}{22}{\psi_1}{\f_2}{\f_2}{\psi_1}{\cO_5}{\f_6}{\f_1^{\half}}{\f_2}{\f_2}{\f_1^{\half}}{\cO_{5}^{\half,\half}}{\f_6} ={}&{} -\frac{i \left(l_5+\frac{1}{2}\right)}{l_5+1}, \\
\opeFuncDecomp{\uniq 1}{22}{\psi_1}{\f_2}{\f_2}{\psi_1}{\cO_5}{\f_6}{\f_1^{\half}}{\f_2}{\f_2}{\f_1^{\half}}{\cO_{5}^{-\half,-\half}}{\f_6} ={}&{} -\frac{i \left(\Delta _5-\frac{3}{2}\right) \left(\De_{251}+l_5-\frac{3}{2}\right)^2}{4 \left(\Delta _5-2\right)
	\left(\Delta _5+l_5-2\right) \left(\Delta _5+l_5-1\right)},
\\
\opeFuncDecomp{\uniq 3}{12}{\psi_1}{\f_2}{\f_2}{\psi_1}{\cO_5}{\f_6}{\f_1^{-\half}}{\f_2}{\f_2}{\f_1^{\half}}{\cO_{5}^{\half,\half}}{\f_6} ={}&{} \frac{i \left(l_5+\frac{1}{2}\right) \left(l_5-\De_{125}+\frac{5}{2}\right)}{\left(\Delta _6-1\right)\left(l_5+1\right)},
\\
\opeFuncDecomp{\uniq 3}{12}{\psi_1}{\f_2}{\f_2}{\psi_1}{\cO_5}{\f_6}{\f_1^{-\half}}{\f_2}{\f_2}{\f_1^{\half}}{\cO_{5}^{-\half,-\half}}{\f_6} ={}&{}  \frac{i \left(\De_{152}+l_5-\frac{3}{2}\right) \left(\De_{251}+l_5-\frac{3}{2}\right)}{4 \left(\Delta_5-\frac{3}{2}\right)^{-1}\left(\Delta _5-2\right) \left(\Delta
	_6-1\right) \left(\Delta _5+l_5-2\right)}\nn
\\
&\qquad\x\frac{\left(\De_1+\De_2+\De_5+l_5-\frac{9}{2}\right)}{\Delta _5+l_5-1}\;,
\\
\opeFuncDecomp{\uniq 3}{21}{\psi_1}{\f_2}{\f_2}{\psi_1}{\cO_5}{\f_6}{\f_1^{-\half}}{\f_2}{\f_2}{\f_1^{\half}}{\cO_{5}^{\half,\half}}{\f_6} ={}&{}  \frac{i\left(\De_{125}+l_5-\frac{3}{2}\right)}{\Delta _6-1},
\\
\opeFuncDecomp{\uniq 3}{21}{\psi_1}{\f_2}{\f_2}{\psi_1}{\cO_5}{\f_6}{\f_1^{-\half}}{\f_2}{\f_2}{\f_1^{\half}}{\cO_{5}^{-\half,\half}}{\f_6} ={}&{} \frac{i \left(\Delta _5-\frac{3}{2}\right)
	\left(l_5+\frac{1}{2}\right)  
	\left(l_5-\De_1-\De_2-\De_5+\frac{11}{2}\right)
}{4 \left(\Delta _5-2\right)
	\left(\Delta _6-1\right) \left(l_5+1\right) \left(l_5-\Delta _5+2\right)}
\nn\\
&\qquad \x \frac{\left(l_5-\De_{152}+\frac{5}{2}\right)\left(l_5-\De_{251}+\frac{5}{2}\right) }{-\Delta _5+l_5+3},
\eea 
\normalsize

We would like to emphasize two points. Firstly, as there is not a unique way to write \equref{eq: OPE data in terms of scalar 6j symbol}, the statement that the $\cK$ in \equref{eq: K coefficients for 2f2s with an exchange of a scalar t channel} are the only nonzero coefficients for $\<\psi\f\f\psi\>$ with an exchange of a scalar $\f_6$ in the t-channel should be understood for a particularly chosen decomposition in \equref{eq: OPE data in terms of scalar 6j symbol}. One can of course change the decomposition, which would then require a new set of $\cK$ coefficients. For example, we used a set of $\cK$ coefficients with $\cO_6$ held constant in \equref{eq: K coefficients for 2f2s with an exchange of a scalar t channel}; another set with $\cO_5$ held constant instead can be immediately obtained via \equref{eq: relation between J coefficients for 5 and 6 exchange}.\footnote{It should be noted that not all different decompositions are related to each other via symmetries. For example, $\cO_4'=\f_1^{\half}$ whereas $\cO_1'=\f_1^{\pm\half}$ in \equref{eq: K coefficients for 2f2s with an exchange of a scalar t channel}: this follows from fixing $b=\half$ in \equref{eq: J for 2f2s} as we noted after the equation. If we were to fix $b=-\half$ instead, we would then have a set of $\cK$ coefficients with $\cO_4'=\f_1^{-\half}$ and $\cO_1'=\f_1^{\pm\half}$, and these new coefficients are not related to \equref{eq: K coefficients for 2f2s with an exchange of a scalar t channel} in any manifestly symmetric way.} Secondly, we note that the absence of $\cK^{\uniq 1}_{12}$, $\cK^{\uniq 1}_{21}$, $\cK^{\uniq 3}_{11}$, and $\cK^{\uniq 3}_{22}$ is not coincidental: they are forbidden by the parity symmetry as we work in a parity-definite basis.\footnote{\label{footnote: parity of three-point structures}We would like to caution the reader that this statement follows from the symmetries of three-point structures $\<\cO_1\cO_2\cO_3\>^a$ under the transformation $X\rightarrow -X$ in embedding space, hence it is true whether the relevant physical theory has parity symmetry or not, i.e. $\<\cO_1\cO_2\cO_3\>^a$ are merely formal entities and should not be thought of as physical three-point structures.}

\section{$\cK$ Coefficients}
\label{sec: K coefficients}

In this appendix, we present the explicit expression for the $\cK$ coefficients defined in \equref{eq: K coefficients} for $\<\psi\f\f\psi\>$. As there are a different number of three-point tensor structures depending on whether $l_{5,6}=0$, the minimal complete set of nonzero $\cK$ coefficients differs for each case. We already presented the results for $\<\psi\f\f\psi\>$ with $l_6=0$ in \equref{eq: K coefficients for 2f2s with an exchange of a scalar t channel}, so we will detail the $\<\psi\f\f\psi\>$ with $l_6\ne 0$ below. For $\<\psi_1\psi_2\psi_2\psi_1\>$ and $\<\psi\psi\psi\psi\>$, the coefficients become quite lengthy so we do not reproduce them here; please see the \texttt{Mathematica} file of \cite{Albayrak:2020rxh} for their explicit expressions.

For the correlator $\<\psi\phi\phi\psi\>$, the list below constitutes a sufficient set of nonzero $\cK$ coefficients  if the exchanged operator in t-channel is not a scalar.\footnote{For scalar exchange in the t-channel, see \equref{eq: K coefficients for 2f2s with an exchange of a scalar t channel}.} For convenience, we will use a different shorthand notation in this section, i.e.
\be 
\De_{abc}=\De_{a}+\De_{b}+\De_{c}\;,\quad \Delta_{ab}^{c}=\De_{a}+\De_{b}-\De_{c}.
\ee 

The coefficients are:
\footnotesize
\begin{equation*}
	\begin{array}{cc}
		\scalebox{.8}{\opeFuncDecomp{\uniq 2}{11}{\psi_1}{\f_2}{\f_2}{\psi_1}{\cO_5}{\cO_6}{\f_1^{-\half}}{\f_2}{\f_2}{\f_1^{-\half}}{\cO_{5}^{-\half,-\half}}{\cO_6} 
		}& \frac{i
			\left(\Delta _5-\frac{3}{2}\right) \left(\Delta_{125}+l_5-\frac{9}{2}\right){}^2
			\left(\Delta_{15}^{2}+l_5-\frac{3}{2}\right){}^2}{16 \left(\Delta _5-2\right)
			\left(\Delta _6-1\right) l_6 \left(\Delta _5+l_5-2\right) \left(\Delta _5+l_5-1\right)}
		\\
		\scalebox{.8}{\opeFuncDecomp{\uniq 2}{22}{\psi_1}{\f_2}{\f_2}{\psi_1}{\cO_5}{\cO_6}{\f_1^{-\half}}{\f_2}{\f_2}{\f_1^{-\half}}{\cO_{5}^{-\half,\half}}{\cO_6} 
		}& -\frac{i
			\left(\Delta _5-\frac{3}{2}\right) \left(l_5+\frac{1}{2}\right) \left(-\Delta
			_{125}+l_5+\frac{11}{2}\right){}^2 \left(-\Delta_{15}^{2}+l_5+\frac{5}{2}\right){}^2}{16 \left(\Delta _5-2\right) \left(\Delta _6-1\right)
			\left(l_5+1\right) l_6 \left(-\Delta _5+l_5+2\right) \left(-\Delta _5+l_5+3\right)} 
		\\
		\scalebox{.8}{\opeFuncDecomp{\uniq 2}{22}{\psi_1}{\f_2}{\f_2}{\psi_1}{\cO_5}{\cO_6}{\f_1^{-\half}}{\f_2}{\f_2}{\f_1^{-\half}}{\cO_{5}^{\half,-\half}}{\cO_6} 
		}& -\frac{i
			\left(\Delta_{12}^{5}+l_5-\frac{3}{2}\right){}^2}{4 \left(\Delta _6-1\right) l_6} 
		\\
		\scalebox{.8}{\opeFuncDecomp{\uniq 2}{11}{\psi_1}{\f_2}{\f_2}{\psi_1}{\cO_5}{\cO_6}{\f_1^{-\half}}{\f_2}{\f_2}{\f_1^{-\half}}{\cO_{5}^{\half,\half}}{\cO_6} 
		}& \frac{i
			\left(l_5+\frac{1}{2}\right) \left(-\Delta_{12}^{5}+l_5+\frac{5}{2}\right){}^2}{4
			\left(\Delta _6-1\right) \left(l_5+1\right) l_6} 
		\\
		\scalebox{.8}{\opeFuncDecomp{\uniq 3}{12}{\psi_1}{\f_2}{\f_2}{\psi_1}{\cO_5}{\cO_6}{\f_1^{-\half}}{\f_2}{\f_2}{\f_1^{\half}}{\cO_{5}^{-\half,-\half}}{\cO_6} 
		}& \frac{i
			\left(\Delta _5-\frac{3}{2}\right) \left(\Delta_{125}+l_5-\frac{9}{2}\right)
			\left(\Delta_{15}^{2}+l_5-\frac{3}{2}\right) \left(\Delta_{25}^{1}+l_5-\frac{3}{2}\right)}{8 \left(\Delta _5-2\right) \left(\Delta _6-1\right)
			\left(\Delta _5+l_5-2\right) \left(\Delta _5+l_5-1\right)} 
		\\
		\scalebox{.8}{\opeFuncDecomp{\uniq 4}{12}{\psi_1}{\f_2}{\f_2}{\psi_1}{\cO_5}{\cO_6}{\f_1^{-\half}}{\f_2}{\f_2}{\f_1^{\half}}{\cO_{5}^{-\half,-\half}}{\cO_6} 
		}& \frac{i
			\left(\Delta _5-\frac{3}{2}\right) \left(\Delta_{125}+l_5-\frac{9}{2}\right)
			\left(\Delta_{15}^{2}+l_5-\frac{3}{2}\right) \left(\Delta_{25}^{1}+l_5-\frac{3}{2}\right)}{8 \left(\Delta _5-2\right) l_6 \left(\Delta
			_5+l_5-2\right) \left(\Delta _5+l_5-1\right)} 
		\\
		\scalebox{.8}{\opeFuncDecomp{\uniq 3}{21}{\psi_1}{\f_2}{\f_2}{\psi_1}{\cO_5}{\cO_6}{\f_1^{-\half}}{\f_2}{\f_2}{\f_1^{\half}}{\cO_{5}^{-\half,\half}}{\cO_6} 
		}& \frac{i
			\left(\Delta _5-\frac{3}{2}\right) \left(l_5+\frac{1}{2}\right) \left(-\Delta
			_{125}+l_5+\frac{11}{2}\right) \left(-\Delta_{15}^{2}+l_5+\frac{5}{2}\right)
			\left(-\Delta_{25}^{1}+l_5+\frac{5}{2}\right)}{8 \left(\Delta _5-2\right) \left(\Delta
			_6-1\right) \left(l_5+1\right) \left(-\Delta _5+l_5+2\right) \left(-\Delta
			_5+l_5+3\right)} 
		\\
		\scalebox{.8}{\opeFuncDecomp{\uniq 4}{21}{\psi_1}{\f_2}{\f_2}{\psi_1}{\cO_5}{\cO_6}{\f_1^{-\half}}{\f_2}{\f_2}{\f_1^{\half}}{\cO_{5}^{-\half,\half}}{\cO_6}
		}& \frac{i
			\left(\Delta _5-\frac{3}{2}\right) \left(l_5+\frac{1}{2}\right) \left(-\Delta
			_{125}+l_5+\frac{11}{2}\right) \left(-\Delta_{15}^{2}+l_5+\frac{5}{2}\right)
			\left(-\Delta_{25}^{1}+l_5+\frac{5}{2}\right)}{8 \left(\Delta _5-2\right)
			\left(l_5+1\right) l_6 \left(-\Delta _5+l_5+2\right) \left(-\Delta _5+l_5+3\right)} 
		\\
		\scalebox{.8}{\opeFuncDecomp{\uniq 3}{21}{\psi_1}{\f_2}{\f_2}{\psi_1}{\cO_5}{\cO_6}{\f_1^{-\half}}{\f_2}{\f_2}{\f_1^{\half}}{\cO_{5}^{\half,-\half}}{\cO_6}
		}& \frac{i
			\left(\Delta_{12}^{5}+l_5-\frac{3}{2}\right)}{2 \left(\Delta _6-1\right)} 
		\\
		\scalebox{.8}{\opeFuncDecomp{\uniq 4}{21}{\psi_1}{\f_2}{\f_2}{\psi_1}{\cO_5}{\cO_6}{\f_1^{-\half}}{\f_2}{\f_2}{\f_1^{\half}}{\cO_{5}^{\half,-\half}}{\cO_6}
		}& \frac{i
			\left(\Delta_{12}^{5}+l_5-\frac{3}{2}\right)}{2 l_6} 
		\\
		\scalebox{.8}{\opeFuncDecomp{\uniq 3}{12}{\psi_1}{\f_2}{\f_2}{\psi_1}{\cO_5}{\cO_6}{\f_1^{-\half}}{\f_2}{\f_2}{\f_1^{\half}}{\cO_{5}^{\half,\half}}{\cO_6}
		}& \frac{i
			\left(l_5+\frac{1}{2}\right) \left(-\Delta_{12}^{5}+l_5+\frac{5}{2}\right)}{2
			\left(\Delta _6-1\right) \left(l_5+1\right)} 
		\\
		\scalebox{.8}{\opeFuncDecomp{\uniq 4}{12}{\psi_1}{\f_2}{\f_2}{\psi_1}{\cO_5}{\cO_6}{\f_1^{-\half}}{\f_2}{\f_2}{\f_1^{\half}}{\cO_{5}^{\half,\half}}{\cO_6}
		}& \frac{i
			\left(l_5+\frac{1}{2}\right) \left(-\Delta_{12}^{5}+l_5+\frac{5}{2}\right)}{2
			\left(l_5+1\right) l_6} 
		\\
		\scalebox{.8}{\opeFuncDecomp{\uniq 3}{21}{\psi_1}{\f_2}{\f_2}{\psi_1}{\cO_5}{\cO_6}{\f_1^{\half}}{\f_2}{\f_2}{\f_1^{-\half}}{\cO_{5}^{-\half,-\half}}{\cO_6}
		}& \frac{i
			\left(\Delta _5-\frac{3}{2}\right) \left(\Delta_{125}+l_5-\frac{9}{2}\right)
			\left(\Delta_{15}^{2}+l_5-\frac{3}{2}\right) \left(\Delta_{25}^{1}+l_5-\frac{3}{2}\right)}{8 \left(\Delta _5-2\right) \left(\Delta _6-1\right)
			\left(\Delta _5+l_5-2\right) \left(\Delta _5+l_5-1\right)} \\
		\scalebox{.8}{\opeFuncDecomp{\uniq 4}{21}{\psi_1}{\f_2}{\f_2}{\psi_1}{\cO_5}{\cO_6}{\f_1^{\half}}{\f_2}{\f_2}{\f_1^{-\half}}{\cO_{5}^{-\half,-\half}}{\cO_6}
		}& \frac{i
			\left(\Delta _5-\frac{3}{2}\right) \left(1-\Delta _6\right) \left(\Delta
			_{125}+l_5-\frac{9}{2}\right) \left(\Delta_{15}^{2}+l_5-\frac{3}{2}\right) \left(\Delta_{25}^{1}+l_5-\frac{3}{2}\right)}{8 \left(\Delta _5-2\right) \left(\Delta _6-1\right) l_6
			\left(\Delta _5+l_5-2\right) \left(\Delta _5+l_5-1\right)} \\
		\scalebox{.8}{\opeFuncDecomp{\uniq 3}{12}{\psi_1}{\f_2}{\f_2}{\psi_1}{\cO_5}{\cO_6}{\f_1^{\half}}{\f_2}{\f_2}{\f_1^{-\half}}{\cO_{5}^{-\half,\half}}{\cO_6}
		}& \frac{i
			\left(\Delta _5-\frac{3}{2}\right) \left(l_5+\frac{1}{2}\right) \left(-\Delta
			_{125}+l_5+\frac{11}{2}\right) \left(-\Delta_{15}^{2}+l_5+\frac{5}{2}\right)
			\left(-\Delta_{25}^{1}+l_5+\frac{5}{2}\right)}{8 \left(\Delta _5-2\right) \left(\Delta
			_6-1\right) \left(l_5+1\right) \left(-\Delta _5+l_5+2\right) \left(-\Delta
			_5+l_5+3\right)} \\
		\scalebox{.8}{\opeFuncDecomp{\uniq 4}{12}{\psi_1}{\f_2}{\f_2}{\psi_1}{\cO_5}{\cO_6}{\f_1^{\half}}{\f_2}{\f_2}{\f_1^{-\half}}{\cO_{5}^{-\half,\half}}{\cO_6}
		}& \frac{i
			\left(\Delta _5-\frac{3}{2}\right) \left(1-\Delta _6\right) \left(l_5+\frac{1}{2}\right)
			\left(-\Delta_{125}+l_5+\frac{11}{2}\right) \left(-\Delta_{15}^{2}+l_5+\frac{5}{2}\right) \left(-\Delta_{25}^{1}+l_5+\frac{5}{2}\right)}{8
			\left(\Delta _5-2\right) \left(\Delta _6-1\right) \left(l_5+1\right) l_6 \left(-\Delta
			_5+l_5+2\right) \left(-\Delta _5+l_5+3\right)} \\
		\scalebox{.8}{\opeFuncDecomp{\uniq 3}{12}{\psi_1}{\f_2}{\f_2}{\psi_1}{\cO_5}{\cO_6}{\f_1^{\half}}{\f_2}{\f_2}{\f_1^{-\half}}{\cO_{5}^{\half,-\half}}{\cO_6}
		}& \frac{i
			\left(\Delta_{12}^{5}+l_5-\frac{3}{2}\right)}{2 \left(\Delta _6-1\right)}
		\\
		\scalebox{.8}{\opeFuncDecomp{\uniq 4}{12}{\psi_1}{\f_2}{\f_2}{\psi_1}{\cO_5}{\cO_6}{\f_1^{\half}}{\f_2}{\f_2}{\f_1^{-\half}}{\cO_{5}^{\half,-\half}}{\cO_6}
		}& \frac{i
			\left(1-\Delta _6\right) \left(\Delta_{12}^{5}+l_5-\frac{3}{2}\right)}{2 \left(\Delta
			_6-1\right) l_6} \\
		\scalebox{.8}{\opeFuncDecomp{\uniq 3}{21}{\psi_1}{\f_2}{\f_2}{\psi_1}{\cO_5}{\cO_6}{\f_1^{\half}}{\f_2}{\f_2}{\f_1^{-\half}}{\cO_{5}^{\half,\half}}{\cO_6}
		}& \frac{i
			\left(l_5+\frac{1}{2}\right) \left(-\Delta_{12}^{5}+l_5+\frac{5}{2}\right)}{2
			\left(\Delta _6-1\right) \left(l_5+1\right)} 
		\\
		\scalebox{.8}{\opeFuncDecomp{\uniq 4}{21}{\psi_1}{\f_2}{\f_2}{\psi_1}{\cO_5}{\cO_6}{\f_1^{\half}}{\f_2}{\f_2}{\f_1^{-\half}}{\cO_{5}^{\half,\half}}{\cO_6}
		}& \frac{i
			\left(1-\Delta _6\right) \left(l_5+\frac{1}{2}\right) \left(-
			\Delta_{12}^{5}+l_5+\frac{5}{2}\right)}{2 \left(\Delta _6-1\right) \left(l_5+1\right) l_6} \\
		\scalebox{.8}{\opeFuncDecomp{\uniq 1}{22}{\psi_1}{\f_2}{\f_2}{\psi_1}{\cO_5}{\cO_6}{\f_1^{\half}}{\f_2}{\f_2}{\f_1^{\half}}{\cO_{5}^{-\half,-\half}}{\cO_6}
		}& -\frac{i
			\left(\Delta _5-\frac{3}{2}\right) \left(\Delta_{25}^{1}+l_5-\frac{3}{2}\right){}^2}{4
			\left(\Delta _5-2\right) \left(\Delta _5+l_5-2\right) \left(\Delta _5+l_5-1\right)} 
	\end{array}
\end{equation*}
\begin{equation}
	\begin{array}{cc}
		\scalebox{.8}{\opeFuncDecomp{\uniq 2}{22}{\psi_1}{\f_2}{\f_2}{\psi_1}{\cO_5}{\cO_6}{\f_1^{\half}}{\f_2}{\f_2}{\f_1^{\half}}{\cO_{5}^{-\half,-\half}}{\cO_6}
		}& \frac{i
			\left(\Delta _5-\frac{3}{2}\right) \left(\Delta _1+\frac{1}{2} \left(-\Delta
			_6+l_6-1\right)\right) \left(\Delta _1+\frac{1}{2} \left(\Delta _6-l_6-4\right)\right)
			\left(\Delta_{25}^{1}+l_5-\frac{3}{2}\right){}^2}{4 \left(\Delta _5-2\right) \left(\Delta
			_6-1\right) l_6 \left(\Delta _5+l_5-2\right) \left(\Delta _5+l_5-1\right)} \\
		\scalebox{.8}{\opeFuncDecomp{\uniq 1}{11}{\psi_1}{\f_2}{\f_2}{\psi_1}{\cO_5}{\cO_6}{\f_1^{\half}}{\f_2}{\f_2}{\f_1^{\half}}{\cO_{5}^{-\half,\half}}{\cO_6}
		}& \frac{i
			\left(\Delta _5-\frac{3}{2}\right) \left(l_5+\frac{1}{2}\right) \left(-\Delta_{25}^{1}+l_5+\frac{5}{2}\right){}^2}{4 \left(\Delta _5-2\right) \left(l_5+1\right)
			\left(-\Delta _5+l_5+2\right) \left(-\Delta _5+l_5+3\right)} \\
		\scalebox{.8}{\opeFuncDecomp{\uniq 2}{11}{\psi_1}{\f_2}{\f_2}{\psi_1}{\cO_5}{\cO_6}{\f_1^{\half}}{\f_2}{\f_2}{\f_1^{\half}}{\cO_{5}^{-\half,\half}}{\cO_6}
		}& -\frac{i
			\left(\Delta _5-\frac{3}{2}\right) \left(l_5+\frac{1}{2}\right) \left(\Delta
			_1+\frac{1}{2} \left(-\Delta _6+l_6-1\right)\right) \left(\Delta _1+\frac{1}{2}
			\left(\Delta _6-l_6-4\right)\right) \left(-\Delta_{25}^{1}+l_5+\frac{5}{2}\right){}^2}{4
			\left(\Delta _5-2\right) \left(\Delta _6-1\right) \left(l_5+1\right) l_6 \left(-\Delta
			_5+l_5+2\right) \left(-\Delta _5+l_5+3\right)} \\
		\scalebox{.8}{\opeFuncDecomp{\uniq 1}{11}{\psi_1}{\f_2}{\f_2}{\psi_1}{\cO_5}{\cO_6}{\f_1^{\half}}{\f_2}{\f_2}{\f_1^{\half}}{\cO_{5}^{\half,-\half}}{\cO_6}
		}& i \\
		\scalebox{.8}{\opeFuncDecomp{\uniq 2}{11}{\psi_1}{\f_2}{\f_2}{\psi_1}{\cO_5}{\cO_6}{\f_1^{\half}}{\f_2}{\f_2}{\f_1^{\half}}{\cO_{5}^{\half,-\half}}{\cO_6}
		}& -\frac{i
			\left(\Delta _1+\frac{1}{2} \left(-\Delta _6+l_6-1\right)\right) \left(\Delta
			_1+\frac{1}{2} \left(\Delta _6-l_6-4\right)\right)}{\left(\Delta _6-1\right) l_6} \\
		\scalebox{.8}{\opeFuncDecomp{\uniq 1}{22}{\psi_1}{\f_2}{\f_2}{\psi_1}{\cO_5}{\cO_6}{\f_1^{\half}}{\f_2}{\f_2}{\f_1^{\half}}{\cO_{5}^{\half,\half}}{\cO_6}
		}& -\frac{i
			\left(l_5+\frac{1}{2}\right)}{l_5+1} \\
		\scalebox{.8}{\opeFuncDecomp{\uniq 2}{22}{\psi_1}{\f_2}{\f_2}{\psi_1}{\cO_5}{\cO_6}{\f_1^{\half}}{\f_2}{\f_2}{\f_1^{\half}}{\cO_{5}^{\half,\half}}{\cO_6}
		}& \frac{i
			\left(l_5+\frac{1}{2}\right) \left(\Delta _1+\frac{1}{2} \left(-\Delta
			_6+l_6-1\right)\right) \left(\Delta _1+\frac{1}{2} \left(\Delta
			_6-l_6-4\right)\right)}{\left(\Delta _6-1\right) \left(l_5+1\right) l_6}
	\end{array}
\end{equation}
\normalsize

\section{Symmetries of $6j$ Symbols}
\label{sec: symmetries of 6j symbols}

\begin{figure}
	\centering
	\includegraphics[scale=1.4]{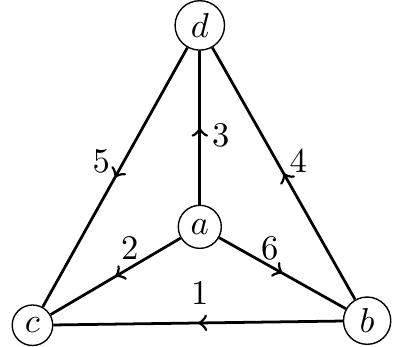}
	\caption[$6j$ symbol written as a tetrahedron in diagrammatic notation]{\label{fig: tetrahedron} Diagrammatic form of the $6j$ symbol $\left(\Psi^{\tl{(s)}cd}_{\tl\cO_5},\Psi^{(t)ab}_{\cO_6}\right)$ as a tetrahedron.}
\end{figure}

By representing the $6j$ symbol as a tetrahedron as in figure \ref{fig: tetrahedron}, we can reveal its symmetries, as was done in \cite{Liu:2018jhs}. Explicitly, we can consider the three transformations as the generators of the symmetry group:
\begin{enumerate}
	\begin{subequations}
		\label{eq: 6j symmetries}
		\item[\textbf{S1}] Rotation around the axis that passes through the vertex $a$ and the center of the triangle $\De bcd$, generated by the permutation $(1\tl 5\tl 4)(\tl 154)(236)(\tl 2\tl 3\tl 6)(bcd)$:
		\be 
		\label{eq: 6j symmetry 1}
		\left\{\begin{aligned}
			\cO_1  \cO_2  \cO_6 \\
			\cO_3  \cO_4  \cO_5
		\end{aligned}\right\}^{cdab} =(-1)^{2l_{\cO_6}} 
		\left\{\begin{aligned}
			\tl\cO_5  \cO_3  \cO_2 \\
			\cO_6  \tl\cO_1  \cO_4
		\end{aligned}\right\}^{dbac}\;.
		\ee 	
		
		\item[\textbf{S2}] Rotation around the axis that passes through the vertex $c$ and the center of the triangle $\De abd$, generated by the permutation $(125)(\tl 1\tl 2\tl 5)(\tl3 46)(3\tl 4\tl 6)(bad)$:
		\be 
		\label{eq: 6j symmetry 2}
		\left\{\begin{aligned}
			\cO_1  \cO_2  \cO_6 \\
			\cO_3  \cO_4  \cO_5
		\end{aligned}\right\}^{cdab} =(-1)^{2l_{\cO_1}}  \left\{\begin{aligned}
			\cO_2  \cO_5  \tl \cO_3 \\
			\tl\cO_4  \cO_6  \cO_1
		\end{aligned}\right\}^{cbda}\;.
		\ee 
		
		\item[\textbf{S3}] Reflection with respect to the plane that passes through the points $c$, $d$, and the midpoint of the line segment $\overline{ab}$, generated by the permutation $(12)(\tl1\tl2)(34)(\tl3\tl4)(6\tl6)(ba)$:
		\be 
		\label{eq: 6j symmetry 3}
		\left\{\begin{aligned}
			\cO_1  \cO_2  \cO_6 \\
			\cO_3  \cO_4  \cO_5
		\end{aligned}\right\}^{cdab} =(-1)^{2l_{\cO_6}}  
		\left\{\begin{aligned}
			\cO_2  \cO_1  \tl\cO_6 \\
			\cO_4  \cO_3  \cO_5
		\end{aligned}\right\}^{cdba}\;.
		\ee 
	\end{subequations}
\end{enumerate}
The overall phases in the front follow from the fermionic nature of the correlators and can be checked explicitly.

The validity of \equref{eq: 6j symmetries} depends on the choice of three point basis. For example, the first two equalities require us to work in a basis which respects the cyclic permutations; i.e., we should have $\<\cO_1\cO_2\cO_3\>^a=\<\cO_2\cO_3\cO_1\>^a=\<\cO_2\cO_3\cO_1\>^a$. Generically, we can always find a basis which respects this property.

The equality \equref{eq: 6j symmetry 3} on the other hand requires the basis to respect inversions, i.e. we should have $\<\cO_1\cO_2\cO_3\>^a=\<\cO_2\cO_1\cO_3\>^a$. We can always choose a basis to respect this \emph{unless} we have $l_{\cO_1}=l_{\cO_2}$. In that case, we can no longer choose two independent bases $\<\cO_1\cO_2\cO_3\>^a$ and $\<\cO_2\cO_1\cO_3\>^a$ to satisfy the required equality; we need the same basis to satisfy this condition. However, if we work in a parity definite basis, all \emph{nonzero} $6j$ symbols will have an even number of parity odd three-point structures, therefore the equality holds. Assuming we are in such a basis, we can use following relations to derive all permutations:
\be 
\textbf{S1}^3=\textbf{S2}^3=\textbf{S3}^2=\textbf{E} \text{ with } (\textbf{S2}*\textbf{S1})^2=(\textbf{S3}*\textbf{S1})^4=(\textbf{S3}*\textbf{S2})^2=\textbf{E}
\ee 
where \textbf{E} is the identity transformation. 

In summary, we can derive \equref{eq: 6j symmetries} and similar identities by considering the inversions and rotations of the tetrahedron and are valid in a parity definite basis with a cyclic property. These conditions are trivially satisfied for external scalars as there is only one three-point structure. 

An interesting set of transformations is the one that does not move the edges $5,6$. There are only three such permutations:
\be 
(12)(\tl 1\tl 2)(3 4)(\tl 3\tl 4)(6 \tl 6)(a b),\\
(5\tl 5)(2 3)(\tl 2\tl 3)(1 4)(\tl 1 \tl 4)(c d),\\
(13)(\tl 1\tl 3)(24)(\tl 2\tl 4)(5\tl 5)(6\tl 6)(ab)(cd),
\ee 
which yields 
\begin{multline}
	\left\{\begin{aligned}
		\cO_1  \cO_2  \cO_6 \\
		\cO_3  \cO_4  \cO_5
	\end{aligned}\right\}^{cdab} =(-1)^{2l_{\cO_6}}  
	\left\{\begin{aligned}
		\cO_2  \cO_1  \tl\cO_6 \\
		\cO_4  \cO_3  \cO_5
	\end{aligned}\right\}^{cdba}
	=(-1)^{2l_{\cO_5}} 
	\left\{\begin{aligned}
		\cO_4  \cO_3  \cO_6 \\
		\cO_2  \cO_1  \tl \cO_5
	\end{aligned}\right\}^{dcab}
	\\
	=(-1)^{2l_{\cO_5}+2l_{\cO_6}} 
	\left\{\begin{aligned}
		\cO_3  \cO_4  \tl\cO_6 \\
		\cO_1  \cO_2  \tl\cO_5
	\end{aligned}\right\}^{dcba}\;.
\end{multline}
But we also know how to relate $\left\{\begin{aligned}
	\cO_1  \cO_2  \cO_6 \\
	\cO_3  \cO_4  \tl\cO_5
\end{aligned}\right\}$ to $\left\{\begin{aligned}
	\cO_1  \cO_2  \cO_6 \\
	\cO_3  \cO_4  \cO_5
\end{aligned}\right\}$, and likewise for $\cO_6$, due to shadow symmetry of the partial waves.

We may also be interested in interchanging $\cO_{5,6}$ in the $6j$ symbol, and this can be achieved with the transformation $\mathfrak{T}=(1\tl 1)(2\tl 4)(3\tl 3)(56)(ad)(bc)$:
\be 
\label{eq: 6j symmetry regarding exchange of 5 and 6}
\left\{\begin{aligned}
	\cO_1  \cO_2  \cO_6 \\
	\cO_3  \cO_4  \cO_5
\end{aligned}\right\}^{cdab} =(-1)^{2\left(l_{\cO_5}+l_{\cO_6}\right)}  
\left\{\begin{aligned}
	\tl\cO_1  \tl\cO_4  \cO_5 \\
	\tl\cO_3  \tl\cO_2  \cO_6
\end{aligned}\right\}^{badc}.
\ee 
Likewise, with the transformation $\mathfrak{T}=(1\tl 3)(2\tl 2)(4\tl 4)(5 \tl6)(ac)(bd)$, we get
\be 
\left\{\begin{aligned}
	\cO_1  \cO_2  \cO_6 \\
	\cO_3  \cO_4  \cO_5
\end{aligned}\right\}^{cdab} =  
\left\{\begin{aligned}
	\tl\cO_3  \tl\cO_2  \tl\cO_5 \\
	\tl\cO_1  \tl\cO_4  \tl\cO_6
\end{aligned}\right\}^{abcd}.
\ee 

	\backmatter

	\bibliography{collectiveReferenceLibrary}{}

	\bibliographystyle{utphys}

\end{document}